\documentclass[nofootinbib,11pt]{report}

\usepackage{graphicx}

\usepackage{rotating}

\usepackage{dcolumn}

\usepackage{amssymb}

\usepackage{amsmath}

\usepackage{bm}% bold math

\usepackage{listings}

\usepackage{fancyhdr}

\usepackage{wrapfig}

\usepackage{subfig}

\usepackage[vcentermath]{youngtab}

\usepackage[francais,english]{babel}

\usepackage[utf8]{inputenc}  

\usepackage{hyperref}

\usepackage{color}

\usepackage{cite}

\definecolor{MyBlue}{rgb}{0.15,0.15,0.70}
\hypersetup{
colorlinks=true,
citecolor=MyBlue,
linkcolor=MyBlue,
urlcolor=MyBlue
}

\usepackage[width=14cm, height=22cm, left=3cm]{geometry}

\newcommand{\beq}{\begin{equation}}
\newcommand{\eeq}{\end{equation}}
\newcommand{\bea}{\begin{eqnarray}}
\newcommand{\eea}{\end{eqnarray}}
\newcommand{\ben}{\begin{enumerate}}
\newcommand{\een}{\end{enumerate}}

\newcommand{\pa}{\partial}

\newcommand{\bo}{\square}
\newcommand{\na}{\nabla}
\newcommand{\ed}{{\rm d}}

\newcommand{\we}{\wedge}
\newcommand{\Lie}{{\cal L}}
\newcommand{\Ord}{{\cal O}}

\newcommand{\ti}{\tilde}
\newcommand{\da}{\dagger}
\newcommand{\Rs}{\mathbb{R}}

\renewcommand\({\left(}
\renewcommand\){\right)}
\renewcommand\[{\left[}
\renewcommand\]{\right]}
\newcommand{\bra}{\langle}
\newcommand{\ket}{\rangle}

\newcommand{\os}{\overset}
\newcommand{\us}{\underset}

\newcommand{\nn}{\nonumber}

\newcommand{\al}{\alpha}
\newcommand{\be}{\beta}
\newcommand{\ga}{\gamma}
\newcommand{\Ga}{\Gamma}
\newcommand{\de}{\delta}
\newcommand{\De}{\Delta}
\newcommand{\ep}{\epsilon}
\newcommand{\vep}{\varepsilon}
\newcommand{\ze}{\zeta}
\newcommand{\et}{\eta}
\newcommand{\te}{\theta}

\newcommand{\ka}{\kappa}
\newcommand{\la}{\lambda}
\newcommand{\La}{\Lambda}
\newcommand{\ro}{\rho}
\newcommand{\si}{\sigma}

\newcommand{\ta}{\tau}
\newcommand{\ph}{\phi}
\newcommand{\vph}{\varphi}
\newcommand{\ch}{\chi}
\newcommand{\om}{\omega}
\newcommand{\Om}{\Omega}

\allowdisplaybreaks

\numberwithin{equation}{section}

\begin{document}

\thispagestyle{fancy}

\fancyhf{}
\fancyhead[LO]{UNIVERSIT\'{E} DE GEN\`{E}VE\\ Département de physique théorique \\ }
\fancyhead[CO]{}
\fancyhead[RO]{FACULT\'E DES SCIENCES\\ Professeur Michele Maggiore \\ }

\begin{center}

\vspace*{2cm}

\Huge{\bf Aspects of Infrared Non-local Modifications of General Relativity}

\vspace{1.5cm}

\LARGE{TH\`ESE}

\vspace{1.5cm}

\Large{présentée à la Faculté des sciences de l'Université de Genève \\ pour obtenir le grade de Docteur ès sciences, mention physique}

\vspace{1cm}

\Large{par}

\vspace{0.5cm}

\Large{\bf Ermis MITSOU}

\vspace{0.5cm}

de

\vspace{0.5cm}

Brot-Dessous (NE)

\vspace{2cm}

Thèse N$^{\circ}$ 4770

\vspace{2cm}

GEN\`EVE

\vspace{0.5cm}

Atelier d'impression ReproMail

\vspace{0.3cm}

2015

\end{center}

\pagebreak

\begin{figure}[h!]
\begin{center}
\includegraphics[width=14.5cm]{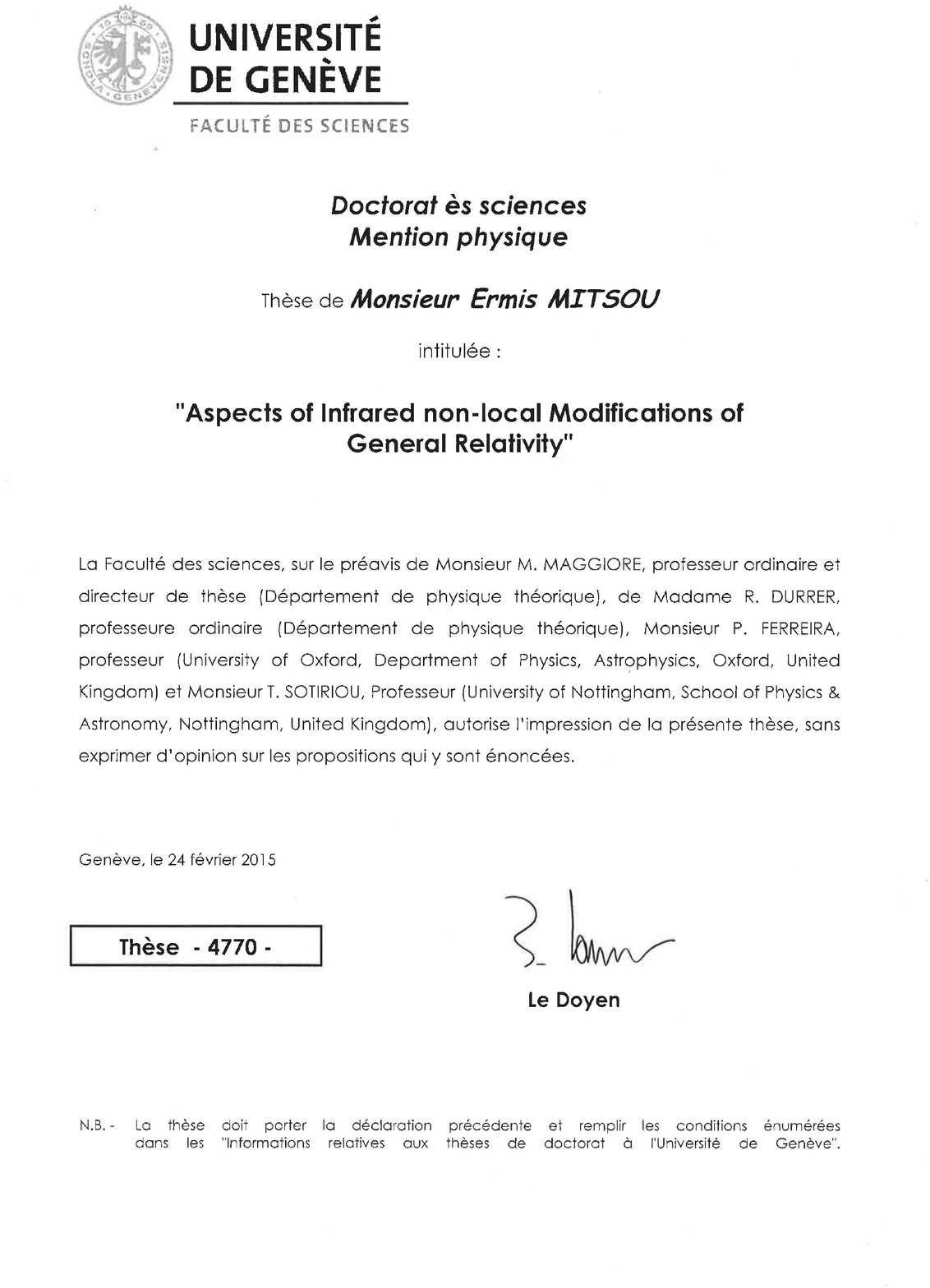} 
\end{center}
\end{figure}

\pagebreak

{\it à Poinpon,} \\
{\it et à tous ceux qui m'ont prêté un crayon et une feuille quand j'en avais besoin...}

\renewcommand{\abstractname}{Abstract}

\begin{abstract}

\begin{otherlanguage}{francais}

Le sujet de recherche abordé dans cette thèse concerne le problème de l'énergie sombre en cosmologie, cette forme d'énergie encore non-identifiée qui domine actuellement la dynamique de notre univers. Nous considérons l'éventualité que ce soit une modification non-locale de la théorie de la Relativité Générale qui soit à l'origine de cet effet. Inspirés par la théorie de gravité massive, nous construisons des théories non-locales dans lesquelles la gravité peut avoir une masse mais où l'invariance sous difféomorphismes n'est pas brisée. Nous nous focalisons sur la cosmologie de ces théories et les confrontons à certaines contraintes observationnelles. Sur un plan plus théorique, nous nous attardons également sur les subtilités de la théorie des champs non-locale, en clarifiant certains malentendus sur la question de stabilité. 

\end{otherlanguage}

\end{abstract}

\renewcommand{\abstractname}{\Large Résumé}

\begin{abstract}

\begin{otherlanguage}{francais}

Dans cette thèse, notre première motivation est de construire une théorie de gravité massive qui soit invariante sous transformations de coordonnées et ne fasse pas appel à une métrique extérieure de référence, ce qui est possible si l'on a recours à des termes non-locaux. Cependant, les contraintes phénoménologiques nous mèneront à des modifications non-locales de la Relativité Générale dans lesquselles la gravité n'est pas forcément massive, mais où la cosmologie reproduit les observations actuelles.

La structure dynamique d'une théorie des champs non-locale présente quelques subtilités par rapport aux théories locales, et ne pas en tenir compte peut nous mener à conclusions erronées. Nous commençons donc par l'étude de la dynamique des théories de jauge massives, linéaires et locales, sous plusieurs angles différents, afin de mettre en avant les propriétés qui ne seront pas exportables dans le cas non-local. Nous en profitons pour discuter un aspect intéressant de la théorie linéaire d'un champ massif de spin-2, qui consiste en une symétrie de jauge cachée dans le secteur scalaire. Elle n'apparaît que lorsque les champs non-dynamiques sont éliminés à travers leur équations du mouvement et, en ce sens, elle correspond à une symétrie de la physique, mais pas à une symétrie de l'action. 

Nous terminons l'étude des théories massives locales linéaires en les reformulant en tant que théories massives non-locales mais invariantes de jauge, à travers le formalisme de St\"uckelberg. Ceci constitue notre premier pas dans les théories non-locales, même si en l'occurrence la non-localité n'est qu'apparante et disparaît avec un choix de jauge approprié. Cependant, la technologie ainsi développée nous permet de définir une théorie d'un champ spin-2 massif linaire réellement non-locale et invariante de jauge. 

Suite à cela nous faisons une pause pour discuter en profondeur les subtilités des théories non-locales susmentionnées. La première est que des équations non-locales et causales ne peuvent pas être obtenues à travers le principe variationnel standard appliqué à une action non-locale, mais qu'il existe cependant un principe variationnel plus général qui fait l'affaire. Ensuite, à travers un processus de localisation, qui consiste à réécrire les équations sous forme locale en introduisant des champs auxiliaires, nous voyons que le contenu dynamique de ces théories est plus large qu'il n'y paraît. Ces champs obéissent des équations dynamiques, mais leurs conditions initiales sont contraintes par le choix de définition de nos opérateurs non-locaux dans la théorie originale. Ce dernier fait implique que nous ne pouvons pas quantifié de manière consistante les théories non-locales et donc que ces dernières ne peuvent être interprétées qu'en tant que théories classiques effectives. 

Le contenu original de cette partie consiste à clarifié une certaine confusion qui a lieu dans la littérature concernant l'impact de ces champs auxiliaires sur la stabilité des solutions d'intérêt. En effet, il se trouve que dans la plupart des modèles non-locaux étudiés, ces champs sont des ``ghosts", c'est-à-dire des champs dont l'énergie cinétique est négative. Cependant, le fait que leur conditions initiales soient contraintes a mené certains à déduire que leur effet sur la stabilité classique est automatiquement nul. Nous montrons que cet argumentation est justement le fruit d'un raisonnement de théorie des champs locale ne s'appliquant pas aux théories non-locales. En conclusion, les champs auxiliaires sont tout autant capables de déstabiliser une solution que n'importe quel champ dynamique non-contraint. Cependant, contrairement au cas quantique, la présence de ``ghosts" n'invalide pas nécessairement la théorie, car les divergences peuvent être assez lentes ou même contrées par des effets non-linéaires. C'est pourquoi, une étude de stabilité classique est nécessaire dans chaque cas.

Une fois ces quelques points clarifiés, nous reprenons la théorie de spin-2 linéaire non-locale que nous avions construite et tentons de la généraliser en une théorie de gravité non-locale, c'est-à-dire, nous construisons des extensions non-linéaires. Pour ce faire, nous empruntons deux procédés différents: un qui se base sur une action non-locale et un qui opère directement au niveau des équations du mouvement à l'aide de projecteurs transverses. Nous obtenons ainsi une classe de modèles non-linéaires que nous soumettons à certaines contraintes phénoménologiques. Celles-ci réduisent les modèles à deux extensions à un paramètre des modèles de Maggiore (M) et de Maggiore - Mancarella (MM) récemment proposés, qui les relient continument à la Relativité Générale avec une constante cosmologique. 

Ces modèles contiennent un  ``ghost" ultra-léger, mais des études numériques récentes et complètes des perturbations cosmologiques montrent que les modèles M et MM sont statistiquement équivalents à $\La$CDM, dans les marges d'erreurs des données actuelles. Cela suggère que les extensions le sont également, puisqu'elles ne font que nous rapprocher de $\La$CDM. Ceci les rends intéressantes, malgré le fait qu'un paramètre de plus diminue le pouvoir prédictif d'une théorie. Pour finir, nous étudions numériquement et analytiquement l'arrière-plan cosmologique de ces modèles. 
 
\end{otherlanguage}

\end{abstract}

\renewcommand{\abstractname}{Remerciements}
\begin{abstract}

\begin{otherlanguage}{francais}

En premier lieu je souhaiterais remercier mon superviseur et directeur de thèse, Michele Maggiore, pour m'avoir offert l'opportunité d'apprendre le métier de chercheur/enseignant en physique théorique. J'ai toujours pu compter sur ses conseils avisés et profiter de son expérience dans toutes les facettes de l'activité académique, tout en bénéficiant d'une grande liberté pour mes recherches personnelles. Au-delà de ses compétences de chercheur, j'ai pu également apprécier son attitude positive ainsi que sa fascination contagieuse pour les mystères de la physique. Je le remercie aussi pour sa lecture attentive de la présente thèse, ses commentaires et ses corrections.

En second lieu j'aimerais remercier les personnes avec qui j'ai eu le plaisir de collaborer ces quatre dernières années, à savoir, Maud Jaccard, Lukas Hollenstein, Stefano Foffa, Yves Dirian et évidemment Michele. Notre interaction a été une composante incontournable de mon doctorat, tant dans son impact sur ma formation et évolution en tant que chercheur, que sur le plan humain. Je pense avoir eu de la chance d'interagir avec toutes ces différentes personnalités qui constitueront sans aucun doute des références de qualité dans mon parcours professionnel. 

Aux membres de notre groupe de cosmologie je veux dire un grand merci pour les échanges, d'ordre académique ou pas, pour les sessions de ``crap-coffee", pour leur humeur joviale et pour l'ambiance chaleureuse qu'ils génèrent dans notre communauté. Je souhaiterais adresser un remerciement particulier à Ruth Durrer, Stefano Foffa et Michele pour leur influence, leurs conseils, ainsi que leur soutien dans mes recherches de travail. Un énorme merci également aux secrétaires du groupe, Cécile Jaggi-Chevalley et Francine Gennai-Nicole, pour leur disponibilité et leur aide dans les tourments administratifs, ainsi qu'à Andreas Malaspinas pour son assistance technique, sa disponibilité et les échanges au deuxième étage de l'école de physique qui durent toujours un peu plus long que prévu. 

J'ai également l'immense plaisir de remercier les hurluberlus du bureau 205 du Pavillon de physique I, David Daverio et Yves Dirian, pour les braves types qu'ils sont, les pauses café-clope qui ont révolutionné la physique, les pizzas toujours trop grasses et leurs caractères bien trempés. A David je dédicace en partie cette thèse, pour son amitié, sa complicité et en souvenir de ces débats interminables et bruyants sur la physique fondamentale, entre-autres. Ce doctorat n'aurait clairement pas été le même sans lui.

Je pense également à mes amis de toujours, Bryan, Chris, Fab, Ivàn, Jon, Kevin et Yannick qui me soutiennent (et me supportent) depuis tant d'années. A mes colocataires, mes ``honey", je leur dis merci pour l'ambiance sereine et agréable qu'ils procurent à notre ``coulouc". Merci à mes parents qui m'ont toujours soutenu dans mes ambitions et m'ont donné les moyens de les réaliser. Je remercie aussi mon frère Olivier pour son soutien et lui dédicace en partie ce travail en gage de reconnaissance de son courage et de sa determination. 

Pour finir, ma plus grande pensée va vers ma femme, mon amour, Julie, qui m'a toujours aimé, soutenu et compris. Je lui suis le plus reconnaissant du monde pour tout cela, ainsi que pour son courage, sa passion de vivre et cette force subtile qui l'habite. Je suis heureux qu'elle nous ait accordé sa confiance pour ce premier pas vers notre avenir. C'est un pas rempli d'espoir malgré l'incertitude des chemins où nous mènent nos ambitions. Ce travail porte les traces de l'énergie qu'elle m'insuffle et c'est pourquoi il lui est naturellement dédicacé en premier lieu. 

\end{otherlanguage}

\end{abstract}

\renewcommand{\abstractname}{Jury de thèse}
\begin{abstract}

\vspace{0.5cm}

\begin{itemize} \itemsep4pt

\item
Professeur Michele Maggiore, Université de Genève, Suisse (directeur de thèse).

\item
Professeure Ruth Durrer, Université de Genève, Suisse.

\item
Professeur Pedro G. Ferreira, Université d'Oxford, Royaume-Uni.

\item
Professeur Thomas Sotiriou, Université de Nottingham, Royaume-Uni.

\end{itemize}

\vspace{1cm}

Je tiens évidemment à remercier les membres du jury pour la considération, la lecture et l'évaluation de la présente thèse, ainsi que pour leurs corrections et leurs suggestions.

\end{abstract}

\setcounter{tocdepth}{1}
\tableofcontents

\chapter{Introduction}

\vspace{-0.1cm}

During my PhD, the research that I have conducted within the group of my PhD advisor Prof. Maggiore has focused on several aspects of the problem of dark energy in late-time cosmology. Here are the resulting publications:
\begin{itemize} \itemsep6pt

\item
{\it ``Stability analysis and future singularity of the} $m^2 R \square ^{-2} R$ {\it model of non-local gravity"} \\
with Yves Dirian \\
JCAP {\bf 10} (2014) 065

\item
{\it ``Cosmological dynamics and dark energy from non-local infrared modifications of gravity''} \\
with Stefano Foffa and Michele Maggiore \\
Int. J. Mod. Phys. A {\bf 29} (2014) 1450116

\item
{\it ``Apparent ghosts and spurious degrees of freedom in non-local theories"} \\
with Stefano Foffa and Michele Maggiore \\
Phys.\ Lett.\ B {\bf 733} (2014) 76-83

\item
{\it ``A non-local theory of massive gravity'}' \\
with Maud Jaccard and Michele Maggiore \\
Phys.\ Rev.\ D {\bf 88} (2013) 044033 

\item
{\it ``Bardeen variables and hidden gauge symmetries in linearized massive gravity}'' \\
with Maud Jaccard and Michele Maggiore \\
Phys.\ Rev.\ D {\bf 87} (2013)  044017 

\item
{\it ``Zero-point quantum fluctuations in cosmology''} \\
with Lukas Hollenstein, Maud Jaccard and Michele Maggiore \\
Phys.\ Rev.\ D {\bf 85} (2012) 124031

\item
{\it ``Early dark energy from zero-point quantum fluctuations''} \\
with Lukas Hollenstein, Maud Jaccard and Michele Maggiore \\
Phys.\ Lett.\ B {\bf 704} (2011) 102-107

\end{itemize}
An important part of this work consisted in the construction and study of a non-local theory of massive gravity and related non-local modifications of General Relativity that would produce a dark energy effect in accordance with observations. This is the subject on which I would like to focus my PhD thesis.

\section{Background}

In the last decades the field of cosmology has witnessed an effervescence which could be compared to the one that permeated particle physics in the 60's and the 70's, resulting in the birth of the Standard Model (SM). As often in science, it is the development of the experimental/observational branch of the discipline that allows the theoretical research to blossom. Indeed, the important activity in observational cosmology during the last two decades turned the discipline into a precise quantitative science, with more and more satellite, balloon and ground-based missions coming to enrich and refine the data pool. This allowed theorists to converge on a six-parameter concordance model, dubbed ``$\La$CDM", whose statistical predictions fit the data within the current error bars. These two factors, the rich/accurate data and the theoretical concordance model, constitute a solid basis for modern cosmology. This is still a very active area of research, as many more missions will take place in the future, thus providing more accurate input that will allow discriminating between models.

An important aspect of the concordance model, on top of the fact that it matches observations in a satisfying way, is that it mostly relies on well-understood physics. Indeed, on one side there is General Relativity (GR), which determines the dynamics of space-time in the presence of matter, and on the other hand there is the SM, which determines the content and microscopic dynamics of that matter. It is remarkable that the combination of these two pillars of modern theoretical physics suffices to describe already many aspects of the observed cosmology.

Nevertheless, there are also important parts of the concordance model which still remain unaccounted for from the theoretical point of view. The two outstanding ones in late-time cosmology are referred to as the ``dark matter'' and ``dark energy'' problems. These are significant extra elements compared to what GR and the SM alone would predict. They have therefore greatly contributed to the enthusiasm for theoretical cosmology and in setting-up further observational missions.

Before we discuss these two issues, let us also briefly mention the other important challenge in cosmology that is the understanding of its very early stages. The currently dominating paradigm, and by far, is the theory of inflation \cite{Guth,Linde} (see \cite{Baumann} for a review), which consists in the universe undergoing a period of accelerated expansion. This is theoretically appealing because it naturally leads to an approximately homogeneous, isotropic and spatially flat universe, as the one we observe. Most importantly, however, it explains the large-scale structure by relating it to primordial quantum fluctuations generated during this inflationary phase.

\subsubsection{Dark matter}

On Earth and solar-system scales the dynamics of GR and the matter content of the SM suffice to explain the observed phenomena, at least at the level of accuracy reached by experiment\footnote{A possible exception to this statement would be the neutrino masses, which are taken to be zero in the SM, while it has been discovered that $m_{\nu} \neq 0$ from measurements of neutrino oscillations.}. Unfortunately, this success story does not apply to larger scales such as the galactic, extragalactic and cosmological ones. 

On astrophysical scales, the rotation curves of galaxies and the motions of galaxies in galaxy clusters cannot be explained by the masses that we see in the telescope. Rather, the observed motions correspond to the gravitational forces one would have had in the presence of a larger amount of non-relativistic matter. On cosmological scales, it seems that non-relativistic matter constitutes nearly 30\% of the critical density today, while the observed baryonic matter, which matches the expected abundance from SM Big-Bang nucleosynthesis, can only account for $\sim 5\%$. 

Therefore, the simplest modification one can think of, that would correct this discrepancy, is to include a speculative type of particle with the following properties. It should not interact (or very weakly) with light, thus making it practically invisible, it should be rather massive so that it scales as non-relativistic matter and also stable on a time-scale of the age of the universe. Cosmological structure formation also suggests that it is non-relativistic at the time at which it decouples from the original plasma, and that its interactions are dominated by gravity. This way that matter can clump into halos, which then provide the necessary gravitational potential for ordinary matter to agglomerate into the galaxies, clusters, filaments we see today\footnote{Indeed, in the absence of that effect, it would have taken longer for ordinary matter to form the large scale structures, in contradiction with observations.}. Furthermore, the fact that no such new particle has been detected in accelerators yet, along with the fact that Big-Bang nucleosynthesis should not be disturbed too much, implies that it should interact very weakly with SM matter. This is what one refers to as ``Cold Dark Matter'', making the last three letters of ``$\La$CDM'', \footnote{``Cold" because it is massive, weakly interacting, and ``Dark'' because it does not interact with light. Note that a more appropriate term would be ``cold transparent matter'' because a dark object does interact with light since it absorbs it. For example, a black hole is ``dark'', dark matter is not, although the name is certainly more catchy.}.

\subsubsection{Dark energy}

Another important effect which is theoretically puzzling lies in the trend of the late-time expansion of the universe. In the late 90's, two independent groups \cite{Retal,Petal} analyzed the light-curves of type Ia supernovae and reported that the data imply an accelerated expansion of the universe at late times. This behaviour has been confirmed by many satellite and ground-based observations and will be further studied by missions planned for the future. The main complementary evidence comes from the Cosmic Microwave Background radiation anisotropies (CMB) and the Baryon Acoustic Oscillations in the large scale structure of matter (BAO)\footnote{It should be noted however that what is actually being measured in all of these three independent observations is the distance-redshift relation $D(z)$, \cite{Durrer}. Thus, the possibility remains that the inferred acceleration is only an apparent effect of physics which influence $D(z)$, \cite{Durrer}.}. 

This observation was surprising because ordinary fluids such as matter and radiation can only produce a decelerating expansion. Indeed, from the second Friedmann equation it follows that acceleration implies a negative pressure $p < - \ro/3$, since the energy density $\ro$ must be positive. In the case of dark matter, although its precise nature still eludes us, the most probable scenario is that it corresponds indeed to some massive particle(s) that could one day be detected in a collider. On the other hand, because of its negative pressure, dark energy seems to lie one step beyond in the scale of mysteriousness. Indeed, its properties are not the ones of a fluid made of standard particles and the speculations about its fundamental nature are much more variable. This discovery was rewarded with the Nobel prize of physics in 2011, given the astonishing implications for our understanding of the universe. 

Clearly, there are two, not mutually exclusive possibilities in order to explain this effect: either one must postulate the existence of a new source on the right-hand side of the Einstein equation that would support this expansion, or one must modify GR in the infra-red so that acceleration is obtained by altering the behavior of gravity itself\footnote{It is interesting to note however that in most cases this distinction may not be clear, as it is often possible to reproduce the phenomenology of modified gravity models with appropriate dark energy sources \cite{KunzSapone}.}. The degrees of freedom or mechanism which are responsible for this late-time acceleration being yet unknown, the community refers to them generically as ``dark energy''. This energy would then account for nearly 70\% of today's total energy of the universe.

\begin{figure}[h!]
\begin{center}
\includegraphics[width=16cm]{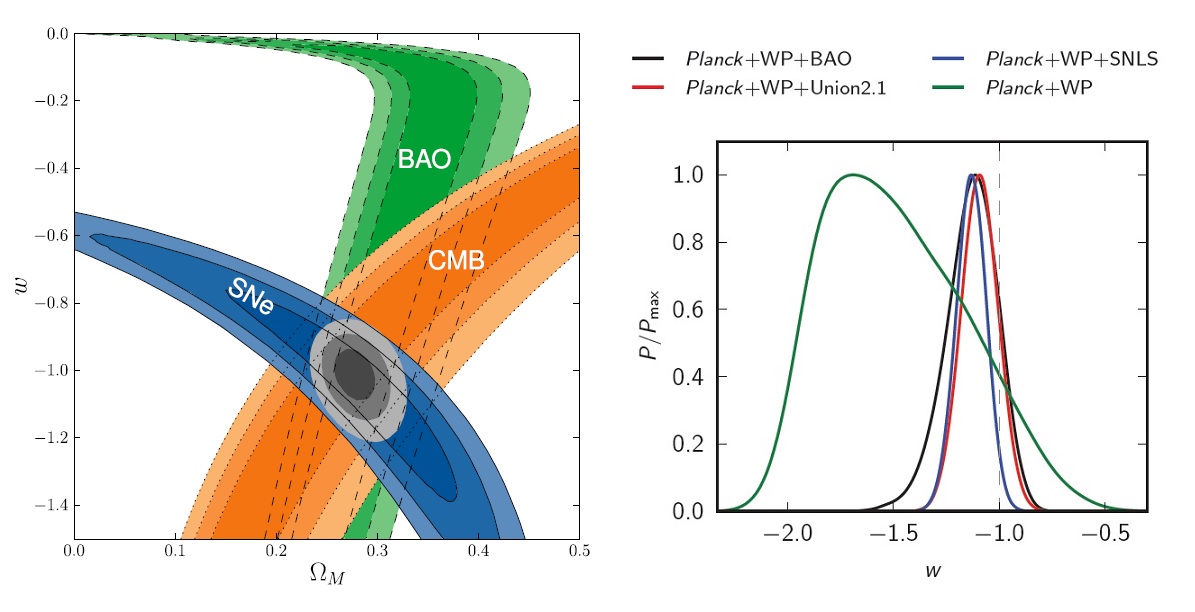} 
\caption{Left panel: the $1\si, 2\si$ and $3\si$ confidence regions of the combined constraints of type Ia supernovae (blue), CMB (orange) and BAO (green), without systematic errors. Plot by Amanullah et al. \cite{Amanullah} using the Union 2 compilation of supernovae, the WMAP7 data for the CMB and the SDSS DR7 and 2dF Galaxy Survey data for the BAO (2010). Right panel: the confidence region for $w$ from the Planck collaboration \cite{Planck} (2013). The combined CMB constraints of Planck and WMAP7 alone (green line), in combination with supernovae data (SNSL in blue and Union 2.1 in red) or BAO data (black). The latter are a combination of SDSS DR7, WiggleZ, BOSS DR9 and 6dF Galaxy Survey data.}
\label{fig:wOm} 
\end{center}
\end{figure}

From the theoretical point of view, quite remarkably, the best dark energy candidate for fitting the data \cite{Planck} is also the simplest term one could think of in the Einstein equation, namely, a positive cosmological constant
\beq \label{eq:EinsteinLa}
G_{\mu\nu} + \La g_{\mu\nu} = 8\pi G T_{\mu\nu} \, .
\eeq
This ``$\La$'' is the one which is found in ``$\La$CDM'' so that the name of the model reflects how it describes the ``dark'' sector. A very revealing plot is the one which combines constraints from type Ia supernovae, CMB and BAO observations, on the $\(w, \Om_M \)$ plane, where $\Om_M$ is the energy fraction corresponding to non-relativistic matter (dark and ordinary) today and $w$ is the equation of state of the dark energy component. Assuming a spatially flat universe we have that the fraction corresponding to dark energy today is $1 - \Om_M$, and also assuming a constant $w$ in time one gets figure \ref{fig:wOm}, \cite{Planck, Amanullah}. Indeed, one directly sees that dark energy makes up approximately 70\% of today's energy budget and is consistent with the time-evolution of a cosmological constant since $w \approx -1$. 

Now if we rather put this $\sim \La$ term on the right-hand side and interpret it as a constant source, we have that 
\beq
\ro_{\La} \equiv T_{tt} = \frac{\La}{8 \pi G} \, , \hspace{1cm}  p_{\La} \equiv T^i_i/3 = -\frac{\La}{8\pi G} \, .
\eeq
Thus, this energy-momentum tensor has a non-diluting (constant) energy density and negative pressure. These are both counter-intuitive properties for fluids made of particles, but might be accounted for if we resort to a more ``microscopic'' interpretation. Indeed, a constant source could typically correspond to the contribution of a potential term $\La \sim V(\bra \hat{\ph} \ket)$ in the quantum effective action of some Higgs-like field in a broken symmetry phase. This kind of dark energy is known as ``quintessence'' and, along with its generalizations (``$K$-essence'', etc.), represent one of the most studied alternatives to the cosmological constant. An important difference with the latter is that $\bra \hat{\ph} \ket$ is not necessarily constant in time and that the new field brings in additional degrees of freedom in cosmological perturbation theory. 

On the other hand, if we interpret (\ref{eq:EinsteinLa}) as a modification of gravity, i.e. on the left-hand side of (\ref{eq:EinsteinLa}), involving just another constant of nature $\La$, then this seems the most economic, conservative and also natural solution. Unfortunately, it is the quantum side of physics which will disagree with this interpretation. In the following section we will review succinctly the main arguments of the so-called ``cosmological constant problem''.

\section{The quantum vacuum problem} \label{sec:quantvacprob}

We may start by noting that the cosmological constant term plays exactly the role of the vacuum energy of field theory on flat space-time. Indeed, the $\La$ term in the Einstein equation corresponds to a constant term in the Einstein-Hilbert action
\beq \label{eq:Laterm}
S = \frac{1}{16\pi G} \int \ed^4 x \, \sqrt{-g} \( R - 2 \La \) \, .
\eeq
In the case $g_{\mu\nu} = \et_{\mu\nu}$ this is just a constant that produces an overall energy shift. This does not mean that a vacuum energy has no observable effects, as is clearly demonstrated by the Casimir effect in QFT for instance, but that only energy {\it differences} are relevant, not absolute values\footnote{See \cite{BMM} for a review of the Casimir effect.}. In GR however, every kind of energy gravitates, since this is what we find by definition on the right-hand side of the Einstein equation, and the physics therefore depends on the absolute value of $\La$. For energies way below the Planck scale, since the interactions with gravitons are heavily suppressed, the gravitational dynamics can be treated in very good approximation semi-classically. This means that gravity can be described classically, but sourced by the vacuum expectation value of quantum matter fields. Formally, we have
\beq
G_{\mu\nu} = 8\pi G \bra0 | T_{\mu\nu}[\hat{\ph}] | 0 \ket \, ,
\eeq
although the vacuum state $| 0 \ket$ may not be unique or easy to define. In any case, the quantum vacuum energy of matter is expected to appear as a cosmological constant on the right-hand side. In QFT on flat space-time, each bosonic field mode brings in a vacuum energy contribution which is formally diverging 
\beq
E_0(\vec{p}) = \frac{1}{2} \, \sqrt{\vec{p}^2 + m^2}\, (2\pi)^3 \de^{(3)}(0) \to \frac{1}{2}\, \sqrt{\vec{p}^2 + m^2}\,L^3
\eeq
and must thus be regularized by inserting some infra-red cut-off length $L$. The total vacuum energy is then the integral over all the modes, which must also be regularized but with an ultra-violet cut-off $\La_{\rm c} \gg L^{-1}$
\beq
E_{\rm vac} \equiv \int^{\La_{\rm c}} \frac{\ed^3 p}{(2\pi)^3}\, E_0(\vec{p}) = \frac{L^3 \La_{\rm c}^4}{16\pi^2} + \dots \, ,
\eeq
where the dots are lower-order terms in $\La_{\rm c}$. Finally, the vacuum energy density is simply
\beq
\ro_{\rm vac} = \frac{E_{\rm vac}}{L^3} = \frac{\La_{\rm c}^4}{16\pi^2} + \dots  \, ,
\eeq
so the infra-red regulator is irrelevant for this ``local'' quantity. This computation can also be performed for the rest of the $T^{\rm vac}_{\mu\nu}$ components and on less trivial backgrounds, if the latter have enough isometries, such as in cosmology for instance. Bearing some subtleties, one gets that $T^{\rm vac}_{\mu\nu} = {\rm const.} \times g_{\mu\nu}$ for the leading order term\footnote{If one uses a cut-off on momentum space then the result for the leading term $\sim \La_{\rm c}^4$ is actually $p_{\rm vac} = \ro_{\rm vac}/3$, whereas if $T^{\rm vac}_{\mu\nu} \sim g_{\mu\nu}$ then one should rather find $p_{\rm vac} = - \ro_{\rm vac}$. The $1/3$ ratio is the one obeyed by radiation and is inconsistent with a constant $\ro$ because then the continuity equation $\dot{\ro} = -3H \( \ro + p\)$ is not satisfied, so this result is in contradiction with general covariance. This apparent problem arises because these cut-off-dependent (``bare'') quantities are not the physical (``renormalized'') quantities. Since the cut-off is imposed on the $3$-momenta, it breaks covariance and thus so does the resulting energy-momentum tensor. The freedom in choosing the counter-terms then allows one to impose the correct relation for the renormalized quantities $p^{\rm ren}_{\rm vac} = - \ro^{\rm ren}_{\rm vac}$. As a matter of fact, had we started with a regularization that preserves covariance, such as dimensional regularization, this is the result we would have obtained. Thus, the apparent $1/3$ ratio is an artefact of our regularization scheme, and the physics cannot depend on it \cite{Maggiore1,HJMM1,HJMM2}.}, so this takes indeed the form of a cosmological constant. 

For fermionic fields, we have the same result but with the opposite sign. Thus, as soon as the number of bosons and fermions is not equal, we have that the ``natural'' value of $\ro_{\rm vac}$ is as high as the cut-off of this theory, from the effective field theory point of view. For the SM, where we know that effective theory to hold at least up to the scale where it has been tested ($\La_{\rm c} \sim$ TeV), we have at least $\ro_{\La} \sim {\rm TeV}^4 = 10^{12}\, {\rm GeV}^4$. As a matter of fact, since the SM has more fermionic degrees of freedom than bosonic ones, we should even expect a negative result. What is known as the ``cosmological constant problem" \cite{Weinberg1,Weinberg2} is that what we observe in cosmology is rather a tiny positive value $\ro_{\La} \sim 10^{-47} \, {\rm GeV}^4$, that is, a difference of at least sixty orders of magnitude!

\subsubsection{Renormalization group viewpoint}

Although the above description of the ``quantum vacuum catastrophe'' is probably the standard point of view on the dark energy problem in the community, it must be stressed that it relies more on theoretical hand-waving arguments than experimentally tested physics. Indeed, the vacuum energy is a feature of perturbative QFT whose absolute value is not observable in that theory, i.e. it is not an aspect of the theory which is checkable. Therefore, we do not know if it has any physical validity for us to take into account {\it as such} when generalizing to generally-covariant physics.  Moreover, even in QFT the absolute value of the vacuum energy is an ill-defined notion since one can get rid of it by choosing the so-called ``normal ordering'' for the Hamiltonian operator\footnote{This is usually expressed in terms of creation and annihilation operators, but in terms of $\hat{\ph}$ and its conjugate momentum $\hat{\pi}$ it amounts to adding a singular term $\sim \[ \hat{\ph}(x), \hat{\pi}(x) \]$ in $\hat{H}$ which of course vanishes classically.}, i.e. this issue is related to the ordering ambiguity of quantum mechanics. And this is not the only argument which casts doubt on the effect of vacuum energy within the QFT framework.

Indeed, an important remark is that this is merely a ``naturalness'' argument, not a prediction \cite{Maggiore1,HJMM1,HJMM2,MaggioreQFT}. In QFT the parameters of the Lagrangian cannot be predicted, only their dependence on the probing scale can, i.e. their running under the renormalization group. Thus, one can a priori fix them at any value suggested by experiment at some scale, and only then will their values at other scales be predicted. In the case of the leading part of the cosmological constant, there is no dependence on the probing scale, since it is a constant, and it can thus be chosen arbitrarily small at all scales. The apparent unnaturalness of this choice is then due to the fact that the observed tiny magnitude corresponds to a huge precision compared to the expected value. If what we expect is of order one, then the value we wish to give is of order $10^{-60}$, i.e. 60 digits of precision with respect to the natural scale. The unnaturalness argument thus corresponds to this incredibly fine tunning that must be performed. However, from the renormalization group point of view, only the running is physical, not the absolute values of the cut-off dependent quantities, so the above mentioned fine-tunning is not between physical quantities.

\subsubsection{Effective field theory viewpoint}

So why should one continue taking the cosmological constant problem so seriously? The point is that in the effective field theory viewpoint of QFT \cite{CWZ,CCWZ,Weinberg3,Donoghue1,Donoghue2,Burgess}, which is its modern interpretation, the cut-off-dependent quantities do acquire some physical substance. Indeed, the cut-off scale is usually related to the strong-coupling scale for perturbatively non-renormalizable theories, i.e. the energy at which the perturbative expansion breaks down. For instance, in the case of GR this scale would be the Planck mass. In practical examples of effective theories with known ultra-violet completions, the cut-off is related to the mass of some new particle, which is thus not seen in the effective theory, and which softens the interaction by being produced precisely near the cut-off. This allows us to access higher energy scales perturbatively, but with a larger theory encompassing the heavy particles. This is for example the case of the Higgs field when the effective theory is a massive Yang-Mills theory with fixed mass, or of the $W^{\pm}, Z$ bosons when the effective theory is Fermi's theory of four-fermion weak interactions, or the radial mode in the effective theory of the Goldstone modes of a sigma model. In all these cases, the cut-off of the effective theory is related to the activation of some new degrees of freedom. 

The question that now arises is whether this effective field theory logic applies to vacuum energy. Indeed, by definition, the vacuum has nothing to do with particles nor interaction scales. Thus, as long as we are within the QFT framework, it appears that we should keep adding-up the vacuum energies of higher and higher momenta. This would then end only at a scale where the mathematical description is not QFT anymore\footnote{An analogous case is the theory of fluids, which is an effective theory of space-time fields whose underlying ultra-violet completion is not a field theory but the dynamics of a large number of constituent elements. In that case, one also finds that the orders of magnitude of the parameters of the fluid are related to the fundamental scales arising in the microscopic element interactions.}. We are aware of such a scale, the Planck scale. Indeed, there the graviton interactions are strong and thus the structure of space-time becomes ambiguous, so that the local Minkowski approximation of QFT stops making sense. Thus, from the effective field theory point of view, we get an even larger estimate of the quantum vacuum energy, that is $\ro_{\La} \sim M^4 \sim 10^{76} \, {\rm GeV}^4$, giving a difference of $123$ orders of magnitude with the observed value! 
\\
\\
\indent From the above paragraphs we understand that the issue of the quantum vacuum in GR is not so well defined and is rather complicated, to say the least. Nevertheless, it is always a good theoretical exercise to look for alternative ways to describe a given phenomenon, even when what keeps us from choosing the simplest solution could be a matter of ``semantics''. Moreover, with increasing observational data, these alternatives can be tested. Thus, even if $\La$CDM turns out to still be a good fit in ten or twenty years, the strength of this statement would be much more important if several alternatives had also been considered.

To summarize, the problem of dark energy is two-fold. First one has to come up with a mechanism/argument for taming the quantum vacuum. In most cases, this is achieved only at the cost of making $\ro_{\rm vac}$ vanish exactly (e.g. supersymmetry), unless there is some fine-tunning. If $\ro_{\rm vac} = 0$, then one must also come up with a mechanism for producing some form of dark energy.

\section{Massive gravity}

As already mentioned, in this thesis we are going to explore the possibility of modifying gravity in the infrared in order to account for the dark energy effect, instead of considering some extra source on the right-hand side of the Einstein equation. One of the most studied modifications of the gravitational Lagrangian, motivated by both ultra-violet and infra-red physics, is the one where the Ricci scalar is replaced by an arbitrary function $f(R)$. Among other modifications involving also tensor curvature invariants, this class is distinguished by the fact that it has no ghosts (see \cite{SotiriouFaraoni} for a review). Another much studied model of infrared modified gravity is the Dvali-Gabadadze-Porrati (DGP) brane-world model \cite{DGP}. Although it has been shown to be non-viable, its theoretical by-products, such as the Galileon theory \cite{NRT}, have been instrumental in the development of massive gravity. 

Since GR describes a massless particle, when interpreted as a QFT on flat space-time, the simplest modification one can think of that hopefully alters only the infrared physics is giving a mass to that particle. The resulting theory of ``massive gravity'' has been both an inspiration and a (chronologically) starting point for our work on non-local modifications of gravity, so we find appropriate to summarize some of its important features.

\subsubsection{Expected advantages}

By (Lorentz-invariant) ``massive gravity'' is commonly meant a deformation of GR having the following properties:
\begin{itemize}
\item
In the absence of matter fields, Minkowski space-time is a linearly stable solution.
\vspace{0.2cm}
\item
The theory is Lorentz-invariant over that background.
\vspace{0.2cm}
\item
The spectrum of its linearized QFT over that background is a massive spin-2 particle.
\end{itemize}
It is not surprising that Minkowski space-time plays a privileged role in defining massive gravity, since the notions of particle, and thus mass, are well-defined only through the isometries of that background, i.e. the Poincar\'e group. A formulation of ``massiveness'' which would be applicable to more general backgrounds would involve the notion of gap, that is, that the field quanta have a minimal amount of energy $m > 0$. Classically, whenever the background is symmetric enough so that a dispersion relation of the perturbations $\om(\vec{k})$ can be defined, we would have that $\om(\vec{0}) = m > 0$.

Following the general wisdom of weakly interacting theories on Minkowksi space-time, a mass usually makes the field both insensitive to, and of little influence on, energy-momentum scales obeying $p, E \ll m$. Indeed, this is merely the fact on which effective field theory is based. Extrapolating these assumptions, as such, to the case of a fully non-linear theory of massive gravity would have the following consequences. 

First, massive gravity would be insensitive to a cosmological constant, since the latter is the most extreme example of infrared source. Second, the deceleration of the expansion of the universe should decrease as the background curvature approaches the $m$ scale, since the gravitational interaction would be cut-off at energies lower than $m$. This would suggest that the mass $m$ should be of the order of the Hubble parameter today $H_0$.

Any mechanism that would screen the cosmological constant, or more generally infrared sources, from gravity goes by the name ``degravitation'', an idea that has been first considered independently of any massive theory of gravity \cite{DGS1,DGS2,AHDDG,DHK}. This provides a very elegant resolution of the cosmological constant problem, by revealing that the true question is not why is $\ro_{\rm vac}$ so small, but rather why it affects gravity so little. 

Finally, another expected advantage of massive gravity is that, unlike the cosmological constant, a small value of the graviton mass would be ``technically natural", in the following sense. Indeed, a naive dimensional analysis would first suggest that, under radiative corrections, $\de m^2 \sim \La_{\rm c}^2$, which is not that much of an improvement compared to $\ro_{\rm vac} \sim \La^4_{\rm c}$. However, as in any gauge theory, adding a fixed mass necessarily breaks the gauge symmetry, here diffeomorphisms. In the massless case that symmetry protects the mass from being generated by loop corrections, so as $m^2 \to 0$, the corrections should tend to zero as well. This is the naturalness argument of 't Hooft \cite{'tHooft}, which implies that $\de m^2 \sim m^2$ and thus, by dimensional analysis, $\de m^2 \sim m^2 \log \La_{\rm c}$. In conclusion, the renormalized mass would be close to the bare one even for huge values of $\La_{\rm c}$.

Thus, following these naive expectations for a massive theory, one could obtain both a solution to the cosmological constant problem and possibly a naturally small dark energy. Of course, as stressed, these are hand-waving arguments that have no reason to apply in the case of non-linear theories over non-trivial backgrounds such as GR in cosmology. Nevertheless, they are certainly enough to tickle one's curiosity about what kind of phenomenology a theory of massive gravity would imply. This has indeed been the case recently, as the passed few years have witnessed an important excitement in this area. However, massive gravity has a much longer history that dates back to the late 30's.

\subsubsection{Brief history}

Since Minkowski space-time plays a privileged role in defining massive gravity, in order to conceptually appreciate the theory it is convenient to adopt the particle physics interpretation of GR: the latter is the unique theory, under some reasonable assumptions, of a massless spin-2 particle with consistent interactions \cite{Hinterbichler1, Hinterbichler2}. Indeed, GR can be expressed as a special relativistic gauge theory in terms of the perturbation around Minkowski space-time $h_{\mu\nu} \equiv M/2 \( g_{\mu\nu} - \et_{\mu\nu} \)$
\beq \label{eq:SEHh}
S_{\rm EH} = \int \ed^4 x \[ -\frac{1}{2}\, \pa_{\mu} h_{\nu\ro} \pa^{\mu} h^{\nu\ro} + \pa_{\mu} h^{\mu\nu} \pa^{\ro} h_{\ro\nu} - \pa_{\mu} h^{\mu\nu} \pa_{\nu} h + \frac{1}{2}\, \pa_{\mu} h \pa^{\mu} h + \Ord(\la h, \la^2 h^2, \dots) \pa h \pa h \]  \, ,
\eeq
where the indices are displaced using $\et_{\mu\nu}$, i.e. the special relativistic convention. Here $M \equiv \sqrt{8\pi G}$ (in natural units $\hbar = c = 1$) is the reduced Planck mass and $\la \equiv M^{-1}$ is the reduced Planck length playing the role of the small coupling constant. The diffeomorphisms now act as a non-abelian gauge symmetry on $h_{\mu\nu}$
\bea
\de h_{\mu\nu} & = & - \pa_{\mu} \xi_{\nu} - \pa_{\nu} \xi_{\mu} - \Lie_{\xi} h_{\mu\nu} \nn \\
 & = & - \pa_{\mu} \xi_{\nu} - \pa_{\nu} \xi_{\mu} - \xi^{\ro} \pa_{\ro} h_{\mu\nu} - h_{\ro\nu} \pa_{\mu} \xi^{\ro} - h_{\mu\ro} \pa_{\nu} \xi^{\ro}   \, ,
\eea
whose ``global'' subgroup\footnote{That is, the subgroup inducing a homogeneous transformation for $h_{\mu\nu}$.} are the isometries of Minkowski $\pa_{(\mu} \xi_{\nu)} = 0$, i.e. the Poincar\'e group. This is a derivatively coupled effective field theory whose cut-off, or strong-coupling scale, is given by the Planck scale. 

Since $h_{\mu\nu}$ is a two-tensor one can form two Lorentz-invariant quadratic combinations to form a mass term, these being $h_{\mu\nu}h^{\mu\nu}$ and $h^2$. At the linearized level, the only combination which yields a linearly stable theory was found by Fierz and Pauli (FP), in 1939, to be \cite{FierzPauli}
\beq \label{eq:FPmass}
S_{\rm FP} = -\frac{m^2}{2} \int \ed^4 x \( h_{\mu\nu} h^{\mu\nu} - h^2 \) \, .
\eeq
The linear theory describes a massive spin-2 excitation, so by one of Wigner's theorems, there are five degrees of freedom. Any other mass term will necessarily introduce a sixth degree of freedom which is a Lorentz scalar but is also a ghost, i.e. it has a negative kinetic energy and thus makes the total energy unbounded from below.

Quite later, in 1970, it was independently realized by van Dam and Veltman \cite{vanDamVeltman}, and Zakharov \cite{Zakharov}, that unlike spin-1 massive gauge theories, the spectrum of the spin-2 one is discontinuous in the massless limit, a feature that is known as the ``vDVZ'' discontinuity. Indeed, inverting the quadratic form of the graviton Lagrangian to obtain the propagator and saturating it with conserved sources one gets
\beq
T^{\mu\nu}(-k) D_{\mu\nu\ro\si}(k) T^{\ro\si}(k) = T^{\mu\nu}(-k)  \[ - \frac{i}{k^2 + m^2} \( \frac{1}{2} \, \et_{\mu\ro} \et_{\nu\si} +  \frac{1}{2} \, \et_{\mu\si} \et_{\nu\ro} - \frac{1}{3}\, \et_{\mu\nu} \et_{\ro\si}  \)  \]T^{\ro\si}(k) \, , 
\eeq
whereas in the massless case the last factor is $1/2$ instead of $1/3$. This implies that in the massless limit one obtains the GR result plus an extra scalar pole, i.e. a ``fifth force'' between the sources
\beq  \label{eq:vDVZprop}
\lim_{m \to 0} T^{\mu\nu}(-k) D_{\mu\nu\ro\si}(k) T^{\ro\si}(k) = {\rm GR} + \frac{1}{6}\, \ti{T}^{\mu\nu}(-k)  \[ - \frac{i}{k^2} \, \et_{\mu\nu} \et_{\ro\si}  \] \ti{T}^{\ro\si}(k) \, .
\eeq
This means that however small the mass may be, there will be $\Ord(1)$ differences with GR. For instance, if one fixes the normalization of $M$ by requiring the correct Newtonian limit, then the bending of light by a massive object deviates by $25\%$ from the GR prediction \cite{Hinterbichler1,Hinterbichler2}. Moreover, if the limit $m \to 0$ is not continuous, the argument that makes the mass ``natural'' under radiative corrections does not necessarily hold anymore. Most importantly however, this discontinuity suggests that giving a mass to gravity does not only modify its infra-red behaviour!

Nevertheless, this is an artefact of the linearized theory and no discontinuity appears if one considers the fully non-linear kinetic term. In 1972 Vainshtein \cite{Vainshtein} computed the spherically symmetric stationary solution perturbatively, both close to and far away from the source. In the latter case, he found that the zero-order part was not the Schwarzschild solution, a mark of the vDVZ discontinuity, and that the expansion parameter was $r_V/r$, with $r_V \equiv \( m^{-4} M^{-2} M_S \)^{1/5}$ now known as the ``Vainshtein radius'' ($M_S$  is the mass of the source). This implies that the region of validity of this solution $r > r_V$ is pushed to infinity in the massless limit since then $r_V \to \infty$. Moreover, as one approaches from infinity, the non-linearities become important at $r_V$. On the other hand, close to the source the expansion parameter is $r/r_V$ and the zero-order part is the Schwarzschild solution\footnote{A solution extending to all of space-time and matching the two asymptotic behaviours has been very difficult to find and its existence was first established numerically only in 2009 \cite{BabichevDeffayetZiour}. See \cite{BabichevDeffayet} for an introduction to the Vainshtein mechanism and the modern approach to the subject.}. Thus, GR {\it is} recovered close to the source and in the massless limit, but this cannot be seen in a perturbative expansion from the linear regime (far away from the source). This is now known as the ``Vainshtein mechanism'' and consists in the discontinuity of the linearized theory being ``cured'' by strong non-linear effects. The fifth force that appears in the propagator (\ref{eq:vDVZprop}) is indeed present in the linear regime, but is then screened by non-linear effects at small scales. 

Soon after Vainshtein's work, still in 1972, Boulware and Deser showed \cite{BoulwareDeser} that, unlike non-linear spin-1 gauge theories, considering the fully non-linear kinetic term of GR while keeping only the FP quadratic potential  reactivated the sixth ghost mode which was precisely avoided with the FP tuning (\ref{eq:FPmass}) in the linearized theory. Three decades later, in 2002, it was shown that this could still make sense as an effective field theory of an interacting massive graviton \cite{ArkaniHamedGeorgiSchwartz}. Indeed, the ghost's mass lies above the cut-off $\La_5 = (m^4 M)^{1/5}$ and the later is parametrically larger than $m$. However, for a mass of the order of the Hubble scale today $m \sim H_0$ one gets the very large scale $\La_5^{-1} \sim 10^{11}$ km, i.e. way larger than the millimeter scale down to which gravity has been tested. By adding higher powers to the Fierz-Pauli potential one can push the cut-off down to $\La_3 \equiv (m^2 M)^{1/3}$, giving $\La_3^{-1} \sim 10^3$ km, which is however still quite large \cite{ArkaniHamedGeorgiSchwartz}. Moreover, around a heavy source the effective theory breaks down at a distance that is parametrically larger than $\La^{-1}$ and also $r_V$, so that one has no access to the region where GR is recovered \cite{ArkaniHamedGeorgiSchwartz}.

The resolution of the ghost problem came only in 2010 in the works of de Rham, Gabadadze and Tolley (dRGT) \cite{deRhamGabadadze, deRhamGabadadzeTolley} which showed, in some special limit, that adding appropriately tuned higher-order terms in the potential removes the ghost at all orders in perturbation theory\footnote{Moreover, this special structure of the potential has been shown to be stable enough under quantum corrections, in the sense that it does deviate from its ghost-free form, but that the resulting ghost has a mass lying above the cut-off\cite{deRhamHeisenbergRibeiro}.}. Shortly after that, it was shown that the degree of freedom count is indeed five without having considered any limit and non-perturbatively \cite{HassanRosen1, HassanRosen2}, \footnote{Note however that this does not necessarily imply that the Minkowski solution is stable in the fully non-linear theory. Indeed, it is already a remarkably difficult task to demonstrate this in the case of GR \cite{ChristodoulouKlainerman}.}. Another advantage is that in the presence of a heavy source with mass $M_S$, the corresponding Vainshtein radius $r_V \sim \( m^{-2} M^{-2} M_S  \)^{1/3}$ is now larger than the distance at which the effective theory breaks down, so that there exists a region where GR is recovered \cite{Hinterbichler1,Hinterbichler2, deRham}. 

Unfortunately however, the cut-off is still $\La_3$, although it has been argued that the actual region of validity of the theory could extend to higher energies \cite{deRham}. Most importantly, it turns out that the theory admits only approximately (spatially flat) homogeneous and isotropic solutions \cite{DAdRDGPT} (for non-trivial $a(t)$), an important drawback for cosmology. One can have spatially open, or Bianchi type anisotropic solutions, but these are plagued by ghost instabilities \cite{DFGLM1,DFGLM2}. Even so, the successful construction of an effective theory of a massive graviton with the above properties is a remarkable theoretical achievement. 

A review and discussion of the theoretical and phenomenological properties of the dRGT theories, can be found in the reviews \cite{Hinterbichler1, Hinterbichler2, deRham}. In the present thesis, the aspect of massive gravity which interests us is of a more conceptual nature. Indeed, when trying to express this theory in terms of the full metric $g_{\mu\nu}$ one inevitably ends up with $\et_{\mu\nu}$ in the mass term as well, since the latter is not generally covariant. This leads to the following conceptual issues.

\subsubsection{Conceptual shortcomings of the dRGT approach}

The first source of discomfort is of course the lack of invariance. To deal with it one can still reinterpret the theory as a generally covariant one where there exists a privileged set of coordinates in which the tensor $\et$ takes the form $\et = {\rm diag} \( -1, 1, \dots, 1 \)$. A related alternative, which practically amounts to the same situation, would be to consider this trivial metric $\et$ as a dynamical field as well through a version of the so-called ``St\"uckelberg trick''. One introduces four auxiliary scalars $\ph^a$ through the replacement
\beq
\et_{\mu\nu} \to \et_{ab}\, \pa_{\mu} \ph^a \pa_{\nu} \ph^b \, ,
\eeq
so that now $\et_{\mu\nu}$ does transform like a tensor (while $\et_{ab}$ is an ``internal'' metric) and takes its trivial form in the $x^{\mu} = \de^{\mu}_a \ph^a$ coordinates. 

The St\"uckelberg trick is often cited as the prime example that any theory can be made gauge-invariant by simply introducing auxiliary fields patterned on the gauge transformation, a fact which is obviously true. However what cannot be retrieved after breaking diffeomorphism invariance with a mass is one of the founding principles of the theory: relativity. Indeed, the theory may be generally covariant but there exists a privileged set of coordinates, a preferred frame of reference, the one in which $\et_{\mu\nu}$ becomes trivial. It must be emphasized that this preferred frame is determined at the theory level, i.e. it is independent of the specific solution we are interested in. This should be contrasted with the {\it dynamically} privileged frames that arise in many situations, such as the rest frame of the CMB in cosmology, or the rest-frames of the sun in solar-system physics.

Another source of conceptual discomfort is the problem of choice: why $\et$? Indeed, in principle one could, and actually one does \cite{HassanRosenSchmidtMay,GuaratoDurrer}, consider other choices for this ``reference metric'', which is usually denoted by $f_{\mu\nu}$, \footnote{In this case Minkowski space-time is not guaranteed to exist as a stable solution. If the background is $\bar{g}_{\mu\nu} \neq \et_{\mu\nu}$, and not necessarily $f_{\mu\nu}$, then the field $h_{\mu\nu}$ transforms homogeneously only when the diffeomorphism generator $\xi^{\mu}$ is a Killing vector of $\bar{g}_{\mu\nu}$. Thus, the global space-time transformations are not the Poincar\'e group any more and the notion of a massive particle becomes ill-defined.}. But even if the phenomenology privileges one of these metrics, we would still be left with a ``God-given'' non-dynamical field. One way to solve this issue is bimetric gravity, first proposed in \cite{IshamSalamStrathdee} and recently extended to a ghost-free theory of massive bigravity \cite{HassanRosen3, vSSMEMH}, in which case one considers an Einstein-Hilbert kinetic term for the reference metric as well, making it dynamical and restoring explicit general covariance and relativity. A second dynamical metric opens a whole new window for the above mentioned conceptual issues and actually does exhibit a stable flat Friedmann-Lema\^itre-Robertson-Walker (FLRW) solution \cite{CCNP}. This has also been an active area of research lately, but unfortunately it is seems hard to obtain models where all perturbations are bounded on the backgrounds of interest \cite{CCP,LagosFerreira,CDGM}.

The above considerations lead us to wonder whether there might be a way to construct a theory of massive gravity in terms of a single metric $g_{\mu\nu}$ that is both explicitly covariant and privileges no reference frame. It turns out that this is possible, but that the price to pay is the loss of space-time locality.

\section{Non-local gravity}

A non-local theory is a theory in which the equations of motion are not differential but integro-differential, with both space and time integrations. Therefore, the dynamics of the field at $x$ do not only depend on the values of this field in the infinitesimal neighborhood of $x$, but on a finite or infinite region of space-time. In particular, in the case of time non-localities the corresponding physics exhibit memory effects. Since the field value  at $t + \ed t$ depends on the field values on a finite past interval $[t_i, t]$, the field ``remembers'' its history. Here we will restrict to non-local operators that are the inverses of some differential operators. Then, general covariance will imply that space and time non-localities come together.

Non-local modifications of GR have been considered in the early attempts to construct degravitating mechanisms \cite{AHDDG,DHK}. Moreover, they also appear from loop corrections to the quantum effective action for the metric, i.e. the action for the expectation value $\bra \hat{g}_{\mu\nu} \ket$ \cite{Jordan,CalzettaHu,BGVZ,DonoghueElMenoufi,TsamisWoodard}. Based on this justification, phenomenological non-local modifications of GR have already been considered as possible explanations of dark energy, with \cite{DeserWoodard1} being the pioneering one. More generally, non-local effects may appear in many classical effective descriptions where dissipative effects or subsystems are considered \cite{Galley,GalleyTsangStein}. 

In our work during my PhD we have first started by trying to construct a generally-covariant theory of massive gravity at the price of non-locality \cite{JaccardMaggioreMitsou2}, based on an earlier construction \cite{DHK, Dvali} which rather focused on its degravitation properties. The corresponding cosmology not being viable, we proceeded with the study of non-local modified gravity models that are still controlled by a fixed mass parameter, but in which the graviton remains massless \cite{FoffaMaggioreMitsou2,DirianMitsou}. These theories contain ghost modes, i.e. fields with negative kinetic energy, and we have spent some time understanding their effects both at the classical and quantum levels \cite{JaccardMaggioreMitsou2, DirianMitsou, FoffaMaggioreMitsou1}. Independent of the work in which I have been involved, the group has been very productive on the phenomenological analysis of these models \cite{Maggiore2,KehagiasMaggiore,MaggioreMancarella,DFKKM,CFM,DFKMP}.

\section{Thesis summary}

In this thesis we will describe part of the above-mentioned work and will also try to extend a bit further some of its concepts, constructions and conclusions. In the second chapter, we will start by revisiting linear massive gauge theories, since manipulating them will be important in understanding how to construct and especially analyze non-local theories. In particular, we will see how the field components of these theories split into dynamical/non-dynamical modes and the relation to the constraints of gauge theory, an identification which will be crucial in the non-local case. Part of this analysis will also cover a study that we carried out in \cite{JaccardMaggioreMitsou1} before we started the research on non-local gravity. It concerns a hidden symmetry in massive linearized gravity and the thorough analysis we will perform here will hopefully allow us to understand that feature better. The chapter will end with a non-local formulation of these local theories and a construction of a more general, genuinely non-local, theory of a linear massive graviton, with a scalar mode that is {\it not} necessarily a ghost. The latter part contains unpublished original material.

This will bring us to the subject of non-local field theory, so in the third chapter we will discuss the many subtleties that arise when considering non-localities. Indeed, a first feature is that the variational principle has to be generalized in order to provide causal equations of motion. Moreover, non-local theories cannot be quantized without enlarging their set of solutions in the classical limit, so that they can only be interpreted as classical effective theories. 

Most importantly however, their dynamical structure must be clarified in order to properly settle classical stability issues. This is a subject that has not been treated rigorously enough in some important part of the related literature, in my opinion. An original part of this thesis consists in unveiling the misunderstanding that lies at the origin the confusion. Indeed, as we shall see, one has to separate the notion of degree of freedom and dynamical field (or ``radiative'', ``propagating'' field). Whereas the two notions are equivalent in local field theory, this is no longer true in the presence of non-localities. If some field has its initial conditions constrained, and thus does not represent a degree of freedom, this does not necessarily mean that it does not propagate. 

Then, in the fourth chapter we will come back to the linear non-local theory constructed in chapter 2 and we will try to extend it to a generally-covariant non-local theory of massive gravity. There are two possible procedures, the ``action-based'' one and the ``projector-based'' one, whose resulting theories can be very different. After having constructed a class of models in both cases, we will apply some phenomenological constraints in order to reduce the number of free parameters. For the projector-based model the result will be that the tensor modes cannot be massive, while in the action-based model they can, but the corresponding mass term is irrelevant for the cosmological background. Since this is the part that will interest us here, the action-based model can also be taken with zero tensor mass. What is then left is the mass of the scalar mode, and the two models are one-parameter extensions of the models proposed by Maggiore \cite{Maggiore2} and Maggiore and Mancarella \cite{MaggioreMancarella}. The extensions continuously interpolate between these models and GR with a cosmological constant, so that the phenomenological successes of the former should remain valid for the extended models as well. 

In the last chapter we will analyze the background cosmology, using both numerical simulations and analytical approximations. The analysis of the one-parameter extensions is an original part of this thesis and confirms that they become indistinguishable from $\La$CDM for large values of the extension parameter. We will finish with a discussion of the fact that these solutions are phenomenologically viable, despite the presence of a ghost mode.  

Finally, in the appendix \ref{sec:bitensors} we have tried to provide a more or less rigorous mathematical support for the non-local operators that are invoked in generally-covariant non-local theories. These correspond to the generalization of the integration kernels of Green's theory, which are convolved with functions, to ``bi-tensors'' in differential geometry, that are convolved with tensors. The appendix also contains derivations of the properties of these operators that are most useful to us. For the reader who is less interested in these technicalities, rest assured that whenever some property or definition will be used, on top of referring to sections of this appendix we will also give lighter explanations that should satisfy (but not bore) a more physically-oriented mind. 

I acknowledge the use of Mathematica and especially of the ``xACT'' package for symbolic tensor computations \cite{xAct}.

\section{Notation \& conventions} \label{sec:convnnot}

We work on a $D$-dimensional manifold $\cal M$, also define $d \equiv D - 1$ and we focus on the case $D \geq 4$. The manifold $\cal M$ is equipped with a Lorentzian metric $g$, that is, a symmetric covariant tensor of rank $2$ whose component matrix $g_{\mu\nu}$ in some local coordinates has eigenvalues with the sign signature $\( -, +, \dots, + \)$ and thus $g \equiv \det ( g_{\mu\nu} ) \in \Rs^{*-}$. We denote by $\et_{\mu\nu}$ the Minkowski metric $\et = {\rm diag}(-1, 1, \dots, 1)$ and use the convention $\vep_{01\dots d} = - \vep^{01\dots d} = +1$ for the Levi-Civita symbol, so that
\beq
\frac{1}{D!}\, \sqrt{-g}\, \vep_{\mu_1 \dots \mu_D} \, \ed x^{\mu_1} \we \dots \we \ed x^{\mu_D} \equiv \sqrt{-g} \, \ed^D x \, ,
\eeq  
is the volume $D$-form. For the Riemann and Ricci tensors the conventions are
\beq
R^{\ro}_{\,\,\,\si\mu\nu} \equiv \pa_{\mu} \Ga^{\ro}_{\,\,\,\si\nu} - \pa_{\nu} \Ga^{\ro}_{\,\,\,\si\mu} + \Ga^{\ro}_{\,\,\,\al\mu} \Ga^{\al}_{\,\,\,\si\nu} + \Ga^{\ro}_{\,\,\,\al\nu} \Ga^{\al}_{\,\,\,\si\mu} \, , \hspace{0.5cm} R_{\mu\nu} \equiv R^{\ro}_{\,\,\,\mu\ro\nu} \, , \hspace{0.5cm} R \equiv g^{\mu\nu} R_{\mu\nu} \, ,
\eeq
and for the Christoffel symbols
\beq
\Ga^{\ro}_{\,\,\,\mu\nu} \equiv \frac{1}{2}\, g^{\ro\si} \( \pa_{\mu} g_{\nu\ro} +  \pa_{\nu} g_{\mu\ro} -  \pa_{\ro} g_{\mu\nu} \) \, .
\eeq
We use $\bo \equiv g^{\mu\nu} \na_{\mu} \na_{\nu}$ to denote the d'Alembertian and $\De \equiv \pa_i \pa_i$ to denote the Laplacian on flat space-time. The space-time Fourier transform convention is
\beq
\ph(x) = \int \frac{\ed^D k}{(2\pi)^D} \, \ph(k) \, \exp \[ i \et_{\mu\nu} k^{\mu} x^{\nu} \] \, , \hspace{1cm} \ph(k) = \int \ed^D x \, \ph(x) \, \exp \[ -i \et_{\mu\nu} k^{\mu} x^{\nu} \]  \, ,
\eeq
so for consistency the spatial Fourier transform is
\beq
\ph(\vec{x}) = \int \frac{\ed^d k}{(2\pi)^d} \, \ph(\vec{k}) \, \exp \[ i \vec{k} \cdot \vec{x} \] \, , \hspace{1cm} \ph(\vec{k}) = \int \ed^d x \, \ph(\vec{x}) \, \exp \[ -i \vec{k}\cdot \vec{x} \]  \, .
\eeq
We use natural units $\hbar = c = 1$ and also the following reduced Planck masses $M \equiv \( 8 \pi G \)^{-1/2}$ and $\ti{M} \equiv (16 \pi G)^{-1/2}$, which are actually masses only in $D = 4$.

\chapter{Linear massless/massive gauge theories} \label{ch:linearmassgauge}

In this chapter we propose to study the massive and massless theories of spin-1 and spin-2 fields through several approaches, each one of them providing a complementary viewpoint. As already mentioned in the introduction, the notions of degree of freedom and of dynamical field are not equivalent in non-local field theory. It is therefore important to first understand their equivalence in local field theory, and especially gauge theory, where not all fields propagate. We will thus see, in many different ways, how the field content splits into dynamical and non-dynamical fields and how this is related to the degrees of freedom of the theory. This will then allow us to understand the spectrum of non-local gauge theories, without making any confusion between the constraints that are due to non-locality and the ones that are due to gauge symmetry. Finally, this analysis will also bring us useful by-products that will allow us to construct linear non-local massive spin-2 gauge theories. 

Although our main interest is in gravity and thus the spin-2 field, the spin-1 case will be very helpful in facilitating our intuition and argumentation. Indeed, it shares many properties with the spin-2 case, but at the same time has less fields, thus simplifying our analysis. On top of this, the spin-1 theory stands as exceptional, regarding some important properties, when compared with higher spin theories $s \geq 2$. Thus, the study of the spin-1 case will turn out to be essential in contrasting with some peculiarities of the spin-2 case.  

For the kinetic term of the theory, in each case, we will consider the only one that is stable, i.e. the one that exhibits the highest gauge symmetry. These are the kinetic terms of electrodynamics and of linearized GR. For the mass terms however we will consider the most general quadratic Lorentz-invariant potential, which in the case of the spin-2 field usually activates a ghost mode. Indeed, that ghost will be a recurrent subject in this thesis, so it is important that we include these actions as well in our study. Moreover, considering this general case will lead us to the definition of projectors that are going to be very useful for constructing a genuinely non-local ghost-free theory. This chapter is based on, and extends, the following papers \cite{JaccardMaggioreMitsou2,JaccardMaggioreMitsou1}.

\section{Technical preliminaries}

\subsection{Inverse differential operators} \label{eq:diffop}

In this chapter we will consider only spatially localized fields, that is, fields which tend to zero sufficiently fast at infinity and which can therefore be represented by their spatial Fourier transform. On this space of fields the operator $\De - m^2$, where $\De \equiv \pa_i \pa_i$ is the Laplacian, is negative-definite, as is obvious in its Fourier representation. It has therefore zero kernel when acting on fields whose values and first spatial derivatives tend to zero at spatial infinity. This means that it admits a unique (right and left) inverse $\( \De - m^2 \)^{-1}$, and actually a unique power $\( \De - m^2 \)^{\ka}$ for $\ka \in \Rs$, which can again be obtained through its Fourier representation. These operators commute among themselves and with spatial derivatives.

These nice properties do not generalize to the Klein-Gordon operator $L \equiv \bo - m^2$ because it has a non-trivial kernel, the vector space generated by the plane-wave solutions (see appendix \ref{sec:Greensfunc} for detailed properties). It therefore admits more than one right-inverse $L L^{-1} = {\rm id}$ and no left-inverse in general. The space of inverses is parametrized by the elements of the kernel since any two inversions are related by a homogeneous solution
\beq
L \[ L^{-1} (\ph) - L'^{-1} (\ph) \] = 0 \, .
\eeq
Thus, if one picks a $L^{-1}$ once and for all, all other inversions are found by adding a homogeneous solution, as we know from calculus. Here we will denote by ``$L^{-1}$'' the inverses of $L$ that are also $\Rs$-linear operators
\beq
L^{-1} \( \al \ph + \be \ph' \) = \al L^{-1} \ph + \be L^{-1} \ph' \hspace{1cm}  \al, \be = {\rm const} \in \Rs \, ,
\eeq
which must be contrasted with the general inverse operator which is affine
\beq
L^{-1}_{\rm gen.}(\ph) = L^{-1} \ph + \psi \, , \hspace{1cm} L \psi = 0 \, , 
\eeq
with $\psi$ independent of $\ph$. The operators $L^{-1}$ can then be represented by the convolution with a Green's distribution
\beq \label{eq:Lm1phG}
(L^{-1} \ph)(x) = \int \ed^D y\, G(x,y) \, \ph(y) \, , \hspace{1cm}  L_x G(x,y) = \de^{(D)}(x-y) \, ,
\eeq
which by Poincar\'e covariance must be of the form $G(x,y) = G(x-y)$. The quantity $i G$ is also called a ``propagator'' depending on the context. The different choices of $L^{-1}$ now correspond to the different time boundary conditions of $G(x)$, which in turn correspond to the time boundary conditions of $(L^{-1}\ph)(x)$, \footnote{Given the set of fields we consider, the spatial boundary conditions are zero at infinity.}. 

Two Green's functions are of particular relevance for physics on flat space-time, the retarded one in classical field theory and the Feynman one in perturbative QFT. Imposing trivial initial conditions
\beq
\lim_{x^0 \to - \infty} G(x) = 0 \, , \hspace{1cm} \lim_{x^0 \to - \infty} \pa_{x^0} G(x) = 0 \, , \label{eq:retpropBC}
\eeq
gives the retarded propagator
\beq \label{eq:retardedpropnoannex}
G_{\rm r}(x) = \lim_{\ep \to 0^+} \int \frac{\ed^D k}{(2\pi)^D} \frac{\exp \( i \et_{\mu\nu} k^{\mu} x^{\nu} \)}{(k^0 + i \ep)^2 - \vec{k}^2 - m^2} \, ,
\eeq
while imposing no positive-frequency ingoing waves and no negative-frequency outgoing waves
\bea
\lim_{x^0 \to - \infty} G(x) & = & \int \frac{\ed^d k}{(2\pi)^d} \int_{-\infty}^0 \frac{\ed k^0}{2\pi}\, a(k)\, \exp\[ i \et_{\mu\nu} k^{\mu} x^{\nu} \] \, , \\
\lim_{x^0 \to +\infty} G(x) & = & \int \frac{\ed^d k}{(2\pi)^d} \int_0^{\infty} \frac{\ed k^0}{2\pi}\, a(k)\, \exp\[ i \et_{\mu\nu} k^{\mu} x^{\nu} \] \, , \label{eq:FeynpropBC}
\eea
gives the Feynman propagator
\beq
G_{\rm F}(x) = \lim_{\ep \to 0^+} \int \frac{\ed^D k}{(2\pi)^D} \frac{\exp \( i \et_{\mu\nu} k^{\mu} x^{\nu} \)}{- k^2 - m^2 + i \ep} \, .
\eeq
Indeed, by writing (\ref{eq:Lm1phG}) in Fourier space, and using the converging contour integrals with the residue theorem, we get that $L^{-1} \ph$ obeys the above mentioned boundary/initial conditions in each respective case. The domains of definition of the corresponding operators $L^{-1}_{\rm r}$ and $L^{-1}_{\rm F}$ are the fields obeying the same boundary conditions as $G$ in each respective case. On their respective domains of definition, both operators commute with partial derivatives and are also left-inverses\footnote{See appendices \ref{sec:leftinv} and \ref{sec:commrel} where we show this for $\bo^{-1}_{\rm r}$ in real space and on arbitrary globally hyperbolic space-times. It can also be worked-out in Fourier space for both $\bo^{-1}_{\rm r}$ and $\bo^{-1}_{\rm F}$, since if the Fourier representation gives a finite result, i.e. if the operators are defined, then it is obvious that they commute with the derivatives and are also left-inverses.}. In practice the $L^{-1}_{\rm r}$ may act after some derivatives, in which case it is convenient to have a stronger condition for its applicability. At the bottom of appendix \ref{sec:Greensfunc} we provide such a condition which we call ``having finite past''. Loosely speaking, it amounts to $\ph$ being non-zero only after a finite time.

The retarded Green's function arises in situations where one wants to solve a sourced equation
\beq \label{eq:solLphJ}
L \ph = J \, , \hspace{1cm} \ph = \int \ed^D y\, G_{\rm r}(x-y) \, J(y) \, ,
\eeq
in a causal way, i.e. such that $\ph(x)$ depends only on $J(x')$ with $x'$ in the past light-cone of $x$. This is indeed the case as we can see by the real space representation in $D = 4$ given in equation (\ref{eq:Gr4DMink}) of appendix \ref{sec:Greensfunc}. Flipping the sign of $\ep$ in (\ref{eq:retardedpropnoannex}) amounts to flipping the sign of $x^0$, after having redefined $k^0 \to -k^0$, so this gives us the advanced propagator $G_{\rm a}$ which is supported on the future light-cone and is thus anti-causal. We thus have
\beq
G_{\rm r}(-x^0, \vec{x}) = G_{\rm a}(x^0, \vec{x}) \, ,
\eeq
while $G_{\rm r}(x^0, \vec{x})$  is symmetric under the individual sign flip of spatial arguments. In perturbative QFT it is rather the Feynman propagator which is relevant because it is the one that arises in the computation of the scattering amplitudes. More precisely, it represents the particles of $\ph$ which mediate the interaction between sources $J$ at different space-time points. To see this one can invoke the corresponding action
\beq \label{eq:XX}
S = \lim_{\ep \to 0^+} \int \ed^D x \[ \frac{1}{2}\, \ph \( \bo - m^2 + i \ep \) \ph - \ph J \] \, ,
\eeq
which has been regularized with an $\ep$ factor that ensures the convergence of the corresponding path integral. Thus, unitarity of $e^{i S}$ forces upon us this choice for the sign of $\ep$. We then have that by integrating-out $\ph$
\beq \label{eq:intoutphFprop}
\int_{\cal B} D \ph \, e^{i S} \sim \exp \[ \frac{i}{2} \int \ed^D x \, J G_{\rm F} J \] \, .
\eeq
Differentiating twice (\ref{eq:XX}) with respect to the source one gets that the Feynman propagator is the two-point function
\beq
\bra 0 | \hat{\ph}^{\da}(k) \hat{\ph}(k) | 0 \ket = - \frac{i}{k^2 + m^2 - i \ep} \, .
\eeq
Actually, this path integration has been performed a bit formally since we have not specified its boundary conditions $\cal B$. However, these are already fixed for consistency reasons and there are several instructive ways to see this that will be useful for us at some point later on. First, note that the path integral is dominated by the classical solutions, which in this case are given by free wave-packets at infinity (where $J = 0$) with dispersion relation
\beq
k^0 = \pm \( \sqrt{m^2 + \vec{k}^2} - i \ep \) \, .
\eeq
Thus, positive-frequency modes diverge at past infinity, while negative-frequency modes diverge at future infinity. This means that the only boundary conditions for which the path integral makes sense around classical solutions are the Feynman ones (\ref{eq:FeynpropBC}), i.e. only negative-frequency waves at past infinity and only positive-frequency waves at future infinity. Conversely, if one imposes these boundary conditions but sets $\ep = 0$, then the result of integrating-out $\ph$ is the Feynman propagator. One can also understand these boundary conditions from the point of view of the canonical quantization. One simply needs
\beq
\bra 0 | T \dots | 0 \ket \sim \int D \ph \, \dots \, e^{i S[\ph]} \, ,
\eeq
where $| 0 \ket$ is the vacuum state at past infinity and $\bra 0|$ is the one at future infinity. We then have that $a |0 \ket = 0$, where $a$ is the free annihilation operator corresponding to the amplitude of the modes with positive frequency, while $\bra 0 | a^{\da} = 0$, where $a^{\da}$ is the creation operator corresponding to the amplitude of the modes with negative frequency.

Finally, note that since $G_{\rm F}(k)$ is a function of $k^2$, we have that $G_{\rm F}(x)$ is symmetric under the individual flip of any of its arguments, so it is symmetric under time-reversal in particular. As a consequence it has both retarded $\sim \te(x^0-y^0)$ and advanced $\sim \te(y^0-x^0)$ parts. This is expected because in a scattering process the information of the whole interval $t \in ]-\infty, \infty[$ is required, so that for finite $t$ the dependence is acausal.

\subsection{Degrees of freedom, dynamical and non-dynamical fields} \label{sec:dofprop}

In non-local theories the question of degrees of freedom of a theory can be a subtle issue, so it is important that we define clearly the words we will be using. The number of degrees of freedom of a field theory, denoted by $N_{\rm f}$, is the number of initial field configurations that we are free to choose in order to evolve the system uniquely in time. In the theories we are going to study below we will find two types of fields. The ``dynamical''  (or ``radiative'') ones are those obeying a second-order equation in time
\beq \label{eq:dyn}
\( \bo - m^2 \) \ph = J \, ,
\eeq
while the ``non-dynamical'' (or ``non-radiative'') ones are those that obey a purely spatial differential equation
\beq \label{eq:nondyn}
\( \De - m^2 \) \ph = J \, .
\eeq
In the dynamical case (\ref{eq:dyn}) the solution for a $\ph$ which is solely excited by $J$ takes the form (\ref{eq:solLphJ}). This means that, by measuring $\ph$ at some $x$, one can deduce some information about the excitations of $J$ at some other $x'$ (as long as $x$ is in the future light-cone of $x'$). We thus say that the field ``propagates'' the information of the source. This is how one can gain information about a distant object, by detecting the waves it emits in some {\it dynamical} field. Going even further, this is how two ``sources'' at different space-time points are going to interact through the ``force'' mediated by $\ph$. Note that this scenario does not focus on the initial conditions that would have been given to $\ph$. These are actually trivial since $\ph$ is solely excited by the source. Thus, the forces that are present in the theory correspond to the dynamical fields, independently of whether these are degrees of freedom or not. Finally, since the dynamical fields induce poles in the propagator, and ``propagate'' the information of sources, one can equivalently refer to them as ``propagating'' fields.

In the non-dynamical case (\ref{eq:nondyn}) the equation seems to be in conflict with relativity since it is not Lorentz invariant and implies an action at a distance, i.e. $\ph$ reacts instantaneously to the source $J$. As we will see however, in these cases, either $\ph$ will not be physically observable (gauge-dependent), or it will itself be a spatially non-local functional of the fundamental fields. In the latter case the measurement of $\ph$ is spatially non-local to begin with and can thus not be performed at a single time, so there is no contradiction with relativity. In that case, the information of the source does not propagate but is instead communicated simultaneously, to an unphysical or non-local field. Thus, non-dynamical fields do not allow us to gain local information on the source's dynamics nor do they mediate any interaction. 

Now, in the dynamical case, we have that one needs to provide the initial conditions $\ph(t_i, \vec{x})$ and $\dot{\ph}(t_i, \vec{x})$ on $\Rs^d$ in order to evolve the field in time, so that it corresponds to $N_{\rm f} = 2$. In the non-dynamical case we have that the field is totally determined by the source at every time and, in particular, at the initial condition surface, so that $N_{\rm f} = 0$. In the dynamical case the solutions for $J = 0$ are linear superpositions of plane-waves, whose vector space is isomorphic to the initial data space, while in the non-dynamical case the source-free solution is $\ph = 0$.

It therefore seems obvious that, if one denotes the number of dynamical fields by $N_{\rm d}$, then $N_{\rm f} = 2 N_{\rm d}$, \footnote{If the dynamical equations where of order $n$ in the time-derivatives, this would give $N_{\rm f} = n N_{\rm d}$.}. This appears as a trivial statement in local field theory, but does not hold at all for non-local theories. It is thus important to stress in advance that the notion of dynamical field and degree of freedom should be considered separately.

\section{Standard Lagrangian approach} \label{sec:stdLagfor}

\subsection{Spin 1} \label{sec:stdLagforA}

\subsubsection{Massive}

So let us start by considering the case of massive electrodynamics, that is, the Proca action
\beq \label{eq:ProcaAction}
S \equiv \int \ed^D x \[ -\frac{1}{4}\, F_{\mu\nu} F^{\mu\nu} - \frac{1}{2}\, m^2 A_{\mu} A^{\mu} + A_{\mu} j^{\mu} \] \, , \hspace{1cm} F_{\mu\nu} \equiv \pa_{\mu} A_{\nu} - \pa_{\nu} A_{\mu} \, ,
\eeq
where $j^{\mu}$ is a conserved external source, i.e. $\pa_{\mu} j^{\mu} = 0$, and the mass parameter $m$ breaks the U$(1)$ gauge symmetry\footnote{In realistic cases where $j^{\mu}$ is also made of fundamental fields, the argument that the $m = 0$ action is gauge-invariant because $\pa_{\mu} j^{\mu} = 0$ no longer holds. Indeed, conservation equations can only hold for some field configurations, namely the on-shell ones, whereas a symmetry should hold for {\it all} field configurations in the action. There are then two possibilities. Either the $A_{\mu} j^{\mu}$ term corresponds to non-minimal couplings to other fields through $F_{\mu\nu}$, in which case it is itself gauge-invariant, or it emerges through minimal couplings that involve the covariant derivative $\na \equiv \pa - i A$, in which case its variation is compensated by a non-trivial variation of $A$-independent terms. Then, because of that gauge symmetry, by Noether's theorem for local symmetries we have that $\pa_{\mu} j^{\mu} = 0$ on-shell. In the massive case, if the matter sector is unchanged, then we still have a global U$(1)$ symmetry and it is thus Noether's theorem for global symmetries which implies $\pa_{\mu} j^{\mu} = 0$.}
\beq \label{eq:gsA}
\de A_{\mu} = - \pa_{\mu} \te \, .
\eeq
The equations of motion are
\beq \label{eq:ProcaEOM0}
\pa_{\mu} F^{\mu\nu} - m^2 A^{\nu} = -j^{\nu} \, , 
\eeq
and taking the divergence one gets
\beq \label{eq:ProcaEOMdiv}
m^2 \pa_{\mu} A^{\mu} = 0 \, ,
\eeq
so we can rewrite them as
\beq \label{eq:ProcaEOM}
\( \bo - m^2 \) A_{\mu} = - j_{\mu} \, , \hspace{1cm}  \pa_{\mu} A^{\mu} = 0 \, .
\eeq
Thus, as soon as $m \neq 0$, and therefore the gauge symmetry is lost, the usual Lorentz gauge condition of massless electrodynamics $\pa_{\mu} A^{\mu} = 0$  appears as the scalar part of the equations of motion. The latter along with the $\mu = 0$ components of the Klein-Gordon equation imply that $A_0$ is non-dynamical
\beq \label{eq:A0sol}
\( \De - m^2 \) A_0 = \pa_i \dot{A}_i - j_0 \, , \hspace{1cm}  \dot{A}_0 = \pa_i A_i \, ,
\eeq
and that its initial conditions are totally determined in terms of the ones of $A_i$ and $j_0$. We are then left with
\beq \label{eq:ProcaEOMi}
\( \bo - m^2 \) A_i = - j_i \, ,
\eeq
that is, $d$ unconstrained fields transforming in the vector representation of SO$(d)$ and obeying a massive Klein-Gordon equation. This amounts to $N_{\rm f} = 2 N_{\rm d} = 2d$ degrees of freedom, corresponding to the initial conditions of $A_i$ and $\dot{A}_i$. In $d = 3$ this gives $N_{\rm d} = 3$.

\subsubsection{Massless}

In the case where $m = 0$, we have the gauge symmetry (\ref{eq:gsA}), so the Lorentz gauge $\pa_{\mu} A^{\mu} = 0$ can be reached by performing a gauge transformation, the result being again (\ref{eq:A0sol}) and (\ref{eq:ProcaEOMi}), but with $m = 0$. Now however these equations have a residual gauge symmetry given by the gauge parameters satisfying $\bo \te = 0$. To see what we can do with it, we can consider the general solution of the divergence of (\ref{eq:ProcaEOMi}) 
\beq \label{eq:solpaiAi}
\pa_i A_i = \ph^{\rm hom} - \bo_{\rm r}^{-1} \pa_i j_i \, , 
\eeq
where $\ph^{\rm hom}$ is a homogeneous solution $\bo \ph^{\rm hom} = 0$. Remember that for the action of $\bo_{\rm r}^{-1}$ to be defined the source $\pa_i j_i$ must have finite past. Using the residual gauge transformation on that equation we get
\beq
\pa_i A_i - \De \te = \ph^{\rm hom} - \bo_{\rm r}^{-1} \pa_i j_i \, .
\eeq
It is thus possible to cancel $\ph^{\rm hom}$ by choosing
\beq
\te = - \De^{-1} \ph^{\rm hom} \, ,
\eeq
so that $\pa_i A_i$ is totally determined by the source and its initial conditions are thus fixed. The degrees of freedom are therefore the $N_{\rm f} = 2(d-1)$ components of the transverse part $A^{\rm t}_i$, i.e. $\pa_i A^{\rm t}_i = 0$, and its first derivatives.

It may appear however that the longitudinal part $\pa_i A_i$ is still a dynamical field, since it obeys a dynamical equation $\bo \pa_i A_i = \pa_i j_i$, even though it does not correspond to a degree of freedom. This would be in contradiction with $N_{\rm f} = 2N_{\rm d}$. As it turns out, this is only an artefact of our choice of gauge, which is the natural one from the point of view of the massive theory, since then $\pa_{\mu} A^{\mu} = 0$ holds continuously with $m \to 0$. Indeed, one can always introduce a $\bo^{-1}_{\rm r}j$ term in the gauge parameter to make it appear as a source of a gauge-dependent component. We can therefore choose a different gauge to start with, such as the one which precisely eliminates the longitudinal mode
\beq
\pa_i A_i = 0 \, .
\eeq
This choice is more natural from the Hamiltonian point of view, as we will see soon. The equation of motion of $A_0$ then reads
\beq
\De A_0 = -j_0 \, ,
\eeq
and we have that the divergence of the equation of $A_i$ is automatically satisfied. We thus have that the initial conditions of both $A_0$ and $\pa_i A_i$ are fixed {\it and} that these fields are non-dynamical. We can therefore conclude that in the massless theory we have indeed $N_{\rm f} = 2 N_{\rm d} = 2(d-1)$, which for $d = 3$ gives $N_{\rm d} = 2$.

\subsection{Spin 2} \label{sec:stdLagforh}

Let us know consider linearized GR along with the most general quadratic potential
\bea
S & = & \int \ed^D x \[ -\frac{1}{2}\, \pa_{\mu} h_{\nu\ro} \pa^{\mu} h^{\nu\ro} + \pa_{\mu} h^{\mu\ro} \pa^{\nu} h_{\nu\ro} - \pa_{\mu} h^{\mu\nu} \pa_{\nu} h + \frac{1}{2}\, \pa_{\mu} h \pa^{\mu} h \right. \nn \\
 & & \left. \hspace{1.2cm}  - \frac{1}{2}\, m^2 \( h_{\mu\nu} h^{\mu\nu} - (1 + \al) h^2 \) + h_{\mu\nu} T^{\mu\nu} \] \nn \\
 & \equiv & \int \ed^D x \[ \frac{1}{2}\, h_{\mu\nu} {\cal E}^{\mu\nu\ro\si} h_{\ro\si}  - \frac{1}{2}\, m^2 \( h_{\mu\nu} h^{\mu\nu} - (1 + \al) h^2 \) + h_{\mu\nu} T^{\mu\nu} \] \, ,  \label{eq:FPaction}
\eea 
where ${\cal E}$ is known as the ``Lichnerowicz operator''
\beq \label{eq:Lichne}
{\cal E}^{\mu\nu\ro\si} \equiv \( \et^{\mu(\ro} \et^{\si)\nu} - \et^{\mu\nu} \et^{\ro\si} \) \bo - \et^{\mu(\ro} \pa^{\si)} \pa^{\nu} - \et^{\nu(\ro} \pa^{\si)} \pa^{\mu} + \et^{\mu\nu} \pa^{\ro} \pa^{\si} + \et^{\ro\si} \pa^{\mu} \pa^{\nu}  \, .
\eeq
The Fierz-Pauli theory corresponds to the choice $\al = 0$. Here $T^{\mu\nu}$ is some external conserved source $\pa_{\mu} T^{\mu\nu} = 0$ and the mass term breaks the following linear gauge symmetry
\beq \label{eq:gsh}
\de h_{\mu\nu} = - \pa_{\mu} \xi_{\nu} - \pa_{\nu} \xi_{\mu} \, .
\eeq 
The equations of motion are
\beq \label{eq:FPEOM}
\( \bo - m^2 \) h_{\mu\nu} - \et_{\mu\nu} \( \bo - (1+\al) m^2 \) h - \pa_{\mu} \pa^{\ro} h_{\ro\nu} - \pa_{\nu} \pa^{\ro} h_{\ro\mu} + \et_{\mu\nu} \pa^{\ro} \pa^{\si} h_{\ro\si} + \pa_{\mu} \pa_{\nu} h = -T_{\mu\nu} \, ,
\eeq
their divergence is
\beq \label{eq:divFPEOM}
m^2 \( \pa^{\mu} h_{\mu\nu} - (1+\al) \pa_{\nu}  h \) = 0 \, ,
\eeq
and their trace is
\beq
(D-2) \( \pa_{\mu} \pa_{\nu} h^{\mu\nu} - \bo h \) + \( (1+\al) D - 1 \) m^2 h = -T \, .
\eeq
Taking the double divergence, we can simplify the trace equation to
\beq \label{eq:trhimp}
(D-2) \al \bo h + \( (1+\al) D - 1 \) m^2 h = -T \, ,
\eeq
which is a dynamical equation for $h$ only when $\al \neq 0$. In the following sections we will see that in this case the kinetic term of $h$ has the wrong sign with respect to the rest of the fields, so that $h$ is a ghost. For the moment, we can already say that if $\al = - (D-1)/D$ then $h$ is massless, because the mass term solely depends on the traceless part $h_{\mu\nu} - \et_{\mu\nu} h /D$, while if $\al < -(D-1)/D$ then $h$ is also a tachyon. In particular, for $\al = -1/2$ it is a tachyon with mass $- m^2$. On the other hand, if $\al = 0$, then (\ref{eq:trhimp}) becomes an algebraic equation for $h$ and the latter gets totally determined by the source
\beq \label{eq:hep0}
h = -\frac{1}{d m^2} \, T \, ,
\eeq
so that it is no longer a degree of freedom, nor a dynamical field. Another peculiarity of this choice for $\al$ is that the double divergence of the equation of motion, i.e. the divergence of (\ref{eq:divFPEOM}), is gauge-invariant\footnote{It is actually the linearization of the Ricci scalar.}. This suggests that, although $m \neq 0$, there is some kind of leftover gauge symmetry in the equations of motion, in contrast with massive electrodynamics where both the equation and its divergence are not gauge-invariant. However, the equations of motion are not invariant under any gauge transformation (\ref{eq:gsh}), even a pure-scalar one $\xi_{\mu} = \pa_{\mu} \te$. We will understand this point better in the following sections, when we will have the appropriate technology at our disposal. For the moment we can simply note the interesting fact that for a pure-scalar transformation 
\beq
\de h_{\mu\nu} = - 2\pa_{\mu} \pa_{\nu} \te \, ,
\eeq
the action varies by
\beq
\de S \sim \te \( \pa_{\mu} \pa_{\nu} h^{\mu\nu} - (1+ \al)\bo h \) + \al \( \bo \te \)^2  \, ,
\eeq
so for $\al = 0$ this is proportional to the divergence of (\ref{eq:divFPEOM}) and therefore vanishes for on-shell $h_{\mu\nu}$ configurations\footnote{This corresponds to the well-known fact that, when $h_{\mu\nu}$ takes the form $h_{\mu\nu} = \pa_{\mu} \pa_{\nu} \ph$ for some function $\ph$, the Fierz-Pauli mass term is a total derivative. The generalization of this property to terms of cubic and higher order in $\pa_{\mu} \pa_{\nu} \ph$ gives
rise to the Galileon family of operators \cite{NRT}.}. Thus, if $h_{\mu\nu}$ is a solution, then
\beq \label{eq:appsymSh}
S[h_{\mu\nu}] = S[h_{\mu\nu} - 2 \pa_{\mu} \pa_{\nu} \te] \, .
\eeq
Clearly, something special happens at the Fierz-Pauli point $\al = 0$, although it is not a gauge symmetry. So let us start with this $\al = 0$ case. 

\subsubsection{Massive $\al = 0$}

Using (\ref{eq:hep0}) and (\ref{eq:divFPEOM}) the system of equations simplifies to
\bea
\( \bo - m^2 \) h_{\mu\nu} & = & -T_{\mu\nu} + \frac{1}{d} \[ \et_{\mu\nu} T - \frac{\pa_{\mu} \pa_{\nu}}{m^2} \,  T \] \, , \label{eq:FP1} \\
\pa_{\mu} h^{\mu\nu} - \pa^{\nu} h & = & 0 \, , \label{eq:FP2} \\
h & = & -\frac{1}{d m^2} \, T \, . \label{eq:FP3}
\eea 
The latter allows us to fix $h_{00}$
\beq
h_{00} = h_{ii} + \frac{1}{d m^2} \, T \, .
\eeq
The $i$ component of (\ref{eq:FP2}), along with the $0i$ component of (\ref{eq:FP1}), fix $h_{0i}$
\beq \label{eq:intch0i}
\( \De - m^2 \) h_{0i} = \pa_j \dot{h}_{ij} - T_{0i}  \, , \hspace{1cm} \dot{h}_{0i} = \pa_j h_{ij} + \frac{1}{d m^2} \, \pa_i T \, .
\eeq
Using the $0$ component of (\ref{eq:FP2}) and the trace of the spatial part of (\ref{eq:FP1}), we get
\beq \label{eq:intchii}
\( \De - m^2 \) h_{ii} = \pa_i \pa_j h_{ij} - T_{00}  \, , \hspace{1cm}  \dot{h}_{ii} = \pa_i h_{i0} \, .
\eeq
We finally split $h_{ij}$ into its trace $h_{ii}$ and traceless part $\ti{h}_{ij}$, and isolate $h_{00}, h_{0i}, h_{ii}$ in the above equations
\bea
\( \frac{d-1}{d}\,\De - m^2 \) h_{00} & = & \pa_i \pa_j \ti{h}_{ij} - T_{00} + \frac{1}{d}\, T - \frac{d-1}{d^2 m^2}\, \De T \, , \\
\( \De - m^2 \) h_{0i} & = & \pa_j \ti{h}_{ij} - T_{0i} +  \( (d-1)\, \De - d m^2 \)^{-1} \pa_i \( \pa_j \pa_k \ti{h}_{jk} - \pa_j T_{0j} \) \, , \\
\( \De - m^2 \) \dot{h}_{0i} & = &  \( \De - m^2 \) \( \pa_j \ti{h}_{ij} + \frac{1}{d m^2} \, \pa_i T \) + \frac{1}{d}\,\pa_i \( \pa_j \pa_k \ti{h}_{jk} - T_{00} \) \, , \\
\( \frac{d-1}{d}\,\De - m^2 \) h_{ii} & = & \pa_i \pa_j \ti{h}_{ij} - T_{00}  \, , \\
\( \De - m^2 \) \dot{h}_{ii} & = & \pa_i \pa_j \ti{h}_{ij} - \pa_i T_{0i} +  \( (d-1)\, \De - d m^2 \)^{-1} \De \( \pa_j \pa_k \ti{h}_{jk} - \pa_j T_{0j} \)  \, , \nn \\
\eea
so the corresponding initial conditions are determined by the ones of $\ti{h}_{ij}$ and $T_{\mu\nu}$ and these fields are non-dynamical. We are thus left with only $\ti{h}_{ij}$ being unconstrained, obeying a massive Klein-Gordon equation (the spatial traceless part of (\ref{eq:FP1}))
\beq
\( \bo - m^2 \) \ti{h}_{ij} = -\ti{T}_{ij} - \frac{1}{d m^2} \( \pa_i \pa_j - \frac{1}{d}\, \de_{ij} \De \) T  \, ,
\eeq
and transforming in the tensor representation of SO$(d)$. We thus have $N_{\rm f} = 2 N_{\rm d} = D^2 - D - 2$, which in $D = 4$ gives $N_{\rm d} = 5$.

\subsubsection{Massive $\al \neq 0$}

Let us now move on to the $\al \neq 0$ case. As we have seen already in the $\al = 0$ case, the equations for the $h_{\mu\nu}$ components can easily become lengthy in the process of spotting the non-dynamical fields and their precise form is not particularly illuminating. This is even worse here because of the undetermined $\al$ parameter. We therefore propose to simply sketch the procedure for generic $\al$ and then give the precise equations for the case $\al = -1/2$ which is considerably simpler. In the subsequent sections where the method of analysis will be more suited, we will treat the generic case explicitly to see that it is not qualitatively different from $\al = -1/2$. 

So let us sketch the procedure for the generic case. In the $\al = 0$ case, the trace equation eliminated $h_{00}$, so we were able to use the divergence equation to eliminate $h_{0i}$ and $h_{ii}$. Here, since the trace is dynamical, we have that either $h_{00}$, or $h_{ii}$ will remain dynamical. More precisely, using the $0$ component of (\ref{eq:divFPEOM}) and the appropriate combination of the $00$ component of (\ref{eq:FPEOM}) and (\ref{eq:trhimp}), we find non-dynamical equations for $h_{00}$ and $\dot{h}_{0i}$ which fix the initial conditions in terms of the ones of $h_{ii}$ and $h_{0i}$. We can then use the $0i$ component of (\ref{eq:divFPEOM}) along with the $0i$ component of (\ref{eq:FPEOM}) to do the same for $h_{0i}$. We are then left with the $ij$ component of (\ref{eq:FPEOM}) in an appropriate combination with the $00$ component and (\ref{eq:trhimp}), which yield dynamical equations for the unconstrained fields $h_{ij}$. This therefore corresponds to $N_{\rm f} = 2 N_{\rm d} =  D^2 - D$, or $N_{\rm d} = 6$ when $D = 4$, i.e. the trace $h_{ii}$ (or equivalently the Lorentz trace $h \equiv h_{ii} - h_{00}$) is now part of the dynamical spectrum. 

In particular, for $\al = -1/2$, equations (\ref{eq:FPEOM}), (\ref{eq:divFPEOM}) and (\ref{eq:trhimp}) can be brought to the simple form
\beq \label{eq:al12nFP}
\( \bo - m^2 \) \bar{h}_{\mu\nu} = -T_{\mu\nu}  \, , \hspace{1cm} \pa_{\mu} \bar{h}^{\mu\nu} = 0 \, , 
\eeq 
where
\beq \label{eq:hbardef}
\bar{h}_{\mu\nu} \equiv h_{\mu\nu} - \frac{1}{2}\, \et_{\mu\nu} h \, .
\eeq
Using the second equation along with the $0\mu$ component we get
\beq
\(\De - m^2 \) \bar{h}_{0i} = \pa_j \dot{\bar{h}}_{ij} - T_{0i} \, , \hspace{1cm} \dot{\bar{h}}_{0i} = \pa_j \bar{h}_{ij} \, ,
\eeq
and
\beq
\( \De - m^2 \) \bar{h}_{00} = \pa_i \pa_j \bar{h}_{ij} - T_{00} \, .
\eeq 
These fields are thus non-dynamical and their initial conditions are fixed in terms of the ones of $h_{ij}$ and $T_{\mu\nu}$. We are thus left with $h_{ij}$ obeying
\beq
\( \bo - m^2 \) \bar{h}_{ij} = -T_{ij} \, ,
\eeq
so $N_{\rm f} = 2 N_{\rm d} = D^2 - D$, and in particular $N_{\rm d} = 6$ for $D = 4$.

\subsubsection{Massless}

Let us now pass to the $m = 0$ case. First remember that (\ref{eq:divFPEOM}) is a possible choice of gauge only if $\al \neq 0$, since otherwise its divergence is gauge-invariant. We therefore have that the $m = 0$ case follows from the massive $\al \neq 0$ case by simply setting $m \to 0$, although now (\ref{eq:divFPEOM}) is obtained by a gauge transformation, as in the spin-1 theory. We work in the gauge corresponding to $\al = -1/2$ so that our equations are (\ref{eq:al12nFP}) with $m = 0$. Again, as in electrodynamics, there is a residual gauge symmetry given by the gauge parameters that satisfy $\bo \xi^{\mu} = 0$. The divergence and trace of the spatial part of (\ref{eq:al12nFP}) read
\beq
\bo \pa_j \bar{h}_{ij} = -\pa_j T_{ij} \, , \hspace{1cm} \bo \bar{h}_{ii} = -T_{ii} \, ,
\eeq
and their solutions take the form
\beq
\pa_j \bar{h}_{ij} = \ph_i^{\rm hom} - \bo_{\rm r}^{-1}\pa_j T_{ij} \, , \hspace{1cm} \bar{h}_{ii} = \ph^{\rm hom} - \bo_{\rm r}^{-1} T_{ii} \, , 
\eeq
where $\bo \ph_i^{\rm hom} = 0$ and $\bo \ph^{\rm hom} = 0$ are homogeneous solutions. We can then perform a residual gauge transformation 
\beq
\pa_j \bar{h}_{ij} - \De \xi_i - \pa_i \dot{\xi}_0 = \ph_i^{\rm hom} - \bo_{\rm r}^{-1}\pa_j T_{ij} \, , \hspace{1cm} \bar{h}_{ii} + (d-2) \pa_i \xi_i - d \dot{\xi}_0 = \ph^{\rm hom} - \bo_{\rm r}^{-1} T_{ii} \, , 
\eeq
and we see that we can kill the homogeneous solutions with the choice 
\bea
\dot{\xi}_0 & = & -\frac{1}{2(d-1)} \[  (d-2) \De^{-1} \pa_i \ph_i^{\rm hom} + \ph^{\rm hom} \] \, , \\
\xi_i & = & - \De^{-1} \[ \ph_i^{\rm hom} + \frac{1}{2(d-1)}\, \pa_i \( (d-2) \De^{-1} \pa_j \ph_j^{\rm hom} + \ph^{\rm hom} \) \] \, .
\eea
Therefore, $\pa_j \bar{h}_{ij}$ are $\bar{h}_{ii}$ are fully determined by the source and thus carry no degrees of freedom. The only unconstrained components are the spatial transverse-traceless part $\bar{h}^{\rm tt}_{ij}$, i.e. $\pa_j \bar{h}^{\rm tt}_{ij} = 0$ and $\bar{h}_{ii}^{\rm tt} = 0$, whose equation of motion is the spatial transverse-traceless part of (\ref{eq:al12nFP})
\beq
\bo \bar{h}_{ij}^{\rm tt} = -T_{ij}^{\rm tt} \, ,
\eeq
and correspond to $N_{\rm f} = d^2- d - 2$ degrees of freedom. As in the spin-1 case, the fact that $h_{ii}$ and $\pa_i h_{ij}$ are apparently dynamical is a gauge artefact. By starting all over again but rather considering the gauge 
\beq
\pa_i \bar{h}_{ij} = 0 \, ,
\eeq
we find indeed that they both obey non-dynamical equations and thus have that $N_{\rm f} = 2 N_{\rm d}$, with $N_{\rm d} = 2$ in the $D = 4$ case.

\section{Canonical formalism} \label{sec:canform}

The most rigorous way to perform the degree of freedom count and to study the stability of a theory is through the canonical formalism (see for instance \cite{Hinterbichler1,Hinterbichler2,BoulwareDeser,HassanRosen1,HassanRosen2} for the case of massive gravity). It is also the most suited way to unambiguously see that $N_{\rm f} = 2N_{\rm d}$ for gauge theories. Here we assume that the reader has the basic knowledge of constrained Hamiltonian systems, i.e. Dirac's algorithm, weak equality\footnote{Weak equality ``$\approx$'' holds for ``$=$ up to the addition of terms that are zero on the constrained hypersurface''.}, first/second class constraint terminology\footnote{A constraint is ``first class'' if its Poisson bracket with any other constraint and the Hamiltonian is weakly zero. A constraint that is not first class is called ``second class''.} etc. \footnote{See for instance \cite{HenneauxTeitelboim} for details on this formalism.}. 

\subsection{Spin 1}

\subsubsection{Massive}

Since $A_0$ has no kinetic term $\sim \dot{A}_0^2$ in (\ref{eq:ProcaAction})
\beq \label{eq:SAdp1}
S = \int \ed^D x \[ \frac{1}{2}\, \dot{A}^2_i - \frac{1}{4}\, F_{ij} F_{ij} - \frac{1}{2}\, m^2 A_i^2 + A_i j_i  - \dot{A}_i \pa_i A_0 + \frac{1}{2} \( \pa_i A_0 \)^2 + \frac{1}{2}\, m^2 A_0^2 - A_0 j_0 \] \, , 
\eeq
we Legendre transform only with respect to $\dot{A}_i$. The conjugate momenta (the electric field) read
\beq
E_i \equiv \frac{\pa L}{\pa \dot{A}_i} = \dot{A}_i - \pa_i A_0 \, , 
\eeq
so that the action in canonical form is
\beq
S = \int \ed^D x \[ E_i \dot{A}_i - {\cal H}[E_i, A_i, A_0] \] \, ,
\eeq
where
\beq \label{eq:HProca}
{\cal H}[E_i, A_i, A_0] = \frac{1}{2}\, E_i^2 + \frac{1}{4}\, F_{ij} F_{ij} + \frac{1}{2}\, m^2 A_i^2 - A_i j_i - A_0 \( \pa_i E_i - j_0 \) - \frac{1}{2}\, m^2 A_0^2 \, ,
\eeq
is the Hamiltonian density. Since $A_0$ is an auxiliary non-dynamical field, it can be integrated-out in order to provide a clearer picture of the dynamics, i.e. it can be replaced with the solution of its own equation of motion
\beq
{\cal H}[E_i, A_i] = \frac{1}{2}\, E_i^2 + \frac{1}{4}\, F_{ij} F_{ij} + \frac{1}{2}\, m^2 A_i^2 + \frac{1}{m^2} \(\pa_i E_i \)^2 - A_i j_i - \frac{2}{m^2}\, \pa_i E_i\, j_0 + \Ord(j^2)  \, .
\eeq
It is then clear that we have $N_{\rm f} = 2N_{\rm d} = 2d$ degrees of freedom forming two vectors $A_i$ and $E_i$ under SO$(d)$.

\subsubsection{Massless}

In the case $m = 0$, we have to go back to (\ref{eq:HProca}) and observe that $A_0$ becomes a Lagrange multiplier enforcing the Gauss constraint 
\beq
{\cal G} \equiv \pa_i E_i - j_0 = 0 \, .
\eeq
We now enter Dirac's constraint formalism so let us define the Poisson bracket
\beq
\left\{ \Ord, \Ord' \right\} \equiv  \int \ed^d x \[ \frac{\de \Ord}{\de A_i} \frac{\de \Ord'}{\de E_i} - \frac{\de \Ord'}{\de A_i} \frac{\de \Ord}{\de E_i} \] \, ,
\eeq 
and let us also smear the phase space functions of interest
\beq
A[f] \equiv \int \ed^d x\, f_i A_i \, , \hspace{1cm} E[g] \equiv \int \ed^d x\, g_i E_i \, , \hspace{1cm} G[A_0] \equiv \int \ed^d x \, A_0\, {\cal G} \, , \hspace{1cm} H \equiv \int \ed^d x\, {\cal H} \, ,
\eeq
so that time-evolution is given by\footnote{Note that the second term here is needed because $\Ord$ can depend on the source which has its own time-dependence. The $\pa_t$ operator will of course not act on the smearing fields $f_i$, $g_i$ and $A_0$.}
\beq \label{eq:Orddot}
\dot{\Ord} = - \left\{ H, \Ord \right\} + \pa_t \Ord \, .
\eeq
We then have that, for a conserved source, ${\cal G}$ is first class
\beq
\dot{G}[A_0] = 0 \, , \hspace{1cm} \left\{ G[A_0], G[A'_0] \right\} = 0
\eeq
so $A_0$ is not determined by the equations of motion and $G[A_0]$ generates abelian gauge transformations on phase space
\beq
\de A[f] = -\left\{ G[A_0], A[f] \right\} = -\int \ed^d x \, f_i \pa_i A_0 \, , \hspace{1cm} \de E[g] = -\left\{ G[A_0], E[g] \right\} = 0 \, ,
\eeq
which for $A_i$ and $E_i$ translate into
\beq
\de A_i = - \pa_i A_0 \, , \hspace{1cm} \de E_i = 0 \, .
\eeq
This implies that $\pa_i A_i$ is pure-gauge, the simplest example being the Coulomb gauge $\pa_i A_i = 0$. Along with ${\cal G} = 0$, we thus get that the longitudinal parts of $A_i$ and $E_i$ are non-dynamical, leaving only the transverse parts as the $N_{\rm f} = 2(d-1)$ degrees of freedom of the theory. Moreover, here we can clearly see why $N_{\rm f} = 2 N_{\rm d}$. Indeed, the fields with constrained initial conditions $A_0$ and $\pa_i A_i$ appear as a Lagrange multiplier $A_0$, which is thus totally arbitrary and in fact represents the gauge parameter, and a canonical couple $\pa_i A_i, \pa_i E_i$ subject to a constraint (spatial differential equation) and a gauge transformation on phase space. Thus both $A_0$ and $\pa_i A_i$ are non-dynamical and thus $N_{\rm f} = 2 N_{\rm d}$. Finally, note that in both the massive and massless cases, the quadratic part of the (gauge-fixed for $m=0$) Hamiltonian is positive definite, so these theories are stable.

\subsection{Spin 2}

\subsubsection{Massive $\al \neq 0$}

Since the $h_{0\mu}$ components have no kinetic term in (\ref{eq:FPaction}), we first remove all time-derivatives that act upon them by integrating by parts
\bea
S & = & \int \ed^D x \[ \frac{1}{2}\( \dot{h}_{ij}^2 - \dot{h}_{ii}^2\) - \frac{1}{2} \( \pa_i h_{jk} \)^2 + \( \pa_i h_{ij} \)^2 - \pa_i h_{ij} \pa_j h_{kk} + \frac{1}{2} \( \pa_i h_{jj} \)^2 \right. \nn \\
 & & \left. \hspace{1.2cm} - 2\, \dot{h}_{ij} \pa_i h_{j0} + 2\, \dot{h}_{ii} \pa_j h_{j0}  + \pa_i h_{ij} \pa_j h_{00} - \pa_i h_{jj} \pa_i h_{00} + 2\, \pa_{[i} h_{j]0} \pa_i h_{j0} \right. \nn \\
 & & \left. \hspace{1.2cm} - \frac{1}{2}\, m^2 \( h_{ij}^2 - (1+\al) h_{ii}^2 + 2(1+\al) h_{00} h_{ii} - 2 h_{0i}^2 - \al h_{00}^2 \) \right. \nn \\
 & & \left. \hspace{1.2cm} + h_{00} T_{00} - 2h_{0i} T_{0i} + h_{ij} T_{ij} \] \, , \label{eq:Shdp1}
\eea
and then Legendre transform only with respect to $\dot{h}_{ij}$. The conjugate momenta read
\beq \label{eq:pi0}
\pi_{ij} \equiv \frac{\pa L}{\pa \dot{h}_{ij}} = \dot{h}_{ij} - \de_{ij} \dot{h}_{kk} - 2\, \pa_{(i} h_{j)0} + 2\, \de_{ij} \pa_k h_{k0} \, , 
\eeq
and the inversion gives
\beq \label{eq:hofpi}
\dot{h}_{ij} = \pi_{ij} - \frac{1}{d-1} \de_{ij}  \pi_{kk} + 2\, \pa_{(i} h_{j)0}   \, ,
\eeq
so that the action in canonical form reads
\beq
S = \int \ed^D x \[ \pi_{ij} \dot{h}_{ij} - {\cal H}[h_{ij}, \pi_{ij}, h_{00}, h_{0i}] \] \, ,
\eeq
and the Hamiltonian density is
\bea 
{\cal H}[h_{ij}, \pi_{ij}, h_{00}, h_{0i}] & \equiv & \frac{1}{2} \( \pi_{ij}^2 - \frac{1}{d-1}\, \pi_{ii}^2 \) + \frac{1}{2} \( \pa_i h_{jk} \)^2 - \( \pa_i h_{ij} \)^2 + \pa_i h_{ij} \pa_j h_{kk} - \frac{1}{2} \( \pa_i h_{jj} \)^2  \nn \\
 & & + \frac{1}{2}\, m^2 \( h_{ij}^2 - (1+\al) h_{ii}^2 - 2 h_{0i}^2 - \al h_{00}^2 \) - h_{ij} T_{ij} \label{eq:linHam} \\
 & & + h_{00} \( \pa_i \pa_j h_{ij} - \De h_{ii} + (1+\al) m^2 h_{ii} - T_{00} \) + 2 h_{0i} \( \pa_j \pi_{ij} + T_{0i} \)   \, .\nn 
\eea
We see that $h_{0i}$ is an auxiliary field that appears quadratically whatever the value of $\al$, so we can integrate it out as we did for $A_0$ in the spin-1 case to get
\bea 
{\cal H}[h_{ij}, \pi_{ij}, h_{00}] & = & \frac{1}{2} \( \pi_{ij}^2 - \frac{1}{d-1}\, \pi_{ii}^2 \) + \frac{1}{m^2}  \( \pa_i \pi_{ij} \)^2   + \frac{1}{2} \( \pa_i h_{jk} \)^2 - \( \pa_i h_{ij} \)^2 + \pa_i h_{ij} \pa_j h_{kk}  \nn \\
 & & - \frac{1}{2} \( \pa_i h_{jj} \)^2 + \frac{1}{2}\, m^2 \( h_{ij}^2 - (1+\al) h_{ii}^2 - \al h_{00}^2 \) \label{eq:h0iintout}  \\
 & & + h_{00} \( \pa_i \pa_j h_{ij} - \De h_{ii} + (1+\al) m^2 h_{ii} - T_{00} \)  - h_{ij} T_{ij} + \frac{2}{m^2}\, \pa_i \pi_{ij} T_{0j} + \Ord(T^2) \, .  \nn
\eea
Now, for $\al \neq 0$ we have that $h_{00}$ is also a quadratic auxiliary field, so we can integrate it out as well
\bea
{\cal H}[h_{ij}, \pi_{ij}] & = & \frac{1}{2} \( \pi_{ij}^2 - \frac{1}{d-1}\, \pi_{ii}^2 \) + \frac{1}{m^2} \( \pa_i \pi_{ij} \)^2   + \frac{1}{2} \( \pa_i h_{jk} \)^2 - \( \pa_i h_{ij} \)^2 + \pa_i h_{ij} \pa_j h_{kk} - \frac{1}{2} \( \pa_i h_{jj} \)^2 \nn \\
 & & + \frac{1}{2 m^2 \al} \[ \pa_i \pa_j h_{ij} - \De h_{ii} + (1+\al) m^2 h_{ii}  \]^2 +\frac{1}{2}\, m^2 \( h_{ij}^2 - (1+\al) h_{ii}^2  \)     \\
 & & \nn \\
 & & - h_{ij} T_{ij} - \frac{1}{m^2 \al} \( \pa_i \pa_j h_{ij} - \De h_{ii} + (1+\al) m^2 h_{ii} \) T_{00} + \frac{2}{m^2}\, \pa_i \pi_{ij} T_{0j}  + \Ord(T^2) \, . \nn
\eea
To see the instability in this setting we can harmonically decompose $\pi_{ij}$
\beq \label{eq:harmdecpi}
\pi_{ij} \equiv \frac{1}{d}\, \de_{ij} \pi + \( \pa_i \pa_j - \frac{1}{d}\, \de_{ij} \De \) l + \pa_{(i} v_{j)} + t_{ij} \, , \hspace{1cm} \pa_i v_i = t_{ii} = 0 \, , \hspace{0.5cm} \pa_i t_{ij} = 0  \, , 
\eeq
and trade $l$ for the more convenient variable
\beq
\Pi \equiv \pi + (d-1) \, \De l  \, ,
\eeq
to get that the part of $\cal H$ which is quadratic in $\pi_{ij}$ in the scalar sector reads
\beq
{\cal H}_{\Ord \(\pi_{\rm scal.}^2\)} = \frac{1}{2d(d-1)}\, \Pi^2 + \frac{1}{d^2 m^2} \( \pa_i \Pi \)^2 - \frac{1}{d(d-1)}\, \pi \Pi \, .
\eeq
We see that $\cal H$ is thus not positive-definite, or that by completing the square there is a negative-definite term. Since this occurs at the level of the conjugate momenta, we have that the corresponding mode is a ghost. The degrees of freedom are the $h_{ij}$ and $\pi_{ij}$ components, that is, a total of $N_{\rm f} = 2 N_{\rm d} = d^2 + d = D^2 - D$.

\subsubsection{Massive $\al = 0$}

So let us go back to (\ref{eq:h0iintout}) and move on to the $\al = 0$ case where $h_{00}$ becomes a Lagrange multiplier enforcing the constraint
\beq
{\cal C}_t \equiv \pa_i \pa_j h_{ij} - \( \De - m^2 \) h_{ii} - T_{00} = 0 \, .
\eeq
Defining the Poisson bracket
\beq
\left\{ \Ord, \Ord' \right\} \equiv  \int \ed^d x \[ \frac{\de \Ord}{\de h_{ij}} \frac{\de \Ord'}{\de \pi_{ij}} - \frac{\de \Ord'}{\de h_{ij}} \frac{\de \Ord}{\de \pi_{ij}} \] \, ,
\eeq 
and the smeared observables
\beq
h[f] \equiv \int \ed^d x \, f_{ij} h_{ij} \, , \hspace{0.5cm} \pi[g] \equiv \int \ed^d x\, g_{ij} \pi_{ij} \, , \hspace{0.5cm} C_t[h_{00}] \equiv \int \ed^d x \, h_{00}\, {\cal C}_t \, , \hspace{0.5cm} H \equiv \int \ed^d x\, {\cal H}
\eeq
we get that ${\cal C}_t$ is second class (using (\ref{eq:Orddot}) and for a conserved source)
\beq \label{eq:Cprime}
\dot{C}_t[h_{00}] = -\int \ed^d x \, h_{00} \, {\cal C}' \equiv C'[h_{00}] \, , \hspace{1cm} {\cal C}' \equiv \pa_i \pa_j \pi_{ij} + \frac{1}{d-1}\, m^2 \pi_{ii} + \pa_i T_{0i} \, ,
\eeq
so it is a priori not conserved under time evolution. To repair this, we must therefore consider ${\cal C}'$ as an additional (secondary) constraint and append a term $q\, {\cal C}'$ to the total Hamiltonian density $\cal H$
\beq
{\cal H} \to {\cal H} + q\, {\cal C}' \, ,
\eeq
with $q$ a Lagrange multiplier. Now, demanding that ${\cal C}'$ be conserved fixes $h_{00}$
\beq
\dot{C}'[q] \sim {\cal C}_t + d m^2 \( h_{00} - h_{ii} \) - T   \approx 0  \, ,
\eeq
so we can choose
\beq
h_{00} = h_{ii} + \frac{1}{d m^2} \, T  \, ,
\eeq
which is nothing but (\ref{eq:hep0}), and the Hamiltonian density now reads
\bea 
{\cal H}[h_{ij}, \pi_{ij}, q] & = & \frac{1}{2} \( \pi_{ij}^2 - \frac{1}{d-1}\, \pi_{ii}^2 \) + \frac{1}{m^2}  \( \pa_i \pi_{ij} \)^2 \label{eq:Hh00la} \\
 & &  + \frac{1}{2} \( \pa_i h_{jk} \)^2 - \( \pa_i h_{ij} \)^2 + \frac{1}{2} \( \pa_i h_{jj} \)^2   + \frac{1}{2}\, m^2 \( h_{ij}^2 + h_{ii}^2 \)  \nn \\
 & & + q \, {\cal C}' - h_{ij} T_{ij} - h_{ii} T_{00} + \frac{1}{d m^2} \, {\cal C}_t T + \frac{2}{m^2}\, \pa_i \pi_{ij} T_{0j} + \Ord(T^2) \, . \nn
\eea
We must finally demand that ${\cal C}_t$ be conserved under time-evolution with this new Hamiltonian
\beq
\dot{C}_t \sim {\cal C}' - \frac{d m^4}{d-1}\, q \approx 0 \, ,
\eeq
which ends up fixing $q = 0$. The resulting Hamiltonian preserves the constraints ${\cal C}_t, {\cal C}'$ under time-evolution as long as they are imposed on the initial conditions. These constraints kill precisely the degree of freedom which causes the instability in the $\al \neq 0$ case and make the Hamiltonian positive-definite. Indeed, by performing a harmonic decomposition of $\pi_{ij}$ (\ref{eq:harmdecpi}) and also $h_{ij}$
\beq
h_{ij} \equiv  \frac{1}{d}\, \de_{ij} \ph + \( \pa_i \pa_j - \frac{1}{d}\, \de_{ij} \De \) \la + \pa_{(i} \be_{j)} + \ta_{ij} \, , \hspace{1cm} \pa_i \be_i = \ta_{ii} = 0 \, , \hspace{0.5cm} \pa_i \ta_{ij} = 0 \, , 
\eeq 
we can actually solve the constraints explicitly and get that they relate the traces to the longitudinal parts
\bea
\(\frac{d-1}{d}\, \De - m^2 \) \ph & = & \frac{d-1}{d}\, \De^2 \la - T_{00} \, , \\
\(\frac{d-1}{d}\, \De + m^2 \) \pi & = & -(d-1) \(\frac{d-1}{d}\, \De^2 l + \pa_i T_{i0} \) \, .
\eea
Actually, as in the $\al \neq 0$ case, one can use more convenient combinations instead of the longitudinal modes
\beq \label{eq:PhiPi}
\Phi \equiv \De \la - \ph \, , \hspace{1cm} \Pi \equiv \pi + (d-1)\, \De l  \, ,
\eeq
so that, using the constraints, we can express $\ph, \la, \pi, l$ in terms of $\Phi$ and $\Pi$. We get
\beq
\ph = -\frac{1}{m^2} \[ \frac{d-1}{d}\, \De \Phi - T_{00} \] \, , \hspace{1cm} \De \la = - \frac{1}{m^2} \[ \frac{d-1}{d}\, \De \Phi - m^2 \Phi - T_{00} \] \, .
\eeq
and
\beq
\pi = -\frac{d-1}{m^2} \[ \frac{1}{d} \, \De \Pi + \pa_i T_{i0} \] \, , \hspace{1cm}  \De l = \frac{1}{m^2} \[ \frac{1}{d}\, \De \Pi + \frac{1}{d-1}\, m^2 \Pi + \pa_i T_{i0} \] \, .
\eeq
and the action reads
\beq \label{eq:SHFP}
S = \int \ed^D x \[ \frac{1}{d}\,\Pi \dot{\Phi} + \frac{1}{2}\, \pa_i v_j \pa_i \dot{\be}_j + t_{ij} \dot{\ta}_{ij} - {\cal H} \] \, ,
\eeq
where
\beq
{\cal H} = {\cal H}_{\rm scalar} + {\cal H}_{\rm vector} + {\cal H}_{\rm tensor} \, ,
\eeq
with\footnote{The harmonic variables of the source $\ro, p, q, \si, q_i, \si_i, \si_{ij}$ are defined in (\ref{eq:Tharmdec}) and the conservation equation in terms of them reads (\ref{eq:consharmh}).}
\bea \label{eq:HFP}
{\cal H}_{\rm scalar} & = & \frac{1}{2d(d-1)}\, \Pi^2 + \frac{d-1}{2d}\, \Phi \( m^2 - \De \) \Phi - \frac{1}{d m^2}\, \Pi \De q + \frac{d-1}{d} \, \Phi \( \ro - \De \si \) \nn \\
 & & + \frac{d-1}{d m^2}\, \Phi \De \( p + \frac{d-1}{d}\, \De \si \) + \Ord(T^2) \nn \\
{\cal H}_{\rm vector} & = & \frac{1}{4 m^2}\, \pa_i v_j \( m^2 - \De \) \pa_i v_j + \frac{1}{2}\, m^2 \( \pa_i \be_j \)^2 + \frac{1}{m^2}\, v_i \De q_i + \frac{1}{2}\, \be_i \De \si_i + \Ord(T^2)  \nn \\
{\cal H}_{\rm tensor} & = &  \frac{1}{2}\, t_{ij}^2 + \frac{1}{2}\, \ta_{ij} \( m^2 - \De \) \ta_{ij} - \ta_{ij} \si_{ij}  + \Ord(T^2) \, .
\eea
The quadratic part is positive definite and the theory is thus stable, with $N_{\rm f} = 2 N_{\rm d} = d^2+d-2 = D^2 - D - 2$ degrees of freedom. These correspond to $\Phi, \be_i, \ta_{ij}$ and their conjugate momenta.

\subsubsection{Massless}

We can finally proceed to the massless case where $h_{0i}$ becomes a Lagrange multiplier as well. We must therefore go back to (\ref{eq:linHam}) with $m = 0$ and define the constraint imposed by $h_{0i}$ as
\beq
{\cal C}_i \equiv \pa_j \pi_{ij} - T_{0i} \, , \hspace{1cm} C_s[h_{0i}] \equiv \int \ed^d x\, h_{0i} \, {\cal C}_i \, .
\eeq
Observe that the secondary constraint ${\cal C}'$ of Fierz-Pauli theory (\ref{eq:Cprime}) actually reduces to $\pa_i {\cal C}_i$ in the $m \to 0$ limit. The only non-trivial Poisson bracket arises in
\beq
\dot{C}_t[h_{00}] = - C_s[\pa_i h_{00}] \approx 0 \, , 
\eeq
so the system is now first class and the $h_{0\mu}$ are not determined by the equations of motion. Rather, they serve as the gauge parameters of the gauge transformations generated by $C_t$ and $C_s$ on phase space
\bea
\de h[f] & = & -\left\{ C_t[h_{00}],  h[f] \right\} - \left\{ C_s[h_{0i}],  h[f] \right\} = - \int \ed^d x\, f_{ij} \, \pa_i h_{j0} \, , \\
\de \pi[g] & = & -\left\{ C_t[h_{00}],  \pi[g] \right\} - \left\{ C_s[h_{0i}],  \pi[g] \right\} = \int \ed^d x\, \pi_{ij} \( \pa_i\pa_j - \de_{ij} \De \) h_{00} \, ,
\eea
which for $h_{ij}$ and $\pi_{ij}$ imply
\bea \label{eq:gshcan}
\de h_{ij} = -\pa_{(i} h_{j)0} \, , \hspace{1cm}  \de \pi_{ij} = \( \pa_i\pa_j - \de_{ij} \De \) h_{00} \, .
\eea
These can be used to fix the gauge to
\beq
\pa_i h_{ij} = 0 \, , \hspace{1cm} \pi_{ii} = 0 \, ,
\eeq
which, along with ${\cal C}_t = 0$ and ${\cal C}_i = 0$, imply that $h_{ij}$ and $\pi_{ij}$ are both transverse-traceless. The degrees of freedom of the theory are thus $N_{\rm f} = 2N_{\rm d} = d^2 - d - 2$. The Hamiltonian density in this gauge is positive-definite
\beq
{\cal H} = \frac{1}{2} \, \pi_{ij}^2 + \frac{1}{2} \( \pa_i h_{jk} \)^2 + \Ord(T) \, ,
\eeq
so the theory is stable. 

We can now note that the combinations $\Phi$ and $\Pi$ defined in (\ref{eq:PhiPi}), that were used in the treatment of the massive theory, are actually invariant under (\ref{eq:gshcan})\footnote{This will become clear in the next section where we will deduce the transformations of the harmonic variables under the gauge symmetry.}. We thus have that the scalar sector of the FP theory $\al = 0$, once the second class constraints are solved, becomes invariant under the transformations generated by the {\it massless constraints} $C_t, C_s$. It is important to note however that not all of these constraints appear in the massive theory and, for those who do, they are second class for $m \neq 0$. This means they do not correspond to gauge symmetries, since there is no totally undetermined field playing the role of the gauge parameter. Thus, by discussing the transformations of the scalars in the massive theory we are actually comparing objects in two different theories. 

Nevertheless, the fact that these modes are gauge-invariant is again a property of the scalar sector of FP theory alone, since this is not the case of the vector sector, where $v_i$ is gauge-invariant but $\be_i$ is not, and it is also not the case in massive electrodynamics. Moreover, it is also not the case for the ghost scalar when $\al \neq 0$, so this has all the characteristics of the issue that was discussed in the previous section: on the FP point $\al = 0$ there is something that looks like a gauge symmetry but that is actually not. 

Another interesting feature we can already see here is the vDVZ discontinuity of the $\al = 0$ theory. Indeed, sending $m \to 0$ in (\ref{eq:HFP}) effectively neutralizes the vector modes but the scalar mode remains, that is, one more dynamical field that in the $m = 0$ case.

It seems that the use of harmonic variables has helped our understanding of this apparent symmetry issue and has generally made the dynamics of the theory more transparent. Unfortunately, in the canonical formalism the action is a bit too crowded because of the presence of the conjugate momenta, so this is still not the optimal way to understand the theory. We therefore now propose to use harmonic variables, but in the Lagrangian formulation.

\section{Harmonic formalism} \label{sec:harmdec}

In this section we consider the $d$-harmonic decomposition, but at the level of the Lagrangian formulation. This will allow us to explore the above mentioned ``residual gauge symmetry'' of Fierz-Pauli theory, but it will also make the dynamical structure of the theory more transparent. Moreover, this formalism is also easily applicable in the case of a de-Sitter background. It will thus allow us to understand in a different language a number of results in the literature on the degrees of freedom of massive gravity over de-Sitter. This section is based on original work from our group \cite{JaccardMaggioreMitsou1}.

To briefly introduce the harmonic decomposition, let us start by noting that at each space-time point $x$, the field components form irreducible representations of SO$(d)$, i.e. $A_0(x)$ is a SO$(d)$-scalar, $A_i(x)$ a SO$(d)$-vector, the traceless part of $h_{ij}(x)$ is a SO$(d)$-tensor and so on. If we now consider the full group of isometries of $d$-dimensional space, i.e. the Euclidean group ISO$(d)$ of rotations and translations, then a field is no longer seen as an infinite collection of independent SO$(d)$ representations, but as a finite collection of irreducible representations of ISO$(d)$. For instance, we have that inside of $A_i$ there hides a scalar under SO$(d)$, that is, $\pa_i A_i$. We can therefore split $A_i$ into its scalar part $\pa_i A_i$ and its transverse vector part $A^{\rm t}_i$, obeying $\pa_i A_i^{\rm t} = 0$, which obviously do not mix under translations, nor under rotations since the latter commute with $\pa_i$. For tensors one can analogously decompose the traceless part of $h_{ij}$ into a scalar, a transverse vector and a transverse-traceless tensor. 

We will therefore refer to ``$d$-scalars'', ``$d$-vectors'' and ``$d$-tensors'' for these irreducible representations of ISO$(d)$, while the irreducible representations of SO$(d)$ will be referred to as ``SO$(d)$-vectors'' and ``SO$(d)$-tensors''. Note that $d$-vectors and $d$-tensors are thus automatically transverse. The basic advantage of the harmonic decomposition in our analysis lies in the following fact: the massless dynamical fields form the highest possible irreducible representation of ISO$(d)$, while the massive ones form the highest possible representation of SO$(d)$. This formalism is thus ideal for observing the activation of modes by mass.

\subsection{Spin 1}

We start by splitting $A_i$ and $j_i$ into longitudinal and transverse parts
\bea
A_0 \equiv \psi \, , \hspace{1cm} A_i \equiv \pa_i \la + \be_i \, , \hspace{1cm}  \pa_i \be_i = 0 \, , \\
j_0 \equiv -\ro \, , \hspace{1cm}  j_i \equiv \pa_i \si + \si_i \, , \hspace{1cm}  \pa_i \si_i = 0 \, , \label{eq:jharmdec}
\eea
with the inverse map being
\beq
\la = \De^{-1} \pa_i A_i \, , \hspace{1cm} \be_i = P_{ij} A_j \, ,
\eeq
and so on for $j_i$, where $P_{ij}$ is the projector on the subspace of $d$-vector fields (transverse SO$(d)$-vectors)
\beq
P_{ij} \equiv \de_{ij} - \pa_i \De^{-1} \pa_j \, , \hspace{1cm} P_i^{\,\,k} P_k^{\,\,\,j} = P_i^{\,\,j} \, , \hspace{1cm} \pa_i P_{ij} = 0 \,  .
\eeq
Note that the harmonic variables are therefore spatially non-local combinations of the original fields. In terms of the harmonic variables the gauge transformation (\ref{eq:gsA}) reads
\beq
\de \psi = - \dot{\te} \, , \hspace{1cm} \de \la = - \te \, , \hspace{1cm} \de \be_i = 0 \, ,
\eeq
so that $\be_i$ is gauge-invariant, while $\la$ and $\psi$ can combine to form the gauge-invariant combination
\beq
\Psi \equiv \psi - \dot{\la} \, .
\eeq
On the other hand, current conservation $\pa_{\mu} j^{\mu} = 0$ translates into
\beq \label{eq:consharmA}
\dot{\ro} = -\De \si \, ,
\eeq
and this equation will be implicitly used every time some $\sim \dot{\ro}$ term appears. We then get that the action (\ref{eq:ProcaAction}) can be written as
\beq
S = \int \ed^D x \[ -\frac{1}{2}\, \pa_{\mu} \be_i \pa^{\mu} \be_i - \frac{1}{2}\, m^2 \be_i \be_i + \frac{1}{2} \, \pa_i \Psi \pa_i \Psi + \frac{1}{2}\, m^2 \( \psi^2 - \pa_i \la \pa_i \la \) + \ro \Psi + \be_i \si_i \] \, ,
\eeq
where we consider $\la, \be_i, \psi$ as the independent fields, while $\Psi$ is just a shorthand notation for the combination $\psi - \dot{\la}$. Here one might be tempted to use $\Psi$ as an independent field instead of $\psi$ or $\la$, but should refrain from doing so. This is because $\Psi$ depends on time-derivatives of the original fields $A_{\mu}$. This in turn implies that the initial conditions of $\Psi$ are not determined, since they require the knowledge of the initial value of $\ddot{\la}$. So keep in mind that one can only consider field redefinitions that preserve the initial data for the Cauchy problem to remain well-posed.

\subsubsection{Massless}

This subtlety being mentioned, the first thing to notice in the above action is that for $m = 0$ it is explicitly gauge-invariant since it depends only on the gauge-invariant quantities $\Psi$ and $\be_i$. The latter obeys a massless Klein-Gordon equation
\beq
\bo \be_i = -\si_i \, ,
\eeq
and thus constitutes the $2(d-1)$ degrees of freedom of the theory. The equation of motion of $\psi$ is the Poisson equation
\beq
\De \Psi = \ro \, ,
\eeq
while the equation of motion of $\la$ is the time-derivative of it. Pay attention to the way in which gauge-invariance neutralizes the longitudinal mode $\la$ in this setting. The latter {\it does} have a kinetic term $\sim \dot{\la}^2$ in the action, which would naively make it dynamical, but the fact that it enters only through the combination $\Psi$ and that the latter ultimately obeys a purely spatial equation makes $\la$ non-dynamical. Therefore, in the massless case, it turns out that we {\it can} effectively consider $\Psi$ as an independent variable and vary the action with respect to it because the initial conditions of $\la$ are pure-gauge so the initial data of $\Psi$ {\it are} defined. This will no longer be true in the massive theory. 

Until now, the spatial differential equations we obtained always concerned gauge-dependent fields, so that we did not need to worry about questions of instantaneous response to a source. Here, we are witnessing an equation that involves only spatial derivatives for $\Psi$, which is a gauge-invariant variable. As anticipated in section \ref{sec:dofprop}, we see however that $\Psi$ is a spatially non-local functional of the original fields, so that it cannot be measured instantaneously to begin with.  

\subsubsection{Massive}

Turning on the mass $m \neq 0$, we first see that the gauge-invariant variables are not sufficient to describe the mass term since the latter breaks the gauge symmetry. This means that the equation of $\la$ will not be implied by the one of $\psi$ any more and therefore that its time-derivatives will now make it a dynamical field. The equation of motion of $\be_i$ is now a massive Klein-Gordon equation
\beq
\( \bo - m^2 \) \be_i = -\si_i \, ,
\eeq
while the ones of $\psi$ and $\la$ read
\beq \label{eq:EOMApsila}
\( \De - m^2 \) \psi = \ro + \De \dot{\la} \, , \hspace{1cm} \( \pa_t^2 + m^2 \) \la = \si + \dot{\psi} \, .
\eeq
Then, isolating $\dot{\psi}$ in the latter and plugging the result in the time-derivative of the equation of $\psi$, we get
\beq
\( \bo - m^2 \) \la = - \si \, ,
\eeq
so $\la$ corresponds to the additional $2$ degrees of freedom one gets when $m \neq 0$. On the other hand, solving for $\ddot{\la}$ in its own equation of motion and plugging the result in the time-derivative of the equation of $\psi$, we get that $\psi$ is non-dynamical and that its initial conditions are totally determined by the ones of the other fields
\beq
\( \De - m^2 \) \psi = \ro + \De \dot{\la} \, , \hspace{1cm} \dot{\psi} = \De \la  \, .
\eeq
One could therefore integrate-out $\psi$, and redefine the longitudinal modes by a spatially non-local operation\footnote{Note that the square-root is real because $\De \( \De - m^2 \)^{-1}$ is positive-definite as it can be seen by using its Fourier representation.}
\beq
\ti{\la} \equiv \sqrt{ \De \( \De - m^2 \)^{-1}}\, \la \, , \hspace{1cm} \ti{\si} \equiv \sqrt{ \De \( \De - m^2 \)^{-1}}\, \si \, ,
\eeq
to find a spatially local action for the dynamical fields only
\beq
S = \int \ed^D x \[ -\frac{1}{2}\, \pa_{\mu} \be_i \pa^{\mu} \be_i - \frac{1}{2}\, m^2 \be_i \be_i + \be_i \si_i  + m^2 \( - \frac{1}{2}\, \pa_{\mu} \ti{\la}\pa^{\mu} \ti{\la} - \frac{1}{2}\, m^2 \ti{\la}^2 + \ti{\la}\, \ti{\si} \) + \Ord(j^2)\] \, .
\eeq
Now only the dynamical fields appear in the action. This was already the case in the canonical formalism after having integrated-out $A_0$, but the advantage here is that the action is analytic in $m^2$ so that one can study the $m \to 0$ limit unambiguously. Note that the new dynamical mode one gets in the massive case (here $\la$ or $\ti{\la}$) is not gauge-invariant, as one would expect in a massive theory, and disappears in the $m \to 0$ limit.

Finally, observe that, after having eliminated the non-dynamical field, the dynamical ones come with the Klein-Gordon kinetic terms, even though they are not representations of the Lorentz group. This is because Poincar\'e invariance implies the standard relativistic dispersion relation $E^2 = m^2 + \vec{p}^2$ for the dynamical fields.

\subsection{Spin 2}

We start by defining the harmonic variables
\bea
h_{00} & \equiv & \psi \, , \nn \\
h_{0i} & \equiv & \pa_i \ch + \ch_i  \, ,  \\
h_{ij} & \equiv & \frac{1}{d}\,\de_{ij} \, \ph + \( \pa_i \pa_j - \frac{1}{d}\, \de_{ij} \De \) \la + \pa_{(i} \be_{j)} + \ta_{ij} \, , \nn \\
T_{00} & \equiv & \ro \, , \nn \\
T_{0i} & \equiv & -\pa_i q - q_i \, , \label{eq:Tharmdec} \\
T_{ij} & \equiv & \de_{ij} \, p + \( \pa_i \pa_j - \frac{1}{d}\, \de_{ij} \De \) \si + \pa_{(i} \si_{j)} + \si_{ij}  \, , \nn
\eea
where
\beq
\pa_i \ch_i = \pa_i \be_i = \ta_{ii} = \pa_i q_i = \pa_i \si_i = \si_{ii} = 0 \, , \hspace{1cm} \pa_i \ta_{ij} = \pa_i \si_{ij} = 0 \, ,
\eeq
while the inverse relation is
\bea
\ch & = & \De^{-1} \pa_i h_{0i} \, , \\
\ch_i & = & P_{ij} h_{0j} \, , \\ 
\ph & = & h_{ii} \, , \\
\la & = & -\frac{1}{d-1}\,\De^{-1} \[ h_{ii} - d \De^{-1} \pa_i \pa_j h_{ij} \] \, , \\
\be_i & = & 2 \De^{-1} P_{ij} \pa_k h_{jk}  \, , \\
\ta_{ij} & = & P_{ijkl} h_{kl} \, ,
\eea
and so on for the components of $T_{\mu\nu}$, where $P_{ijkl}$ is the projector on the subspace of $d$-tensors (transverse-traceless SO$(d)$-tensors)
\beq
P_{ij}^{\,\,\,\,kl} \equiv P_{(i}^{\,\,\,k} P_{j)}^{\,\,\,l} - \frac{1}{d-1}\, P_{ij} P^{kl} \, , \hspace{0.5cm} P_{ij}^{\,\,\,\,nm} P_{nm}^{\,\,\,\,\,\,\,kl} = P_{ij}^{\,\,\,\,kl} \, , \hspace{0.5cm} \pa_i P_{ijkl} = 0 \, , \hspace{0.5cm} P_{iikl} = 0 \, .
\eeq
Decomposing the gauge parameter as well
\bea \label{eq:xiAB}
\xi_0 = A \, , \hspace{1cm} \xi_i = \pa_i B + B_i \, , \hspace{1cm} \pa_i B_i = 0 \, ,
\eea
we get that the gauge transformation (\ref{eq:gsh}) reads
\bea
\de \psi & = & - 2\dot{A} \, , \\
\de \ch & = & - A - \dot{B} \, , \\
\de \ch_i & = & - \dot{B}_i \, , \\
\de \ph & = & - 2 \De B \, , \\
\de \la & = & - 2 B \, , \\
\de \be_i & = & - 2 B_i \, , \\
\de \ta_{ij} & = & 0 \, ,
\eea
so one can form the following independent gauge-invariant combinations
\beq
\Psi \equiv \psi - 2 \dot{\ch} + \ddot{\la} \, , \hspace{1cm} \Phi \equiv \De \la - \ph  \, , \hspace{1cm} \Xi_i \equiv  \ch_i - \frac{1}{2}\, \dot{\be}_i  \, ,
\eeq
known as ``Bardeen potentials'' \cite{Bardeen}, of which $\Phi$ is already known from the previous section. Finally, the conservation equation $\pa_{\mu}T^{\mu\nu} = 0$ gives
\beq \label{eq:consharmh}
\dot{\ro} = -\De q \, , \hspace{1cm} \dot{q} = -p - \frac{d-1}{d}\, \De \si \, , \hspace{1cm} \dot{q}_i = -\frac{1}{2}\, \De \si_i \, ,
\eeq
and, as in the spin-1 case, these will be implicitly used whenever we have a time-derivative acting on a source component in the subsequent computations. The action (\ref{eq:Shdp1}) in terms of these variables reads
\bea
S & = & \int \ed^D x \[ \frac{d-1}{d} \( - \frac{1}{2}\, \dot{\Phi}^2 + \frac{d-2}{2 d} \( \pa_i \Phi \)^2 + \pa_i \Phi \pa_i \Psi \) + \( \pa_i \Xi_j \)^2 - \frac{1}{2}\, \pa_{\mu} \ta_{ij} \pa^{\mu} \ta_{ij} \right. \nn \\
 & & \left. \hspace{1.2cm} - \frac{1}{2}\, m^2 \( \frac{d-1}{d} \( -\Phi^2 + 2 \Phi \De \la \) + 2\( \De \la - \Phi \) \psi - \al \(\Phi + \psi - \De \la\)^2\right. \right. \nn \\
 & & \left. \left. \hspace{1.2cm} - 2 \( \pa_i \ch \)^2 - 2 \ch_i^2 + \frac{1}{2} \( \pa_i \be_j\)^2 + \ta_{ij}^2  \)  + \Psi \ro - \Phi p + 2 \Xi_i q_i + \ta_{ij} \si_{ij}  \] \, . \label{eq:Sharmh}
\eea
As in the case of electrodynamics, note that for $m = 0$ only gauge-invariant quantities appear, thus making the symmetry manifest. Again, we cannot consider $\Psi$ and $\Xi_i$ as independent variables with respect to which we could vary the action because they contain time-derivatives of the original fields and their initial conditions are thus not defined. This is however not the case of $\Phi$, so we choose to consider $\psi, \ch, \Phi, \la, \ch_i, \be_i, \ta_{ij}$ as the independent fields, while $\Psi$ and $\Xi_i$ are mere shorthand notations.

\subsubsection{Massless}

So let us start with the massless case $m = 0$ and compute the equations of motion. In the $d$-scalar sector, the ones of $\psi$ and $\Phi$ read
\beq \label{eq:PhiPsim0EOM}
\De \Phi = \frac{d}{d-1}\,\ro \, , \hspace{1cm} \ddot{\Phi} - \frac{d-2}{d} \, \De \Phi - \De \Psi = \frac{d}{d-1}\, p \, , 
\eeq
respectively, while the ones of $\ch$ and $\la$ are the first and second time-derivative of the former. To simplify the second equation we note that by taking the double time-derivative of the first one and using (\ref{eq:consharmh}) we get
\beq
\De \ddot{\Phi} = \frac{d}{d-1}\, \ddot{\ro} = -\frac{d}{d-1}\, \De \dot{q} = \frac{d}{d-1}\, \De \( p + \frac{d-1}{d}\, \De \si \) \, ,
\eeq
so that the equation of $\Phi$ actually gives
\beq \label{eq:Psim0EOM}
\De \Psi = -\frac{d-2}{d-1}\, \ro + \De \si \, .
\eeq
In the vector sector we have the equation of motion of $\ch_i$
\beq
\De \Xi_i = q_i \, ,
\eeq
and the one of $\be_i$ which is its time-derivative, while finally for the tensor modes
\beq
\bo \ta_{ij} = -\si_{ij} \, .
\eeq
Therefore, the Bardeen variables $\Phi, \Psi, \Xi_i$ are physical but non-dynamical fields, thus leaving the $d^2 - d - 2$ components of $\ta_{ij}$ and $\dot{\ta}_{ij}$ as the only degrees of freedom/dynamical fields of the theory. 

\subsubsection{Massive}

Let us now turn on the mass $m \neq 0$ in which case the equations of motion of $\ch, \la, \be_i$ are no longer implied by the ones of $\Phi, \psi$ and $\ch_i$. Since there are a lot of variables now, it is not particularly illuminating to work at the level of the equations of motion. Rather, we can do directly as we did in the end of the spin-1 case, that is, to integrate-out at the level of the action the manifestly non-dynamical modes, i.e. those without time-derivatives. For notational simplicity, we will consider the $d$-scalar, $d$-vector and $d$-tensor sectors separately. As far as the last two are concerned, the procedure and properties are exactly analogous to the ones of the $d$-scalar and $d$-vector modes in massive electrodynamics. For the $d$-tensor sector there is nothing to do, we simply have that it becomes massive
\beq
S_{\rm tensor} = \int \ed^D x \[ - \frac{1}{2}\, \pa_{\mu} \ta_{ij} \pa^{\mu} \ta_{ij} - \frac{1}{2}\, m^2 \ta_{ij}^2 + \ta_{ij} \si_{ij} \] \, .
\eeq
For the $d$-vector sector the non-dynamical field is $\ch_i$, so integrating it out in (\ref{eq:Sharmh}) and using the spatially non-local redefinition
\beq
\ti{\be}_i \equiv \sqrt{\De \( \De - m^2 \)^{-1}} \, \be_i \, , \hspace{1cm} \ti{\si}_i \equiv \sqrt{\De \( \De - m^2 \)^{-1}} \, \si_i \, , 
\eeq
we get the spatially local action
\beq
S_{\rm vector} = \frac{m^2}{2} \int \ed^D x \[ - \frac{1}{2}\, \pa_{\mu} \ti{\be}_i \pa^{\mu} \ti{\be}_i - \frac{1}{2}\, m^2 \ti{\be}_i^2 + \ti{\be}_i \ti{\si}_i  \] \, .
\eeq
As for the longitudinal mode in the spin-1 case, we have that the $d$-vector mode activated by the mass $\be_i$ is not gauge-invariant and smoothly disappears in the $m \to 0$ limit. The novel feature in the spin-2 case, as already anticipated in the previous sections, lies in the $d$-scalar sector. We can start by integrating-out $\ch$ to get
\bea
S_{\rm scal.} & = & \int \ed^D x \[ \frac{d-1}{d} \( - \frac{1}{2}\, \dot{\Phi}^2 - \frac{d-1}{d m^2} \( \pa_i \dot{\Phi} \)^2 + \frac{d-2}{2 d} \( \pa_i \Phi \)^2 + \pa_i \Phi \pa_i \( \psi + \ddot{\la} \) \) \right.  \\
 & & \left. \hspace{1.2cm} - \frac{1}{2}\, m^2 \( \frac{d-1}{d} \( -\Phi^2 + 2 \Phi \De \la \) + 2\( \De \la - \Phi \) \psi - \al \(\Phi + \psi - \De \la\)^2  \) \right. \nn \\
 & & \left. \hspace{1.2cm} + \psi \ro - \Phi p + \frac{2(d-1)}{d m^2}\, \Phi \De \( p + \frac{d-1}{d}\, \De \si \) + \De \la \( p + \frac{d-1}{d}\, \De \si \) + \Ord(T^2)  \] \, . \nn
\eea
At this point, it is convenient to trade $\psi$ for the trace
\beq
h \equiv h_{\mu}^{\,\,\,\mu} = - \psi - \Phi + \De \la \, ,
\eeq
in which case we have
\bea
S_{\rm scal.} & = & \int \ed^D x \[ \frac{d-1}{d} \( - \frac{1}{2}\, \dot{\Phi}^2 - \frac{d-1}{d m^2} \( \pa_i \dot{\Phi} \)^2 - \frac{d+2}{2 d} \( \pa_i \Phi \)^2 - \pa_i \Phi \pa_i h - \De  \Phi \(\ddot{\la} + \De \la \)  \) \right. \nn  \\
 & & \left. \hspace{1.2cm} - \frac{1}{2}\, m^2 \( \frac{d+1}{d}\, \Phi^2 + 2 \( \De \la \)^2 - \frac{2(d+1)}{d}\,\Phi \De \la + 2 h \( \Phi - \De \la \) - \al h^2  \) \right. \nn \\
 & & \left. \hspace{1.2cm} - h \ro - \Phi \( \ro + p \) + \frac{2(d-1)}{d m^2}\, \Phi \De \( p + \frac{d-1}{d}\, \De \si \) + \De \la \( \ro + p + \frac{d-1}{d}\, \De \si \) \right. \nn \\
 & & \left. \hspace{1.2cm}  + \Ord(T^2)  \] \, .  \label{eq:Sscalh} 
\eea
We next integrate-out $h$, i.e. we solve the equation of motion of $h$
\beq \label{eq:hsolharm}
h = - \frac{1}{\al m^2} \(  \frac{d-1}{d}\, \De \Phi - m^2 \Phi + m^2 \De \la - \ro \) \equiv - \frac{1}{\al m^2}\, G  \, ,
\eeq
plug it back inside the action. Choosing the above defined $G$ and $\ti{\Phi} \equiv \Phi + G/m^2$ as the independent fields instead of $\left\{ \Phi, \la \right\}$ we get a diagonal action
\bea
S_{\rm scal.} & = & \frac{d-1}{d} \int \ed^D x \[  - \frac{1}{2}\, \pa_{\mu} \ti{\Phi} \pa^{\mu} \ti{\Phi}  - \frac{1}{2}\, m^2 \ti{\Phi}^2  - \ti{\Phi} \( \ro - \De \si \)  \right. \label{eq:SscalhG} \\
 & & \hspace{2.2cm} \left. + \frac{1}{m^4} \( \frac{1}{2}\, \pa_{\mu} G \pa^{\mu} G  - \frac{1 + d\( 1+ 1/\al \)}{2(d-1)}\, m^2 G^2 - \frac{m^2}{d-1}\,G \( \ro - d p \) \)+ \Ord(T^2) \] \, .  \nn
\eea
Note that $G$ is proportional to the on-shell trace $h$, so in particular it is a Lorentz scalar on-shell. As a check, we can compare its equation of motion 
\beq
\( \bo + \frac{1 + d\( 1+ 1/\al \)}{d-1}\, m^2 \) G = -\frac{m^2}{d-1} \( \ro - d p \) \, ,
\eeq
with (\ref{eq:trhimp}), using (\ref{eq:hsolharm}), and see that they match exactly. We have thus shown what we had claimed in section \ref{sec:stdLagfor}, i.e. it is indeed the trace which is the unstable mode and, more precisely, it is a ghost with mass
\beq \label{eq:mghost}
m_{\rm ghost}^2 = \frac{1 + d\( 1+ 1/\al \)}{d-1}\, m^2 \, .
\eeq
Unlike the case of the $d$-scalar sector or massive electrodynamics, here the action is non-analytic in both $m^2$ and $\al$, if our fields are combinations of the $h_{\mu\nu}$ that are analytic in these parameters. In the $\al \to 0$ limit, with $m$ kept fixed, we see that $m_{\rm ghost}$ diverges, while the coupling to the source remains constant, so we effectively have $G = 0$ and thus also $\ti{\Phi} = \Phi$. It is also instructive to see how this condition appears when working directly at the $\al = 0$ point. So let us go back at the level of (\ref{eq:Sscalh}) where now $h$ is a Lagrange multiplier. Integrating it out will therefore result in fixing another field, which we choose to be $\la$. The equation of motion of $h$ is then simply $G = 0$ and, plugging this inside the action we are indeed left with\footnote{After deriving this result we were informed by S. Deser (private communication) that  
a similar form was obtained in an old and little known paper \cite{DeserTrubatch}. Interestingly enough, this paper appeared in 1966, that is 14 years before the introduction of gauge-invariant variables by Bardeen \cite{Bardeen} in cosmological perturbation theory.}
\beq \label{eq:Sscalh}
S_{\rm scal.} = \frac{d-1}{d} \int \ed^D x \[  - \frac{1}{2}\, \pa_{\mu} \Phi \pa^{\mu} \Phi - \frac{1}{2}\, m^2 \Phi^2 - \Phi \( \ro - \De \si \) + \Ord(T^2) \] \, .
\eeq
At this stage we can make a series of remarks. First, we have now reached in this formalism the same conclusion we did in the previous section, namely, that the scalar degree of freedom in FP theory is gauge-invariant and it does not go away in the $m \to 0$ limit. Second, what we have gained here with respect to the canonical formalism is a clearer picture of the whole $\(m^2, \al\)$ plane provided by (\ref{eq:SscalhG}). Indeed, we are now able to see more clearly the fact that the vDVZ discontinuity as $m \to 0$ arises only for $\al = 0$. If we go back to (\ref{eq:SscalhG}) and take $m \to 0$ with $\al \neq 0$ fixed, we get
\beq
\ti{\Phi} \equiv \Phi + \frac{1}{m^2}\, G \to \frac{1}{m^2} \, G \, ,
\eeq
so that the kinetic terms in the action cancel out and we retrieve the same number of degrees of freedom as in the massless theory. We now understand that the vDVZ discontinuity and the fact that the FP $d$-scalar mode is gauge-invariant are intimately related features. Indeed, in the $m \to 0$ limit we retrieve the gauge symmetry so that only gauge-invariant combinations can survive. For instance, in the case of electrodynamics, we have that the longitudinal mode must disappear since it is not gauge-invariant. Here, for $\al \neq 0$ we have that the $d$-scalar modes are not gauge-invariant and must thus disappear in the $m \to 0$ limit. On $\al = 0$ however, since $\Phi$ survives the $m \to 0$ limit, it must be gauge-invariant.

\subsubsection{Hidden gauge symmetry}

It is now the appropriate moment for discussing the fact that FP theory seems to have something that looks like a symmetry but which is not quite it. This was the novel result of our paper \cite{JaccardMaggioreMitsou1}, where (\ref{eq:Sscalh}) was derived, and we have therefore elaborated on the physical significance of $\Phi$ being gauge-invariant.  

Already from section \ref{sec:stdLagforh} we know that, {\it for on-shell configurations} $h_{\mu\nu}$, the action is invariant under gauge transformations of the form $\xi_{\mu} = \pa_{\mu} \te$ (see eq. (\ref{eq:appsymSh})), but not the equations of motion. The fact that this holds for on-shell configurations in equivalent to the fact that here some equations of motion have to be used, i.e. the ones of the non-dynamical fields, in order to get the invariance. The additional information we gain here with respect to section \ref{sec:stdLagforh} is that the sector of the equations of motion which corresponds to the dynamical field $\Phi$ is also gauge-invariant, a feature which is not visible when we work with $h_{\mu\nu}$. Moreover, the transformation considered in section \ref{sec:stdLagforh} was one-dimensional, whereas here we have two gauge parameters in the $d$-scalar sector of $\xi_{\mu}$ (\ref{eq:xiAB}), that is, $A$ and $B$. Trading the former, for $\bar{A} \equiv A - \dot{B}$, we can write the corresponding gauge parameter (\ref{eq:xiAB})
\beq \label{eq:gshAB}
\xi_{\mu} = \bar{A} \de_{\mu}^0 + \pa_{\mu} B \, ,
\eeq
so that $B$ corresponds to the $\te$ parameter considered in section \ref{sec:stdLagforh}, while $A$ parametrizes the additional transformation under which  (\ref{eq:Sscalh}) is invariant.

Now observe that, if we perform the gauge transformation (\ref{eq:gshAB}) at the level of the original action, we get a new action that depends on the gauge parameters
\beq
S_{A,B}[h_{\mu\nu}] = S[h_{\mu\nu}] + \De S[h_{\mu\nu}, A, B] \, ,
\eeq
where $\De S \neq 0$ for general $h_{\mu\nu}$, so that this is not a gauge symmetry. If we decompose $h_{\mu\nu}$ harmonically, we have that $S_{A,B}$ will correspond to $S$ with $\psi, \ch, \Phi, \la$ replaced by
\beq
\psi - 2\dot{A} \, , \hspace{1cm} \ch - A - \dot{B} \, , \hspace{1cm} \Phi \, , \hspace{1cm} \la - 2 B \, , 
\eeq
respectively. Then, since these are simply the original variables that have been shifted, integrating-out the non-dynamical ones will automatically yield again (\ref{eq:Sscalh}), i.e. whatever the values of $A$ and $B$. Thus, although the actions $S_{A,B}$ are not the same, they do reduce to the same action once the non-dynamical fields are integrated-out. In conclusion, although the action is {\it not} invariant under the $A,B$ transformation, the physics {\it is}. It is in this sense that this is a ``hidden'' gauge symmetry.

\subsection{de-Sitter background} \label{eq:de-Sitterharm}

As a final display of the power of the harmonic formalism, let us apply it to the case where the background space-time is de-Sitter and see whether it is still a gauge-invariant field which propagates in the $d$-scalar sector for $\al = 0$. This is not guaranteed a priori because on flat space-time we concluded that it was the vDVZ discontinuity which was responsible for this and, as it turns out, there is no discontinuity on a de-Sitter background \cite{Hinterbichler1,Hinterbichler2}. 

It is convenient to work in the following coordinates
\beq
g_{00} = -1 \, , \hspace{1cm} g_{0i} = 0 \, , \hspace{1cm} g_{ij} = a^2 \de_{ij} \, , \hspace{1cm} a \equiv e^{H t} \, ,
\eeq
where $a(t)$ is the scale factor and $H$ the (constant) Hubble parameter. We consider directly the case case of the linear massive spin-2 field and obtain its action by linearizing the Einstein-Hilbert action with cosmological constant 
\beq
\La \equiv \frac{d(d-1)}{2}\, H^2 \, ,
\eeq
around the corresponding de-Sitter solution. Also appending a FP mass term and a linear source this gives
\bea
S & = & \int {\rm d}^D x \, a^d \[ -\frac{1}{2} \, \na_{\mu} h_{\nu\ro} \na^{\mu} h^{\nu\ro} + \na_{\mu} h_{\nu\ro} \na^{\nu} h^{\mu\ro} - \na_{\mu} h^{\mu\nu} \na_{\nu} h + \frac{1}{2} \, \na_{\mu} h \na^{\mu} h \right. \nn  \\
 & & \left. \hspace{1.6cm} + d H^2 \( h_{\mu\nu} h^{\mu\nu} - \frac{1}{2} h^2 \) - \frac{1}{2}\, m^2 \( h_{\mu\nu} h^{\mu\nu} - h^2 \) + h_{\mu\nu} T^{\mu\nu} \]  \, , 
\eea
where the non-vanishing components of the Christoffel symbols $\Ga^{\ro}_{\,\,\,\mu\nu}$ are
\beq
\Ga^i_{\,\,j0} = H \de^i_j  \, , \hspace{1cm} \Ga^0_{\,\,ij} = H g_{ij} \, .
\eeq
For $m = 0$, we have the gauge symmetry
\beq \label{eq:gshdS}
\de h_{\mu\nu} = - \na_{\mu} \xi_{\nu} - \na_{\nu} \xi_{\mu} \, ,
\eeq
provided the source satisfies the background-covariant conservation equation $\na_{\mu} T^{\mu\nu} = 0$, while for $m^2 = (d-1) H^2$ we have a one-dimensional gauge symmetry
\beq \label{eq:gshdSpm}
\de h_{\mu\nu} = - \na_{\mu} \na_{\nu} \te - g_{\mu\nu} H^2  \te \, ,
\eeq
provided the source satisfies
\beq \label{eq:consdSte}
\na_{\mu} \na_{\nu} T^{\mu\nu} + H^2 T = 0 \, .
\eeq
The latter case is known as the ``partially massless'' theory because the gauge symmetry eliminates the $d$-scalar mode. It is convenient to define the following field strength \cite{DeserWaldron1}
\beq
F_{\mu\nu\ro} \equiv \na_{\mu} h_{\nu\ro} -  \na_{\nu} h_{\mu\ro} \, , \hspace{1cm} F_{\mu} \equiv g^{\nu\ro} F_{\mu\nu\ro} \, ,
\eeq 
which is invariant under (\ref{eq:gshdSpm}) and in terms of which the action becomes
\beq \label{eq:dSF2rep}
S = \int \ed^D x \, a^d \[ - \frac{1}{4} \( F_{\mu\nu\ro} F^{\mu\nu\ro} - 2 F_{\mu} F^{\mu} \) - \frac{1}{2}\, M^2 \( h_{\mu\nu}h^{\mu\nu} - h^2 \) - h_{\mu\nu} T^{\mu\nu} \] \, ,
\eeq
where
\beq
M^2 \equiv m^2 - (d-1) H^2 \, ,
\eeq
is precisely zero for the partially massless theory. This representation is quite elegant from the point of view of the partially massless theory $M = 0$ because it exhibits many analogies with electrodynamics: there is a one-dimensional gauge symmetry, the theory can be written as the square of some gauge-invariant field strength and there is a cohomological chain structure between the gauge parameter $\te$, the field $h_{\mu\nu}$ and the field strength $F_{\mu\nu\ro}$ \cite{DeserWaldron1,Hinterbichler3}. Then, $M$ appears as the mass that will break this symmetry and activate the $d$-scalar mode. In particular, we can already anticipate that for $M^2 < 0$ the theory will be unstable \cite{DeserWaldron2, Higuchi}.

Here we will focus on the $d$-scalar sector of the theory only, since this is where the exotic features lie, and we will neglect the source for simplicity. In defining and using harmonic variables we must now pay attention to the fact that the position of the spatial indices matters, i.e. they are displaced using $g_{ij}$, so for instance
\beq
\De \equiv g^{ij} \pa_i \pa_j = a^{-2} \pa_i \pa_i \, ,
\eeq
The definitions of the harmonic variables are the same, except for the spatial sectors whose natural generalization is 
\bea
h_{ij} & \equiv & \frac{1}{d}\, g_{ij} \, \ph + \( \pa_i \pa_j - \frac{1}{d}\, g_{ij} \De \) \la  \, , \\
T_{ij} & \equiv & g_{ij} \, p + \( \pa_i \pa_j - \frac{1}{d}\, g_{ij} \De \) \si   \, .
\eea
Note that this changes only the definitions of $\ph$ and $p$. The inverse relation now reads
\bea
\ch & = & \De^{-1} \pa^i h_{0i} \, , \\
\ph & = & h_i^i \, , \\
\la & = & -\frac{1}{d-1}\,\De^{-1} \[ h_i^i - d \De^{-1} \pa^i \pa^j h_{ij} \] \, , 
\eea
where $P_i^{\,\,j}$ is the same as before but one must use $g_{ij}$ to displace its indices now. The gauge transformation (\ref{eq:gshdS}) reads
\bea
\de \psi & = & - 2 \dot{A} \, , \\
\de \ch & = & - A - \dot{B} + 2 H B \, , \\
\de \ph & = & - 2 \De B + 2 d H A \, , \\
\de \la & = & - 2 B \, , 
\eea
so the Bardeen variables are 
\bea
\Psi & \equiv & \psi - \( 2 \dot{\ch} - \ddot{\la} + 2 H \dot{\la} \) \, , \\
\Phi & \equiv & \Phi_0 - d H \( 2 \chi - \dot{\la} + 2 H \la \)  \, ,
\eea
where $\Phi_0 \equiv \De \la - \ph$ is the $\Phi$ of Minkowski space-time. We see that now both combinations include time-derivatives of the original variables so that none of these can be taken as a fundamental field since their initial conditions are undetermined. In particular, this seems to imply that the $d$-scalar degree of freedom on the FP point $\al = 0$ will not be gauge-invariant.

Nevertheless, one must note that the Bardeen variables are the only gauge-invariant combinations (up to combinations among themselves) that are {\it local in time} in the harmonic variables. This is certainly convenient, although not at all a physical requirement. In fact, as we will see in a moment, if we abandon this property we get access to gauge-invariant combinations that do not suffer from the above initial condition problem. 

We can now write down the $d$-scalar part of the action
\bea
S_{\rm scal.} & = & \frac{d-1}{d}  \int \ed^D x \, a^d \[ - \frac{1}{2} \( \dot{\Phi} - d H \Psi \)^2 + \frac{d-2}{2 d} \, \pa_i \Phi \pa^i \Phi + \pa_i \Phi \pa^i \Psi \right. \\
 & & \left. \hspace{1.6cm} - \frac{1}{2}\, m^2 \( -\Phi_0^2 + 2 \Phi_0 \De \la  + \frac{2d}{d-1} \( \De \la - \Phi_0 \) \psi  - \frac{2d}{d-1}\, \pa_i \ch \pa^i \chi \)  \] \, .  \nn
\eea
Again, when $m = 0$ we see that only gauge-invariant combinations appear and we retrieve the flat space-time result for $H \to 0$. For $H \neq 0$, the second and third terms of the first line can be rewritten in a convenient way
\beq
- \frac{1}{d H} \int \ed^D x\, a^d \pa_i \Phi \,\pa^i \( \dot{\Phi} - d H \Psi \) = \int \ed^D x\, a^d \[ \frac{d-2}{2 d} \, \pa_i \Phi \pa^i \Phi + \pa_i \Phi \pa^i \Psi \] \, ,
\eeq
where we have used the fact that 
\beq
\pa_i \Phi \,\pa^i \dot{\Phi} = \frac{1}{2}\, a^{-2} \pa_t \( \pa_i \Phi \)^2 \, ,
\eeq
and then integrated by parts the time derivative. Observe also that 
\beq
K \equiv \dot{\Phi} - d H \Psi = \dot{\Phi}_0 - d H \psi  \, ,
\eeq
so that this combination actually only depends on $\Phi_0$ and $\psi$. We can then consider $\psi, \ch, \Phi_0, \la$ as our independent variables and integrate by parts here and there to finally get
\bea
S_{\rm scal.} & = & \frac{d-1}{d}  \int \ed^D x \, a^d \[ - \frac{1}{2}\, K^2 - \frac{1}{d H}\, \pa_i \( \Phi_0 - 2 d H \chi \) \pa^i K \right. \nn \\
 & & \left. \hspace{2.6cm}  - \De \la \( \dot{K} + d H K + m^2 \Phi_0 + \frac{d m^2}{d-1}\, \psi \) \right.  \nn \\
 & & \left. \hspace{2.6cm}   + \frac{1}{2}\, m^2 \( \Phi_0^2  + \frac{2d}{d-1} \Phi_0 \psi + \frac{2d}{d-1}\, \pa_i \ch \pa^i \chi \) \] \, . 
\eea
We start by integrating-out $\chi$
\bea
S_{\rm scal.} & = & \frac{d-1}{d}  \int \ed^D x \, a^d \[ - \frac{1}{2} K^2 - \frac{d-1}{d m^2}\, \pa_i K \pa^i K - \frac{1}{d H}\, \pa_i \Phi_0 \,\pa^i K \right. \nn \\
 & & \left. \hspace{2.6cm}  - \De \la \( \dot{K} + d H K + m^2 \Phi_0 + \frac{d m^2}{d-1}\, \psi \)   + \frac{1}{2}\, m^2 \( \Phi_0^2  + \frac{2d}{d-1} \Phi_0 \psi \)  \] \, .   \nn 
\eea
We now rescale our fields
\beq \label{eq:rescaling}
\left\{ \psi, \Phi_0, \la \right\} \to a^{-(d-1)} \left\{ \psi, \Phi_0, \la \right\} \, ,
\eeq
and trade $\psi$ for the new variable
\beq
\psi' \equiv \psi + \frac{d-1}{d} \, \Phi_0 \, ,
\eeq
so that
\beq
K \to a^{-(d-1)} \( \dot{\Phi}_0 - d H \psi' \) \equiv a^{-(d-1)} K'\, , 
\eeq
and get
\bea
S_{\rm scal.} & = & \frac{d-1}{d}  \int \ed^D x \, a^{-(d-2)} \[ - \frac{1}{2} K'^2 - \frac{d-1}{d m^2}\, \pa_i K'\, \pa^i K' - \frac{1}{d H}\, \pa_i \Phi_0 \,\pa^i K' \right. \nn \\
 & & \left. \hspace{2.6cm}  - \De \la \( \dot{K}' + H K' + \frac{d m^2}{d-1}\, \psi' \)   + \frac{1}{2}\, m^2 \( - \Phi_0^2  + \frac{2d}{d-1} \Phi_0 \psi' \)  \] \, .   \nn 
\eea
We can now trade $\Phi_0$ for the new variable
\beq \label{eq:Omdef}
\Om(t) \equiv \Phi_0(t) - d H \int^t_{t_i} \ed t' \, \psi'(t') + \frac{(d-1)H}{M^2} \[ \dot{\Phi}_0(t_i)  - d H \psi'(t_i) + H \Phi_0(t_i) \] \, ,
\eeq
where $t_i$ is the time at which the initial conditions are given and we have omitted the $\vec{x}$ dependence for notational simplicity. The choice of the time-independent term will be justified later. As already suggested above, this is a non-local generalization of $\Phi_0$ to de-Sitter space-time, for which the Cauchy problem is well-defined. Indeed, it is gauge invariant, as we will show in a moment, and the initial data $\left\{ \Om(t_i), \psi'(t_i) \right\}$ are in bijection with $\left\{ \Phi_0(t_i), \psi'(t_i) \right\}$ 
\bea  
\Om(t_i) & = & \frac{m^2}{M^2}\, \Phi_0(t_i) + \frac{(d-1)H}{M^2} \[ \dot{\Phi}_0(t_i) - d H \psi'(t_i) \]  \, , \\
\dot{\Om}(t_i) & = & \dot{\Phi}_0(t_i) - d H \psi'(t_i) \, ,
\eea
contrary to the Bardeen variable $\Phi$. We can then invert this to get
\beq   \label{eq:phiofOm}
\Phi_0(t) = \Om(t) + d H \int^t_{t_i} \ed t' \, \psi'(t') - \frac{(d-1)H}{m^2} \[ \dot{\Om}_0(t_i) + H \Om(t_i) \] \, , 
\eeq 
so, although $K' = \dot{\Om}$, performing this replacement in the action will yield non-local terms because of the presence of undotted $\Phi_0$'s. However, after integrating-out $\la$ and solving for $\psi'$, we get that the latter becomes a total time derivative
\beq
\psi' = -\frac{d-1}{d m^2} \( \ddot{\Om} + H \dot{\Om} \) \, ,
\eeq
so that (\ref{eq:phiofOm}) becomes local
\beq 
\Phi_0(t) = \frac{M^2}{m^2}\, \Om(t) - \frac{(d-1)H}{m^2} \, \dot{\Om}(t)  \, .
\eeq
We now understand that the time-independent piece in (\ref{eq:Omdef}) was chosen precisely such that it cancels the one arising in the above integral. We are thus left with an action for $\Om$ alone which, after many integrations by parts, gives
\beq \label{eq:SofOmresc}
S_{\rm scal.} = \frac{d-1}{d} \frac{M^2}{m^2} \int \ed^D x \, a^{-(d-2)} \[ -\frac{1}{2}\, \pa_{\mu} \Om \, \pa^{\mu} \Om - \frac{1}{2}\, M^2 \Om^2 \] \, .
\eeq 
If we rescale back in order to obtain the volume form $\sqrt{-g} = a^d$ for the integration measure
\beq
\Om \to a^{d-1} \Om \, ,
\eeq
we get
\beq \label{eq:SofOm}
S_{\rm scal.} = \frac{d-1}{d} \frac{M^2}{m^2} \int \ed^D x \, \sqrt{-g} \[ -\frac{1}{2}\, \pa_{\mu} \Om \, \pa^{\mu} \Om - \frac{1}{2}\, m^2 \Om^2 \] \, ,
\eeq 
so this has also the effect of replacing $M^2$ with $m^2$ in the mass term. The dynamical mode in the $d$-scalar sector is therefore $\Om$, which in terms of the original fields reads
\bea
\Om(t) & \equiv & \Phi_0(t) - d H a^{-(d-1)} (t) \int^t_{t_i} \ed t' \, a^{d-1} (t') \( \psi(t') + \frac{d-1}{d}\, \Phi_0(t') \) \label{eq:Omofh} \\
 & & + \( \frac{a(t_i)}{a(t)} \)^{d-1} \frac{(d-1)H}{M^2} \[ \dot{\Phi}_0(t_i)  - d H \psi'(t_i) + H \Phi_0(t_i) \] \nn \\
 & = & a^{-(d-1)} (t) \int^t_{t_i} \ed t' \, a^{d-1} (t') K(t')  + \( \frac{a(t_i)}{a(t)} \)^{d-1} \[ \frac{m^2}{M^2}\, \Phi_0(t_i) + \frac{(d-1)H}{M^2} \, K(t_i) \]  \, .  \nn
\eea
Under a gauge transformation (\ref{eq:gshdS}) we have that $K$ is gauge-invariant so 
\beq
\de \Om(t) \sim \de\Phi_0(t_i) \sim A(t_i) \, ,
\eeq
and thus $\Om$ is gauge-invariant if we set $A(t_i) = 0$. Note that this restriction is by no means a loss of symmetry since it concerns only a subset of measure zero of the gauge parameters. One can still use such an $A(t)$ to trivialize the time-evolution of a field mode. Thus, after having integrated-out the non-dynamical fields, the $d$-scalar sector remains gauge-invariant even in de-Sitter space. 

A very elegant feature of our result (\ref{eq:SofOm}) is that it renders the dependence of the spectrum on $M$ quite transparent. $\Om$ becomes non-dynamical when $M \to 0$, in which case we reach the partially massless theory with gauge symmetry (\ref{eq:gshdSpm})\footnote{Since the mapping between $\Om$ and $\Phi_0$ is singular as $M \to 0$, we should actually check this result by working directly on the $M = 0$ point, in which case integrating out $\la$ to fix $\psi'$ gives $S_{\rm scal.} = 0$.}, while for $M^2 < 0$ that mode becomes a ghost. The stability condition $M^2 \geq 0$ is known as the ``Higuchi bound'' \cite{Higuchi}. We also see that the ``natural variables'' with respect to the interpretation of $M^2$ as being the mass of the partially massless theory are the rescaled ones, since it is for these fields that $M$ appears as the mass (\ref{eq:SofOmresc}) and for which $\Om$ involves no $a$ in its definition (\ref{eq:Omdef}).

\section{Propagator} \label{sec:propagator}

The dynamical content and stability of a linear theory can also be deduced by looking at its propagator. Moreover, since the propagator is an essential building block of perturbative QFT, it is important to be able to ``read from it'' this important information of the theory. We will not write explicitly the $\ep$ prescription here since it depends on whether one is interested in classical or quantum propagation. It will be however useful to use some QFT language, e.g. the number of dynamical fields becomes the number of particle polarizations/states.

\subsection{Spin 1}

Writing the Proca action (\ref{eq:ProcaAction}) in the form
\beq
S = \int \ed^D x \[ \frac{1}{2}\, A_{\mu} {\cal K}^{\mu\nu} A_{\nu} - A_{\mu} j^{\mu} \] \, ,
\eeq
we can identify the quadratic structure
\beq \label{eq:quadstructA}
{\cal K}^{\mu\nu} = \et^{\mu\nu} \( \bo - m^2 \) - \pa^{\mu} \pa^{\nu}  \, .
\eeq
The propagator is defined by
\beq
{\cal K}^{\mu\ro} D_{\ro\nu} = i \de_{\mu}^{\nu} \, ,
\eeq
whose solution is
\beq
D_{\mu\nu}(k) = - \frac{i}{k^2 + m^2} \( \et_{\mu\nu} + \frac{k_{\mu}k_{\nu}}{m^2} \) \, .
\eeq
The exchange of a photon between two vertices in the computation of a scattering amplitude will then be controlled by the saturated propagator
\beq \label{eq:satpropA}
j^{*\mu}(k) D_{\mu\nu}(k) j'^{\nu}(k) = j^{*\mu}(k)  \[ - \frac{i}{k^2 + m^2} \, \et_{\mu\nu} \] j'^{\nu}(k) \, ,
\eeq
where here $j^{\mu}, j'^{\mu}$ either represent external on-shell sources, in which case conservation implies $k_{\mu} j^{\mu}(k) = k_{\mu} j'^{\mu}(k) = 0$, or parts of a Feynman diagram to which the photon is attached, in which case it is the Ward identity\footnote{The Ward identity is usually presented as a direct consequence of gauge symmetry and it can therefore appear as a surprise that it still holds in the massive case. However, note that one can also derive the identity by simply using the operator equation $\pa_{\mu} \hat{A}^{\mu} = 0$, which is valid in the massive case, when computing correlation functions with on-shell external momenta $\pa_{\mu} \bra 0 | T\left\{ \hat{A}_{\mu}(x) \dots \right\} | 0 \ket = 0$. Thus, the Ward identity still holds in massive electrodynamics, not because $\pa_{\mu} A^{\mu}$ contains no propagating degrees of freedom as in the massless case, but because $\pa_{\mu} A^{\mu}$ is simply zero on-shell.} which implies these equations. In the classical case, the saturated propagator is what controls the interaction mediated by the electromagnetic field in the perturbative equations of motion of the fields present in the source.

In the massive case we have as many possible inversions of ${\cal K}_{\mu\nu}$ as with $\bo - m^2$ because of the homogeneous solutions. These are parametrized by all the possible linear superposition amplitudes $a_i(\vec{k})$ (belonging to some space of integrable functions), that are functions on $\Rs^d$. Going to the $m = 0$ case enlarges that kernel dramatically because now it also includes all the pure-gauge solutions $A_{\mu} = \pa_{\mu} \te$, parametrized by a function $\te$ on $\Rs^D$. Thus, on top of the pole contour prescription, which can be translated into a prescription on initial/final conditions, one must also give a prescription for picking a preferred gauge, i.e. one must add a gauge-fixing term. The usual Lorentz-invariant choice is 
\beq \label{eq:gfA}
S_{\rm gf} = - \frac{1}{2\xi} \int \ed^D x\, \( \pa_{\mu} A^{\mu} \)^2 \, ,
\eeq
so that
\beq
{\cal K}^{\mu\nu} = \et^{\mu\nu}  \bo - \( 1 - \frac{1}{\xi} \) \pa^{\mu} \pa^{\nu} \, ,
\eeq
is invertible and\footnote{For non-linear theories the gauge-fixing term breaks the unitarity of the $S$-matrix and one must also include Faddeev-Popov fields to restore it.}
\beq \label{eq:propm0A}
D_{\mu\nu}(k) = - \frac{i}{k^2} \( \et_{\mu\nu} - (1-\xi)\, \frac{k_{\mu}k_{\nu}}{k^2} \) \, .
\eeq
What matters for the gauge-fixing term to be valid is that the saturated propagator must be independent of it because the physics cannot depend on a choice of gauge. Since $k_{\mu} j^{\mu} (k) = 0$, which is also a consequence of gauge symmetry when $m = 0$, we have indeed the $\xi$-independent result
\beq
j^{*\mu}(k) D_{\mu\nu}(k) j'^{\nu}(k) = j^{*\mu}(k)  \[ - \frac{i}{k^2} \, \et_{\mu\nu} \] j'^{\nu}(k)  \, .
\eeq
Comparing with (\ref{eq:satpropA}) we note that the interaction between two sources mediated by the photon is continuous in the $m \to 0$ limit. At the same time however, we know that the massive photon has $d$ polarizations, while the massless one has $d-1$ polarizations. To understand how a discontinuity in this number can be consistent with a continuous limit at the propagator level, we decompose $j^{\mu}$ and $j'^{\mu}$ harmonically (\ref{eq:jharmdec}) which in Fourier space gives
\beq
j_0 = -\ro \, , \hspace{1cm} j_i(k) = i k_i \si(k) + \si_i(k) \, , \hspace{1cm} k_i \si_i(k) = 0 \, ,
\eeq
and similarly for $j'^{\mu}$. We then restrict to tree-level diagrams and sources with ``mass'' $m_{\rm s}^2 = - k^2$, so that $m_{\rm s}$ is the ``mass'' of the virtual photon that is being exchanged. For instance, in the case where the source is made of minimally coupled electrons and positrons we have that $m_{\rm s} \geq 2 m_e$. We also consider the case $m_{\rm s} > m$ so that we do not have to deal with the complications of resonances\footnote{Demanding heavier sources $m_{\rm s} > m$ and no loops implies that the virtual photon can never be on-shell, i.e. it is never a ``real'' photon. Alternatively, if $m_{\rm s} < m$, then the process in which the photon is on-shell would be kinematically allowed, in which case the propagator would be singular, implying an infinite probability for this process to occur. As in the case of any heavy particle, the resolution of this apparent problem comes by noting that the heavy particle becomes unstable precisely when $m_{\rm s} < m$, since it can then disintegrate into the source's particles. By the optical theorem, we then have that the imaginary part of the vacuum polarization diagram becomes non-zero. Since that diagram is responsible for shifting the mass $m$ under radiative corrections in the propagator, we get that the poles of the {\it renormalized} propagator have a non-vanishing imaginary part. Thus, the case $k^2 = -m_{\rm ren}^2$, where $m_{\rm ren}$ is the renormalized mass of the photon, is not a singularity of the renormalized propagator but rather the maximum of the so-called ``Breit-Wigner'' resonance.}. We can then write the conservation equation (\ref{eq:consharmh}) as
\beq
\ro = \frac{i \vec{k}^2}{\sqrt{m_{\rm s}^2 + \vec{k}^2}}\, \si \, ,
\eeq
and similarly for $j'^{\mu}$, so that the saturated propagator reads
\beq \label{eq:harmsatprop}
\left. j^{*\mu}(k) D_{\mu\nu}(k) j'^{\nu}(k) \right|_{k^2=-m_{\rm s}^2} = -\frac{i}{m^2 - m^2_{\rm s}} \[ \frac{m_{\rm s}^2 \vec{k}^2}{m_{\rm s}^2 + \vec{k}^2}\, \si^*(k) \si'(k) + \si^*_i (k) \si'_i (k)  \]   \, .
\eeq
The first term in the square bracket represents the exchange of the longitudinal photons between the longitudinal modes of the sources, while the second term corresponds to the exchange of a transverse photon between the transverse modes of the sources. We can now focus on the case where $m_{\rm s}^2 \to m^2$ from above, so that the photon gets close to being real. It can therefore be considered as an external photon that has been ``produced'' by $j'^{\mu}$ at $t \to - \infty$ and then ``detected'' by its interaction with $j^{\mu}$ at $t \to \infty$.

This shows how the continuity in the saturated propagator can be reconciled with the discontinuity in the dynamical fields of the photon: the longitudinal information is simply proportional to $m^2$ for real photons and thus smoothly decouples in the massless limit. It is therefore not enough to look at the unsaturated propagators to deduce the number of dynamical fields in the theory, one must also make use of the conservation equation of the source, which brings in the mass dependence.

Note that the source components that appear are the ones that are being propagated so that counting them gives us a {\it lower bound} on the number of dynamical fields $N_{\rm d}$. In the massive case we have $\si_i$ and $\si$, that is $N_{\rm d} \geq d$, while in the massless limit the longitudinal part $\si$ smoothly decouples and becomes unobservable and we are thus left with $N_{\rm d} \geq d-1$. Here these inequalities are saturated, as we know. We will see however that this is not always the case for non-local theories in the presence of ghosts. 

Finally, as far as stability is concerned, we have that the saturated propagator (\ref{eq:harmsatprop}) is the one of a healthy scalar times a positive-definite scalar product of $j_i$ and $j'_i$, so that this theory is stable.

\subsection{Spin 2}

Let us start by identifying the quadratic structure of (\ref{eq:FPaction}) 
\beq
{\cal K}^{\mu\nu\ro\si} \equiv {\cal E}^{\mu\nu\ro\si} - m^2 \( \et^{\mu(\ro} \et^{\si)\nu} - (1+\al)\et^{\mu\nu} \et^{\ro\si} \) \, , 
\eeq
where $\cal E$ was defined in (\ref{eq:Lichne}). The propagator
\beq
{\cal K}^{\mu\nu\al\be} D_{\al\be\ro\si} = i\de^{\mu}_{(\ro} \de^{\nu}_{\si)} \, ,
\eeq
is given by
\bea
D_{\mu\nu\ro\si}(k) & = & - \frac{i}{k^2 + m^2} \[ \frac{1}{2} \( \et_{\mu\ro} \et_{\nu\si} + \et_{\mu\si} \et_{\nu\ro} \)  - \frac{1}{d}\( 1 - \frac{\al}{d}\,\frac{k^2 + m^2}{\mu^2} \)\, \et_{\mu\nu} \et_{\ro\si} \right. \nn \\
 & & \left. \hspace{2.5cm} + \frac{1}{2} \( \et_{\mu\ro} \, \frac{k_{\nu} k_{\si}}{m^2} + \et_{\mu\si} \, \frac{k_{\nu} k_{\ro}}{m^2} + \et_{\nu\ro} \, \frac{k_{\mu} k_{\si}}{m^2} + \et_{\nu\si} \, \frac{k_{\mu} k_{\ro}}{m^2} \)  \right. \nn \\
 & & \left. \hspace{2.5cm} - \frac{1+2\al}{d} \( \et_{\mu\nu}\, \frac{k_{\ro} k_{\si}}{\mu^2} + \et_{\ro\si}\, \frac{k_{\mu} k_{\nu}}{\mu^2} \) + \frac{(1 + 2 \al)(d-1)}{d}\, \frac{k_{\mu} k_{\nu} k_{\ro} k_{\si}}{m^2 \mu^2} \] \, , \nn
\eea
where
\beq
\mu^2 \equiv m^2 - \al \( \frac{d-1}{d}\, k^2 - \frac{d+1}{d}\, m^2 \) \, .
\eeq
In the saturated propagator with conserved sources the terms with uncontracted $k_{\mu}$'s drop
\bea
T^{*\mu\nu} D_{\mu\nu\ro\si} T'^{\ro\si} & = & - \frac{i}{k^2 + m^2} \[ T^*_{\mu\nu} T'^{\mu\nu}  - \frac{1}{d}\( 1 - \frac{\al}{d}\,\frac{k^2 + m^2}{\mu^2} \) T^* T' \] \nn \\
 & = & - \frac{i}{k^2 + m^2} \[ T^*_{\mu\nu} T'^{\mu\nu}  - \frac{1}{d}\, T^* T' \] - \frac{\al}{d^2}\, \frac{i}{\mu^2} \, T^* T' \, . \label{eq:satprophm}
\eea
Note that this neatly splits into the Fierz-Pauli propagator $\al = 0$ plus an extra scalar propagator which can be written as
\beq
- \frac{i}{ - k^2 + m^2_{\rm ghost}} \, \frac{T^* T'}{d(d-1)}  \, ,
\eeq
with $m^2_{\rm ghost}$ given precisely by (\ref{eq:mghost}). Indeed, this is the pole corresponding to the ghost since the ``kinetic'' part $\sim k^2$ in the denominator comes with the wrong sign and the corresponding source is the trace $T$. In the case $\al = -1/2$ we have that $m_{\rm ghost}^2 = -m^2$, so that this becomes also a tachyon, but then the tensor structure becomes the one of the massless theory
\beq \label{eq:al12satprop}
T^{*\mu\nu} D_{\mu\nu\ro\si} T'^{\ro\si} = - \frac{i}{k^2 + m^2} \[ T^*_{\mu\nu} T'^{\mu\nu}  - \frac{1}{d-1}\, T^* T' \] \, . 
\eeq
Let us now focus on the case $\al = 0$. As we did for the spin-1 case, we can again perform the harmonic decomposition of the sources (\ref{eq:Tharmdec}) and use the conservation equations (\ref{eq:consharmh}) with ``source mass'' $m_{\rm s}^2 = - k^2$
\beq 
\ro =  \frac{i\vec{k}^2}{\sqrt{m_{\rm s}^2 + \vec{k}^2}}\,  q \, , \hspace{1cm} q = -\frac{i}{\sqrt{m_{\rm s}^2 + \vec{k}^2}} \( p - \frac{d-1}{d} \,\vec{k}^2  \si \) \, , \hspace{1cm} q_i = \frac{1}{2}\,\frac{i\vec{k}^2}{\sqrt{m_{\rm s}^2 + \vec{k}^2}}\,  \si_i \, ,
\eeq
to get
\beq \label{eq:hlocpropM}
T^{*\mu\nu} D_{\mu\nu\ro\si} T'^{\ro\si} = - \frac{i}{m^2 - m_{\rm s}^2} \[ x^{\da} M  x' + \frac{1}{2}\, \frac{m_{\rm s}^2 \vec{k}^2}{m_{\rm s}^2 + \vec{k}^2}\, \si^*_i \si'_i + \si^*_{ij} \si'_{ij} \] \, .
\eeq
where $x \equiv \( p, \frac{d-1}{d}\,\vec{k}^2 \si \)$ and $M$ is a $2\times 2$ matrix with eigenvalues
\beq
\la_- = 0 \, , \hspace{1cm} \la_+ = \frac{(d^2 - 2d + 2) \vec{k}^4 + 2d \vec{k}^2 m_{\rm s}^2 + d^2 m_{\rm s}^4}{d (d-1) (m_{\rm s}^2 + \vec{k}^2)^2} > 0 \, ,
\eeq
so that there is only one pole corresponding to the $d$-scalar sector and with the correct sign, as expected. We see that the $d$-vector and $d$-tensor sectors are the exact analogues of the $d$-scalar and $d$-vector sectors of electrodynamics (\ref{eq:harmsatprop}). Considering the limits $m_{\rm s}^2 \to m^2 \to 0$ we get that the $\si_i$ part smoothly decouples and we are left with only $\si_{ij}$. In the $d$-scalar sector however we have the vDVZ discontinuity since
\beq
\la_+ \to \frac{d^2 - 2d + 2}{d (d-1)} \neq 0 \, ,
\eeq
so this pole remains. We can compare this with the case $m = 0$ where, because of the gauge symmetry, we must add a gauge-fixing term in the action in order to invert the quadratic structure. The usual Lorentz-invariant choice is
\beq
S_{\rm gf} = - \frac{1}{\xi} \int \ed^D x\, \pa_{\mu} \bar{h}^{\mu\nu} \pa^{\ro} \bar{h}_{\ro\nu} \, , 
\eeq
where $\bar{h}_{\mu\nu}$ has been defined in (\ref{eq:hbardef}), in which case one has
\bea
{\cal K}^{\mu\nu\ro\si} & = & \( \et^{\mu(\ro} \et^{\si)\nu} - \( 1 - \frac{1}{2\xi} \) \et^{\mu\nu} \et^{\ro\si} \) \bo  \\
 & &  - \( 1 - \frac{1}{\xi} \) \( \et^{\mu(\ro} \pa^{\si)} \pa^{\nu} + \et^{\nu(\ro} \pa^{\si)} \pa^{\mu} - \et^{\mu\nu} \pa^{\ro} \pa^{\si} - \et^{\ro\si} \pa^{\mu} \pa^{\nu} \) \, , \nn
\eea
with inverse
\bea
D_{\mu\nu\ro\si} & = & - \frac{i}{k^2} \[  \frac{1}{2} \( \et_{\mu\ro} \et_{\nu\si} + \et_{\mu\si} \et_{\nu\ro} \)  - \frac{1}{d-1}\, \et_{\mu\nu} \et_{\ro\si} \right.  \\
 & & \left. \hspace{1.6cm} + \frac{1}{2}\, (1- \xi) \( \et_{\mu\ro} \, \frac{k_{\nu} k_{\si}}{m^2} + \et_{\mu\si} \, \frac{k_{\nu} k_{\ro}}{m^2} + \et_{\nu\ro} \, \frac{k_{\mu} k_{\si}}{m^2} + \et_{\nu\si} \, \frac{k_{\mu} k_{\ro}}{m^2} \) \]\, .  \nn
\eea
Comparing the saturated one 
\beq
T^{*\mu\nu} D_{\mu\nu\ro\si} T'^{\ro\si} = - \frac{i}{k^2 + m^2} \[ T^*_{\mu\nu} T'^{\mu\nu}  - \frac{1}{d-1}\, T^* T' \] =  - \frac{i}{m^2 - m_{\rm s}^2} \, \si^*_{ij} \si'_{ij}  \, .
\eeq
with (\ref{eq:satprophm}) for $\al = 0$, we see that the discontinuity lies in the factor in front of the $\sim T^* T'$ term which is $1/d$ instead of $1/(d-1)$. This difference is what is precisely needed in order to cancel the $d$-scalar pole. Finally, here too we can see that the $m \to 0$ and $\al \to 0$ limits do not commute. Indeed, taking $m \to 0$ while keeping $\al \neq 0$ fixed we get that
\beq
1 - \frac{\al}{d}\,\frac{k^2 + m^2}{\mu^2} \to \frac{d}{d-1} \, ,
\eeq
so (\ref{eq:satprophm}) becomes the massless propagator, which is independent of $\al$, and there is thus no vDVZ discontinuity.

\section{St\"uckelberg formalism} \label{sec:Stuck}

In using massive theories so far we have encountered two conceptually disturbing features. First, the gauge symmetry is broken and, second, the number of degrees of freedom is discontinuous in the $m \to 0$ limit as it suddenly jumps from $2d$ to $2(d-1)$. The St\"uckelberg trick \cite{Stueckelberg, Hinterbichler1,Hinterbichler2} is an elegant way of killing those two birds with one stone at the level of the action, and with explicit Lorentz covariance. As in the case of propagators, it shows that the degrees of freedom do not change discontinuously as $m \to 0$, but that some of them simply decouple.

\subsection{Spin 1} \label{sec:StuckA}

The so-called ``St\"uckelberg trick'' amounts to introducing auxiliary fields in a way which is patterned on the gauge transformation itself. In the case of massive electrodynamics we have (\ref{eq:gsA}) so one substitutes
\beq
A_{\mu} \to A_{\mu} + \frac{1}{m} \, \pa_{\mu} \ph \, ,
\eeq
in (\ref{eq:ProcaAction}), where $\ph$ is the ``St\"uckelberg field''. Since this technically has the form of a gauge transformation, only the mass term varies and we have that the Proca action becomes
\beq
S[A] \to S[A,\ph] = \int \ed^D x \[ -\frac{1}{4}\, F_{\mu\nu} F^{\mu\nu} - \frac{1}{2}\, m^2 A_{\mu} A^{\mu} - \frac{1}{2}\, \pa_{\mu} \ph \pa^{\mu} \ph- m\, A^{\mu} \pa_{\mu} \ph + A_{\mu} j^{\mu}  \] \, . 
\eeq
By construction, this action is invariant under the gauge transformation
\beq
\de A_{\mu} = - \pa_{\mu} \te \, , \hspace{1cm} \de \ph = m \te \, ,
\eeq
so $\ph$ is a redundant (pure-gauge) field. The equations of motion of $A_{\mu}$ and $\ph$ are, respectively,
\beq \label{eq:AstuEOM}
\pa_{\mu} F^{\mu\nu} - m^2 A^{\nu} = - j^{\nu} + m\, \pa^{\nu} \ph  \, , \hspace{1cm} \bo \ph = -m\pa_{\mu} A^{\mu} \, ,
\eeq
and we see that the latter is nothing but the divergence of the former. The gauge in which $\ph = 0$ is called the ``unitary gauge'', in which case one recovers the equations of Proca theory. However, the advantage of having $\ph$ around is to keep imposing the gauge condition on the gauge field, and by choosing this condition appropriately, $\ph$ can then be interpreted as carrying the information of the longitudinal degrees of freedom that are activated in the massive theory. To see this, let us proceed to two different gauge-fixing scenarios. 

We first choose to impose the Lorentz gauge $\pa_{\mu} A^{\mu} = 0$ so that, along with the equation of motion of $A_0$, we can fix the initial conditions of the latter
\beq 
\( \De - m^2 \) A_0 = \pa_i \dot{A}_i + m \dot{\ph} - j_0 \, , \hspace{1cm}  \dot{A}_0 = \pa_i A_i \, .
\eeq
We are then left with the equations
\beq
\( \bo - m^2 \) A_i = -j_i + m \pa_i \ph \, , \hspace{1cm}  \bo \ph = 0 \, .
\eeq
Now we see that, as in the {\it massless} case in (\ref{sec:stdLagforA}), we also have a residual gauge symmetry given by the $\te$ obeying $\bo \te = 0$. However, since the $A_i$ obey a massive Klein-Gordon equation, we cannot use such a $\te$ to kill the homogeneous solution of $\pa_i A_i$ as in the massless case. Rather, we can use $\te$ to set $\ph = 0$, so that this amounts to choosing the unitary gauge. Thus, with the Lorentz gauge the St\"uckelberg field cannot represent the longitudinal mode since it obeys a massless Klein-Gordon equation.

Another initial choice of gauge is $\pa_{\mu} A^{\mu} = - m \ph$, in which case the conditions on $A_0$ read
\beq 
\( \De - m^2 \) A_0 = \pa_i \dot{A}_i + m \dot{\ph} - j_0 \, , \hspace{1cm}  \dot{A}_0 = \pa_i A_i + m \ph \, ,
\eeq
and the leftover equations are
\beq
\( \bo - m^2 \) A_i = -j_i  \, , \hspace{1cm}  \( \bo - m^2 \) \ph = 0 \, .
\eeq
We have again a residual gauge symmetry but it is now parametrized by the $\te$ obeying $(\bo - m^2) \,\te = 0$. We can thus choose either to set $\ph = 0$ using such a $\te$, or to eliminate the homogeneous solution of $\pa_i A_i$ as in (\ref{sec:stdLagforA}). In the latter case, it is therefore $\ph$ which survives and represents the degrees of freedom associated with the longitudinal part $\pa_i A_i$,  while $A_{\mu}$ contains $2(d-1)$ degrees of freedom as in the massless case. Thus, the interpretation of $\ph$ depends on the choice of gauge one makes. 

Nevertheless, the interpretation in which $\ph$ represents the 2 degrees of freedom of the longitudinal part is the most appealing because it survives in the $m \to 0$ limit. Indeed, for $m = 0$ we have that $\ph$ becomes gauge-invariant and thus an unambiguous degree of freedom. We are then left with massless electrodynamics plus a scalar, totaling $N_{\rm f} = 2N_{\rm d} = 2d$. The important feature is that $A_{\mu}$ and $\ph$ are now decoupled, so if we focus on the dynamics of $A_{\mu}$ then $\ph$ is unobservable. Just as we saw when studying the propagators, the longitudinal modes do not propagate in the $A_{\mu}$ field anymore.

\subsection{Spin 2}

In the spin-2 case we must pattern the introduction of the St\"uckelberg field on (\ref{eq:gsh})
\beq
h_{\mu\nu} \to h_{\mu\nu} + \frac{1}{m} \( \pa_{\mu} A_{\nu} + \pa_{\nu} A_{\mu} \) \, , 
\eeq
in (\ref{eq:FPaction}) to get
\bea
S & = & \int \ed^D x \[ \frac{1}{2}\, h_{\mu\nu} {\cal E}^{\mu\nu\ro\si} h_{\ro\si} - \frac{1}{2}\, F_{\mu\nu} F^{\mu\nu} + 2\al \( \pa_{\mu} A^{\mu} \)^2   - \frac{1}{2}\, m^2 \( h_{\mu\nu} h^{\mu\nu} - (1 + \al) h^2 \) \right. \nn \\
 & & \left. \hspace{1.2cm}  - 2m \( h^{\mu\nu} \pa_{\mu} A_{\nu} - (1+\al) h \pa_{\mu} A^{\mu} \) + h_{\mu\nu} T^{\mu\nu} \] \, , \label{eq:AhAStuck}
\eea 
where as usual $F_{\mu\nu} \equiv \pa_{\mu} A_{\nu} - \pa_{\nu} A_{\mu}$, so that the gauge symmetry is restored
\beq
\de h_{\mu\nu} = - \pa_{\mu} \xi_{\nu} - \pa_{\nu} \xi_{\mu} \, , \hspace{1cm} \de A_{\mu} = m \xi_{\mu} \, .
\eeq
Note that for $\al = 0$ the equation of motion of $A_{\mu}$ takes the form of the equation of massless electrodynamics with an $h_{\mu\nu}$-dependent source, so it is invariant under the U$(1)$ transformation (\ref{eq:gsA}). This means that $A_{\mu}$ represents $2(d-1)$ degrees of freedom, while the difference between Fierz-Pauli theory and the massless theory is $D^2 - D - 2 - (d^2 - d - 2) = 2d$, so if we take the $m \to 0$ limit now we are still discontinuous in the number of degrees of freedom. We can thus perform a second St\"uckelberg trick on this field in order to acquire the U$(1)$ symmetry as well. We replace
\beq
A_{\mu} \to A_{\mu} \to \frac{1}{m} \, \pa_{\mu} \ph \, ,
\eeq 
to get
\bea
S & = & \int \ed^D x \[ \frac{1}{2}\, h_{\mu\nu} {\cal E}^{\mu\nu\ro\si} h_{\ro\si} - \frac{1}{2}\, F_{\mu\nu} F^{\mu\nu} + 2\al \( \pa_{\mu} A^{\mu} \)^2 + \frac{2\al}{m^2} \( \bo \ph \)^2  \right. \nn \\
 & & \left. \hspace{1.2cm} - \frac{1}{2}\, m^2 \( h_{\mu\nu} h^{\mu\nu} - (1 + \al) h^2 \) - 2m \( h^{\mu\nu} \pa_{\mu} A_{\nu} - (1+\al) h \pa_{\mu} A^{\mu} \) \right. \nn \\
 & &  \left. \hspace{1.2cm} - 2 \( h^{\mu\nu} \pa_{\mu} \pa_{\nu} \ph - (1+\al) h \bo \ph \) + \frac{4\al}{m}\, \pa_{\mu} A^{\mu} \bo \ph  + h_{\mu\nu} T^{\mu\nu} \] \, , \nn \\
 \label{eq:ShStuck}
\eea 
which has the gauge symmetry
\beq \label{eq:StuckU1}
\de h_{\mu\nu} = 0 \, , \hspace{1cm} \de A_{\mu} = -\pa_{\mu} \te \, , \hspace{1cm} \de \ph = m \te \, .
\eeq
For $\al \neq 0$ we see that we have a higher derivative theory for $\ph$ which means that it carries a healthy and a ghost-like degree of freedom. Indeed, one can integrate-in a second scalar $\psi$ to lower the derivative order by replacing\footnote{The original action is then obtained by integrating-out $\psi$.}
\beq
\frac{2\al}{m^2} \( \bo \ph \)^2 \to -2\al \( \pa_{\mu} \psi \pa^{\mu} \ph + \frac{1}{2}\, m^2 \psi^2 \) \, ,
\eeq
and then diagonalize the $\ph, \psi$ kinetic sector to find there is a ghost\footnote{This is why any other kinetic term than $F_{\mu\nu}F^{\mu\nu}$ for a vector field implies a ghost by the way.}. 

As for the limit $m \to 0$, the cases $\al = 0$ and $\al \neq 0$ must be considered separately as always. In the former case we have that $A_{\mu}$ decouples, while we still have terms $\sim \pa h \pa \ph$. We must thus diagonalize the $h_{\mu\nu}$ and $\ph$ kinetic sectors by redefining
\beq
h'_{\mu\nu} = h_{\mu\nu} - \frac{2}{d-1}\, \et_{\mu\nu} \ph \, ,
\eeq
to get
\beq
S = \int \ed^D x \[ \frac{1}{2}\, h'_{\mu\nu} {\cal E}^{\mu\nu\ro\si} h'_{\ro\si} - \frac{1}{2}\, F_{\mu\nu} F^{\mu\nu}   - \frac{2d}{d-1}\, \pa_{\mu} \ph \pa^{\mu} \ph + h'_{\mu\nu} T^{\mu\nu} + \frac{2}{d-1}\, \ph T \] \, .  \nn
\eeq 
We see that although $A_{\mu}$ has totally decoupled, the scalar $\ph$ remains coupled to the source and is gauge-invariant under (\ref{eq:StuckU1}). Thus, $\ph$ still interacts with the system and this is the way the vDVZ discontinuity manifests itself in this formalism. For the $\al \neq 0$ case, there is no U$(1)$ gauge symmetry in the equation of $A_{\mu}$ to begin with, so the latter already represents the $2d$ degrees of freedom that are activated by the mass. We therefore do not need to introduce the St\"uckelberg scalar and can take the $m \to 0$ limit at the level of the $h_{\mu\nu}, A_{\mu}$ action (\ref{eq:AhAStuck}), to get that $A_{\mu}$ decouples, leaving us with the massless theory for $h_{\mu\nu}$.

\section{Non-local formulation} \label{sec:nonlocformu}

Another advantage of the St\"uckelberg formalism is that it can serve as an intuitive starting point for constructing non-local gauge theories. Here we follow closely the procedure introduced in \cite{Dvali, DHK} and also used in our paper \cite{JaccardMaggioreMitsou2}.

\subsection{Spin 1}

Let us start by solving in a causal way the equation of motion of the St\"uckelberg field $\ph$ (\ref{eq:AstuEOM})
\beq \label{eq:Astusol}
\ph = \ph^{\rm hom} - m \bo_{\rm r}^{-1} \pa_{\mu} A^{\mu} \, ,
\eeq
where $\ph^{\rm hom}$ is a homogeneous solution $\bo \ph^{\rm hom} = 0$ and $\pa_{\mu} A^{\mu}$ must have finite past for this equation to make sense. For notational simplicity, unless specified otherwise, from now on we will only write ``$\bo^{-1}$'' to denote the retarded inversion of $\bo$.

Since we know that $\pa_{\mu} A^{\mu}$ is not physical, demanding that it has finite past is not too much of a restriction. It would have been way more dramatic if we imposed this condition on all of $A_{\mu}$, because this would exclude free wave-packet solutions since these extend arbitrarily far into the past. We can now proceed and plug (\ref{eq:Astusol}) inside the equation for $A_{\mu}$ to get
\beq \label{eq:intoutph}
\pa_{\mu} F^{\mu\nu} - m^2 {\cal P}^{\nu}_{\,\,\,\mu} A^{\mu} = - j'^{\nu} \, , 
\eeq
where we have a new conserved source
\beq \label{eq:Asourcep}
j'^{\mu} \equiv j^{\mu} - m\, \pa^{\mu} \ph^{\rm hom} \, , \hspace{1cm} \pa_{\mu} j'^{\mu} = 0 \, ,
\eeq
and we have defined the operator
\beq \label{eq:PMinkA}
{\cal P}_{\mu}^{\,\,\,\nu} \equiv \de_{\mu}^{\nu} - \pa_{\mu} \bo^{-1} \pa^{\nu} = \de_{\mu}^{\nu} - \bo^{-1} \pa_{\mu} \pa^{\nu} \, ,
\eeq
which has the following nice properties. It is a projector
\beq
{\cal P}_{\mu}^{\,\,\,\ro} {\cal P}_{\ro}^{\,\,\,\nu} = \de_{\mu}^{\nu} - 2 \pa_{\mu} \bo^{-1} \pa^{\nu} + \pa_{\mu} \bo^{-1} \bo \bo^{-1} \pa^{\nu} = {\cal P}_{\mu}^{\,\,\,\nu} \, ,
\eeq
where we have used the fact that $\bo^{-1}$ is a right inverse of $\bo$, the projected field $A^{\rm T}_{\mu} \equiv {\cal P}_{\mu}^{\,\,\,\nu} A_{\nu}$ is $D$-transverse\footnote{This is not a surprise since the right-hand side of (\ref{eq:intoutph}) is transverse.}
\beq
\pa^{\mu} A^{\rm T}_{\mu} = \pa^{\mu} A_{\mu} - \bo \bo^{-1} \pa^{\nu} A_{\nu} = 0 \, ,
\eeq
and, under a gauge transformation (\ref{eq:gsA}) where the gauge parameter $\te$ has finite past, varies as
\beq \label{eq:ATvar}
\de A^{\rm T}_{\mu} = -\pa_{\mu} \te + \pa_{\mu} \bo^{-1}\bo \te = 0 \, .
\eeq
Indeed, since $\bo^{-1}$ acts on $\bo \te$, it only makes sense for $\bo \te$ with finite past, which implies that $\te$ has finite past and also that $\bo^{-1} \bo = {\rm id}$. This condition on the gauge parameter is reminiscent of the condition we encountered on the initial conditions of the gauge parameter on de-Sitter space-time. Again, this does not exclude the possibility of using $\te$ to neutralize a field mode, so it does not diminish the gauge symmetry in any sense. We thus have that $A^{\rm T}_{\mu}$ is gauge-invariant for all practical purposes.  

Going back at (\ref{eq:intoutph}) we see that we have reached a gauge invariant description of massive electrodynamics with no extra field, but at the price of non-locality. This may a priori sound a bit surprising because we know that this non-local theory is equivalent to a local one. This means that the physics of (\ref{eq:intoutph}) cannot be non-local, i.e. the prediction of the value of some physical observable at $x$ should still only depend on the data in its infinitesimal past light-cone neighbourhood. This is indeed the case because by going to the Lorentz gauge $\pa_{\mu} A^{\mu} = 0$ the equations become local. Thus, non-locality is only an artefact of explicit gauge-invariance and actually affects only the pure-gauge modes. The mass term can therefore be understood as the obstruction to having simultaneously both manifest locality and gauge-invariance.

\subsubsection{Where are the degrees of freedom?}

Let us now try to count the degrees of freedom using (\ref{eq:intoutph}). We choose the Lorentz gauge $\pa_{\mu} A^{\mu} = 0$ so that we retrieve the equation of motion of Proca theory (\ref{eq:ProcaEOM}), but with $j'^{\mu}$ instead of $j^{\mu}$, i.e. we have the homogeneous solution of $\ph$ that is still around. This amounts to as many different sources as $\ph$ has initial data, so we might be worried that our non-local trick might have inserted additional degrees of freedom into the system. Of course there is no miracle, and $\ph^{\rm hom}$ is eliminated by the residual gauge symmetry one has in the St\"uckelberg formalism. Indeed, the equations being
\beq
\( \bo - m^2 \)A_{\mu} = -j_{\mu} + m\,\pa_{\mu} \ph^{\rm hom} \, , \hspace{1cm}  \pa_{\mu} A^{\mu} = 0 \, ,
\eeq
we can transform with $\te$ such that $\bo \te = 0$ to get
\beq
\( \bo - m^2 \)A_{\mu} + m^2 \pa_{\mu} \te = - j_{\mu} + m\,\pa_{\mu} \ph^{\rm hom} \, , \hspace{1cm}  \pa_{\mu} A^{\mu} = 0 \, .
\eeq
Since $\bo\ph^{\rm hom} = 0$ as well, we can choose $\te = m^{-1}\ph^{\rm hom}$ and retrieve Proca theory exactly. Indeed, remember from section \ref{sec:StuckA} that in the $\pa_{\mu} A^{\mu} = 0$ gauge, $\ph$ cannot represent the longitudinal mode because it is massless $\bo \ph = 0$, so fully gauge-fixing can only result in the unitary gauge $\ph = 0$. This shows us that we could have avoided keeping track of $\ph^{\rm hom}$ in the above computations since at the end of the day this ``freedom'' is pure-gauge. In the St\"uckleberg formalism if we set $\pa_{\mu} A^{\mu} = 0$, then we still have a residual gauge-symmetry. In the non-local formalism with $\ph^{\rm hom} = 0$ if we set $\pa_{\mu} A^{\mu} = 0$ we have the Proca equations and thus no residual gauge symmetry.

Nevertheless, we also saw in section \ref{sec:StuckA} that if we rather choose the gauge $\pa_{\mu} A^{\mu} = - m \ph$, then $\ph$ obeys $(\bo-m^2)\,\ph = 0$, so its homogeneous solution could be interpreted as carrying the longitudinal degrees of freedom of the theory. However, here if $\ph$ were to carry the plane wave solutions of the longitudinal mode, then the gauge choice $\pa_{\mu} A^{\mu} = - m \ph$ would not be admissible because $\pa_{\mu} A^{\mu}$ would not have finite past.

We therefore conclude that the St\"uckelberg fields cannot represent the mode that is activated by the mass in this non-local formulation and thus one can safely set $\ph^{\rm hom} = 0$. From now on $j'^{\mu} = j^{\mu}$ and we will also neglect the homogeneous solutions when integrating-out the St\"uckelberg fields in the spin-2 case. Indeed, there too the homogeneous solutions of the St\"uckelberg fields will be massless so that they cannot represent the dynamical fields of the theory. They ultimately correspond to the residual gauge freedom of  the St\"uckelberg formalism.

\subsubsection{Filtered response to linear sources}

The non-local equation of motion (\ref{eq:intoutph}), although quite elegant, can be simplified even more if we restrict to the case where all of $A_{\mu}$ has finite past and thus so does $j_{\mu}$. This is the case where one is interested in the production of electromagnetic waves by a source with finite past, i.e. when any radiation at future infinity is entirely due to $j^{\mu}$. Then, one can write
\beq
A_{\mu} = \bo^{-1} \bo A_{\mu} \, ,
\eeq
so that (\ref{eq:intoutph}) reads
\beq \label{eq:nonlocA}
\( 1 - \frac{m^2}{\bo} \) \pa_{\mu} F^{\mu\nu} = -j^{\nu} \, .
\eeq
In this particular case, we have access to a new interpretation of the mass term as a high-pass filter \cite{Dvali,DHK,Hinterbichler1,Hinterbichler2}. Indeed, going to ``Fourier space''\footnote{This is actually not really possible for the time coordinate since $A_{\mu}$ will in general not vanish at future infinity because of the waves generated by the source. One should rather use a Laplace transform for $t$ since the support of $A_{\mu}$ is bounded in the past.} and neglecting the pole contour prescription, we have
\beq
-\( 1 + \frac{m^2}{k^2} \) i k_{\mu} F^{\mu\nu} = - j^{\nu} \, ,
\eeq   
which can be inverted to give
\beq
k_{\mu} F^{\mu\nu} = -\frac{i k^2}{m^2 + k^2}\, j^{\nu} \, .
\eeq   
Now the left-hand side is the kinetic term of ordinary massless electrodynamics, but the source is multiplied by a filter which modulates its intensity. Indeed, for $k^2 \ll m^2$, i.e. for high frequencies and large wave-lengths, the source of $A_{\mu}(k)$ becomes $\sim k^2$. This is the degravitation analogue for electrodynamics, which ``screens'' the large scale behaviour of the source \cite{Dvali, DHK}. 

It is important to stress one more time that equation (\ref{eq:nonlocA}) is valid only when studying the response to an external source. More precisely, (\ref{eq:nonlocA}) only makes sense if $\pa_{\mu} F^{\mu\nu}$ has finite past, which excludes ingoing radiation at past infinity since that radiation does not obey $\pa_{\mu} F^{\mu\nu} = 0$ because of the mass. Therefore, (\ref{eq:nonlocA}) cannot be taken as a classical model covering every feature of massive electrodynamics. For a full description of the theory, with the constraint of past infinity applying only on non-dynamical fields (here $\pa_{\mu} A^{\mu}$), one needs to consider (\ref{eq:intoutph}).

\subsubsection{Propagators using projectors}

The computation of the propagator in a massive but yet gauge-invariant setting is very instructive, especially in the light of this projector formalism. We can first rewrite (\ref{eq:nonlocA}) as
\beq
\( \bo - m^2 \) {\cal P}_{\mu}^{\,\,\,\nu} A_{\nu} = - j_{\mu} \, ,
\eeq
so that the operator which must be inverted is
\beq
{\cal K}^{\mu\nu} = \( \bo - m^2 \) {\cal P}^{\mu\nu} \, .
\eeq
As in the massless case, the gauge invariance of the equation is reflected in the fact that $\cal K$ is proportional to a projector. It gives zero on pure-gauge modes, which means a non-trivial kernel, which means that it is not uniquely invertible. In section \ref{sec:propagator} we have used the standard method for inverting such operators, which is to introduce a gauge-fixing term that will not affect the saturated propagator. In the spirit of the projector formalism developed here, there is actually a natural way of privileging an inverse that is also easily computable. Indeed, we can note that the space in which ${\cal K}_{\mu\nu}$ lives is the space of transverse operators and that ${\cal P}_{\mu}^{\nu}$ is the identity element. Thus, as long as we restrict to this subspace, the inversion relation becomes
\beq
{\cal K}^{\mu\ro} D_{\ro\nu} = i {\cal P}_{\nu}^{\mu} \, ,
\eeq
and admits a unique transverse inverse (up to the homogeneous solution/initial conditions ambiguity)
\beq \label{eq:propprojA}
D_{\mu\nu} = -\frac{i}{k^2 + m^2} \( \et_{\mu\nu} - \frac{k_{\mu} k_{\nu}}{k^2} \) \, .
\eeq
Not surprisingly, in the massless case this corresponds to the Landau gauge $\xi = 0$ in (\ref{eq:propm0A}). This is the only choice that cannot be expressed through a gauge fixing term (\ref{eq:gfA}) precisely because it is the only choice which imposes transversality $\pa_{\mu} A^{\mu} = 0$, instead of breaking it. In any case, as already noted, since the source is conserved the physically relevant term is the one with no uncontracted $k_{\mu}$'s. In the spin-2 case however, there will be a whole one-parameter family of transverse operators, so this construction will be very useful.

\subsection{Spin 2}

The equations of motion of (\ref{eq:ShStuck}) are
\bea
{\cal E}_{\mu\nu\ro\si} h^{\ro\si} - m^2 \( h_{\mu\nu} - (1+\al)\et_{\mu\nu} h \)  & = & -T_{\mu\nu} + 2m \( \pa_{(\mu} A_{\nu)} - (1+\al)\et_{\mu\nu} \pa_{\ro} A^{\ro} \) \label{eq:StuhhEOM} \nn \\
 & & + 2\( \pa_{\mu} \pa_{\nu} \ph - (1+\al)\et_{\mu\nu} \bo \ph \) \, , \\
\pa_{\mu} F^{\mu\nu} - 2\al \pa^{\nu} \pa_{\mu} A^{\mu} & = & - m j^{\nu} + \frac{2\al}{m}\, \pa^{\nu} \bo \ph \, , \label{eq:StuhAEOM}\\
\al \bo^2 \ph & = & \frac{m^2}{2} \pa_{\mu} j^{\mu} - \al m \bo \pa_{\mu} A^{\mu} \, , \label{eq:StuhfiEOM}
\eea
for $h_{\mu\nu}$, $A_{\mu}$ and $\ph$, respectively, and we find convenient to define the quantity
\beq
j^{\nu} \equiv \pa_{\mu} h^{\mu\nu} - (1+\al)\pa^{\nu} h \, .
\eeq
Again, note that each one of these equations is the divergence of the previous one. For $\al \neq 0$, we can solve for $\ph$
\beq
\ph = \frac{m^2}{2\al} \, \bo^{-2} \pa_{\mu} j^{\mu} - m \bo^{-1}\pa_{\mu} A^{\mu}  \, ,
\eeq
where, as anticipated in the spin-1 case, the homogeneous solution $\bo^2 \ph^{\rm hom} = 0$ can be safely set to zero since it cannot represent a massive mode and is thus ultimately pure-(residual)gauge. Remember that this expression for $\ph$ makes sense only if $\pa_{\mu} j^{\mu}$ and $\pa_{\mu} A^{\mu}$ have finite past. Plugging this inside the equation of $A_{\mu}$ we get
\beq \label{eq:StuhAEOMfiintout}
\pa_{\mu} F^{\mu\nu} = - m {\cal P}^{\nu}_{\,\,\,\mu} j^{\mu} \, ,
\eeq
where every term is independently transverse. Now this equation is gauge-invariant so we must fix the gauge in order to solve it. We choose $\pa_{\mu} A^{\mu} = 0$, invert $\bo$ and then add a pure-gauge term to get the general solution. This gives, setting again to zero any homogeneous solution,
\beq \label{eq:StuckhAsol}
A_{\mu} = - m \bo^{-1} {\cal P}_{\mu}^{\,\,\,\nu} j_{\nu}+ \pa_{\mu} \te \, . 
\eeq
To perform this inversion we now also need $j^{\mu}$ to have finite past, not just its divergence. This is again ok because $j^{\mu}$ does not represent dynamical fields since it is actually zero in the original formulation (\ref{eq:divFPEOM}). Plugging the solution of $A_{\mu}$ in the one of $\ph$ we get
\beq \label{eq:solphAintout}
\ph = \frac{m^2}{2\al} \, \bo^{-2} \pa_{\mu} j^{\mu}  - m \te \, .
\eeq
where we have used the fact that $\te$ has finite past since $\pa_{\mu} A^{\mu} = \bo \te$ has finite past. Now that both $A_{\mu}$ and $\ph$ are expressed in terms of $h_{\mu\nu}$ we can plug them in the equation of motion of the latter to get
\beq \label{eq:nonlocalno0}
{\cal E}_{\mu\nu\ro\si} h^{\ro\si} - m^2 {}_{\al}{\cal P}_{\mu\nu\ro\si} h^{\ro\si} = -T_{\mu\nu}  \, , 
\eeq
where
\bea
_{\al} {\cal P}_{\mu\nu}^{\,\,\,\,\,\,\ro\si} & \equiv & \de_{(\mu}^{\ro} \de_{\nu)}^{\si} - (1+\al)\, \et_{\mu\nu} \et^{\ro\si} \( 1 - \frac{1+\al}{\al}\, \bo^{-1} \bo \) - \( \de_{(\mu}^{\ro} \pa_{\nu)} \bo^{-1} \pa^{\si} + \de_{(\mu}^{\si} \pa_{\nu)} \bo^{-1} \pa^{\ro} \) \nn \\
 & &  + (1 + \al) \,\et^{\ro\si} \[ 2 \pa_{(\mu} \bo^{-1} \pa_{\nu)} - \frac{1+2\al}{\al}\, \pa_{\mu} \pa_{\nu} \bo^{-1} \] - \frac{1+\al}{\al}\, \et_{\mu\nu} \bo^{-1} \pa_{\ro} \pa_{\si}  \nn \\
 & &  + \frac{1+2\al}{\al} \, \pa_{\mu} \pa_{\nu} \bo^{-2} \pa^{\ro} \pa^{\si}  \, . \label{eq:unsimpproj}
\eea
Although we have expressed this such that $\bo^{-1}$ acts separately on $\bo h$ and $\pa_{\mu} \pa_{\nu} h^{\mu\nu}$, this requires only that $j^{\mu}$ has finite past to converge. Given the complexity of this structure, here we will directly focus on the case where all of $h_{\mu\nu}$, and thus $T_{\mu\nu}$, has finite past, so that we can commute all these operators at will. The result is then very elegant since it can be expressed in terms of the vector projectors
\beq
_{\al} {\cal P}_{\mu\nu}^{\ro\si} = {\cal P}_{(\mu}^{\ro} {\cal P}_{\nu)}^{\si} + \frac{1+\al}{\al} \,{\cal P}_{\mu\nu}{\cal P}^{\ro\si} \, , 
% & = & \de_{(\mu}^{\ro} \de_{\nu)}^{\si} + \frac{1+\al}{\al}\, \et_{\mu\nu} \et^{\ro\si} - \frac{1}{\bo} \( \de_{(\mu}^{\ro} \pa_{\nu)}\pa^{\si} + \de_{(\mu}^{\si} \pa_{\nu)}\pa^{\ro} \) \nn \\
% & & - \frac{1+\al}{\al \bo} \( \et_{\mu\nu} \pa^{\ro}\pa^{\si} + \et^{\ro\si} \pa_{\mu}\pa_{\nu} \) +  \frac{1+2\al}{\al} \frac{1}{\bo^2} \, \pa_{\mu} \pa_{\nu} \pa^{\ro} \pa^{\si}  \, , 
\eeq
As anticipated earlier, here we have that $_{\al} {\cal P}$ is a one-parameter family of operators making the tensor on which they act transverse
\beq
\pa^{\mu} {}_{\al} {\cal P}_{\mu\nu}^{\ro\si} h_{\ro\si} = 0  \, .
\eeq
and also gauge-invariant under (\ref{eq:gsh}) for $\xi_{\mu}$ with finite past. It is convenient to switch to another parametrization, namely 
\beq
a = 1 + d\(1 + 1/\al\) \, , \hspace{1cm} \al = \frac{d}{a - d - 1} \, ,
\eeq
and define
\bea
_a {\cal P}_{\mu\nu}^{\ro\si} & \equiv & _0 {\cal P}_{\mu\nu}^{\ro\si} + a\, {}_s {\cal P}_{\mu\nu}^{\ro\si}  \\
 & = & \de_{(\mu}^{\ro} \de_{\nu)}^{\si} - \frac{1-a}{d}\, \et_{\mu\nu} \et^{\ro\si} - \frac{1}{\bo} \( \de_{(\mu}^{\ro} \pa_{\nu)}\pa^{\si} + \de_{(\mu}^{\si} \pa_{\nu)}\pa^{\ro} \) \nn \\
 & & + \frac{1-a}{d \bo} \( \et_{\mu\nu} \pa^{\ro}\pa^{\si} + \et^{\ro\si} \pa_{\mu}\pa_{\nu} \) +  \( 1  - \frac{1-a}{d} \) \frac{1}{\bo^2} \, \pa_{\mu} \pa_{\nu} \pa^{\ro} \pa^{\si} \, , \label{eq:aprojfam}
\eea
where
\beq \label{eq:sPdef}
_0 {\cal P}_{\mu\nu}^{\ro\si} \equiv {\cal P}_{(\mu}^{\ro} {\cal P}_{\nu)}^{\si} - \frac{1}{d}\, {\cal P}_{\mu\nu} {\cal P}^{\ro\si} \, , \hspace{1cm} {}_s {\cal P}_{\mu\nu}^{\ro\si} \equiv  \frac{1}{d}\, {\cal P}_{\mu\nu} {\cal P}^{\ro\si} \, .
\eeq
To avoid confusing ${}_s {\cal P}$ with ${}_a {\cal P}$ where $a = s$, let us stress that the letter ``$s$'' will be exclusively used in order to denote the second operator in (\ref{eq:sPdef}). Now observe that $_0 {\cal P}$ and $_s {\cal P}$ are orthogonal projectors 
\beq
{}_0{\cal P}^2 = {}_0{\cal P} \, , \hspace{1cm} {}_s{\cal P}^2 = {}_s{\cal P} \, , \hspace{1cm} {}_0 {\cal P} {}_s {\cal P} = 0 \, ,
\eeq
on the subspaces of transverse-traceless and transverse-pure-trace tensors, respectively. Indeed, $_0 {\cal P}^{\mu}_{\,\,\,\mu\ro\si} h^{\ro\si} = 0$ so the latter is also invariant under linearized local conformal transformations
\beq \label{eq:gsct}
\de h_{\mu\nu} = \et_{\mu\nu} \te \, ,
\eeq
for $\te$ with finite past. The obvious advantage of the $a$ parametrization is that now the linear combination and product of two such operators follow the simple rules
\beq \label{eq:Paverprod}
\al \,{}_a {\cal P} + \be\, {}_b {\cal P} = \(\al + \be\) {}_{\frac{\al a+\be b}{\al+\be}} {\cal P} \, , \hspace{1cm} {}_a {\cal P} - {}_b {\cal P} = \( a - b \){}_s {\cal P} \, , \hspace{1cm}  {}_a {\cal P} \, {}_b {\cal P} = {}_{ab} {\cal P} \, ,
\eeq
so ${}_a {\cal P}$ is not a projector unless $a = 0$ or $1$. In the latter case, we have the projector on the subspace of transverse tensors $_1 {\cal P} = {}_0 {\cal P} + {}_s {\cal P}$. Thus, $_1 {\cal P}$, $_0 {\cal P}$ and $_s {\cal P}$ are the identity elements of the space on which they project. 

In terms of $\al$ the choice $a = 0$ corresponds to $\al = - d/(d+1)$, which is the value for which the mass of the ghost (\ref{eq:mghost}) vanishes. Indeed, since the ghost is the trace $h$, it is consistent that the mass term in that case is traceless. Interestingly enough, the projector $a = 1$ corresponds to the value $\al = -1$. From now on, every time we assign a numerical value to the argument of $\cal P$ it will be with respect to the ``$a$'' parametrization (\ref{eq:aprojfam}).

Now note that the Lichnerowicz operator (\ref{eq:Lichne}) takes the form ${\cal E} = \bo\, {}_{1-d} {\cal P}$, which corresponds to $\al = -1/2$. Indeed, this is the only ${\cal P}$ that has no $\sim \bo^{-2}$ term, so it is the only case where $\bo {\cal P}$ is a local second-order transverse operator. Therefore, in the case $\al = -1/2$, we can rewrite the equation in a compact form analogous to (\ref{eq:nonlocA})
\beq \label{eq:nonlochnFP}
\( 1 - \frac{m^2}{\bo} \){\cal E}_{\mu\nu\ro\si} h^{\ro\si} = -T_{\mu\nu} \, , \hspace{1cm} \al = -\frac{1}{2}  \, ,
\eeq
which is the result found in \cite{Dvali, DHK, JaccardMaggioreMitsou2}, \footnote{Note that in \cite{Dvali, DHK} the authors erroneously concluded that this theory propagates only the $d$-tensor part of $h_{\mu\nu}$, i.e. it has the same dynamical content as the massless theory, because it has the same tensor structure (adding a gauge-fixing term and inverting one finds that the saturated propagator is indeed (\ref{eq:al12satprop})). Their argument is that one has precisely integrated-out the St\"uckelbergs which correspond to the $d$-vector and $d$-scalar modes, so that the latter do not appear in this equation. As we have seen, this is not true because the St\"uckelbergs do not represent the dynamical fields that are activated by the mass. Moreover, it is not the tensor structure of the propagator alone which determines the dynamical content, otherwise the latter would be the same in massless and massive electrodynamics. As we have also seen, the presence of the mass is important, because it will affect the conservation equation of the source in Fourier space. Indeed, as we pointed out in \cite{JaccardMaggioreMitsou2}, by expressing the saturated propagator (\ref{eq:al12satprop}) in terms of the harmonic variables of the conserved sources, we get (\ref{eq:hlocpropM}) with $M$ having both a positive and a negative eigenvalue (the ghost pole). We then have that $M \to 0$ as $m_{\rm s} \to m \to 0$ so that we have no vDVZ discontinuity, as expected. However, for $m \neq 0$, all the independent components of the source are present and thus so are all the dynamical fields of the local theory.}. Not surprisingly, for this value of $\al$ we also have that, according to (\ref{eq:mghost}),
\beq
m_{\rm ghost}^2 = - m^2 \, ,
\eeq
so that the ghost mode is also a tachyon with the same magnitude of mass as the spin-2 modes. To understand why this happens, note that the differential operator corresponding to this equation is
\beq
{\cal K}^{\mu\nu\ro\si} = \( \bo - m^2 \) {}_{1-d} {\cal P}^{\mu\nu\ro\si} \, .
\eeq
Since it is transverse but not traceless, the appropriate identity for the inversion is
\beq
{\cal K}^{\mu\nu\al\be} D_{\al\be\ro\si} = i\, {}_1 {\cal P}^{\mu\nu}_{\ro\si} \, , 
\eeq
and thus, using the product rule (\ref{eq:Paverprod}) the propagator is trivial to compute
\beq
D_{\mu\nu\ro\si} = - \frac{i}{k^2 + m^2}\, {}_{\frac{1}{1-d}} {\cal P}_{\mu\nu\ro\si}  \, .
\eeq
We see that, because ${\cal E} \sim {\cal P}$, all the poles are at $k^2 = -m^2$, with the ghost mode having the wrong overall sign, but the same magnitude for the mass. Conversely, this is why the rest of the $\al \neq 0$ cases cannot be written as $(\bo - m^2) {}_a {\cal P}$ for some $a$, because the mass of the ghost is not $m$ any more. 

To conclude the $\al \neq 0$ case (\ref{eq:nonlocalno0}), note that in the $m \to 0$ limit we are left with the massless theory. Thus, as expected, there is no discontinuity. Moreover, as in the spin-1 case, the non-locality is ``pure-gauge'' since one can fix the gauge
\beq
\pa_{\mu} \( h^{\mu\nu} - (1+\al) \et^{\mu\nu} h \) = 0 \, ,
\eeq
which remember is possible for $\al \neq 0$, to get 
\beq
{\cal P}_{\mu\nu\ro\si} h^{\ro\si} = h_{\mu\nu} - (1+\al) \et_{\mu\nu} h \, ,
\eeq
and thus the local equation we started with.

\subsubsection{Fierz-Pauli point}

We now pass to the Fierz-Pauli case. We can first observe that the value $\al = 0$ corresponds to a diverging $a$ so that the $\cal P$ operators are not well defined in this limit. However, one should note that now the action is linear in $\ph$ and its equation of motion (\ref{eq:StuhfiEOM}) is
\beq
\pa_{\mu} j^{\mu} \equiv \pa_{\mu} \pa_{\nu} h^{\mu\nu} - \bo h = 0 \, ,
\eeq
to which we will refer as the ``scalar equation''. For $h_{\mu\nu}$ with finite past this is equivalent to $_s {\cal P} \cdot h = 0$, so if the scalar equation holds then $_a {\cal P} \cdot h = {}_0 {\cal P} \cdot h$ and we may still use the projectors. Since now $\pa_{\mu} j^{\mu} = 0$, the equation of motion of $A_{\mu}$ (\ref{eq:StuhAEOM}) has a transverse right-hand side and can be solved as before. The result is then plugged inside (\ref{eq:StuhhEOM}) and $\te$ simply redefines $\ph$ again. In order to determine the latter, we can then take the trace of that equation and isolate $\ph$, to get
\beq
\ph = -\frac{1}{2}\, \bo^{-1} \[ m^2 h + \frac{1}{d}\, T \] \, ,
\eeq
where we have used $\pa_{\mu} j^{\mu} = 0$ and have put to zero the homogeneous solution since it is massless. Plugging this back inside the equation we get the following system
\bea
{\cal E}_{\mu\nu\ro\si} h^{\ro\si} - m^2 {}_0 {\cal P}_{\mu\nu\ro\si} h^{\ro\si} & = & -T^{\rm TT}_{\mu\nu} \, , \label{eq:EPhFP} \\
\pa_{\mu} \pa_{\nu} h^{\mu\nu} - \bo h & = & 0 \, .
\eea
where now the source has changed and is actually the traceless-transverse part of $T_{\mu\nu}$
\beq
T^{\rm TT}_{\mu\nu} \equiv T_{\mu\nu} - \frac{1}{d} \( \et_{\mu\nu} T - \frac{\pa_{\mu} \pa_{\nu}}{\bo}\, T \) \equiv {}_0 {\cal P}_{\mu\nu}^{\ro\si} T_{\ro\si} \, , 
\eeq
thus satisfying
\beq
\pa^{\mu} T^{\rm TT}_{\mu\nu} = 0 \, , \hspace{1cm} T^{\rm TT} = 0 \, .
\eeq
Now note that the scalar equation is just the trace of (\ref{eq:EPhFP}), so that it is not independent and can be dropped. This might appear disturbing because then we are left with the left-hand side of the theory $a = 0$, which is {\it not} the Fierz-Pauli one $\al = 0$, and the corresponding propagator thus has an extra ghost pole. However, when we saturate it with $T^{\rm TT}_{\mu\nu}$ we retrieve indeed the saturated Fierz-Pauli propagator in terms of $T_{\mu\nu}$. Thus, in this formulation the modification of the source is very relevant. The fact that the Fierz-Pauli theory has one less dynamical field is now reflected in the fact that $h_{\mu\nu}$ ``sees'', and thus propagates, one less component of the source. Another advantage of this formulation is that now the reason for the vDVZ discontinuity at $\al = 0$ is obvious, the source remains $T^{\rm TT}$ as $m \to 0$. 

Another option, is to keep the scalar equation and use it to have $_a {\cal P} \cdot h = {}_0 {\cal P} \cdot h$ and thus ${\cal E} \cdot h = \bo {}_0 {\cal P} \cdot h$, to finally get the following system
\bea
\( 1 - \frac{m^2}{\bo} \){\cal E}_{\mu\nu\ro\si} h^{\ro\si} & = & - T^{\rm TT}_{\mu\nu} \, , \\
\pa_{\mu} \pa_{\nu} h^{\mu\nu} - \bo h & = & 0 \, . \label{eq:nonlocFP}
\eea
The first equation is precisely what we have found for the $\al = -1/2$ case (\ref{eq:nonlochnFP}), but now it is the additional scalar equation which makes the whole difference. It cannot be obtained through a gauge transformation and is responsible for killing the ghost. 

Again, since the theory we started with is local, non-locality can only be a pure-gauge effect, although this time this may be a bit less obvious to show because the source term is non-local as well. This is why the source must be part of the gauge-fixing condition
\beq
\pa_{\mu} h^{\mu\nu} = -\frac{1}{d m^2}\, \bo \pa^{\nu} T  \, .
\eeq
Indeed, with this the scalar equation becomes the equation fixing the trace (\ref{eq:FP3}) and, using this to express the source non-locality in terms of $h$, we can arrange the terms to get (\ref{eq:FP1}). Eq. (\ref{eq:FP2}) is then found by taking the divergence of (\ref{eq:FP1}) and using (\ref{eq:FP3}).

\subsubsection{Extra gauge symmetry}

Using again that all $_a {\cal P}$ act the same on $h_{\mu\nu}$, yet another interesting formulation of the Fierz-Pauli non-local equations (\ref{eq:nonlocFP}) is
\bea
\( \bo - m^2 \) {}_0 {\cal P}_{\mu\nu\ro\si} h^{\ro\si} & = & - T^{\rm TT}_{\mu\nu} \, , \\
\pa_{\mu} \pa_{\nu} h^{\mu\nu} - \bo h & = & 0 \, .
\eea
The advantage here is that the first equation is invariant under linearized local conformal transformations (\ref{eq:gsct}), and consistently traceless on both sides. However, this is not the case of the scalar equation. We can thus ``lift'' Fierz-Pauli theory to a non-local gauge theory with one more gauge symmetry
\beq \label{eq:nonlocFP2}
\( \bo - m^2 \) {}_0 {\cal P}_{\mu\nu\ro\si} h^{\ro\si} = - T^{\rm TT}_{\mu\nu} \, , 
\eeq
and now interpret the scalar equation as a gauge condition that is reached using (\ref{eq:gsct}) with
\beq
\te = - \frac{1}{d} \( h - \bo^{-1}\pa_{\mu} \pa_{\nu} h^{\mu\nu} \) \, .
\eeq
This is a very elegant result because now the ghost mode is also neutralized by a gauge symmetry. Indeed, in the spin-1 case we had $N_{\rm d} = d$ because there are $D$ fields, one gauge symmetry and no residual symmetry because of the mass. In the spin-2 case we have $D^2$ fields, $D$ gauge symmetries in general, so that we are left with $N_{\rm d} = D^2 - D$, except in the $\al = 0$ case where an extra gauge symmetry reduces that number by one.

Now the differential operator corresponding to (\ref{eq:nonlocFP2}) is
\beq
{\cal K}^{\mu\nu\ro\si} = \( \bo - m^2 \) {}_{0} {\cal P}^{\mu\nu\ro\si} \, .
\eeq
Since it is both transverse and traceless, the appropriate identity for the inversion is
\beq
{\cal K}^{\mu\nu\al\be} D_{\al\be\ro\si} = i \,{}_0 {\cal P}^{\mu\nu}_{\ro\si} \, , 
\eeq
and thus, using the product rule (\ref{eq:Paverprod}) the propagator reads
\beq
D_{\mu\nu\ro\si} = - \frac{i}{k^2 + m^2}\, {}_0 {\cal P}_{\mu\nu\ro\si}  \, .
\eeq 
Saturating it, one finds the Fierz-Pauli result, i.e. (\ref{eq:satprophm}) with $\al = 0$. This formulation provides us with yet another point of view on the vDVZ discontinuity. Indeed, in the massless theory we saw that the only projector for which $\bo {\cal P}$ is local is the $a = 1-d$ one. This gives $\sim {}_{\frac{1}{1-d}}{\cal P}$ for the propagator and the following tensor structure for the saturated one
\beq
\sim \et_{\mu(\ro} \et_{\si)\nu} - \frac{1}{d-1}\, \et_{\mu\nu} \et_{\ro\si} \, .
\eeq  
On the other hand, Fierz-Pauli theory, because of the extra gauge symmetry that is needed to kill the ghost in the non-local formulation, must have $_0 {\cal P}$ as its differential operator, and thus the tensor structure for the saturated propagator is
\beq
\sim \et_{\mu(\ro} \et_{\si)\nu} - \frac{1}{d}\, \et_{\mu\nu} \et_{\ro\si} \, .
\eeq

\subsection{New non-local theory} \label{sec:newmasstheor}

In the case of electrodynamics, the uniqueness of the projector makes the non-local formulation of Proca theory the only stable non-local theory of a massive vector field. In the tensor case, the presence of {\it two} independent projectors, $_0 {\cal P}$ and $_s {\cal P}$ defined in (\ref{eq:sPdef}), allows us to construct more healthy models than the ones that are obtained from local theories. In particular, as we will see in this thesis, one can construct a novel, genuinely non-local linear theory, that includes the trace scalar but with {\it no} ghost poles in the propagator. This is possible if we also modify non-locally the kinetic term, so it will not correspond to simply adding a non-local mass term to linearized GR. 

To construct that theory, we take full advantage of the projector formalism developed above to write an equation in which the tensor and scalar modes are diagonalized
\beq \label{eq:projeq}
\( \bo - m_g^2 \) {}_0 {\cal P}_{\mu\nu\ro\si} h^{\ro\si} + \( z \bo - m_s^2 \) {}_s {\cal P}_{\mu\nu\ro\si} h^{\ro\si} = -T_{\mu\nu} \, ,
\eeq
so that each one of them can have its own mass. The $z$ factor will be useful in tracking ghost-like behaviour. Now since by definition ${}_0 {\cal P}  + z \, {}_s {\cal P} \equiv {}_z {\cal P}$, the only case in which the kinetic part is local, and thus coincides with  linearized GR, is 
\beq \label{eq:GRkinterm}
z = 1-d \, .
\eeq
To study the stability and particle content of these theories let us compute the corresponding propagator. Because of the scalar sector we have that the differential operator
\beq
{\cal K}^{\mu\nu\ro\si} \equiv \( \bo - m_g^2 \) {}_0 {\cal P}_{\mu\nu\ro\si} + \( z \bo - m_s^2 \) {}_s {\cal P}_{\mu\nu\ro\si}  \, ,
\eeq
is transverse but not traceless, so that the appropriate identity element for the inversion is 
\beq
{\cal K}^{\mu\nu\al\be} D_{\al\be\ro\si} = i \,{}_1 {\cal P}^{\mu\nu\ro\si} \, ,
\eeq
and the solution is (using the product rule (\ref{eq:Paverprod}))
\beq
D_{\mu\nu\ro\si} = - \frac{i}{k^2 + m^2}\, {}_0 {\cal P}_{\mu\nu\ro\si} - \frac{i}{z k^2 + m^2} \, {}_s {\cal P}_{\mu\nu\ro\si} \, .
\eeq
Saturating it with conserved sources we get
\beq \label{eq:propaeqbmn0}
T^{*\mu\nu} D_{\mu\nu\ro\si} T'^{\ro\si} = - \frac{i}{k^2 + m_g^2} \( T^*_{\mu\nu} T'^{\mu\nu} - \frac{1}{d}\, T^* T' \) - \frac{1}{d}\,\frac{i}{z k^2 + m_s^2} \, T^* T'  \, , 
\eeq
which is the Fierz-Pauli propagator with mass $m_g$ plus a healthy scalar propagator, for $z > 0$, with mass $m_s/\sqrt{|z|}$. Thus, the first term in (\ref{eq:projeq}) describes the massive SO$(d)$-tensor modes, while the second term describes the massive trace mode. This is a remarkable advantage compared to local massive spin-2 theory, where that extra scalar can only be a ghost. In our formalism, instead of having to fight to kill that extra mode allowed by the diffeomorphism symmetry, we have the opportunity to simply let it participate in the dynamics since we can choose $z$ freely. Moreover, its mass is also free, instead of being determined by the one of the tensor modes. Note also that for $m_g \neq 0$ this is not a scalar-tensor theory, nor a bigravity theory in disguise, where the scalar or the second metric would have been integrated-out. Indeed, in scalar-tensor theories the graviton is not massive, while in bigravity theories there is also a massless graviton. 

We thus have that stability requires $z > 0$, as it could have been expected from (\ref{eq:projeq}). This means however that, if we want the kinetic term to be the one of GR (\ref{eq:GRkinterm}), then the scalar is a ghost. The exception is when both masses are zero, in which case that mode is neutralized by the residual gauge symmetry of linearized GR. Thus, as in Fierz-Pauli theory, continuity with GR at $m_i \to 0$ can only be achieved in the presence of a ghost. Conversely, any ghost-free massive theory will have a discontinuity, at the linearized level at least. 

This can be easily seen by considering the massless limit $m_g \to 0$ in the saturated propagator. So let us rewrite the latter as 
\bea 
T^{*\mu\nu} D_{\mu\nu\ro\si} T'^{\ro\si} & = & - \frac{i}{k^2 + m_g^2} \( T^*_{\mu\nu} T'^{\mu\nu} - \frac{1}{d-1}\, T^* T' \) \nn \\
 & & - \frac{1}{d(d-1)}\, \frac{i}{k^2 + m_g^2}\, T^* T' - \frac{1}{d}\,\frac{i}{z k^2 + m_s^2} \, T^* T'   \, , \label{eq:mto0newprop}
\eea
so that the first term reduces to the GR result in the $m_g \to 0$ limit. We see that we are left with the usual vDVZ discontinuity of the Fierz-Pauli propagator, representing the gauge-invariant combination of the two $d$-scalars in $h_{ij}$, plus the massive scalar mode. Taking also $m_s \to 0$, we see that only in the case (\ref{eq:GRkinterm}) does one obtain linearized GR, but then the massive theory has a ghost. 

There is however an important difference with FP theory regarding that discontinuity. Here the discontinuity is already visible at the level of the equations of motion (\ref{eq:projeq}), since we do not retrieve the massless local equations in the $m_g ,m_s \to 0$ limit, for $z \neq 1-d$. On the other hand, in FP theory the action tends to the massless one in the $m \to 0$ limit. The reason for this difference is the presence of projectors, and thus gauge-invariance. Indeed, thanks to the projectors the tensor structure ${\cal K}^{\mu\nu\ro\si}$ in the equations of motion (\ref{eq:projeq}) is identical\footnote{Up to Klein-Gordon operators.} to the structure of the propagator (\ref{eq:propaeqbmn0}). Because of this, any discontinuity in the latter must also arise in the former. In FP theory on the other hand, the tensor structure ${\cal K}^{\mu\nu\ro\si}$ in the action and the one in the propagator $D_{\mu\nu\ro\si}$ are not at all the same and one can thus have a discontinuity in the latter that does not show up in the former.

\subsubsection{Genuine non-locality}

Let us now try to turn (\ref{eq:projeq}) into a system of local equations by fixing the gauge. The choice which makes the $_a {\cal P}$ operator local and involves only local operators is
\beq \label{eq:gfaz}
\pa_{\mu} \( h^{\mu\nu} - \frac{1-a}{D-a}\, \et^{\mu\nu} h \) = 0 \, ,
\eeq
which is accessible since $(1-a)/(D-a) \neq 1$. For generic masses $m_g$ and $m_s$ this gauge does not make the equation local, whatever the choice of $a$, so the system is genuinely non-local. The only exception is when $m^2_s = z m^2_g \equiv z m^2$ because then (\ref{eq:projeq}) can be expressed in terms of a single $\cal P$ operator
\beq
\( \bo - m^2 \) {}_z {\cal P}_{\mu\nu\ro\si} h^{\ro\si} = -T_{\mu\nu} \, ,
\eeq
and we can fix the (\ref{eq:gfaz}) gauge with $a = z$ to get the local system
\bea
\( \bo - m^2 \) \( h_{\mu\nu} - \frac{1-z}{D-z}\, \et_{\mu\nu} h \) & = & - T_{\mu\nu} \, , \label{eq:mz1} \\
\pa_{\mu} \( h^{\mu\nu} - \frac{1-z}{D-z}\, \et^{\mu\nu} h \) & = & 0 \, . \label{eq:mz2}
\eea
This is reminiscent of the situation in local massive spin-2 equations, because (\ref{eq:mz2}) looks like the divergence of (\ref{eq:mz1}). Upon close inspection however, we observe that the analogy does not hold because here the divergence of (\ref{eq:mz1}) implies that $\pa_{\mu} h^{\mu\nu} - \frac{1-z}{D-z}\, \pa^{\nu} h$ is a free dynamical field, not zero. Because of this, these equations do {\it not} derive from the local action
\beq
S = \int \ed^D x \[ \frac{1}{2}\, h_{\mu\nu}\( \bo - m^2 \)  \( h^{\mu\nu} - \frac{1-z}{D-z}\, \et^{\mu\nu} h \) + h_{\mu\nu} T^{\mu\nu} \] \, ,
\eeq 
which describes an obviously unstable theory since it does not have the GR tuning in the kinetic sector. Therefore, even in the case of local gauge-fixed equations, the theory does not derive from a local action and we thus have genuine non-locality.

In the case of local theories, the fact that one could localize the equations by fixing the gauge was a consequence of the fact that the integrated-out fields where pure-gauge. It therefore seems that, if we now wish to localize the above equations by integrating-in some auxiliary fields, the latter will not be pure-gauge, so that these theories cannot be obtained by some St\"uckelberg-ed local theory. This is not a surprise, since we know Proca and Fierz-Pauli theories to be the only ghost-free local theories of spin-1 and spin-2 dynamics, respectively.

\chapter{Subtleties of non-local field theory}

Now that we have reached the subject of non-local field theory, it is important that we discuss some peculiar features that distinguish it from local field theory. This chapter is based on, and extends, \cite{JaccardMaggioreMitsou2,DirianMitsou,FoffaMaggioreMitsou1}.

\section{Non-local actions} \label{sec:nonlocactGa}

\subsection{Schwinger-Keldysh formalism} \label{eq:SKform}

The first point is that causal non-local equations of motion cannot derive from the strict application of the variational principle on some non-local action. Indeed, say we wish to vary an action containing a term of the form
\beq
\int \ed^D x \, \ph \bo_{\rm r}^{-1} \psi  = \int \ed^D x\, \ed^D y \, \ph(x) G_{\rm r}(x,y) \psi(y)  \, ,
\eeq  
where ``r'' denotes the retarded Green's function. The variation with respect to $\ph$ will provide a causal equation of motion
\beq \label{eq:GrSnoncausalex}
\int \ed^D y \, G_{\rm r}(x,y) \psi(y) = \( \bo_{\rm r}^{-1} \psi \)(x) \, ,
\eeq
but the variation with respect to $\psi$ will involve the ``transposed'' Green's function $G^T_{\rm r}(x,y) \equiv G_{\rm r}(y,x) \equiv G_{\rm a}(x,y)$, which is thus the advanced one
\beq
\int \ed^D y \, G_{\rm r}(y,x) \ph(y) = \( \bo_{\rm a}^{-1} \ph \)(x) \, ,
\eeq
so that this equation is anti-causal. In the case $\ph = \psi$, such as in the kinetic terms that would correspond to the non-local theories we constructed, one would rather get the term
\beq
\int \ed^D y \( G_{\rm r}(x,y) + G_{\rm r}(y,x) \) \ph(y) = \( \bo_{\rm r}^{-1} \ph + \bo_{\rm a}^{-1} \ph \)(x) \, ,
\eeq
i.e. the retarded function is effectively symmetrized inside the action. This is a direct consequence of the time-reversal and time-translational symmetries, i.e. the physics that derives from an action is reversible and invariant under time-translations. Conversely, if the equations of motion are non-local but causal, then there is an arrow of time and they can therefore not derive from an action. This is why causal non-local equations encompass for example dissipative/non-conservative systems \cite{Galley, GalleyTsangStein} and systems with memory. Yet another way to understand this is by noting that, although one uses initial conditions to evolve the equations, the variation of the action is performed by fixing boundary conditions in time. This is clearly non-local data and thus the result will in general depend on the whole time-interval, with the only exception being for local actions \cite{Galley}. 

Therefore, non-local equations of motion appear to be of less fundamental significance since they cannot derive from an action and thus cannot be understood as the saddle point approximation of some path integral. Nevertheless, one should remember that this is actually not the rigorous connection between quantum mechanics and classical equations. Rather, the equation of motion of a classical field $\ph$ has physical relevance because it can be understood as the $\hbar \to 0$ limit of the equation of motion of some expectation value $\bra \hat{\ph} \ket(t) \equiv \bra \Psi | \hat{\ph}(t) | \Psi \ket$ of the corresponding operator $\hat{\ph}$, for some fixed state $\Psi$. The evolution of $\bra \hat{\ph} \ket(t)$ is governed by the quantum effective action $\Ga$ and, as it turns out, in interacting theories $\Ga$ is indeed non-local because of the non-local nature of quantum corrections \cite{Jordan,CalzettaHu,BGVZ,DonoghueElMenoufi,TsamisWoodard}\footnote{More precisely, in perturbative QFT the propagator $\sim \( k^2 + m^2 \)^{-1}$ corresponds to a non-local operator $\( \bo - m^2 \)^{-1}$ in real space, so the loop corrections will in general be non-local. For scales $k^2 \ll m^2$ however one can expand
\beq
\frac{1}{k^2 + m^2} = \frac{1}{m^2} \( 1 - \frac{k^2}{m^2} + \Ord(k^4) \) \, ,
\eeq
in which case the corresponding real-space corrections are a series of local, but higher-derivative operators. In the presence of massless particles however, such as in the case of gravity for example, the propagator becomes non-analytic in $k^2$ around $k^2 = 0$, so these corrections are non-local at all scales.}. So non-locality is not such an exotic feature when one is interested in realistic equations of motion deriving from some underlying QFT and, as a matter of fact, non-local terms $\sim \bo^{-1}$ even dominate in the infra-red. So how can these equations be causal?

The important point is to realize that $\Ga$ is not an action in the usual sense of an integral over all of space-time and thus it is a somewhat modified variational principle that allows us to extract physically sensible equations of motion. Indeed, the effective action $\Ga$ we are discussing here, which we will denote by ``$\Ga_{\rm in-in}$'', should not be confused with the better known quantum effective action $\Ga_{\rm in-out}$ that is used in the computation of scattering amplitudes and is an action of the usual form $\int_{t_i}^{t_f} L(t)$. In order to clearly distinguish the two, let us first describe $\Ga_{\rm in-out}$. In that case one is interested in $S$-matrix elements $\bra \Psi_{\rm out} | \Psi_{\rm in} \ket$ where the ket is a state at the initial time $t_i$ and the bra is a state at final time $t_f$. Therefore, the path integral representation of this quantity involves the integral of the Lagrangian
\beq
\bra \Psi_{\rm out} | \Psi_{\rm in} \ket \sim \int \( \prod_{t \in [t_i, t_f]} \ed \ph(t) \) \Psi^*_{\rm out} [\ph(t_f)] \Psi_{\rm in}[\ph(t_i)]  \, e^{i \int_{t_i}^{t_f} \ed t \, L[\ph(t')]} \, ,
\eeq
over the whole time interval $[t_i, t_f]$. The quantum effective action $\Ga_{\rm in-out}[\vph]$, where $\vph(t) \equiv \bra \Psi_{\rm out} | \hat{\ph}(t) | \Psi_{\rm in} \ket$, is then the Legendre transform of the generating functional 
\beq
W_{\rm in-out}[J] = -i \log \int \( \prod_{t \in [t_i, t_f]} \ed \ph(t) \) \Psi^*_{\rm out} [\ph(t_f)] \Psi_{\rm in}[\ph(t_i)]  \, e^{i \int_{t_i}^{t_f} \ed t \( L[\ph(t')] - J(t') \ph(t') \)} \, ,
\eeq
where $J$ is an external linear source. Although the equations of motion of $\Ga_{\rm in-out}$ provide the time-evolution of $\vph(t)$ for $J = 0$, by construction, $\Ga_{\rm in-out}$ is mostly used for its property of being the generating functional of 1PI diagrams. Indeed, the equations of motion of $\vph(t)$ are not very relevant because they are acausal, since the sum over paths will depend on both what happens before {\it and} after $t$. Moreover, if one works with vacuum-to-vacuum amplitudes on backgrounds with non-trivial evolution, as is in the case of cosmology for instance, then the initial vacuum is not proportional to the final vacuum\footnote{Or the latter is not even known.} and $\bra 0_{\rm out} | \hat{\ph} | 0_{\rm in} \ket$ is not even real\footnote{This is why $\Ga_{\rm in-out}$ can be used for computing the lowest order quantum corrections to a potential $V(\vph)$ on flat space-time, because then $| 0_{\rm out} \ket \sim | 0_{\rm in} \ket$ and one can restrict to the cases $\ph = {\rm const}$ where the time-non-locality is irrelevant \cite{PeskinSchroeder}.}. Thus, this $\vph$ usually lacks physical interpretation by not being an eigenvalue of the operator $\hat{\ph}$ and intrinsically non-local in its definition.

In order to get causal equations of motion for some real field one rather needs to consider the quantum effective action for an expectation value $\bra \hat{\ph} \ket(t) \equiv \bra \Psi_{\rm in} | \hat{\ph}(t) | \Psi_{\rm in} \ket$, i.e. with both the ket and the bra being the same state defined at $t_i$, \footnote{As explained in \cite{Barvinsky2}, even in the case of scattering amplitudes what is physically observable is not the amplitude, but the corresponding probability
\beq
|\bra \Psi_{\rm out} | \Psi_{\rm in} \ket |^2 = \bra \Psi_{\rm in} | \( |\Psi_{\rm out} \ket\bra \Psi_{\rm out} | \)| \Psi_{\rm in} \ket \, ,
\eeq
which also takes the form of an expectation value of some operator.}. Now however the path integral is constructed in a different way and we enter the so-called ``in-in'' or ``Schwinger-Keldysh'' or ``closed time-path'' formalism \cite{Schwinger,BakshiMahanthappa1,BakshiMahanthappa2,Keldysh,Jordan,CalzettaHu,Vilkovisky,Barvinsky1}. In the scattering case, we had that 
\beq
\bra \Psi_{\rm out} | \hat{\ph}(t) | \Psi_{\rm in} \ket \sim \int \( \prod_{t \in [t_i, t_f]} \ed \ph(t) \) \Psi^*_{\rm out} [\ph(t_f)] \ph(t) \Psi_{\rm in}[\ph(t_i)]  \, e^{i \int_{t_i}^{t_f} \ed t \, L[\ph(t)]} \, ,
\eeq
because one must connect $| \Psi_{\rm in} \ket$ from $t_i$ to $\hat{\ph}$ at $t$ and then the latter to $\bra \Psi_{\rm out} |$ at $t_f$. In the case of $\bra \Psi_{\rm in} | \hat{\ph}(t) | \Psi_{\rm in} \ket$ we connect $| \Psi_{\rm in} \ket$ from $t_i$ to $\hat{\ph}$ at $t$, but then we have to connect the latter back to $\bra \Psi_{\rm in} |$ at $t_i$, i.e. by going backwards in time. This gives
\bea
\bra \Psi_{\rm in} | \hat{\ph}(t) | \Psi_{\rm in} \ket & \sim & \int \( \prod_{t' \in [t_i, t]} \ed \ph_+(t) \) \( \prod_{t' \in [t_i, t]} \ed \ph_-(t) \) \Psi^*_{\rm in} [\ph_-(t_i)] \ph(t) \Psi_{\rm in}[\ph_+(t_i)]  \\
 & & \hspace{1cm} \times \de \( \ph_+(t) - \ph_-(t) \) \exp \[i \int_{t_i}^t \ed t' \, L[\ph_+(t')] + i \int_t^{t_i} \ed t' \, L[\ph_-(t')] \] \, . \nn 
\eea
It is now obvious that the dynamics of $\bra \hat{\ph} \ket(t)$ can only depend on the physics in the time-interval $[t_i, t]$ so that its evolution must be causal. The corresponding quantum effective action $\Ga_{\rm in-in}$ will then be the Legendre transform of the generating functional
\bea
W_{\rm in-in}[J_+, J_-] & = & -i\log \int \( \prod_{t' \in [t_i, t]} \ed \ph_+(t) \) \( \prod_{t' \in [t_i, t]} \ed \ph_-(t) \) \Psi^*_{\rm in} [\ph_-(t_i)] \Psi_{\rm in}[\ph_+(t_i)]  \label{eq:Winin} \\
 & &  \times \de \( \ph_+(t) - \ph_-(t) \) \exp \[i \int_{t_i}^t \ed t' \(  L[\ph_+(t')] - L[\ph_-(t')]  - \ph_+ J_+ + \ph_- J_-  \) \] \, , \nn
\eea
and will thus depend on two fields $\Ga_{\rm in-in}[\vph_+, \vph_-]$, the one representing $\vph$ on $[t_i,t]$, going forward in time $\vph_+$ and the one representing $\vph$ on $[t, t_i]$, going backwards $\vph_-$. Concretely, 
\beq
\Ga_{\rm in-in}[\vph_+, \vph_-;t] = \int_{t_i}^t \ed t' \(  L[\ph_+(t')] - L[\ph_-(t')] \)  + \Ord(\hbar) \, ,
\eeq
where $L$ is the fundamental Lagrangian and the quantum corrections will typically mix the two sectors precisely because of non-locality. For instance, we may find terms of the form\footnote{In general one finds arbitrary powers of different Green's functions, but always such that the corresponding integration kernel is zero when its second argument is outside the past light-cone of its first argument.}
\beq \label{eq:Gaininex}
\int_{t_i}^t \ed t' \, \ed t'' \, \vph_+(t') G_{\rm r}(t',t'') \vph_-(t'') \, ,  
\eeq
where $G_{\rm r}$ is the retarded Green's function. Note that $\vph_+(t')$ is indeed causally propagated forward in time to $\vph_-(t'')$, since the latter occurs in front of it in this bended time-line. As in the scattering case, the variational principle is now a direct consequence of the relation between $\Ga$ and $W$. By construction
\beq
\frac{\de \Ga_{\rm in-in}}{\de \vph_+(t')} = -J_+(t') \, , \hspace{1cm} \frac{\de \Ga_{\rm in-in}}{\de \vph_-(t')} = J_-(t') 
\eeq
so for vanishing external source we get that the variation of $\Ga_{\rm in-in}$ is zero. The additional requirement here is that one must evaluate these equations at $t$ where the two functions coincide by definition $\vph_+(t) \equiv \vph_-(t) \equiv \vph(t)$. Since $\vph_+$ is ``going forward in time'' it will obey a causal equation, while since $\vph_-$ ``goes backward in time'' it will obey an anti-causal equation. It is thus the equation for $\vph_+$ which is relevant for us, while the one of $\vph_-$ is its time-reversed copy. Applying this variational principle to the example given above (\ref{eq:Gaininex}) we get that the corresponding term in the action is indeed causal $\bo^{-1}_{\rm r} \vph$. 

One should also note that the boundary conditions of this variational principle are given at the extremities of the time-line, which here correspond to simply $t_i$ but for two fields $\vph_{\pm}$. Thus, for the field $\vph$ at the end of the application of the variational principle, these are nothing but the initial conditions. Therefore, this is a variational principle that relies on fixing initial data instead of boundary data. Going back to section \ref{eq:diffop}, remember that the Feynman propagator is the $\bo^{-1}$ corresponding to the boundary conditions of the ``in-out'' path integral with $|\Psi_{\rm in} \ket = | 0_{\rm in} \ket$ and $| \Psi_{\rm out} \ket = |0_{\rm in} \ket$. It is symmetric $(\bo^{-1}_{\rm F})^T = \bo^{-1}_{\rm F}$ and thus privileges no time direction, consistent with the fact that the boundary conditions of the path integral are defined at both past and future infinity. Here we see that the retarded propagator is the $\bo^{-1}$ of the ``in-in'' path integral for $|\Psi_{\rm in} \ket = | 0_{\rm in} \ket$, where one fixes initial conditions instead of boundary conditions and where the arrow of time is explicit. Indeed, for a scalar field in (\ref{eq:Winin}) one must insert a $i \ep \ph_+^2$ factor in $L[\ph_+]$ and a $- i\ep \ph_-^2$ factor in $L[\ph_-]$ for the path integral to converge. For the classical solutions $\vph$, which dominate the path integral, this imposes no ingoing positive frequency modes at past infinity, through $\ph_+$, and no negative frequency modes at past infinity again, through $\ph_-$, so these effectively become the boundary conditions of the retarded Green's function (\ref{eq:retpropBC}).

Finally, note that the above construction holds only for theories for which the fundamental Lagrangian is local, with the non-localities in $\Ga$ being due to quantum corrections. This is because in constructing the path integral one must first pass through the canonical formalism and the latter does not exist in the non-local case precisely because of time non-locality. Nevertheless, the ``in-in'' action and the corresponding variational principle can be taken independently of their quantum origin as a well-defined action-based formulation for classical non-local field theory. As a matter of fact, such a construction has also been used from the purely classical point of view in order to enlarge the scope of action-based mechanics to include dissipative systems as well \cite{Galley,GalleyTsangStein}. In particular, this has allowed for a generalization of Noether's theorem that provides the variation of the charges in terms of the dissipative part of the action \cite{GalleyTsangStein}.

\subsection{Formal action} \label{sec:formalaction}

An interesting observation about the issue that was raised in the previous section is that the whole problem revolves around the type of Green's function that will appear in the equations of motion. Apart from that, the equations one would derive using the standard variational principle on some $S_{\rm in-out}$ or with the modified variational principle applied on some $S_{\rm in-in}$, would be formally the same. Since the usual $S_{\rm in-out}$ action is simpler and closer to our habits, it would be very convenient if we could use it anyway, even if we have to rely on purely formal manipulations. Indeed, we could for instance decide that all $\bo^{-1}$ occurrences inside the action are formal, i.e. undetermined linear inverses of $\bo$. Then, once the equations of motion have been computed, one should turn all the $\bo^{-1}$ into retarded ones by hand. This is in fact a standard way of proceeding (see \cite{TsamisWoodard, DeserWoodard1, Barvinsky2,  DeserWoodard2} and references therein).

Since the difference of the convolution with two different $\bo^{-1}$ is a homogeneous solution, we can give a meaning to this formal action as a functional on the quotient space of fields modulo homogeneous solutions of $\bo$. In this space the kernel of $\bo$ is trivial, by construction, and thus the equivalence class $[\bo^{-1}]$ is unique. In the case of the equations of motion however, where homogeneous solutions matter, one has to choose the appropriate representative $[\bo^{-1}]$ that suits for sensible physics, i.e. $\bo^{-1}_{\rm r}$.

Now note that treating all the $\bo^{-1}$ as equivalent during the variation implies some important simplifications. For instance, this means that we can effectively integrate $\bo^{-1}$ by parts. Indeed
\bea
\int \ed^D x \, \ph(x) \bo^{-1} \psi(x) & \equiv & \int \ed^D x \, \ed^D y\,  \ph(x) G(x,y) \psi(y) \nn \\
 & = & \int \ed^D x \, \ed^D y \, \psi(y)  G^T(y,x) \ph(x) \nn \\
 & = &  \int \ed^D y\,  \psi(y) \(\bo^{-1}\)^T \ph(y) \nn \\
 & \equiv & \int \ed^D y\, \psi(y) \bo^{-1} \ph(y) \, , 
\eea    
since the transposed $\(\bo^{-1}\)^T$ is also a right-inverse $\bo \(\bo^{-1}\)^T = {\rm id}$ (see appendix \ref{sec:convGdef} for the case $\bo^{-1}$). A related simplification is the fact that now $\bo^{-1}$ is also a left-inverse $\bo^{-1} \bo \equiv {\rm id}$ since, from appendix \ref{sec:leftinv}, we know that $\bo^{-1} \bo$ is the identity up to a homogeneous solution. As an example, the formal action corresponding to the non-local equation (\ref{eq:projeq}) reads
\beq \label{eq:formalactionex}
S = \frac{1}{2} \int \ed^D x \[  h_{\mu\nu} \( \( \bo - m_g^2 \) {}_0 {\cal P}^{\mu\nu\ro\si} + \( z\bo - m_s^2 \) {}_s {\cal P}^{\mu\nu\ro\si} \) h_{\ro\si} + h_{\mu\nu} T^{\mu\nu} \] \, ,
\eeq
where the $\bo^{-1}$ inside the projectors are formal. Finally, note that integrating-out fields to get non-local formulations can now be performed at the level of this formal action.

\subsection{Non-local path integral} \label{eq:nonlocpertQFT}

In section \ref{eq:SKform}, the obstruction to the existence of a ``in-out'' action for some causal non-local equations was traced back to the fact that $G_{\rm r}$ is not symmetric under time-reversal. However, this is not the case of its Feynman cousin $G_{\rm F}$ and it is the latter that appears in the path integral for scattering amplitudes, i.e. the ``in-out'' case with $|\Psi_{\rm in} \ket = | 0_{\rm in} \ket$ and $\bra\Psi_{\rm out} | = \bra 0_{\rm out} |$. Thus, there is no need for formal manipulations in writing down such a path integral for our non-local theories. 

For instance, we can now literally integrate-out the St\"uckelbergs of the local theories, i.e. by integrating over them in the path integral\footnote{Of course, for quadratic fields, this has precisely the effect of replacing the fields by the solution to their equation of motion, although with the Feynman prescription if some $\bo - m^2$ has been inverted in the process.}. More precisely, we can start with the path integral of the original local theory, perform the St\"uckelberg trick, insert a gauge-fixing term for the gauge field, and then integrate-out the St\"uckelberg field to get a non-local theory. For example, for Proca theory, this procedure gives
\beq
\int D \ph \, e^{i S[A, \ph, j]} \sim \exp  i \int \ed^D x \[ \frac{1}{2}\, A_{\mu} \( \bo - m^2 + i \ep \) {\cal P}_{\rm F}^{\mu\nu} A_{\nu} - \frac{1}{2\xi} \( \pa_{\mu} A^{\mu} \)^2 + A_{\mu} j^{\mu} \] \, ,
\eeq
where, as we know from section \ref{eq:diffop}, it is the Feynman inversion of $\bo$ which arises in the transverse projector $\cal P$. Contrary to the case of classical physics, where the retarded prescription is lost inside the path integral because of symmetrization, here there is no inconsistency since the Feynman propagator is symmetric. The equations of motion of this action are acausal, but the scattering amplitudes are the ones of Proca theory, by construction. This is simply a local QFT with a field that has been integrated-out. Indeed, the two-point function can be computed by further integrating-out $A_{\mu}$ and taking the double functional derivative with respect to the source. One gets
\beq
\bra 0 | \hat{A}^{\da}_{\mu}(k) \hat{A}_{\nu}(k) |0 \ket = - \frac{i}{k^2 + m^2 - i \ep}\, \et_{\mu\nu} + \( \dots \) k_{\mu}k_{\nu} \, ,
\eeq
whose physical part is thus the same as the propagator (\ref{eq:propprojA}) with the Feynman $\ep$ prescription. Moreover, note that the presence of $\bo^{-1}_{\rm F}$ does not constrain the fields more than in the local case, since the boundary conditions of the path integral are the ones for which $\bo^{-1}_{\rm F}$ is defined anyways.

In local QFT one usually integrates out a dynamical field when one is not interested in the scattering amplitudes containing the associated particles in the ``in'' and ``out'' states. The important question now is whether one can proceed in the same way for the genuinely non-local theories, i.e. without having a corresponding local action for them. Indeed, in the case of the non-local formulation of Proca theory, we were sure that the non-local path integral was not pathological because it simply amounted to the one of a local theory with some integrated-out field. To make sense of a path integral corresponding to the non-local spin-2 theories introduced in the previous section we should first find some local formulation, and study its own quantization.

\section{Localization}

In the case of local equations, as discussed in section \ref{sec:dofprop} and as shown in the case of linear massive gauge theories, we have that each dynamical field brings in two degrees of freedom corresponding to its initial value and the one of its time-derivative. In non-local field theory this rule does not hold anymore, and properly understanding the consequences of this fact is very important if we wish to settle stability issues. Of course, the notion of dynamical field may seem a bit ambiguous when non-localities are around, so we must first express the theory in a way where this terminology is well-defined. Our argumentation will be much more transparent if we parallel it with a simple example highlighting the important features. Consider the following non-local equation for some field $\ph$ with source $J$
\beq \label{eq:pheqex}
\bo \ph - m^4 \bo^{-1}_{\rm  r} \ph = J \, .
\eeq
This of course makes sense only if $\ph$ has finite past, but we can also decide that time starts at some finite $t_i$, in which case the initial conditions of $\ph$ could be chosen freely\footnote{What one should not do in this case however, is consider the times $t < t_i$ because for them the Green's function will be advanced.}. In any case, for our purposes it will not matter whether the initial conditions of $\ph$ are constrained for consistency or not. Equation (\ref{eq:pheqex}), although quite clear to understand, is an integro-differential equation and thus not very transparent as far as the dynamical content is concerned. It is therefore very convenient to introduce an auxiliary field $\psi$ which we define by
\beq \label{eq:psidefex}
\psi \equiv m^2 \bo_{\rm r}^{-1} \ph \, ,
\eeq
to get that the equation now takes a local form
\beq \label{eq:phipsieqex}
\bo \ph = m^2 \psi + J \, .
\eeq
One must then supplement it with the equation satisfied by $\psi$ which, by construction, is a dynamical equation
\beq \label{eq:psieqex}
\bo \psi = m^2 \ph  \, .
\eeq
Observe that this appears as the inverse of the operation of ``integrating-out'', so we may say that we have ``integrated-in'' $\psi$. However, if we now reverse-engineer and integrate-out $\psi$, then the most general solution of (\ref{eq:psieqex}) reads
\beq
\psi = \psi^{\rm hom} + m^2 \bo_{\rm r}^{-1} \ph \, ,
\eeq 
where $\bo \psi^{\rm hom} = 0$ is a homogeneous solution. Note that this is (\ref{eq:psidefex}) only in the case $\psi^{\rm hom} = 0$ and, in particular, we must have $\psi \to 0$ if $m \to 0$. Since the set of homogeneous solutions is isomorphic to the set of initial conditions, the definition of $\psi$ (\ref{eq:psidefex}) constrains its initial conditions to be zero at $t \to - \infty$ if $\ph$ has finite past, or at $t_i$ if this is when we start the convolution in (\ref{eq:pheqex}). 

In any case, we have that $\psi$ is a dynamical field, i.e. it obeys a second-order equation in time, but does not represent degrees of freedom of the theory, i.e. its initial conditions are not free to choose (see section \ref{sec:dofprop} for a reminder on these definitions). Such fields are thus commonly referred to as a ``spurious degrees of freedom'' in the literature. However, as we will see later, their effect on the physics will be far from being ``spurious'', so we will avoid this terminology. We will rather refer to such fields as ``constrained dynamical fields''. For the moment, note that the local equations (\ref{eq:phipsieqex}) and (\ref{eq:psieqex}), {\it subject to the constraints on the initial data of} $\psi$, have exactly the same solutions as (\ref{eq:pheqex}), by construction. They thus provide a more transparent point of view on the physics, since we are certainly more used to working with local equations.

\subsubsection{Understanding $N_{\rm f} \neq 2 N_{\rm d}$}

We thus have that the number of dynamical fields in (\ref{eq:pheqex}), both constrained and unconstrained, is $N_{\rm d} = 2$, while the number of degrees of freedom is $N_{\rm f} = 2$, so the local field theory rule $N_{\rm f} = 2 N_{\rm d}$ does not hold. To understand where the constraints on $\psi$ come from observe in (\ref{eq:psidefex}) that the information of the initial data of $\psi$ amounts to the information of the initial data of the Green's function in $\bo^{-1}$ and therefore to the choice of inversion $\bo^{-1}$. Thus, this additional data that suddenly pop up were actually here all along. They were determining the choice of $\bo^{-1}$ we were using, while now they are expressed as initial conditions of some auxiliary field.

Another way to understand this is by noting that if we do consider an arbitrary $\psi^{\rm hom}$ the effect is that the source is shifted
\beq
J \to J +  m^2 \psi^{\rm hom} \, ,
\eeq
as we already saw when we were integrating-out the St\"uckelebrgs in section \ref{sec:nonlocformu}. Since adding a homogeneous part can be interpreted as changing the Green's function in $\bo^{-1}$, considering a $\psi^{\rm hom} \neq 0$ can be interpreted as a different choice of $\bo^{-1}$, \footnote{More precisely, since by construction $\psi \to 0$ if $\ph \to 0$, we would have that $\psi^{\rm hom}$ is a linear functional of $\ph$ and the new $\bo^{-1}$ can thus still be written as the convolution with a Green's function.}.

Whatever the way we choose to see this, the conclusion is that different initial conditions of $\psi$ correspond to different choices of $\bo^{-1}$ in the original non-local theory and thus different original theories. This implies that the initial data of $\psi$ are theory-level data, in contrast with the initial conditions of regular dynamical fields which represent different solutions of the same theory. Thus, the unconstrained theory of $\ph$ and $\psi$ represents many more theories than (\ref{eq:pheqex}), one for every choice of $\psi^{\rm hom}$.

\subsubsection{Local action and diagonalization}

We can now pass to the action corresponding to these equations
\beq  \label{eq:phlocSex}
S = \int \ed^D x \[ \frac{1}{2}\, \ph \bo \ph + \frac{1}{2}\, \psi \bo \psi - m^2 \ph \psi - \ph J\] \, ,
\eeq
which could have also been obtained by integrating-in $\psi$ directly in the formal action of (\ref{eq:psidefex}) 
\beq \label{eq:phSex}
S = \int \ed^D x \[ \frac{1}{2}\, \ph \bo \ph - \frac{m^4}{2}\, \ph \bo^{-1} \ph - \ph J \] \, .
\eeq
Now one can diagonalize (\ref{eq:phlocSex}) to get
\beq
S = \int \ed^D x \[ \frac{1}{2}\, \ph_+ \( \bo - m^2 \) \ph_+ + \frac{1}{2}\, \ph_- \( \bo + m^2 \) \ph_- - \frac{1}{\sqrt{2}}\( \ph_+ + \ph_-\) J\] \, ,
\eeq
where $\ph_{\pm} \equiv \( \ph \pm \psi \)/\sqrt{2}$, so $\ph_-$ is a tachyon. This could have been directly deduced by looking at the propagator of the non-local theory (\ref{eq:pheqex}) or (\ref{eq:phSex})
\beq
D(k) = -\frac{i}{k^2 - \frac{m^4}{k^2}} = \frac{1}{2} \( - \frac{i}{k^2 + m^2} - \frac{i}{k^2 - m^2} \) \, ,
\eeq
which indeed reflects the spectrum of the localized theory. The constraint on the initial conditions of $\psi$ translates into equal initial conditions for $\ph_+$ and $\ph_-$. In particular, if $m \to 0$ then this gives $\ph_+ = \ph_-$ at all times, since they obey the same equation. This is consistent with the fact that if $m \to 0$ then $\psi \to 0$. 

As already mentioned, by this ``localization'' procedure we obtain a bijective map between the solutions of the non-local equation (\ref{eq:psidefex}) and the solutions of a trivial local field theory, as long as we carefully take into account the constraints on the initial conditions. The dynamical content is therefore clearly a healthy scalar field and a tachyonic one. Thus, non-local field theories ``hide'' constrained dynamical fields.

\subsubsection{Localization versus gauge theory constraints} \label{sec:nonlocgauconst}

It is now {\it very} important to understand that this kind of constraint on the initial conditions has nothing to do with the constraints that arise in local gauge theories. Indeed, one of the reasons for spending so much time analyzing linear local gauge theories was to clearly see how one obtains $N_{\rm f} = 2N_{\rm d}$, i.e. how the constrained fields are necessarily non-dynamical and vice-versa. As we have seen in more than one way, the constraints of gauge theory are encoded within the action, i.e. the latter is all we need to deduce them. This is most obvious in the canonical formalism, where half of the constraints are the equations of motion of components that are Lagrange multipliers, while the other half can be imposed thanks to the arbitrariness of these Lagrange multipliers in the rest of the equations of motion. It is thus the structure of the action itself, which is ultimately due to the presence of the gauge symmetry, which constraints the initial conditions of some fields {\it and} automatically makes them non-dynamical. Here on the other hand the constraints on $\psi$ are not the consequence of some equation of motion, symmetry, or any other particular structure. They are constraints that simply follow by the definition of $\psi$ as a shortcut notation for a fixed functional of $\ph$ and must be appended to the action\footnote{As we saw, this is nothing but the information of the ``retardedness'' of the Green's function, which was also appended to our formal action.}. It is therefore important not to confuse constraints that are due to some gauge symmetry of the theory, and constraints that are due to localization, especially when we deal with non-local gauge theories.

\subsection{Quantization} \label{sec:quantization}

Now that we have found a way of reformulating a non-local theory in terms of a local, but constrained, theory, we can address the issue that was raised in section \ref{eq:nonlocpertQFT}, namely, of whether one gets a sensible QFT by simply plugging a genuinely non-local action inside a path integral without asking any further questions. We see that the problem of non-locality, which kept us from defining a canonical quantization, has now translated into the problem of implementing, somehow, the constraints of the auxiliary fields at the quantum level.  

So let us simply consider a local action with constrained boundary data \'a la Feynman, since we work in an ``in-out'' framework and thus compute $\bra 0_{\rm out} | T \dots | 0_{\rm in} \ket$. In general the constraints will not concern specific fields in the diagonalized action, but rather linear combinations of their boundary data. Translating these into the constraints on the creation operators and thus on the particles, they will generally amount to projections on some Hilbert subspace. Thus, constraining this external particle information corresponds to considering only a sub-block of the $S$-matrix, i.e. not all the possible ``in'' and ``out'' states. In the simplest case where the constraints impose Feynman boundary conditions on a single field, this translates into zero corresponding particles on external legs. However, since the field is dynamical its propagator will appear in the internal lines. Let us call the corresponding particles ``auxiliary''.

Now, if the $S$-matrix is in block-diagonal form and the constraints correspond to choosing one of these blocks, then the evolution will be unitary. Starting with no auxiliary particles in the initial state, no such particles are produced in the final state and thus probability is conserved in this subspace. This is precisely what happens in non-abelian gauge theories where one introduces the Faddeev-Popov particles in order to guarantee that if we start with no longitudinally polarized gauge bosons these will not be produced in the final state. However, in that case, it is the gauge-symmetric structure of the theory, ultimately leading to the BRST symmetry, which implies this highly non-trivial result \cite{PeskinSchroeder}. Here there is no such structure for the auxiliary localizing fields\footnote{The only exception are precisely the non-local formulations of local theories since then the localizing fields are the St\"uckelbergs that are pure-gauge.}, so the $S$-matrix will generally not be in block diagonal form. Thus, the auxiliary particles will be produced in the ``out'' state and not taking into account these states will mean that the evolution is not unitary. Put differently, part of the probability will ``leak'' in final states that are not part of the physical Hilbert subspace. For more complicated constraints on the initial and final states, analogous unitarity problems will necessarily occur. 

One possibility for avoiding this conclusion could be that the auxiliary particles are much heavier than the energies at which we are interested, so that they cannot be produced in the final states and evolution is unitary. Indeed, this is what happens in effective field theories, where some heavy field $\Phi$ has been integrated-out
\beq
e^{i S_{\rm eff}[\ph]} \sim \int D \Phi \, e^{i S[\ph, \Phi]} \, ,
\eeq
with $S_{\rm eff}$ providing a unitary evolution in the subspace of zero $\Phi$ particles at low energies. Unfortunately however, in this case one usually has that the non-local operator is of the form $\( \bo - m^2\)^{-1}$, since the integrated-out mode is massive. By definition then, the effective theory is valid ($S_{\rm eff}$ is unitary) only up to the cut-off $\La < m^2$. Then, for such scales $p, E < \La$ we can expand
\beq
\( \bo - m^2\)^{-1} = - \frac{1}{m^2} \( 1 + \frac{\bo}{m^2} + \dots \) \, ,
\eeq
so that the effective theory cannot be non-local. 

We can thus conclude that, if we take the localized theory as the ``fundamental one'' and try to quantize it, then we have to consider all the dynamical fields on equal footing. There is no way in which the constraints that we impose classically may be somehow implemented in the quantum context without spoiling unitarity. Then, considering the classical limit of this QFT will result in the unconstrained localized equations of motion, thus representing more solutions than the ones of the original non-local theory. In conclusion, it makes no sense quantizing a non-local action. This is why the non-local models proposed in the literature are usually interpreted as the quantum effective action $\Ga$ of some underlying local fundamental action $S$, or as any other type of classical effective action.  

Finally, we can now answer the question raised in section \ref{eq:nonlocpertQFT}, of whether one could simply plug a genuinely non-local action inside a path integral and start computing scattering amplitudes. We argued that in the case of massive electrodynamics this was justified because it simply amounts to integrating-out a field in a local theory. Here we see that in general, the would-be integrated-out fields, i.e. the localizing auxiliary fields, must be deconstrained in any quantization scheme that preserves unitarity. Thus, the quantum theory will not have the non-local theory as its classical limit, but a larger theory. The case of massive electrodynamics, or of Fierz-Pauli theory, is special, in that the integrated-out fields are pure-gauge (St\"uckelbergs) and thus do not correspond to particles in the local theory anyways.

The bottom-line here is that all non-local models should be understood as classical theories, that are therefore entirely determined by their equations of motion. This is going to be understood in the rest of the thesis.

\section{Constrained dynamical fields and classical stability} \label{sec:classstab}

A question of prime importance is whether a constrained dynamical field may destabilize a solution of interest. Indeed, in the literature, this special status has been invoked in order to minimize the impact of constrained dynamical ghosts on classical stability \cite{Barvinsky2, DeserWoodard2}. As we will now show, the impact on stability of such modes is the same as the one of ordinary dynamical fields. Nevertheless, note that, in contrast with the quantum context where a ghost is a fatal flaw\footnote{Indeed, in the quantum theory, a ghost gives rise to a negative-energy state, and therefore the vacuum can decay into ghosts plus ordinary (positive-energy) particles, as long as the total energy remains zero. The corresponding decay rate is infinite because the kinematic integral is unbounded, so this instability is fatal. More precisely, putting a cut-off on momenta we get, by dimensional analysis, that the decay probability per unit time and unit volume is $\Ga \sim \La_{\rm c}^4$. This actually holds for ghosts with tachyonic mass, so that the corresponding field oscillates and there is a notion of particle, although with negative-definite energy $E = - \sqrt{\vec{p}^2 + m^2}$. For ghosts with non-tachyonic mass part of the modes are diverging instead of oscillating so in that case one cannot even define particles.}, at the classical level a ghost does not necessarily imply an instability. Indeed, the stability verdict is not obvious in the presence of non-linear effects, as we will see in concrete examples, so each case must be analyzed individually.

\subsubsection{Classical ghost impact}

Loosely speaking, a solution is ``stable'', or at least ``metastable'', if arbitrary small perturbations of its initial conditions yield solutions that are close enough to the original one\footnote{The notions of ``small'' and ``close enough'' are of course subjective since they depend on the choice of a distance in field space and can be taken from either an absolute or a relative point of view.}. Thus, if some field is dynamical but not a degree of freedom, then its initial conditions cannot be perturbed and this may affect the stability verdict. Indeed, if the unstable modes obey an unsourced linear equation, then constraining their initial conditions to zero implies that they vanish at all times and the trivial solution is stable. One could still get away with non-trivial initial data giving diverging solutions since, by linearity, the auxiliary field does not interact with the physically observable ones and thus observable quantities remain bounded.

However, this is unfortunately not at all a realistic example, for all physically relevant theories contain (self-)interactions. In that case, the information of initial conditions becomes irrelevant. Indeed, consider the simplest example where the constrained unstable field has a linear source with compact support in time. As the source is turned on the field responds by taking a non-zero value, and thus when the source is turned off the field evolves as if it had started with non-trivial initial conditions. Moreover, the instability is communicated to the rest of the fields through the interactions, leading to diverging physical observables. Therefore, in the presence of (self-)interactions, there is no difference between constrained or unconstrained dynamical fields, {\it any} dynamical field matters in the classical stability analysis. It is not important whether some field has incoming waves at past infinity or not, these will be anyways generated at future infinity by its interactions. In the example given above, for instance, we have that the tachyonic mode $\ph_-$ makes the $\ph = 0$ solution of the non-local theory unstable. Of course, this would have been the case even if it were not sourced, because the initial conditions are not $\ph_-(t_i) = \dot{\ph}_-(t_i) = 0$, but this example shows how the diagonalization makes the constrained dynamical modes interact with the source as well\footnote{This can be expected whenever the hidden dynamical field has a corresponding pole in the saturated propagator of the non-local theory.}.

\subsubsection{Comparing with other works}

The above argument allows us to understand some weaknesses in the argumentation of \cite{Barvinsky2} and \cite{DeserWoodard2}, which erroneously conclude that the constraints on the initial/boundary conditions of ghosts neutralize their destabilizing power. Let us consider each case separately.

In their pioneering work on non-local modifications of GR for cosmological purposes, Deser and Woodard proposed the following simple formal action \cite{DeserWoodard1} 
\beq \label{eq:DW}
S_{\rm DW} \equiv \frac{1}{16 \pi G} \int \ed^D x \, \sqrt{-g} \[ R + R f(\bo^{-1} R) \] \, .
\eeq
In \cite{DeserWoodard2}, where they analyze its stability, they note that, when localized, the theory has a dynamical ghost when $f$ is non-linear \cite{NOSZ}. As they correctly show, working with the non-local equations, this mode is not a degree of freedom since its initial conditions are fixed. More precisely, this is a phenomenological model in which the $\bo^{-1}_{\rm r}$ that appears in the equations of motion starts its convolution at some finite $t_i$. Thus, the non-local equations of motion become local at $t = t_i$, \footnote{That is, since the non-localities take the form $\int_{t_i}^t \dots$, they all vanish at $t = t_i$.} and, being a gauge theory, some of them will constraint the initial data. As in the linear cases that we have studied, these are nothing but the equations of motion of the time-components $g_{0\mu}$ that are first-order in time-derivatives. In \cite{DeserWoodard2} it is indeed found that the modification does not change this property, so that there are as many constraints on the initial data as in GR for the same field content $g_{\mu\nu}$. Thus, the degrees of freedom are the same as in GR\footnote{In the localized formulation this would have been deduced by simply noting that the initial conditions of the auxiliary scalars vanish at $t_i = 0$. We will see later on a concrete example of this using a similar model.}. From this however the authors infer that the dangerous mode is saved from propagating, because of the gauge structure of GR, and thus that it cannot affect classical stability. 

This statement reveals precisely the confusion that might arise in non-local gauge theory which we discussed in section \ref{sec:nonlocgauconst}, i.e. that one considers all modes whose initial data are constrained as non-dynamical ones. The constraint we have on the auxiliary scalar here is {\it not} a gauge-theory constraint which would automatically make it non-dynamical. Rather, it is a constraint that comes from the fixed choice of inversion $\bo^{-1}$ and thus does {\it not} neutralize that mode. Again, counting degrees of freedom is {\it not} equivalent to counting dynamical fields\footnote{According to the definitions of section \ref{sec:dofprop}.} in non-local field theory. By going to the localized formulation the situation becomes clear. The gauge constraints reduce $g_{\mu\nu}$ to the two dynamical fields of a massless graviton (just as in GR), while the localizing scalars have constrained initial conditions but remain dynamical. We thus have an interacting dynamical ghost that can potentially destabilize the solution of interest. 

In \cite{Barvinsky2} the proposed model is rather
\beq \label{eq:Barvinsky}
S_{\rm B} \equiv \frac{1}{16 \pi G} \int \ed^D x \[ R - \al R_{\mu\nu} L^{-1} G^{\mu\nu}  \] \, ,
\eeq
where $L \equiv \bo + \Ord(R)$. Localizing this action one finds again dynamical ghosts, and it is argued that they do not influence stability because of their fixed boundary data. The author even illustrates this argument with the following example. Consider the simplest local theory and turn it into a non-local one artificially as follows
\beq
S = \int \ed^D x \[ \frac{1}{2}\, \ph \bo \ph - \ph J\] = \int \ed^D x \[ \frac{1}{2}\, (\bo\ph) \bo^{-1} (\bo \ph) - \ph J\] \, . 
\eeq
Then localize by integrating-in another scalar
\beq
S = \int \ed^D x \[ -\frac{1}{2}\, \psi \bo \psi + \psi \bo \ph - \ph J\] \, , \hspace{1cm} \psi = \bo^{-1} \bo \ph \equiv \ph \, ,
\eeq
and diagonalize $\psi = \psi' + \ph$
\beq
S = \int \ed^D x \[ -\frac{1}{2}\, \psi' \bo \psi' + \frac{1}{2}\, \ph \bo \ph - \ph J\]  \, .
\eeq
Of course, this ghost is only an artefact of this procedure. Indeed, its equation of motion is $\bo \psi = 0$ and, for zero initial conditions, we have $\psi = 0$, so integrating it out gives back the original local theory\footnote{Note that this holds also on non-trivial space-times.}. With this example however, the author implies that this apparent ghost is of the same kind that arises in the localization of (\ref{eq:Barvinsky}), and thus that the latter must also be harmless. This is not true because the above example precisely avoids that the ghost couples to the source. In contrast, in the localization of (\ref{eq:Barvinsky}), after diagonalization, the ghost mode {\it does} couple to the source. 

A probable source of confusion is the fact that the author works in Euclidean space, in which case $\bo^{-1}$ is uniquely defined on fields that vanish sufficiently fast at infinity and thus these are the natural constraints for the localizing fields. The fact that these are boundary constraints, instead of initial condition constraints, implies that whatever modulations the constrained field might experience in the bulk, its asymptotic values are zero. However, Wick rotating to Lorentz space-time we get that these boundary conditions turn into Feynman boundary conditions\footnote{Indeed, the trends $\sim e^{\om t_{\rm E}}$ at $t_{\rm E} \to -\infty$ and $\sim e^{-\om t_{\rm E}}$ at $t_{\rm E} \to +\infty$, with $\om > 0$, for the boundary conditions in Euclidean time turn into $\sim e^{i \om t}$ at $t \to -\infty$ and $\sim e^{-i\om t}$ at $t \to +\infty$ in Lorentzian time, i.e. no ingoing positive-frequency waves and no outgoing negative-frequency waves.}, which is not the type of constraints one must impose for causal physics. Rather, using the retarded propagators the constraints apply on the initial conditions, so there is no control on the behaviour of the ghost at future infinity. As we have argued, in the presence of non-linearities, this mode will be generically activated.

\subsubsection{Small summary}

The take-away message here is that the intuitive property $N_{\rm f} = 2 N_{\rm d}$ of local field theory has to be abandoned in the non-local case. There are hidden dynamical fields that appear only after all boxes have been put in the numerator and thus $N_{\rm f} \leq 2 N_{\rm d}$. The fact that their initial conditions are constrained is a consequence of the definite choice of Green's function in the non-local theory. One has to be even more careful in non-local gauge theories where there are two types of constraints that should not be confused: the ones due to the gauge symmetry, which neutralize modes, and the ones due to the localization, which do not affect propagation. From the above paragraphs it is now clear that what matters for realistic physics are the dynamical fields rather than the ones with unconstrained initial conditions. Constrained ghosts and tachyons are thus as dangerous as their unconstrained cousins. We must stress however once more time that, because these theories are classical, the presence of ghosts or tachyons does not necessarily imply an instability, as non-linearities can affect their evolution non-negligibly. Therefore, in the presence of such modes a case-by-case classical stability analysis is required to settle the issue.

\chapter{Non-local gravity}

We are now ready to consider generally-covariant extensions of the non-local field theories introduced in the second chapter. This chapter is based on, and extends, \cite{JaccardMaggioreMitsou2,FoffaMaggioreMitsou2,DirianMitsou}

\subsubsection{Manipulating $\bo^{-1}$ on curved space-time}

Now $\bo^{-1}$ is a right-inverse of $\bo \equiv \na_{\mu} \na^{\mu}$ and therefore depends on the metric field $g_{\mu\nu}$. For the reader who is interested in the mathematical details of this operator on curved space-time we suggest a first look at the appendix \ref{sec:bitensors}. An important property is that now $\bo^{-1}$ mixes the indices of the tensor on which it acts, just like $\bo$ does. It also commutes with the metric, in the sense that 
\beq
\bo^{-1} g_{\mu\nu} X^{\nu} = g_{\mu\nu} \bo^{-1} X^{\nu} \, ,
\eeq
but of course the $\bo^{-1}$ operators on each side of the equation are different since they act on different spaces. Moreover, note that there is more than one operator which reduces to $\bo^{-1}$ on flat space-time. For example, we have $\( \bo - \xi R \)^{-1}$ when acting on scalars, $\( \de_{\mu}^{\nu} \bo - \xi_1 \de_{\mu}^{\nu} R - \xi_2 R_{\mu}^{\nu} \)^{-1}$ when acting on vectors and so on. We will use the notation ``$\ti{\bo}^{-1}$'' for the as yet undetermined generalizations of $\bo^{-1}$.

For the retarded Green's function of $\ti{\bo}$ to be well-defined we need space-time to be globally hyperbolic, so that there exists a global time function which foliates the manifold, notions of past and future infinity, and of course causality. We will therefore assume that this is the case in what follows, even though the metric is a dynamical field, i.e. a field on which we have a priori no control. As it turns out, for the solutions that will interest us in this thesis, the couple $\( {\cal M}, g\)$ will indeed be globally hyperbolic for the time-intervals of interest.

\section{Constructing generally-covariant equations of motion}

We wish to generalize the models constructed in section \ref{sec:newmasstheor} to generally-covariant theories of $g_{\mu\nu}$. Simply generalizing (\ref{eq:projeq}) to an arbitrary background would correspond to the theory of a linear spin-2 field on curved space-time, which is not what we want. Moreover, working with $h_{\mu\nu}$ is not a good idea because the latter now corresponds to the perturbation around some background metric $h_{\mu\nu} \equiv g_{\mu\nu} - \bar{g}_{\mu\nu}$. Not only this would make our equations depend on $\bar{g}_{\mu\nu}$, but it would also make general covariance hard to implement. 

The obvious solution is to consider non-local combinations of curvature invariants of $g_{\mu\nu}$ and match these to (\ref{eq:projeq}) in the linearized limit over Minkowksi space-time. In doing so however the resulting equations are not transverse (under $\na$) in general. For example, say we have a term of the form
\beq \label{eq:exnotransv}
\ti{\bo}^{-1}_{\rm r} G_{\mu\nu} \, ,
\eeq
in our equation. Perturbing around flat space-time to linear order, since $[\pa_{\mu}, \bo^{-1}_{\rm r}] = 0$, we have that this tensor is transverse because $G_{\mu\nu}$ is. On curved space-time however, this is no longer true because $[\na_{\mu}, \bo^{-1}_{\rm r}] \neq 0$. 

The absence of transversality is inconsistent with gauge-invariance. Indeed, the latter implies that some of the components of the field are not determined by the equations of motion, and thus translates into having less equations of motion than the number of field components. This is the case if the equations are identically transverse, since we have $D$ less equations corresponding to the $D$ gauge parameters of the diffeomorphism symmetry. If the equations are not identically transverse, but we do have the gauge symmetry, then the fields that are not pure-gauge are overdetermined. To resolve this problem, one has two options.

\subsection{Projector-based models}

In the previous chapter we have identified the operators ${\cal P}$ (\ref{eq:aprojfam}) that make a tensor transverse. We could thus use these operators here to make the generalized equations transverse by hand, without affecting the linearized limit (if we choose $_1 {\cal P}$). This option has been considered for instance in \cite{JaccardMaggioreMitsou2,Maggiore2,Porrati} and we will refer to such models as ``projector-based models''.

\subsubsection{No closed form}

On flat space-time we were able to construct explicit expressions for the transverse operators ${\cal P}$. Unfortunately, on arbitrary space-times, these operators exist but admit no closed form in general. This is because now the order of the differential operators matters since covariant derivatives do not commute and in particular $[ \na_{\mu}, \bo_{\rm r}^{-1} ] \neq 0$. As shown in appendix \ref{sec:commrel}, already for an Einstein space $R_{\mu\nu} = \ka\, g_{\mu\nu}$, where $\ka$ is a constant, we have
\beq \label{eq:Einsteinnabocom}
\na_{\mu} \bo_{\rm r}^{-1} = \( \bo - \ka \)_{\rm r}^{-1} \na_{\mu} \, .
\eeq
The only case where this is not a problem is for vectors, where one can simply covariantize the original expression (\ref{eq:PMinkA})
\beq 
{\cal P}_{\mu}^{\,\,\,\nu} \equiv \de_{\mu}^{\nu} - \na_{\mu} \bo_{\rm r}^{-1} \na^{\nu} \, ,
\eeq
making only sense on vectors whose covariant divergence has finite past. Indeed, all the properties of this operator, listed below eq. (\ref{eq:PMinkA}), are still valid and their demonstrations go exactly the same since we did not use $[ \pa_{\mu}, \pa_{\nu} ] = 0$ nor $[ \pa_{\mu}, \bo_{\rm r}^{-1} ] \neq 0$ to derive them. It is therefore a projector on the transverse subspace
\beq
\na^{\mu} A^{\rm T}_{\mu} = 0 \, , \hspace{1cm} A^{\rm T}_{\mu} \equiv {\cal P}_{\mu}^{\,\,\,\nu} A_{\nu}
\eeq
and is invariant under U$(1)$ gauge transformations whose parameter has finite past\footnote{Indeed, as shown in appendix \ref{sec:leftinv}, the property $[\bo, \bo_{\rm r}^{-1}] = 0$ for fields with finite past still holds for globally hyperbolic space-times.}. The only difference with the flat space-time case is that now ${\cal P}_{\mu\nu}$ is not symmetric because $[ \na_{\mu}, \bo_{\rm r}^{-1} ] \neq 0$. 

To understand the obstruction in constructing closed forms for transverse operators ${\cal P}$ of higher rank on generic space-times, let us first see how one could proceed for the vector case. We can start by defining the action of $\cal P$ through an auxiliary field $A$
\beq
A^{\rm T}_{\mu} \equiv A_{\mu} - \na_{\mu} A \, ,
\eeq
obeying
\beq
\bo A = \na_{\mu} A^{\mu} \, .
\eeq
Then, solving for $A$ using the retarded $\bo^{-1}$ one retrieves the definition of ${\cal P}_{\mu}^{\,\,\,\nu} A_{\nu}$. Note that this looks very much like the localization procedure since the initial conditions of $A$ are fixed to zero at past-infinity by the use of $\bo^{-1}_{\rm r}$. This is not a surprise, since a transverse operator is necessarily non-local and the above procedure amounts to localizing it by integrating in $A$.

Now let us try the above construction for ${\cal P}$ in the case of symmetric two-tensors. We can again define
\beq \label{eq:STdef}
h^{\rm T}_{\mu\nu} \equiv h_{\mu\nu} - \na_{(\mu} h_{\nu)}  \, ,
\eeq
where the $D$ components of $h_{\mu}$ obey the $D$ equations
\beq  \label{eq:Smueqp}
\bo h_{\mu} + \na_{\nu} \na_{\mu} h^{\nu} = 2 \na^{\nu} h_{\nu\mu} \, ,
\eeq
or alternatively
\beq \label{eq:Smueq}
\bo h_{\mu} + \na_{\mu} \na_{\nu} h^{\nu} + R_{\mu\nu} h^{\nu} = 2 \na^{\nu} h_{\nu\mu} \, .
\eeq
%could invert the full operator $\de_{\nu}^{\mu} \bo + \na_{\nu} \na^{\mu}$ in (\ref{eq:Smueqp})
To solve for $h_{\mu}$ one must first solve for $\na_{\mu} h^{\mu}$ which, on flat space-time, would be achieved by taking the double-divergence of (\ref{eq:STdef}). Doing this on arbitrary space-time and rearranging the covariant derivatives in a convenient way we get
\beq
\bo \na_{\mu} h^{\mu} + R_{\mu\nu} \na^{\mu} h^{\nu} + \frac{1}{2}\, h^{\mu} \na_{\mu} R = \na_{\mu} \na_{\nu} h^{\mu\nu} \, .
\eeq
We now see that $\na_{\mu} h^{\mu}$ cannot be expressed in terms of $h_{\mu\nu}$ on arbitrary space-times, hence the obstruction for the construction of a closed form for ${\cal P}$. Rather, it seems that one can proceed only in the case of an Einstein space-time $R_{\mu\nu} = \ka\, g_{\mu\nu}$, with $\ka$ a constant
\beq
\na_{\mu} h^{\mu} = \( \bo + \ka \)_{\rm r}^{-1} \na_{\mu} \na_{\nu} h^{\mu\nu} \, .
\eeq
Plugging this back inside (\ref{eq:Smueq}) allows us to express $h_{\mu}$ in terms of $h_{\mu\nu}$,
\beq
h_{\mu} = 2 \( \bo + \ka \)_{\rm r}^{-1} \na^{\nu} h_{\nu\mu} - \( \bo + \ka \)^{-1}_{\rm r} \na_{\mu} \( \bo + \ka \)^{-1}_{\rm r} \na_{\nu} \na_{\ro} h^{\nu\ro} \, ,
\eeq
so plugging this result inside (\ref{eq:STdef}) finally gives
\bea
h^{\rm T}_{\mu\nu} & = & h_{\mu\nu} - 2\na_{(\mu|} \( \bo + \ka \)_{\rm r}^{-1} \na^{\ro} h_{\ro|\nu)} \nn \\
 & & + \na_{(\mu} \( \bo + \ka \)^{-1}_{\rm r} \na_{\nu)} \( \bo + \ka \)^{-1}_{\rm r} \na_{\ro} \na_{\si} h^{\ro\si} \equiv {}_1 {\cal P}_{\mu\nu}^{\,\,\,\,\,\,\ro\si} h_{\ro\si} \, .
\eea
Indeed, specializing to flat space-time, one can then recognize the action of $_1 {\cal P}$ as defined in (\ref{eq:aprojfam}). Note that under a gauge transformation
\beq \label{eq:gshna}
\de h_{\mu\nu} = - \na_{\mu} \xi_{\nu} - \na_{\nu} \xi_{\mu} \, ,
\eeq
by the defining equations (\ref{eq:STdef}) and (\ref{eq:Smueqp}), we have that $\de h_{\mu} = -2 \xi_{\mu}$ and $h_{\mu\nu}^{\rm T}$ is invariant. Thus, as in the case of flat space-time, $h_{\mu\nu}^{\rm T}$ is both transverse and gauge-invariant. 

One can then generalize the whole one-parameter family of transverse operators (\ref{eq:aprojfam}). The transverse-traceless projector can be constructed analogously by defining
\beq
h^{\rm TT}_{\mu\nu} \equiv \ti{h}_{\mu\nu} - \na_{(\mu} h_{\nu)} + \frac{1}{D}\, g_{\mu\nu} \na_{\ro} h^{\ro} \, , \hspace{1cm} \ti{h}_{\mu\nu} \equiv h_{\mu\nu} - \frac{1}{D}\, g_{\mu\nu} h \, ,
\eeq
and
\beq
\bo h_{\mu} + \frac{D-2}{D}\, \na_{\mu} \na_{\nu} h^{\nu} + R_{\mu\nu} h^{\nu} = 2 \na^{\nu} \ti{h}_{\nu\mu} \, .
\eeq
Solving for $h_{\mu}$ on an Einstein space-time one then gets
\bea
h^{\rm TT}_{\mu\nu} & \equiv & \ti{h}_{\mu\nu} - 2 \na_{(\mu|} \( \bo + \ka \)_{\rm r} \na^{\ro} \ti{h}_{\ro|\nu)} +  \frac{1}{d}\, g_{\mu\nu}  \( \bo + \frac{d+1}{d}\,\ka \)_{\rm r} \na_{\ro} \na_{\si} \ti{h}^{\ro\si}  \nn \\
 & & +  \frac{d-1}{d}\, \na_{(\mu} \( \bo + \ka \)_{\rm r} \na_{\nu)} \( \bo + \frac{d+1}{d}\,\ka \)_{\rm r} \na_{\ro} \na_{\si} \ti{h}^{\ro\si} \equiv {}_0 {\cal P}_{\mu\nu}^{\,\,\,\,\,\,\ro\si} h_{\ro\si} \, ,
\eea
which reduces to the action of $_0 {\cal P}$ on flat space-time (\ref{eq:aprojfam}). Again, $h^{\rm TT}_{\mu\nu}$ is invariant under both (\ref{eq:gshna}) and $\de h_{\mu\nu} = - g_{\mu\nu} \te$, which is the generalization of (\ref{eq:gsct}). Finally, note that $_1 {\cal P}$ and $_0 {\cal P}$ are $\Rs$-linear operators even when they cannot be described in closed form, as is easy to check using their definitions involving the auxiliary fields. Thus, the projector on the transverse-pure-trace part can be defined using (\ref{eq:sPdef})
\beq
h^{\rm TpT}_{\mu\nu} \equiv h^{\rm T}_{\mu\nu} - h^{\rm TT}_{\mu\nu} \, ,
\eeq
and the generalization of $_a {\cal P} \cdot h$ is
\beq \label{eq:genaP}
_a h^{\rm T}_{\mu\nu} \equiv h^{\rm TT}_{\mu\nu} + a h^{\rm TpT}_{\mu\nu} \, .
\eeq

\subsubsection{Origin of the obstruction}

The origin of this limitation to Einstein space-times can be traced back to the ``pathology'' of linear higher spin theories \cite{Fronsdal} of not being able to preserve their gauge symmetries on backgrounds that are not Einstein \cite{AragoneDeser1,AragoneDeser2,BvHdWvN,deWitFreedman}, \footnote{Simply put, unlike in the case of differential forms, the presence of symmetric pairs of indices when $s \geq 2$ forces the use of $\na_{\mu}$ in the action. This in turn implies that the gauge symmetry also depends on $\na$ and can therefore not be achieved on arbitrary space-times.}. Indeed, in the vector case $s=1$, the Maxwell action generalizes straightforwardly to arbitrary background
\beq
S = \int \ed^D x\, \sqrt{-g}\[ -\frac{1}{4} \,  g^{\mu\nu} g^{\ro\si} F_{\mu\ro} F_{\nu\si} + A_{\mu} j^{\mu} \] \, ,
\eeq
which is still U$(1)$-symmetric, and the equations of motion are thus covariantly transverse (for a covariantly conserved source)
\beq
\na_{\nu} F^{\mu\nu} = - j^{\mu} \, ,
\eeq
since
\beq
\na_{\mu} \na_{\nu} F^{\mu\nu} = \na_{[\mu} \na_{\nu]} F^{\mu\nu} = R_{\mu\nu} F^{\mu\nu} = 0 \, .
\eeq
This implies that they can be written as a differential operator composed with the transverse projector ${\cal P}_{\mu}^{\,\,\nu}$ acting on $A_{\mu}$. Indeed
\bea
\bo A^{\rm T}_{\mu} - R_{\mu}^{\nu} A_{\nu}^{\rm T} & \equiv & \bo A_{\mu} - \bo\na_{\mu} \bo^{-1} \na_{\nu} A^{\nu} - R_{\mu\nu} A^{\nu} + R_{\mu}^{\nu} \na_{\nu} \bo^{-1} \na_{\ro} A^{\ro} \nn \\
 & = & \bo A_{\mu} - \na_{\mu} \na_{\nu} A^{\nu}  - \[ \bo, \na_{\mu} \] \bo^{-1} \na_{\nu} A^{\nu} - R_{\mu\nu} A^{\nu} + R_{\mu}^{\nu} \na_{\nu} \bo^{-1} \na_{\ro} A^{\ro} \nn \\
 & = & \bo A_{\mu} - \na_{\mu} \na_{\nu} A^{\nu}  - R_{\mu\nu} \na^{\nu} \bo^{-1} \na_{\nu} A^{\nu} - R_{\mu\nu} A^{\nu} + R_{\mu}^{\nu} \na_{\nu} \bo^{-1} \na_{\ro} A^{\ro} \nn \\
 & = & \bo A_{\mu} - \na_{\mu} \na_{\nu} A^{\nu}  - \[ \na_{\mu}, \na_{\nu} \] A^{\nu} - R_{\mu\nu} A^{\nu} \nn \\
 & = & \bo A_{\mu} - \na_{\mu} \na_{\nu} A^{\nu} = \na_{\nu}F^{\mu\nu} \, ,
\eea
so the equation of motion can be written\footnote{In the Proca case the equation of motion in this form is simply modified by $\bo \to \bo - m^2$.}
\beq
\[ \de_{\mu}^{\nu} \bo - R_{\mu}^{\nu} \] A_{\nu}^{\rm T} = -j_{\mu} \, .
\eeq
The Ricci term makes the square bracket commute with the divergence operation, which then gives zero when acting on $A^{\rm T}$. Thus, the existence of a gauge-invariant action is related to the existence of a closed form for the transverse projector that can be read out of the equations of motion. In the case of higher-spin fields, if there existed such a closed form for ${\cal P}$ on arbitrary backgrounds, then one could construct gauge-invariant equations of motion, in closed form, and thus deduce a gauge-invariant action. This is why there exists no closed form for $_a {\cal P}_{\mu\nu}^{\,\,\,\,\,\,\ro\si}$ on arbitrary space-times.

\subsection{Action-based models} \label{sec:actiontrick}

The other possibility for constructing transverse equations of motion, considered for instance in \cite{DeserWoodard1, Barvinsky2, MaggioreMancarella, CKMT}, is to start with a generally-covariant (formal) action. Indeed, say we have such an action for pure gravity
\beq
S = \int \ed^D x \, \sqrt{-g}\, L[g] \, ,
\eeq
where the Lagrangian $L$ is a scalar. Then, performing an infinitesimal (active) diffeomorphism 
\beq
\de g^{\mu\nu} = - \Lie_{\xi} g^{\mu\nu} \equiv - \xi^{\ro} \pa_{\ro} g^{\mu\nu} + g^{\ro\nu} \pa_{\ro} \xi^{\mu} + g^{\mu\ro} \pa_{\ro} \xi^{\nu} = \na^{\mu} \xi^{\nu} + \na^{\nu} \xi^{\mu} \, ,
\eeq
we get
\bea
\de S & = & \int \ed^D x\, \de g^{\mu\nu}\, \frac{\de (\sqrt{-g}\, L)}{\de g^{\mu\nu}} = 2\int \ed^D x \, \sqrt{-g}\, \na^{\mu} \xi^{\nu} \[ \frac{1}{\sqrt{-g}} \frac{\de (\sqrt{-g}\, L)}{\de g^{\mu\nu}} \]  \nn \\
 & = & -2\int \ed^D x \, \sqrt{-g}\, \xi^{\nu} \na^{\mu} \[ \frac{1}{\sqrt{-g}} \frac{\de (\sqrt{-g}\, L)}{\de g^{\mu\nu}} \]\, .
\eea
Since diffeomorphisms are a symmetry of the action we have that $\de S = 0$, for any $g_{\mu\nu}$ and $\xi^{\mu}$, so that 
\beq
\na^{\mu} \[ \frac{1}{\sqrt{-g}} \frac{\de (\sqrt{-g}\, L)}{\de g^{\mu\nu}} \] \equiv 0 \, ,
\eeq
is an identity, independently of whether $S$ is local or not. We thus see that the utility of the formal non-local actions, defined in section \ref{sec:formalaction}, is not only ornamental anymore, it has become a valuable tool in deriving transverse equations of motion. Note that the ad hoc prescription of turning all the $\ti{\bo}^{-1}$ into retarded ones at the end of the variation does not spoil transversality. Indeed, the latter being a local property, it cannot depend on the choice of $\ti{\bo}^{-1}$, since what distinguishes all these operators is non-local information, i.e. the boundary/initial data of the Green's function. All that matters is that $\ti{\bo}^{-1}$ is a right-inverse of $\ti{\bo}$. We can therefore safely apply our variational principle on the formal action.

We stress one more time that formal actions should not be given any physical meaning. Their variation gives rise to non-causal equations of motion, which we make causal by hand afterwards. Moreover, remember that non-local theories are classical theories\footnote{Indeed, as discussed in section \ref{sec:quantization}, one cannot quantize a non-local theory without either enlarging the set of solutions in the classical limit, or losing unitarity.}, so all the information lies in the final, causal, equations of motion. 

Finally, now that $\ti{\bo}$ depends on the metric, we need a formula for the variation of $\ti{\bo}^{-1}$ with respect to $g_{\mu\nu}$ at the level of the formal action. To compute this, we use the same logic as in appendix \ref{sec:commrel}. We apply the variation on $\ti{\bo} \ti{\bo}^{-1} = {\rm id}$ to get
\beq
(\de \ti{\bo}) \ti{\bo}^{-1} + \ti{\bo} \de \ti{\bo}^{-1} = 0 \, ,
\eeq
and then apply $\bo^{-1}$ from the left to isolate the quantity of interest
\beq
\de \ti{\bo}^{-1} \equiv - \ti{\bo}^{-1} (\de \ti{\bo}) \ti{\bo}^{-1} \, .
\eeq
The above equation holds modulo homogeneous solutions, which is indeed the level at which the variation is performed for formal actions.

\subsubsection{Example}

Now that we have all the necessary tools let us work out the simplest example  
\beq \label{eq:actiontrickex}
S = \frac{1}{2} \int \ed^D x  \, \sqrt{-g} \,R \bo^{-1} R \, .
\eeq
Using $\de R = \( R_{\mu\nu} + g_{\mu\nu} \bo - \na_{\mu} \na_{\nu} \) \de g^{\mu\nu}$, integrating by parts at will and sending $\bo^{-1} \to \bo^{-1}_{\rm r}$ at the end we get
\beq \label{eq:nonloceqtrickex}
g_{\mu\nu} R - \na_{\mu} \na_{\nu} \bo^{-1}_{\rm r} R + G_{\mu\nu} \bo_{\rm r}^{-1} R + \frac{1}{2} \( \na_{\mu} \bo^{-1}_{\rm r} R \) \na_{\nu} \bo^{-1}_{\rm r} R - \frac{1}{4}\, g_{\mu\nu} \( \na_{\ro} \bo^{-1}_{\rm r} R \) \na^{\ro} \bo^{-1}_{\rm r} R  \, .
\eeq
Let us now check the transversality of this expression. Taking the divergence and using $[\bo, \na_{\mu}] \ph = R_{\mu\nu} \na^{\nu} \ph$ to simplify the second term
\beq
\bo \na_{\nu} \bo^{-1}_{\rm r} R = R_{\mu\nu} \na^{\nu} \bo^{-1}_{\rm r} R + \na_{\nu} R \, ,
\eeq
we get zero indeed. From this example one thing which is obvious is that the equations of motion of a simple non-local action will usually be rather complicated. There is thus also a practical advantage in describing the model through a formal non-local action, that is, being able to display its information in a compact way. Remember however that, since we have replaced by hand $\bo^{-1} \to \bo^{-1}_{\rm r}$, these are {\it not} the equations of motion of this action, i.e. $\de S \neq 0$ around these solutions.

\subsection{The necessity of considering the scalar mode}

Before we proceed to the construction of the generally-covariant transverse equations, we can already note one limitation of our procedure. Indeed, it appears that the non-local formulation of Fierz-Pauli theory will not be generalizable as wished. Remember that FP theory can be expressed as (\ref{eq:nonlocFP2}) which corresponds to (\ref{eq:projeq}) with $z = 0$, $m_s = 0$ and with the source $T_{\mu\nu}$ replaced by its transverse-traceless part $T_{\mu\nu}^{\rm TT}$. This theory has an extra gauge symmetry (\ref{eq:gsct}) which is responsible for neutralizing the trace mode. In the non-linear context, the natural generalization of this symmetry is the conformal transformation
\beq \label{eq:conftrans}
g_{\mu\nu} \to e^{2 \te} g_{\mu\nu} \, .
\eeq
Thus, in order to keep this field non-dynamical in the non-linear theory we need the latter to be conformally invariant as well. This is however impossible to implement for the following reasons. 

Although we do have building blocks that are generally covariant, the curvature tensors, the only one which is also covariant under (\ref{eq:conftrans}) is the Weyl tensor. A first disadvantage is then that an action made exclusively out of the Weyl tensor could hardly be considered as a deformation of GR. One should then use the $\ti{\bo}^{-1}$ which transforms homogeneously under conformal transformations. For instance, in the case where $\ti{\bo}$ acts on a scalar, i.e. $\ti{\bo} = \bo - \xi R$, the covariant choice is $\xi = (d-1)/4d$ and the transformation is
\beq
\ti{\bo} \to e^{-\frac{D+2}{2}\, \te} \ti{\bo} e^{\frac{D-2}{2}\, \te} \, .
\eeq
Using the conformally-covariant $\ti{\bo}^{-1}$ for tensors of rank $4$, the only action made of the Weyl tensor and $\ti{\bo}^{-1}$ which gives the non-local FP theory in the linearized limit is
\beq
S = \frac{M^2}{2} \int \ed^D x \, \sqrt{-g} \, W_{\mu\nu\ro\si} \frac{1}{\ti{\bo}} \( 1 - \frac{m_g^2}{\ti{\bo}} \) W^{\mu\nu\ro\si} + \Ord(W^3) \, .
\eeq
However, because of the fixed masses, here $M$ and $m_g$, conformal invariance is still not achieved. Indeed, even if the transformation of each term is homogeneous, in the presence of a fixed mass there remains an overall exponential factor $e^{c\te}$, \footnote{These overall factors are not seen in the linearized limit because they multiply second-order terms in the action, or first-order terms in the equations of motion, and thus reduce to $e^{c \te} \to 1$.}. The same happens in a projector-based equation, i.e. the terms that come with different powers of mass do not transform with the same powers of $e^{\te}$, \footnote{The above problems could be resolved if we replace the fixed masses by a scalar field $\ph$ sitting in a non-trivial minimum of its potential and transforming homogeneously under (\ref{eq:conftrans})
\beq
\ph \to e^{\frac{D-2}{2}\, \te} \ph \, .
\eeq
This allows us to use all the curvature invariants, since we can compensate their inhomogeneous transformation with the one of the kinetic term of $\ph$, while at the same time there are no fixed masses and thus no leftover exponential factors under (\ref{eq:conftrans}). The problem now however is that we have one more dynamical field $\ph$ and the gauge symmetry either neutralizes the latter or the scalar mode in $g_{\mu\nu}$, not both. Thus, we still have one more dynamical field than what we started with.}. On top of this problem, note that the coupling to matter should also be made non-local in order to be conformally-invariant, so that the source is $T^{\rm TT}_{\mu\nu}$, yet another challenge. This is why we have also considered the non-local theories that include the trace scalar (\ref{eq:projeq}), but in a healthy way, so that we do not need to implement conformal invariance. From now on we will only consider these models.

\section{Action-based models}

\subsection{Constructing the action}

We now wish to construct an action-based generally-covariant extension of the model (\ref{eq:projeq}) introduced in section (\ref{sec:newmasstheor}). The formal action corresponding to (\ref{eq:projeq}) is
\beq  \label{eq:Slinnonloca}
S = \frac{1}{2} \int \ed^D x \, h_{\mu\nu} \[  \( \bo - m_g^2\) {}_0 {\cal P}^{\mu\nu\ro\si} +  \( z \bo - m_s^2 \) {}_s {\cal P}^{\mu\nu\ro\si} \] h_{\ro\si} \, ,
\eeq
and now $h_{\mu\nu}$ is interpreted as the perturbation of some metric around the Minkowski one
\beq \label{eq:hdef}
h_{\mu\nu} \equiv \frac{M}{2} \( g_{\mu\nu} - \et_{\mu\nu} \) \, ,
\eeq
where $M \equiv (8\pi G)^{-1/2}$ is the reduced Planck mass in $D=4$. The only terms that contribute to the linearized action are those linear and quadratic in curvature. A general enough action to match (\ref{eq:Slinnonloca}) at that order is 
\beq \label{eq:S2R}
S_2 = \frac{M^2}{2} \int \ed^D x\, \sqrt{-g} \[ R + \frac{1}{2}\, R \Ord_1 R - 2 R_{\mu\nu} \Ord_2 R^{\mu\nu} + \frac{1}{2}\, R_{\mu\nu\ro\si} \Ord_3 R^{\mu\nu\ro\si} \]_2 \, ,
\eeq
where the $\Ord_i$ are operators of the form
\beq
\Ord_i = A_i \bo^{-1} + B_i \bo^{-2} \, , 
\eeq
and $A_i, B_i$ are constants. An alternative parametrization that will be useful later is
\beq
S_2 = \frac{M^2}{2} \int \ed^D x\, \sqrt{-g} \[ R + \frac{1}{2}\, R \ti{\Ord}_1 R - 2 R_{\mu\nu} \ti{\Ord}_2 R^{\mu\nu} + \frac{1}{2}\, W_{\mu\nu\ro\si} \Ord_3 W^{\mu\nu\ro\si} \]_2 \, ,
\eeq
where 
\beq \label{eq:Weyldef}
W_{\mu\nu\ro\si} \equiv R_{\mu\nu\ro\si} - \frac{2}{d-1} \( g_{\mu[\ro} R_{\si]\nu} -  g_{\nu[\ro} R_{\si]\mu} \) + \frac{2}{d(d-1)}\, g_{\mu[\ro} g_{\si]\nu} R \, , 
\eeq
is the Weyl tensor and 
\beq
\ti{\Ord}_1 = \Ord_1 - \frac{1}{d(d-1)}\, \Ord_3 \, , \hspace{1cm} \ti{\Ord}_2 = \Ord_2 - \frac{1}{d-1}\, \Ord_3 \, .
\eeq
We can then write (\ref{eq:S2R}) as
\bea
S & = & \frac{1}{2} \int \ed^D x \, h_{\mu\nu} {\cal K}^{\mu\nu\ro\si} h_{\ro\si}  \, ,
\eea
to find
\bea
{\cal K}^{\mu\nu\ro\si} & = & \( 2(\Ord_3 - \Ord_2) +  \bo^{-1} \) \et^{\mu(\ro} \et^{\si)\nu} \bo^2 - \( 2(\Ord_3 - \Ord_2) + \bo^{-1} \) \( \et^{\mu(\ro} \pa^{\si)} \pa^{\nu} + \et^{\nu(\ro} \pa^{\si)} \pa^{\mu} \) \bo  \nn \\
 & & + \( 2(\Ord_2 - \Ord_1) + \bo^{-1} \) \( \et^{\mu\nu} \pa^{\ro} \pa^{\si} + \et^{\ro\si} \pa^{\mu} \pa^{\nu} \) \bo - \( 2(\Ord_2 - \Ord_1) + \bo^{-1} \) \et^{\mu\nu} \et^{\ro\si} \bo^2 \nn \\
 & &  + 2\( \Ord_1 - 2 \Ord_2 + \Ord_3 \) \pa^{\mu} \pa^{\nu} \pa^{\ro} \pa^{\si}   \, . \label{eq:K}
\eea
By diffeomorphism invariance, $\cal K$ is transverse so it must be a combination of $_a {\cal P}$ operators. Equating this to (\ref{eq:Slinnonloca}) we get
\bea
\( 2(\Ord_3 - \Ord_2) + \bo^{-1} \) \bo^2 & = & \bo - m_g^2  \, , \\
\( 2(\Ord_2 - \Ord_1) + \bo^{-1} \) \bo^2 & = & \frac{1}{d} \( \bo - m_g^2 \) - \frac{1}{d} \( z \bo - m_s^2 \)  \, ,
\eea
and the solutions are
\bea
\Ord_1 & = & \[ A_3 + 1 - \frac{D - z}{2d} \] \bo^{-1} + \[ B_3 + \frac{D m_g^2 - m_s^2}{2d} \] \bo^{-2} \, , \nn\\
\Ord_2 & = & A_3 \bo^{-1} + \[ B_3 + \frac{m_g^2}{2} \] \bo^{-2} \, , 
\eea
or alternatively,
\bea
\ti{\Ord}_1 & = & \[ \frac{d^2 - d -1}{d(d-1)}\, A_3 + 1 - \frac{D - z}{2d} \] \bo^{-1} + \[ \frac{d^2 - d -1}{d(d-1)}\,B_3 + \frac{D m_g^2 - m_s^2}{2d} \] \bo^{-2} \, , \nn\\
\ti{\Ord}_2 & = & \frac{d-2}{d-1}\, A_3 \bo^{-1} + \( \frac{d-2}{d-1}\,  B_3  + \frac{m_g^2}{2} \) \bo^{-2} \, . \label{eq:tiO2}
\eea
We have two equations for three operators, which is due to the fact that one can add an arbitrary operator $\Ord$ to all the $\Ord_i$ simultaneously without changing the linearized $S$. This is a consequence of the fact that the Gauss-Bonnet-like combination
\beq
\int \ed^D x \[ R\Ord R -4 R_{\mu\nu} \Ord R^{\mu\nu} + R_{\mu\nu\ro\si} \Ord R^{\mu\nu\ro\si} \] \, ,
\eeq
is a total derivative at the linearized level for all $\Ord$ if $\[ \pa, \Ord \] = 0$. It becomes however non-trivial when $\Ord$ is an inverse differential operator at the non-linear level, even for $D=4$, because then $\[ \na, \bo^{-1} \] \neq 0$. Now that we have expressed the linear action in terms of curvature invariants we can easily generalize it to a fully non-linear theory.

\subsection{Curvature expansion}

There are two types of modifications that can occur in generalizing the above theory. The first one is the same as in the local case, i.e. one can add arbitrary local terms that are higher order in curvature. Since $R_{\mu\nu\ro\si}$ is dimensionful, these terms come with associated mass scales which control the scale at which they influence the physics. Thus, as long as we work at scales larger than the smaller of these masses, the lowest order terms are more than enough. This is the principle of effective field theory, which allows one to consider the most general possible action, compatible with the symmetries of the system, at the energies of interest. 

The second kind of modification is that one can add arbitrary {\it non-local} terms that are higher order in curvature. Unlike their local counterparts, these need not have higher mass dimension. They can actually be dimensionless, such as $\ti{\bo}^{-1} R$ for instance, or even have negative mass dimensions, such as $\ti{\bo}^{-2} R$. This means that their coefficients can have zero dimension, in which case they cannot be neglected for ``natural'' $\Ord(1)$ values, whatever the scale, or positive mass dimension, in which case they dominate the low-energy physics. As a matter of fact, in non-local field theory such power-counting arguments are more limited, because a $\sim \bo^{-1}$ term can dominate at large space-time scales, because of the cumulative effect of the integral, without necessarily having an overall negative mass dimension. 

From the point of view of effective field theory, this is a drawback of non-local field theories, i.e. symmetry alone does not reduce the terms that are relevant for low-energy physics to a finite set. From the point of view of the phenomenologist however, this can be seen as an advantage, since one has many different possibilities for modifying the infrared physics.  

We thus see that by abandoning locality we gain access to way too many non-linear theories and thus need some more input in order to select a given subset. For simplicity we will only consider theories that are second-order in curvature such that there are no terms which do not contribute to the linearized theory. Moreover, we will not consider terms involving derivatives of curvature tensors, such as 
\beq \label{eq:nontrivtermexa}
\(\na_{\mu} \ti{\bo}^{-1} R_{\nu\ro} \)\na^{\nu} \ti{\bo}^{-1} R^{\mu\ro} \, ,
\eeq
for instance. Their inclusion could be very interesting, but as we will argue later, they will not influence our results qualitatively. With these simplifications, we are then left with $A_3$ and $B_3$ as unknown parameters, as well as the operators $\ti{\bo}^{-1}$.

\subsection{Choosing $A_3$ and $B_3$}

\subsubsection{The Ricci model}

From the purely theoretical point of view, the most elegant and simple model is the one with no Riemann tensor terms in the action, i.e. $A_3 = B_3 = 0$. The action can then be conveniently written
\beq \label{eq:RicciS}
S_R = \frac{M^2}{2} \int \ed^D x\, \sqrt{-g} \[R + \frac{1}{2}\, R \[ Z \ti{\bo}^{-1} + m^2 \ti{\bo}^{-2} \] R - m_g^2 R_{\mu\nu} \ti{\bo}^{-2} R^{\mu\nu}   \]   \, ,  
\eeq
where
\beq \label{eq:Z}
Z \equiv \frac{z + d - 1}{2d} \, ,  \hspace{1cm} m^2 \equiv \frac{D m_g^2 - m_s^2}{2d} \, ,
\eeq
and we will refer to it as the ``Ricci'' model. A nice feature of this model is that it shares all the empty space solutions of GR, such as the Schwarzschild and Kerr solutions, whatever the value of the masses. Indeed, since the departure from GR is made of terms quadratic in the Ricci scalar and tensor, we have that every term in the equations of motion will have at least one Ricci tensor or scalar, so that all of them vanish when $R_{\mu\nu} = 0$. This should be contrasted with local massive gravity, where the stationary black hole solution is modified in a non-trivial way, as mentioned in the introduction when we discussed the Vainshtein mechanism.

\subsubsection{The Weyl model}

Remembering that our aim for constructing such theories is to account for dark energy, we should now see what background cosmology has to say about the $A_3$ and $B_3$ parameters. In this context, since the Weyl tensor vanishes for the FLRW metric, only the terms involving the Ricci scalar and Ricci tensor matter. Moreover, for the energy scales of late-time cosmology, the ``past infinity'' of the period of interest is the radiation-dominated era in which case $R = 0$. Thus, in that case $R$ has finite past, while $R_{\mu\nu}$ does not, so it is a natural condition to impose that all $\ti{\bo}^{-1}$ act exclusively on $R$ and $W_{\mu\nu\ro\si}$. If this were not the case, as in the Ricci model, one would have to choose an initial time $t_i$ to begin the convolution with the retarded Green's function. One could then adopt an effective theory point of view and say that at earlier times the energy is above the region of validity of the theory, so that the latter makes sense only for $t > t_i$. Nevertheless, one would still remain with a non-trivial dependence of the history of the universe on that time $t_i$, and with no particular way to privilege a given choice. Most importantly however, in practice the $\sim \ti{\bo}^{-1} R_{\mu\nu}$ terms do not offer a viable cosmological background evolution because they generically give rise to diverging modes \cite{Maggiore2, FoffaMaggioreMitsou1, FerreiraMaroto, ModestoTsujikawa}. Therefore, although the Ricci model may have its theoretical advantages, it is not phenomenologically viable. With the $\ti{\bo}^{-1}$'s acting only on $R$ and the Weyl tensor we avoid these conceptual and practical worries and have a well-defined convolution. 

Another advantage of this prescription is that the beginning of the matter-dominated era marks the beginning of the non-local memory effect since this is when $\ti{\bo}^{-1} R$ starts recording the past. This is a cumulative effect and can become non-negligible at considerably later times. Therefore, in this scenario one obtains an elegant alleviation of the coincidence problem, since dark energy appears as a delayed effect of the matter-radiation transition. This was actually the original motivation for the Deser-Woodard model (\ref{eq:DW}) \cite{DeserWoodard1}, to relate the dark energy scale and timing to an earlier event in the history without having to introduce a new fixed scale. Here we also consider such fixed mass scales but the spirit is the same.  

Given the above considerations, we fix $A_3$ and $B_3$ so that the $R_{\mu\nu}^2$ terms drop in the Weyl representation of the action, i.e. so that $\ti{\Ord}_2 = 0$. Given (\ref{eq:tiO2}), we get
\beq
A_3 = 0 \, , \hspace{1cm} B_3 = - \frac{d-1}{d-2}\, \frac{m_g^2}{2} \, ,
\eeq
and thus
\beq \label{eq:WeylS}
S_W = \frac{M^2}{2} \int \ed^D x\, \sqrt{-g} \[R + \frac{1}{2}\, R \[ Z \ti{\bo}^{-1} - m_R^2 \ti{\bo}^{-2} \] R - \frac{1}{2}\,m_W^2 W_{\mu\nu\ro\si}  \ti{\bo}^{-2} W^{\mu\nu\ro\si}  \]   \, ,  
\eeq
where now
\beq  \label{eq:mR2Weyl}
m_R^2 \equiv  \frac{m_g^2 + (d-2) m_s^2}{2d(d-2)}  \, , \hspace{1cm} m^2_W  \equiv  \frac{d-1}{d-2} \frac{m_g^2}{2} \, .
\eeq
We will refer to this as the ``Weyl'' model. In contrast with the Ricci model, this model does not have the vacuum solutions of GR since the Weyl tensor is precisely the part of the curvature which is non-trivial in this case. Finally, note that both the Ricci (\ref{eq:RicciS}) and the Weyl (\ref{eq:WeylS}) models reduce to GR in the massless limit only if $Z = 0$, which translates into $z = 1-d$ and thus implies that the trace scalar is a ghost, as already noted in section \ref{sec:newmasstheor}.

\subsection{Localization}

Here our expressions will be simpler if we rather use an alternative reduced Planck mass $\ti{M} \equiv (16 \pi G)^{-1/2}$, instead of $M \equiv (8\pi G)^{-1/2}$.

\subsubsection{Weyl model}

Let us first consider the Weyl model (\ref{eq:WeylS}). Since $R$ and $W_{\mu\nu\ro\si}$ are independent components of the Riemann tensor, we have to consider a localizing field for each one of them. One possibility is
\beq
S_W = \int \ed^D x\, \sqrt{-g} \[ \ti{M}^2 R + \ti{M} \ph R + \frac{1}{2 m_R^2} \( \ti{\bo} \ph - \frac{Z}{2}\, \ti{M} R \)^2  + \ti{M} W_{\mu\nu\ro\si}  \ph^{\mu\nu\ro\si} + \frac{1}{2 m_W^2} \( \ti{\bo} \ph_{\mu\nu\ro\si} \)^2\]   \, . \label{eq:WeylSloc} 
\eeq
Indeed, integrating them out using the following solutions
\bea
\ph & = & \ti{M} \(  \frac{Z}{2}\, \ti{\bo}^{-1} R - m_R^2 \ti{\bo}^{-2} R \) \, , \\
\ph_{\mu\nu\ro\si} & = & \ti{M} \( - m_W^2 \ti{\bo}^{-2} W_{\mu\nu\ro\si} \) \, ,
\eea
we retrieve (\ref{eq:WeylS}). It is obvious that $\ph_{\mu\nu\ro\si}$ has the same symmetries as the Weyl tensor
\bea
\ph_{\mu\nu\ro\si} = -\ph_{\mu\nu\si\ro} = -\ph_{\nu\mu\ro\si}  \, , \nn \\
\ph_{\mu\nu\ro\si} + \ph_{\mu\ro\si\nu} + \ph_{\mu\si\nu\ro} = 0 \, , \label{eq:Weylsym}\\
\ph^{\mu}_{\,\,\,\nu\mu\si} = 0 \, , \nn
\eea
corresponding to the Young tableau
\beq
\yng(2,2) \, .
\eeq
Note that (\ref{eq:WeylSloc}) is a higher derivative theory both for the auxiliary fields and for gravity. To gain more insight, let us integrate in two more auxiliary fields in order to lower the derivative order of the $\ph$'s
\bea
S_W & = & \int \ed^D x\, \sqrt{-g} \[ \ti{M}^2 R + \ti{M} \( \phi + \frac{Z}{2}\, \psi \) R - \phi \ti{\bo} \psi - \frac{m_R^2}{2}\,  \psi^2  \right. \label{eq:locS} \\
 & & \left. \hspace{2cm} + \ti{M} W_{\mu\nu\ro\si} \phi^{\mu\nu\ro\si} - \phi_{\mu\nu\ro\si} \ti{\bo} \psi^{\mu\nu\ro\si} - \frac{m_W^2}{2}\,  \psi_{\mu\nu\ro\si} \psi^{\mu\nu\ro\si} \]   \, .  \nn
\eea
The $\psi$'s carry the information of the initial conditions of the second and third time derivatives of the $\ph$'s, so they are also constrained, even though integrating them out does not require inverting $\ti{\bo}$. We see that the action has become linear in the $\phi$'s. Integrating the latter out and choosing the solutions
\beq
\psi = \ti{M}\ti{\bo}^{-1} R \, , \hspace{1cm} \psi_{\mu\nu\ro\si} = \ti{M} \ti{\bo}^{-1} W_{\mu\nu\ro\si} \, ,
\eeq
gives back (\ref{eq:WeylS}). We could have started with this simpler localization\footnote{Indeed, the direct use of Lagrange multipliers to enforce relations among fields is rather the usual procedure \cite{MaggioreMancarella, DFKKM, NOSZ, Koivisto1, Koivisto2, JNOSTZ, BNOS, NO1, NO2, CENO, Kosh}.}, but this might have misled us to think that the initial conditions of the $\phi$'s are arbitrary, since they are not a priori determined by the equations. With this procedure, we see explicitly that actually both the $\psi$'s {\it and} the $\phi$'s are constrained. As a check, note that for $m_R = m_W = 0$ and $Z = 0$ we recover GR. The action becomes linear in the $\psi$'s and thus their equations of motion
\beq
\ti{\bo} \phi = 0\, , \hspace{1cm} \ti{\bo} \phi_{\mu\nu\ro\si} = 0 \, ,
\eeq
imply that the $\ph$'s vanish since they have no homogeneous solution. The action then turns into the Einstein-Hilbert one.

\subsubsection{Ricci model}

Let us now localize (\ref{eq:RicciS}). Although this model is not phenomenologically viable as far as cosmology is concerned, because of the presence of the Ricci tensor, it is interesting to consider it as well for its theoretical properties. Here we can consider a single localizing field $\ph_{\mu\nu}$, since $R$ is the trace of $R_{\mu\nu}$. Going directly to the second-order formulation, we get
\beq
S_R = \int \ed^D x\, \sqrt{-g} \[ \ti{M}^2 R + \ti{M} R_{\mu\nu}\( \phi^{\mu\nu} + \frac{Z}{2}\,  g^{\mu\nu} \psi \) - \phi_{\mu\nu} \ti{\bo} \psi^{\mu\nu} - m_g^2 \psi_{\mu\nu} \psi^{\mu\nu} + \frac{m^2}{2}\, \psi^2  \]   \, .
\eeq

\subsection{Ghosts} \label{eq:ghosts}

At the linearized level integrating in a vector and a scalar would have been sufficient in making the action local. This is because the non-local operators acted on lower-rank tensors such as $\pa^{\nu} h_{\nu\mu}$. Here we see that, since the non-local operators act on curvature invariants, the localization necessarily involves tensors of rank two or more. Thus, the dynamical content of these theories is quite larger. Most importantly however, some of these fields are ghost-like. Indeed, the first hint lies in the fact that, if we diagonalize a term of the form $\ph \bo \psi$, we get
\beq \label{eq:ghostcomb}
\ph \bo \psi = \( \Phi + \Psi \) \bo \( \Phi -\Psi \) = \Phi \bo \Phi - \Psi \bo \Psi \, .
\eeq
This is of course not a rigorous proof because one should first linearize/diagonalize the full action and only then compare the signs of the kinetic terms. However, this procedure is not possible in general without reintroducing non-localities. So let us try in the simplest case.

\subsubsection{The $m_g = 0$ case}

Let us consider the action $m_g = 0$, in which case we only have the auxiliary scalar sector and the Ricci and Weyl models become the same. Then, linearizing over Minkowski space-time which, given the constraints on the scalars, is the solution
\beq
g_{\mu\nu} = \et_{\mu\nu} \, , \hspace{1cm} \phi = \psi = 0 \, ,
\eeq
using (\ref{eq:hdef}), (\ref{eq:Z}), (\ref{eq:mR2Weyl}) and the following redefinitions
\beq
h_{\mu\nu} \equiv h'_{\mu\nu} - \frac{\sqrt{2}}{d-1}\, \et_{\mu\nu} \( \ph + \frac{Z}{2}\, \psi \) \, , \hspace{1cm} \ph \equiv \vph - \( \frac{Z}{2} - \frac{d-1}{d} \) \psi \, ,
\eeq
one gets the diagonal action
\bea
S & = & \int \ed^D x\, \sqrt{-g} \[ \frac{1}{2}\, h'_{\mu\nu} {\cal E}^{\mu\nu\ro\si} h'_{\ro\si} - \frac{d}{d-1}\, \pa_{\mu} \vph \pa^{\mu} \vph - \frac{1}{4d} \( z \, \pa_{\mu} \psi \pa^{\mu} \psi + m^2_{\rm s} \psi^2 \) \right. \nn \\
 & & \left. \hspace{2.2cm}+ h'_{\mu\nu} T^{\mu\nu} - \sqrt{2} \( \frac{1}{d-1}\, \ph + \frac{1}{d}\, \psi \) T  \]   \, .
\eea
Now $\ph$ and $\psi$ also couple to $T_{\mu\nu}$ and thus contribute to the saturated propagator. We thus retrieve the structure of (\ref{eq:mto0newprop}) in the $m_g \to 0$ limit, up to a field normalization, i.e. the auxiliary scalars $\ph$ and $\psi$ correspond to the scalar poles. More precisely, $\ph$ corresponds to the healthy scalar pole which is responsible for the vDVZ discontinuity between the FP propagator and the one of GR when $m_g \to 0$, i.e. it is the longitudinal mode of the massive graviton which does not decouple. On the other hand, $\psi$ corresponds to the trace scalar with mass $m_s$ and is healthy when $z > 0$.

\subsubsection{The $m_g \neq 0$ case}

So what about the auxiliary tensor modes in the $m_g \neq 0$ models? A first argument supporting the presence of ghosts is that there is no particular kinetic structure that would neutralize the time-components which come with the wrong signs. Indeed, the actions of linear tensor theories are ghost-free only in the presence of quadratic combinations that provide a gauge symmetry which kills the ghost modes \cite{Fronsdal}. Here it seems that non-local terms which mix the tensor indices non-trivially, such as the example given in (\ref{eq:nontrivtermexa}), could arrange this situation by providing the necessary structure. As already noted earlier however, it is a notorious problem that higher-spin actions cannot maintain their gauge symmetries on arbitrary backgrounds \cite{AragoneDeser1,AragoneDeser2,BvHdWvN,deWitFreedman}. This is why we did not consider terms such as (\ref{eq:nontrivtermexa}) in our action, because they cannot resolve this ghost problem anyway.  

On top of this issue, which concerns each diagonalized tensor field separately, we also note that in the scalar case the $\sim Z$ term is crucial in making the action ghost-free. Since there is no analogous term in the tensor sector, we expect that the diagonalized fields will exhibit a ghost/non-ghost structure like (\ref{eq:ghostcomb}). The corresponding new poles are indeed also present in the saturated propagator, since the diagonalization will inevitably make the auxiliary fields couple to $T_{\mu\nu}$, it is just that they will add-up with the tensor structure of $h_{\mu\nu}$, \footnote{In the Weyl model the tensor structure will correspond to the one of a $4$-tensor, but since the source is a $2$-tensor, the saturated propagator will reveal the same type of structure as the one of $h_{\mu\nu}$.}. The tensor part of the propagator (\ref{eq:propaeqbmn0}) is thus the sum of these three contributions $h, \ph, \psi$, and only the result has the correct sign. We will therefore have poles with the wrong residue signs for the ghost modes, but since the sum must be healthy, these will necessarily be canceled by healthy poles
\beq \label{eq:propaux}
\sim - \frac{1}{k^2 + m^2} + \frac{1}{k^2 + m^2} \, .
\eeq 
This is why these modes can be missed when working directly at the level of the non-local theory, {\it they simply cancel-out in the propagator}\footnote{This is similar to what happens in Barvinsky's non-local theory (\ref{eq:Barvinsky}) \cite{Barvinsky2}. Indeed, the linearized action is the one of GR, and has thus a healthy propagator, but the non-linear localized action contains an auxiliary tensor $\vph^{\mu\nu}$ on top of the metric, and the latter has obviously ghost modes. We thus have that the propagator of the diagonalized/localized theory has ghost poles that are compensated by healthy ones, as is clear from eq. (28) of \cite{Barvinsky2}. Thus, the ghost propagator simply appears to shift the graviton propagator, canceling it exactly for $\al = 1$.}. This is also why the propagator only provides a lower bound on the number of dynamical fields, because there might by cancellations among the corresponding propagators in the presence of ghosts.

Now note that this cancellation occurs only classically and only at the linearized level in the propagator. More rigorously, classically the retarded $\ep$ prescription for (\ref{eq:propaux}) gives
\beq \label{eq:propauxret}
\lim_{\ep \to 0^+} \[ - \frac{1}{-(k^0 + i\ep)^2 + \vec{k}^2 + m^2} + \frac{1}{-(k^0 + i\ep)^2 + \vec{k}^2 + m^2} \] = 0 \, ,
\eeq 
so we do have a cancellation. Indeed, since $\ep$ displaces the poles in the integral over $k^0$ it is their relative sign that matters for having a retarded response, not the overall sign of the propagator. The prescription is therefore the same for both healthy and ghost fields. 

Quite interestingly, such a cancellation would not occur in a local QFT with a ghost/healthy pair (\ref{eq:propaux}). Indeed, for scattering amplitudes where it is the Feynman propagator that arises, the $\ep$ prescription comes from the modification of the path integral which makes it converge. This means that unitarity forces the choice
\beq
S_{\rm aux. scal.} = \int \ed^D x \[ \frac{1}{2}\, \Phi_1 \( \bo - m^2 + i \ep \) \Phi_1 - \frac{1}{2}\, \Phi_2 \( \bo - m^2 - i \ep \) \Phi_2 \] \, ,
\eeq
for the kinetic terms of the diagonalized auxiliary fields, which in turn translates into
\beq \label{eq:vacdecayima}
\lim_{\ep \to 0^+} \[ - \frac{i}{k^2 + m^2 - i \ep} + \frac{i}{k^2 + m^2 + i \ep} \] = \lim_{\ep \to 0^+} \frac{2\ep}{(k^2 + m^2)^2 + \ep^2} = 2 \pi \de \( k^2 + m^2 \)   \, ,
\eeq 
for the Feynman propagators. Unlike the case of the retarded propagator, now the sign of $\ep$ is always positive and the two terms do not cancel each other. Rather, the result is the real part of the Feynman propagator. Had we chosen the opposite sign for $\ep$ in the propagator for the ghost, we would have lost unitarity but the ghost would have propagated positive energies forward in time, like an ordinary particle \cite{ClineJeonMoore}. 

Coming back to the classical case which involves the retarded propagators, the above argumentation only implies that the corresponding forces between two linear sources will indeed cancel-out. At the fully non-linear level however, these pairs of dynamical ghost/healthy fields will generically have different interactions and will thus be excited by sources in a non-trivial way. We therefore have potential tensor instabilities as soon as $m_g \neq 0$. This leaves only the massless gravity theories $m_g = 0$ as the only potentially ghost-free theories. In the present form these are {\it not} scalar-tensor theories because the scalars are constrained, as also noted in \cite{FoffaMaggioreMitsou1, Barvinsky1, DeserWoodard2, MaggioreMancarella, Kosh, KehagiasMaggiore}, but their dynamical spectrum is the one of a scalar-tensor theory.

\subsubsection{Condensation?}

Note that what the above argumentation tells us is that the Minkowski solution may be perturbatively unstable, nothing more. Indeed, it may very well be the case that there exist other highly symmetric solutions, such as FLRW ones, around which the perturbations are healthy. One then says that the ghosts ``condense'' onto a solution around which the fluctuations have positive-definite kinetic energy, in the same way a tachyon condenses on a non-trivial minimum of the potential. The idea of ghost condensation has already been around for a decade as an interesting mechanism for addressing the dark energy and other cosmological problems \cite{AHCLM}. As in the case of tachyon condensation, one typically needs higher-order terms in the derivatives, which would here correspond to higher order-terms in curvature in the non-local formulation. 

Unfortunately, for the Weyl model, which is the phenomenologically viable one, and for the case of interest where the stable solution is an FLRW solution, ghost condensation is not possible. Indeed, homogeneity and isotropy, along with the symmetries of the $4$-tensors, force the latter to vanish on such space-times. Then, since a ghost field acquires a non-trivial background value when it condenses, by definition, we have that this cannot be the case for the auxiliary $4$-tensors. Thus, we do not believe that the ghost could condense in the Weyl model, unless the stable solution is not homogeneous or isotropic.

\subsection{Stability} \label{eq:classeffthstab}

As already discussed in section \ref{sec:classstab}, the impact of ghosts in classical physics need not be so radical as in the quantum case. Indeed, classical instabilities can be dealt with if they are slow enough to pass phenomenological tests, or if they are stabilized  by background/non-linear effects.

\subsubsection{Non-tachonic ghosts}

If the mass of the ghost is non-tachyonic, we have that the corresponding dispersion relation will be
\beq \label{eq:ghostnontachyon}
\om = \pm \sqrt{\vec{k}^2 - m^2} \, ,
\eeq
so that only the modes at cosmological length-scales $|\vec{k}| < m$ are going to be unstable. Moreover, the maximal frequency of these modes being $\om = m$, we have that the corresponding divergence will manifest itself at cosmological time-scales $\De t \sim m^{-1} \sim H_0^{-1}$, i.e. of the order of the age of the universe. Also, since these modes start at zero, they remain much smaller than one during the whole $\De t$ period in which case our linear analysis is sufficient. Therefore, at scales where these instabilities are observable Minkowski space-time is not the appropriate solution and the solar system/galactic physics are effectively stable. The stability analysis will be important in the context of cosmological perturbation theory where the above dispersion relation argument is not enough anymore, since large space and time scales will be involved. We will come back to this when we will discuss the cosmological phenomenology.

The typical example of such non-tachyonic ghost will be the scalar mode in the case $Z = 0$, where one retrieves GR in the massless limit and thus does not spoil solar system constraints. Indeed, we will see that in this case the viable models are the ones with $m_s^2 > 0$.

\subsubsection{Tachyonic ghost}

Finally, for ghost modes that are also tachyonic, i.e. that obey the dispersion relation
\beq \label{eq:ghosttachyon}
\om = \pm \sqrt{\vec{k}^2 + m^2} \, ,
\eeq
but have a negative energy at the linearized level, there is no divergence in the absence of interactions. Thus, in this case one must also include the non-linearities to pronounce the stability verdict. Tachyonic ghosts are expected in the auxiliary tensor modes since cancellation forces them to come in combinations such as (\ref{eq:propaux}), where it is the overall sign that is wrong.

\section{Projector-based models}

\subsection{Constructing the equations}

We now wish to construct a projector-based generally-covariant extension of (\ref{eq:projeq}). As for the action-based generalizations, here too one has access to a plethora of combinations of curvature invariants and non-local operators. We will again consider only terms that contribute to the linearized equation over Minkowski space-time and no derivatives of curvature invariants. This is a bit more restrictive than in the action-based case since it gives
\beq \label{eq:Pbasedeq}
G_{\mu\nu} + \al \( g_{\mu\nu} R \)^{\rm T} + \be \, {}_a\( \ti{\bo}^{-1} G_{\mu\nu} \)^{\rm T} + \ga \( g_{\mu\nu} \ti{\bo}^{-1} R \)^{\rm T} = 8 \pi G T_{\mu\nu} \, .
\eeq 
Note that for the pure-trace terms $\sim g_{\mu\nu} K$ we have that the transverse part is uniquely defined
\beq
\( g_{\mu\nu} K \)^{\rm TT} = 0 \, , \hspace{1cm} \Leftrightarrow \hspace{1cm} \( g_{\mu\nu} K \)^{\rm T} = \( g_{\mu\nu} K \)^{\rm TpT} \, .
\eeq
For the $\ti{\bo}^{-1} G_{\mu\nu}$ term we have one more free parameter $a$ which corresponds to the choice of transverse operator $_a {\cal P}$. Note that choosing another combination of $R_{\mu\nu}$ and $g_{\mu\nu} R$ instead of $G_{\mu\nu}$ simply amounts to changing $\ga$, thanks to the $\Rs$-linearity of the transverse projectors.

A first remark on this class of models is that they share all the vacuum solutions of GR, just like the action-based Ricci model (\ref{eq:RicciS}). Indeed, if $R_{\mu\nu} = 0$ then, by $\Rs$-linearity of ${\cal P}$, the left-hand side of (\ref{eq:Pbasedeq}) vanishes. Let us now fix the free parameters such that we retrieve (\ref{eq:projeq}) in the linearized limit. Linearizing over Minkowski
\beq
g_{\mu\nu} = \et_{\mu\nu} + 2\la\, h_{\mu\nu} \, , \hspace{1cm} \la \equiv \sqrt{8 \pi G} \, ,
\eeq
and using (\ref{eq:Paverprod}) one gets that (\ref{eq:Pbasedeq}) reads
\beq
\bo \, {}_{(1+ d(2\al-1))} {\cal P}_{\mu\nu}^{\,\,\,\,\,\,\ro\si} h_{\ro\si} + \be \, {}_{(2d\ga/\be -  a (d-1))} {\cal P}_{\mu\nu}^{\,\,\,\,\,\,\ro\si} h_{\ro\si}   = - \la T_{\mu\nu} \, .
\eeq
Using (\ref{eq:Paverprod}), we can also rewrite (\ref{eq:projeq}) as\footnote{With the correct normalization for the source.}
\beq
\bo \, {}_z {\cal P}_{\mu\nu}^{\,\,\,\,\,\,\ro\si} h_{\ro\si} - m_g^2\, {}_{m_s^2/m_g^2} {\cal P}_{\mu\nu}^{\,\,\,\,\,\,\ro\si} h_{\ro\si}   = - \la T_{\mu\nu} \, ,
\eeq
so that, matching the two equations, we get
\beq
1 + d(2\al - 1) = z \, , \hspace{1cm} \be = - m_g^2 \, , \hspace{1cm} 2 d \, \frac{\ga}{\be} - a (d-1) = \frac{m_s^2}{m_g^2} \, .  
\eeq
Keeping $a$ as the free parameter, we then have
\beq
\al  = \frac{z + d - 1}{2} \equiv Z \, , \hspace{1cm} \be = - m_g^2 \, , \hspace{1cm} \ga = - \frac{m_s^2 + a(d-1) m_g^2}{2d} \, ,
\eeq
and thus the generally-covariant extension in terms of $Z, m_s, m_g, a$ reads
\beq \label{eq:projbasedmod}
G_{\mu\nu} + Z \( g_{\mu\nu} R \)^{\rm T} - m_g^2 \, {}_a\( \ti{\bo}^{-1} G_{\mu\nu} \)^{\rm T} - \frac{m_s^2 + a(d-1) m_g^2}{2d} \( g_{\mu\nu} \ti{\bo}^{-1} R \)^{\rm T} = 8 \pi G T_{\mu\nu} \, .
\eeq
As in the case of action-based models, only the $Z = 0$ case reduces to GR in the massless limit, but the price to pay is a scalar ghost in the spectrum. Moreover, as also discussed for the action-based models, the term involving the Einstein tensor in the departure from GR is phenomenologically excluded since it leads to non-viable FLRW solutions \cite{Maggiore2, FoffaMaggioreMitsou1, FerreiraMaroto, ModestoTsujikawa}. This in turn implies that $m_g = 0$, i.e. that the tensor modes are massless. This is in contrast with the action-based models, where the possibility of considering Weyl tensor terms allowed us to have $m_g \neq 0$ without affecting the FLRW solutions\footnote{Of course here too we could use the Weyl tensor but only if we accept derivatives acting on curvature, i.e. terms like $_a \( \ti{\bo}^{-2}\na^{\ro} \na^{\si} W_{\mu\ro\nu\si} \)^{\rm T}$.}. We are thus led to consider the following class of models
\beq \label{eq:ZxiM}
G_{\mu\nu} + \[ Z  g_{\mu\nu} R  - \frac{1}{2d}\, m_s^2 \, g_{\mu\nu} \ti{\bo}^{-1} R \]^{\rm T} = 8 \pi G T_{\mu\nu} \, .
\eeq

\subsection{Localization}

Localizing (\ref{eq:ZxiM}) involves both defining an auxiliary scalar $\psi$ to replace $\ti{\bo}^{-1} R$ and invoking an auxiliary vector $\ph_{\mu}$ for the definition of the transverse part (\ref{eq:STdef}) (\ref{eq:Smueqp}). This gives
\bea
G_{\mu\nu} + Z g_{\mu\nu} R - \frac{1}{2d}\, m_s^2\, g_{\mu\nu} \psi - \na_{(\mu} \ph_{\nu)} & = & 8 \pi G T_{\mu\nu}  \, , \nn \\
\bo \ph_{\mu} + \na_{\nu} \na_{\mu} \ph^{\nu}  & = & 2 Z \na_{\mu} R - \frac{1}{d}\, m^2 \na_{\mu} \psi \, , \nn \\
\ti{\bo} \psi & = & R \, . \label{eq:loceqpb}
\eea
The initial conditions of $\psi$ are determined by its definition
\beq \label{eq:defpsipbmod}
\psi \equiv \ti{\bo}_{\rm r}^{-1} R \, ,
\eeq
and the initial conditions of $\ph_{\mu}$ are similarly determined by the ones of $R$. In the action-based case, the localized action allowed us to gain some insight into the dynamics of the auxiliary fields, i.e. to determine whether some fields are ghost-like or not. The above local equations of motion however do not derive from an action, so such features are less obvious to see here\footnote{To see this, suppose such an action exists. Then, term $\sim \na_{(\mu} \ph_{\nu)}$ in the first equation, which would correspond to the equation of motion of $g_{\mu\nu}$, would be a total derivative $\na_{\mu} \ph^{\mu}$ in the hypothetical action. Thus, such ``friction'' terms cannot derive from an action.}. Nevertheless, one can still detect potentially pathological behaviour. For instance, the vector field $\ph_{\mu}$ does not have the gauge-invariant kinetic term $\na_{\mu} F^{\mu\nu}$.

\section{Solar system constrains}

\subsection{No Vainshtein mechanism}

In local massive gravity, the vDVZ discontinuity is a discontinuity between the action and the propagator, i.e. the former reduces to its GR form in the $m \to 0$ limit, while the latter does not. As we discussed in the introduction however, continuity is restored in the non-linear theory through the Vainshtein mechanism. The strong-coupling scale goes like a negative power of $m$, so that linear perturbation theory breaks down as $m \to 0$, or equivalently at small scales, and thus the propagator does no longer reflect the forces that are present.

In contrast, in all of the above non-local models, thanks to the trivial inversion properties of the linearized projectors, the tensor structure of the propagator (\ref{eq:propaeqbmn0}) is the same as the tensor structure of the linearized action (\ref{eq:Slinnonloca}). Therefore, there is no discontinuity between action and propagator at any point of the parameter plane $\( m_g, m_s \)$ for all $z \neq 0$. Consequently, there is no need for a Vainshtein mechanism and the strong-coupling scale should be the Planck scale $M$. Let us have a closer look at this.

The Vainshtein mechanism is a special case of a more general class of screening mechanisms known as ``$k$-mouflage'' \cite{BabichevDeffayetZiour2}. The latter can occur in scalar-tensor theories where the scalar couples non-minimally to gravity {\it and} has a non-linear kinetic term. The former property is what makes the scalar couple to the source of gravity, after diagonalization, while the latter property is the one responsible for screening it on short distances. Indeed, the higher-order terms in the kinetic term will necessarily involve a mass scale $\La \ll M$, which will correspond to the scale of strong-coupling. Let us now follow the argumentation of \cite{BabichevDeffayet} to see how this screens the scalar force on scales smaller than $\La^{-1}$.

In the case of a scalar-tensor theory a typical non-minimal coupling can be $\sim \ph R$. In the case of local massive gravity the scalar field is the St\"uckelberg scalar, coupling also derivatively to gravity. After diagonalization, the metric to which matter couples becomes of the form 
\beq
h_{\mu\nu} + \al \et_{\mu\nu} \ph \equiv h_{\mu\nu} + \de_{\mu\nu} \, ,
\eeq
where $\al \sim \Ord(1)$ if $h_{\mu\nu}$ and $\ph$ are canonically normalized. We thus have $\de \neq 0$ which corresponds to the difference in the gravitational force felt by matter, i.e. the ``fifth force''. In the diagonalized theory, a typical example for the non-linear kinetic term that leads to $k$-mouflage are the Galileon structures \cite{NRT}, such as
\beq
\frac{1}{2} \, \ph\bo \ph + \frac{1}{2\La^3}\, (\pa\ph)^2 \bo \ph \, .
\eeq
Note that the coupling of $\ph$ to the energy-momentum tensor has the same strength as for the graviton because $\al \sim \Ord(1)$. The equations of motion then read (schematically)
\bea
\pa^2 h + M^{-1} \Ord\(h \pa h \pa h\) & \sim & M^{-1} T \, , \label{eq:pa2hmm} \\
\pa^2\ph + \La^{-3} \Ord \(\pa^4 \ph^2 \) & \sim & M^{-1} T  \, ,
\eea
where the interaction term in (\ref{eq:pa2hmm}) can always be neglected since we work at energies below the Planck scale. We now have the following asymptotic behaviours. At ``large'' scales $\pa \ph \ll \La \ph$, the linear term dominates in the scalar equation so $\pa^2 \ph \sim M^{-1} T \sim \pa^2 h$ and thus $\de \sim \Ord(1)$. At ``small'' scales $\pa \ph \gg \La \ph$, but still $\pa h \ll M h$, \footnote{So that we can neglect non-linearities for $h$.}, it is the non-linear term that dominates, so $\pa^4 \ph^2 \sim \La^3 M^{-1} T \sim \La^3 \pa^2 h$ and thus $\ph \sim \sqrt{\La^3 h /\pa^2}$, which means that now $\de$ is suppressed because $\La/ \pa \ll 1$. Thus, the fifth force is screened. 

In the non-local models we consider here, we see that the localized equations of motion do not have such non-linear kinetic terms in the auxiliary sector. This is why no Vainshtein mechanism takes place and why the strong-coupling scale goes down to the Planck mass. From the theoretical point of view, the absence of the Vainshtein effect is nice because it implies that linear perturbation theory is valid at small scales and for arbitrarily values of the masses. In particular, for $Z = 0$, the solutions of the non-local models, computed as perturbative deformations of the ones of GR, will have an expansion parameter that is analytic in the masses. This feature has been verified for the spherically symmetric static solutions of the models with $Z = \xi = 0$ and $m_g = 0$ \cite{KehagiasMaggiore,MaggioreMancarella}.

\subsection{Constraints on $Z$}

From the phenomenological point of view, the absence of a Vainshtein mechanism implies that the forces that are present on small scales are the ones we read from the propagator (\ref{eq:mto0newprop}) in the massless limit. We then have that the forces corresponding to the two scalar poles (on top of the massless graviton) cancel out only for $Z = 0$, while for $Z \neq 0$ we have a net fifth force which spoils solar system tests. This could have been expected, because $Z$ is a dimensionless parameter and the terms it controls are thus expected to deform GR at all scales, contrary to the terms $\sim m^2_i$, which deform it at the scale $m_i^{-1}$.

It is however interesting to note that, by considering non-linear structures $Z f(\bo^{-1}R)$ in the action-based model one can avoid this conclusion, as shown in \cite{DeserWoodard2} in the context of the Deser-Woodard model (\ref{eq:DW}) \cite{DeserWoodard1}. The argumentation used in \cite{DeserWoodard2} is elegant and will allow us to understand better the effect of the $\sim Z$ terms in our models, so we choose to reproduce it here with some minor adjustments. Let us work in $D=4$ for simplicity. 

First note that homogeneity and isotropy imply that in cosmology the typical time-variation scale of the background is much larger than the gradients of the perturbations. Thus, as far as the action of $\bo$ on the Ricci scalar is concerned, the background dominates. In the standard cosmological history we have that $R$ is always positive so 
\beq
\( \bo - \xi R\)^{-1} R \approx - \( \pa_t^2 + 3 H \pa_t + \xi R \)^{-1} R \, ,
\eeq
is always negative for $\xi \geq 0$. Thus, only the region $x < 0$ of $f(x)$ is relevant for cosmology.  

On the other hand, for solar system physics, the phenomena are non-relativistic and thus the gradients are much more important that the time-derivatives. This is why the standard theoretical tool for solving the Einstein equations in this case is the post-Newtonian expansion, where the small expansion parameter is $v/c$, with $v$ being the typical velocity of the source. We then have that for non-relativistic systems $T_{\mu\nu}$ is dominated by the mass in $\ro \equiv T_{00}$, so that the trace of the Einstein equation reads
\beq
R \approx 8\pi G \ro > 0 \, .
\eeq
For gravitationally bound systems we have the typical profile $\De^{-1} \ro \sim + 1/r$ for the gravitational potential outside the sources. We thus have that $\( \bo - \xi R \)^{-1} R \approx \( \De - \xi R \)^{-1} R$ is positive for $\xi = 0$, but does not have a definite sign for $\xi > 0$, a priori. We must therefore compare the $\De R$ and $\xi R^2$ terms. By dimensional analysis we have that
\beq
\De R \approx 8\pi G \De \ro \approx 8\pi G L^{-2} \ro \, ,
\eeq
where $L$ is the typical size of the bound system. For non-relativistic systems the total mass $M$ dominates the energy density $\ro \approx M/L^3$ and $L$ is way larger than the corresponding Schwarzschild radius $L \gg 2 G M \equiv R_{\rm S}$. Thus, the ratio gives
\beq
\frac{\De R}{\xi R^2} \approx \frac{L}{\xi R_{\rm S}} \gg 1 \, ,
\eeq
for $\Ord(1)$ values of $\xi$, and we conclude that for solar system physics
\beq
\( \bo - \xi R \)^{-1} R > 0 \, .
\eeq
So it is the $x > 0$ part of $f(x)$ which affects the region of GR we do not want to mess with. One should therefore demand that $f(x) \approx 0$ for $x > 0$ in order not to spoil the solar system constraints, which implies in particular $f'(0) \approx 0$. For our models, this means $Z = 0$. 

In the next chapter, we will see that the models where $Z > 1/3$, i.e. the ones where the scalar mode is healthy, actually do not even yield viable cosmological solutions. Thus, from now on we set $Z = 0$ and this implies $z = 1-d$, so that the trace scalar is a ghost (see (\ref{eq:mto0newprop})).

\subsection{The potentially viable models}

We now have a clear picture of which models may provide a viable phenomenology. From the previous section we know that $Z = 0$. This already brings the projector-based model (\ref{eq:ZxiM}) to the form
\beq \label{eq:xiM}
G_{\mu\nu} - \frac{d-1}{2d} m^2 \( g_{\mu\nu} \ti{\bo}^{-1} R \)^{\rm T} = 8 \pi G T_{\mu\nu} \, ,
\eeq
where
\beq \label{eq:m2def}
m^2 \equiv \frac{1}{|z|}\, m_s^2 = \frac{1}{d-1}\, m_s^2  \, ,
\eeq
is the mass of the scalar mode. Eq. (\ref{eq:xiM}) is a one-parameter extension of the model proposed by Maggiore \cite{Maggiore2}, corresponding to the case $\xi = 0$, and we will therefore dub it the ``$\xi$-M model''. The localized form reads
\bea
G_{\mu\nu} - \frac{d-1}{2d}\, m^2\, g_{\mu\nu} \psi - \na_{(\mu} \ph_{\nu)} & = & 8 \pi G T_{\mu\nu} \, , \label{eq:locpbmod1} \\
\bo \ph_{\mu} + \na_{\nu} \na_{\mu} \ph^{\nu}  & = & - \frac{d-1}{d}\, m^2 \na_{\mu} \psi \, ,  \label{eq:locpbmod2} \\
\ti{\bo} \psi & = & R \, . \label{eq:locpbmod3}
\eea
For the action-based Weyl model (\ref{eq:WeylS}) we still have the possibility of considering massive tensor modes $m_g > 0$. In cosmology, this parameter will only affect the perturbations around the FLRW solution, since the background Weyl tensor vanishes. Since from now on we will focus exclusively on the background part of cosmology, we are effectively left with the $m_g = 0$ theory. Thus, the action-based model of interest reads
\beq \label{eq:xiMM}
S = \ti{M}^2  \int \ed^D x\, \sqrt{-g} \[R - \frac{d-1}{4 d}\, m^2 R \ti{\bo}^{-2} R \]   \, ,  
\eeq
where we have again used the mass of the scalar mode $m$. This is a one-parameter extension of the model proposed by Maggiore and Mancarella \cite{MaggioreMancarella}, corresponding to the case $\xi = 0$, so it makes sense to call (\ref{eq:xiMM}) the  ``$\xi$-MM model''. The localized form of (\ref{eq:xiMM}) is
\beq \label{eq:locxiMM}
S = \int \ed^D x\, \sqrt{-g} \[ \ti{M}^2 R + \ti{M} \( \phi + \ti{M}^{-1}\xi \phi \psi \) R - \phi \bo \psi - \frac{d-1}{4 d}\, m^2  \psi^2   \]   \, ,
\eeq
with $\ph$ and $\psi$ obeying
\beq \label{eq:phipsidef}
\ph \equiv - \frac{d-1}{2d}\,  m^2 \ti{M} \( \bo - \xi R \)^{-1}_{\rm r} \psi \, , \hspace{1cm} \psi \equiv \ti{M} \( \bo - \xi R \)^{-1}_{\rm r} R  \, .
\eeq
It is convenient to consider the dimensionless scalars $\phi \to \ti{M} \ph$ and $\psi \to \ti{M} \psi$, so that the equations of motion read
\bea
\( G_{\mu\nu} + g_{\mu\nu} \bo - \na_{\mu} \na_{\nu} \)  \[ 1 + \ph + \xi \ph \psi \] & & \nn \\
+ \na_{(\mu} \ph \na_{\nu)} \psi - \frac{1}{2}\, g_{\mu\nu} \na_{\ro} \ph \na^{\ro} \psi + \frac{d-1}{8 d}\, m^2 g_{\mu\nu} \psi^2 & = & 8\pi G T_{\mu\nu} \,, \label{eq:gttEOM} \\
\( \bo - \xi R \) \ph & = & - \frac{d-1}{2d} \, m^2 \psi \, ,  \label{eq:phEOM} \\
\( \bo - \xi R \) \psi & = & R \, .  \label{eq:psiEOM} 
\eea
From (\ref{eq:locxiMM}) we see that part of the scalar terms induce an effective Planck mass
\beq \label{eq:Meff}
\ti{M}^2 \( 1 + \ph + \xi \ph \psi \) R \equiv \ti{M}_{\rm eff}^2 R \, ,
\eeq
which is not positive-definite. Therefore, gravity becomes unstable as soon as $\ti{M}_{\rm eff}^2 < 0$. 

The Maggiore and Maggiore - Mancarella models, which correspond to the case $\xi = 0$, are currently receiving particular attention \cite{DFKKM, DirianMitsou, CKMT, BLHBP, DFKMP} because their phenomenology seems to privilege them among other non-local models that have been confronted with observations \cite{DeserWoodard1, JaccardMaggioreMitsou2, FoffaMaggioreMitsou2, Maggiore2, KehagiasMaggiore,  DeserWoodard2,  ModestoTsujikawa, Koivisto1, Koivisto2, Woodard, DeffayetWoodard,  DodelsonPark1, DodelsonPark2, NesserisTsujikawa}. Indeed, they have recently passed the constraints of a full Boltzmann/Monte Carlo Markov Chain analysis \cite{DFKMP}, of which they come out as statistically indistinguishable from $\La$CDM, with respect to the current precision of the data. The Maggiore model actually even seems to be slightly privileged. 

The elegance of these models lies in the fact that they are very simple non-local modifications of GR with as many parameters as $\La$CDM, i.e. the mass $m$ plays the role of $\La$. They are therefore very predictive since, once $m$ is fixed such that it reproduces the observed amount of dark energy today, the rest of the physics is determined. It is therefore highly non-trivial that these models can compete with $\La$CDM.

Here we see that, after having narrowed down the set of models to the potentially viable ones, there remains a natural extension of the Maggiore and Maggiore - Mancarella models corresponding to $\bo \to \bo - \xi R$. Considering one more parameter of course degrades predictivity, but it is nevertheless instructive to see what the effect of $\xi$ is.

\section{The effect of $\xi$} \label{sec:deSittersol}

The effect of the $\xi$ parameter is very interesting because for $R \neq 0$ we have that
\beq \label{eq:largexiapprox}
\( \bo - \xi R\)^{-1}_{\rm r} R \approx - \xi^{-1} \, , \hspace{1cm} {\rm if}\,\,\,\, |\xi| \gg |(\bo_{\rm r}^{-1} R)^{-1}| \, .
\eeq
Thus, as soon as $R \neq 0$, the dynamics of these models should be indistinguishable from GR with a cosmological constant $\La \sim m^2$, for large enough $\xi$. If on the other hand $R = 0$, which is the case during RD for the cosmological background for instance, then of course $\ti{\bo}_{\rm r}^{-1} R = 0$ by linearity. 

Not surprisingly, a first effect of $\xi > 0$ is the existence of de-Sitter solutions $G_{\mu\nu} + \La g_{\mu\nu} = 0$. Assuming a constant $R$, we have
\beq \label{eq:dSab}
\La = \frac{d-1}{8d}\, \frac{m^2}{\xi^2} \, , \hspace{1cm} \ph = - 1 \, , \hspace{1cm} \psi = - \frac{1}{\xi} \, ,
\eeq
for the action-based model, and
\beq \label{eq:dSpb}
\La = \frac{d-1}{2d} \, \frac{m^2}{\xi} \, , \hspace{1cm} \ph_{\mu} = 0 \, , \hspace{1cm} \psi = - \frac{1}{\xi} \, ,
\eeq  
for the projector-based one.

\subsubsection{No degravitation}

One of the original motivations for considering non-locality was not only to produce a dark energy effect, but also to degravitate any constant source. The very existence of de-Sitter solutions for $\xi \neq 0$ implies that these sources are not excluded, but the effective $\La$ could still be different from the one that would appear in the equations of motion. Unfortunately, this is not the case. Indeed, for both models, adding a vacuum energy term simply rescales $\La \to \La + \La_{\rm vac}$, so it is not degravitated at all. Note that this argument does not encompass the $\xi = 0$ models, nor the possibility of a dynamical degravitation mechanism, i.e. a time-dependent degravitation in the cosmological context. As we will see however in the next chapter, no such effect will take place.

\chapter{Cosmology}

Here we work in $D=4$ and consider the background cosmology of the action-based $\xi$-MM model (\ref{eq:xiMM}) and the projector-based $\xi$-M model (\ref{eq:xiM}). This chapter is based on, and extends, \cite{FoffaMaggioreMitsou2, DirianMitsou}.

\section{Background equations}

We now consider a flat ($k=0$) FLRW metric in cosmic time
\beq
g_{\mu\nu} \ed x^{\mu} \ed x^{\nu} = - \ed t^2 + a^2(t)\, \ed \vec{x}^2 \, ,
\eeq
so that all fields depend exclusively on time. We will use $x \equiv \log a$ as the time coordinate and denote by a prime the derivative with respect to $x$, so that
\beq
\dot{\ph} = H \ph' \, .
\eeq
The case $k \neq 0$ is also interesting, but we will not consider it both for simplicity and because $k = 0$ is consistent with the present data.

\subsection{Action-based model}

For the equation of $g_{00}$ in (\ref{eq:gttEOM}) we get the modified Friedmann equation
\beq \label{eq:modFried}
H^2 = \frac{8\pi G}{3}\, \si \( \ro + \ro_{\rm DE} \)
\eeq
where $\ro \equiv T_{00}$ and
\beq \label{eq:siroDE}
\si \equiv  \frac{1}{1 + \ph + \ph' + \xi \( \ph \psi + (\ph \psi)' \) + \frac{1}{6}\, \ph' \psi'} \, , \hspace{1cm} \ro_{\rm DE} \equiv \frac{1}{96 \pi G} \, m^2 \psi^2 \, .
\eeq
We see that with this rearrangement the system has turned into a Friedmann equation with a time-dependent Newton's constant and a dynamical dark energy component induced by the mass. Another ``effective Newton's constant'' is (\ref{eq:Meff}) which appears in the localized action (\ref{eq:locxiMM}) and must be monitored since its sign is the one of the kinetic term of gravity. We therefore also define another parameter
\beq \label{eq:tisi}
\ti{\si} \equiv \frac{1}{1 + \ph + \xi \ph \psi} \, .
\eeq
We now go to dimensionless variables
\beq \label{eq:Om}
h \equiv \frac{H}{H_0} \, , \hspace{1cm} \hat{\ro} \equiv \frac{8\pi G}{3 H_0^2}\, \ro  \, , \hspace{1cm} \hat{\ro}_{\rm DE} \equiv \frac{8\pi G}{3 H_0^2}\, \ro_{\rm DE} = \frac{1}{4}\, \mu^2 \psi^2 \, , \hspace{1cm} \mu^2 \equiv \frac{m^2}{9H_0^2} \, , 
\eeq
where the $0$ subscripts denote evaluation at today $x_0 = 0$, so the system of equations is
\bea
h^2 & = & \si \( \hat{\ro} + \hat{\ro}_{\rm DE} \) \, , \label{eq:Friedab} \\
\ph'' + (3 + \ze )\, \ph' + 6\xi(2 + \ze) \ph & = &  3\mu^2 h^{-2} \psi \, , \label{eq:phcEOM} \\
\psi'' + (3 + \ze )\, \psi' + 6\xi(2 + \ze) \psi & = & - 6 (2 + \ze)   \, , \label{eq:psicEOM}
\eea
where for $\ti{\ro}$ we consider a fluid made of matter and radiation
\beq
\hat{\ro} = \hat{\ro}^0_R\, e^{-4x} + \hat{\ro}^0_M\, e^{-3x}  \, ,
\eeq
and
\beq \label{eq:zeab}
\ze \equiv \frac{h'}{h} \os{*}{=} \frac{1}{2} \, \frac{h^{-2} \hat{\ro}' - 3 \mu^2 h^{-2} \psi \( 1 + \xi \psi \) + 4 \ph' \( 1 + \xi \psi \) + \( 1 - 2 \xi \) \ph' \psi' + 4 \xi \ph \( 6 + 6 \xi \ph + \ph' \)}{1 + \( 1 - 6\xi \)\( 1 + \xi \psi \) \ph} \, .
\eeq
where in $*$ we have used the equations of motion to get rid of the second time-derivatives. Given (\ref{eq:phipsidef}), the initial conditions of $\ph$ and $\psi$ are zero at the initial time $t_i$ if the latter is well-inside the RD era since $R_{\rm RD} = 0$
\beq \label{eq:initdata}
\ph(t_i) = \ph'(t_i) = \psi(t_i) = \psi'(t_i) = 0 \, .
\eeq
Now, had we chosen to include the $\sim Z$ term of (\ref{eq:WeylS}), the denominator of $\ze$ would have rather been
\beq
\ze \sim \frac{1}{2\( 1 - 3Z \) + \( 1 - 6 \xi \) \( Z \psi + 2\ph \( 1 + \xi \psi \) \)} \os{\rm RD}{\to} \frac{1}{2\( 1 - 3Z \) } \, .
\eeq
In the case $Z > 1/3$ which corresponds to $z > 0$, and thus to the case where the scalar mode is healthy, we have that $\ze$ has the opposite sign and thus $H$ is growing. Thus, on top of spoiling solar system physics, the $Z$ parameter also spoils the cosmological background solution in the region where it is interesting to consider, i.e. where it makes the scalar healthy. 

Also, observe that the $\si$ factor appears in front of all the energy components, i.e. had we added a ``vacuum'' cosmological constant in the action we would have simply obtained
\beq
\hat{\ro} + \hat{\ro}_{\rm DE} \to \hat{\ro} + \hat{\ro}_{\rm DE} + \hat{\ro}_{\rm vac} \, .
\eeq
From here it is clear that no degravitation of $\hat{\ro}_{\rm vac}$ can be achieved without also degravitating matter and radiation as well. Moreover, what we will observe in the simulations is $\si \geq 1$ at late times, so we will have an enhancement of the source rather than a screening effect. Thus, even in the dynamical context, no degravitation mechanism appears.  

Finally, we would like to spot the variables which characterize conveniently the departure from GR. In $\La$CDM one has that the equation of state parameter of the source can be expressed in terms of $H$. Indeed, one uses the barotropic equation of state $\ro = w p$ and the conservation of energy \beq \label{eq:conseqw}
\dot{\ro} = - 3 H \(\ro + p \) \equiv - 3 H \(1 + w \) \ro \, , \hspace{1cm} \Rightarrow \hspace{1cm} \pa_x \log \ro = - 3 \( 1+ w\) \, ,
\eeq
and replaces $\dot{\ro}$ using the Friedmann equation $\ro \sim H^2$ to find
\beq \label{eq:w}
w = - 1 - \frac{2}{3}\, \ze  \, .
\eeq
In our case, this quantity represents the equation of state of the effective source seen by $H$, namely $\ro_{\rm eff} \equiv \si \( \ro + \ro_{\rm DE} \)$. For RD $(w = 1/3)$, MD $(w=0)$ and de-Sitter $(w = -1)$ phases we get
\beq \label{eq:zeval}
\ze_{\rm RD} = -2 \, , \hspace{1cm} \ze_{\rm MD} = -\frac{3}{2} \, , \hspace{1cm} \ze_{\rm dS} = 0 \, .
\eeq
It will also be interesting to have the equation of state corresponding to $\ro_{\rm DE}$, which we define through the ``conservation equation'' of this effective source (\ref{eq:conseqw})
\beq \label{eq:wDEab}
w_{\rm DE} \equiv -1 - \frac{1}{3}\,\pa_x \log \ro_{\rm DE} = - 1 -\frac{2}{3}\, \frac{\psi'}{\psi} \, .
\eeq

\subsection{Projector-based model}

Homogeneity and isotropy imply that only the $\ph_0$ component of the auxiliary vector is non-zero. It is then convenient to trade it for a new variable (which is {\it not} a scalar)
\beq
\ph \equiv \frac{3}{m^2}\, \dot{\ph}_0 - \psi \, ,
\eeq
whose equation of motion can be found by taking the time-derivative of the $\mu = 0$ part of (\ref{eq:locpbmod2}) and using (\ref{eq:locpbmod3}).
Given the definition of $\psi$ (\ref{eq:defpsipbmod}) and the fact that only retarded Green's functions are invoked in the definition of the transverse part, both $\psi$ and $\ph_0$ have vanishing initial conditions
\beq
\ph_0(t_i) = \ph'_0(t_i) = \psi(t_i) = \psi'(t_i) = 0 \, ,
\eeq
for $t_i$ well-inside the RD phase. Observe that this only implies $\ph(t_i) = 0$. To get the condition on $\ph'(t_i)$ one must evaluate the second-order equation of $\ph_0$, i.e. the $\mu = 0$ part of (\ref{eq:locpbmod2}), at $t_i$, to get that $\ph''_0(t_i) = 0$ as well and thus
\beq
\ph(t_i) = \ph'(t_i) = \psi(t_i) = \psi'(t_i) = 0 \, .
\eeq
Now the modified Friedmann equation, i.e. the $\mu\nu = 00$ part of (\ref{eq:locpbmod1}), takes again the form (\ref{eq:modFried}) with
\beq \label{eq:siroDEpb}
\si = 1 \, , \hspace{1cm}  \ro_{\rm DE} = \frac{1}{24\pi G}\, m^2 \ph \, .
\eeq 
Had we chosen to consider the $\sim Z$ term of (\ref{eq:ZxiM}) we would have rather found
\beq
\si \equiv \frac{1}{1 - 2 Z\( 2 + \ze \)} \, .
\eeq
As in the action-based case, here too the $Z > 1/3$ choice would be problematic. Indeed, in RD we have that $\si = 1$ since $\ze_{\rm RD} = -2$. But if we are supposed to reach a DE phase at late times, i.e. $w \approx -1$, then by (\ref{eq:w}) we have $\ze_{\rm DE} \approx 0$ and thus 
\beq
\si \approx \frac{1}{1 - 4 Z} \, .
\eeq
For $Z > 1/3$ this is negative, so at some point between the two phases $\si^{-1}$ must go through zero, which means that space-time has a singularity $H \to \infty$ before today. As a consequence, the choice of a healthy scalar mode $Z > 1/3$ spoils the background evolution for the projector-based models as well. Note that for $Z = 0$, which is the case of interest, $\si = 1$ so that there is no dynamical degravitation.

Defining again the dimensionless variables (\ref{eq:Om}) but now with
\beq \label{eq:Ompb}
\hat{\ro}_{\rm DE} \equiv \frac{8\pi G}{3 H_0^2}\, \ro_{\rm DE} = \mu^2 \ph \, , 
\eeq
the system of equations becomes
\bea
h^2 & = & \hat{\ro} + \hat{\ro}_{\rm DE} \, , \label{eq:Friedpb} \nn \\
\ph'' + \( 3 - \ze \) \ph' - 3\( 1 + \ze \) \ph & = & - 3 \psi' + 3\( 1 + \ze \) \psi \, , \label{eq:phicEOMpb} \\
\psi'' + \( 3 + \ze \) \psi' + 6 \xi \( 2 + \ze \) \psi & = & - 6\( 2 + \ze \)  \, ,
\eea
and
\beq \label{eq:zepb}
\ze \equiv \frac{h'}{h} = \frac{1}{2}\,\frac{\hat{\ro}' + \mu^2 \ph'}{\hat{\ro} + \mu^2 \ph} \, .
\eeq
Note that $\psi$ has exactly the same equation as in the action-based model (\ref{eq:psicEOM}), i.e. it is the field which localizes $\ti{\bo}^{-1} R$. Finally, we can again define $w_{\rm DE}$ through the ``conservation equation'' of $\ro_{\rm DE}$ to get
\beq \label{eq:wDEpb}
w_{\rm DE} \equiv -1 - \frac{1}{3}\,\pa_x \log \ro_{\rm DE} = - 1 -\frac{1}{3}\, \frac{\ph'}{\ph} \, .
\eeq

\section{Numerical analysis}

\subsubsection{Set-up}

According to the Planck data \cite{Planck}, which {\it assume} $\La$CDM, we have $\hat{\ro}^0_R = 9.21 \times 10^{-5}$ and $\hat{\ro}^0_M = 0.3175$. Since our solutions will be close to $\La$CDM up until today, we will choose these values as well\footnote{For the $\xi = 0$ models a full 
parameter estimation using CMB, BAO and SNe data has been presented in \cite{DFKMP} and the values chosen here are consistent with their results. Since the $\xi > 0$ lie somewhere between the $\xi = 0$ ones and $\La$CDM, these values should be alright for them too.}. The matter-radiation equality then occurs at $x_{\rm eq} \approx -8.15$, with today being $x_0 = 0$. We will start our numerical integration at $x = -40$, that is, well-inside the RD era, so that we can safely impose zero initial conditions on $\ph$ and $\psi$ for both the action-based and projector-based models.

Note that consistency requires $h_0 = 1$, so here this is achieved by tunning $\mu^2$ appropriately. This is analogous to the case of $\La$CDM where one of the energy density components is determined by the defining condition $\sum_i \Om_i^0 = 1$. Here however we do not have the data that determine $\mu^2$ algebraically, since they include the field values today and we only control the initial conditions. Therefore, $\mu^2$ will be determined by successive trials and we will stop when $\log h_0 = \Ord(10^{-6})$. The resulting value will depend on $\xi$, the second parameter of the model, so imposing $h_0 = 1$ actually fixes the relation $\mu^2(\xi)$.

\subsubsection{Data description}

So let us now describe our results that are collectively displayed in the plots and tables of section \ref{sec:plottab}. We have computed the cases $\xi = 2^n$, where 
\beq
n = -\infty, -6, -5, -4, -3, -2, -1, 0 ,1,2 \, .
\eeq
In the plots the color goes from blue to red with increasing $n$, while the $\La$CDM result is given in green for comparison. In figure \ref{fig:dlogH} we have plotted the quantity $\log \( h/h_{\La{\rm CDM}} \)$, where $h_{\La{\rm CDM}}$ is the dimensionless Hubble parameter of $\La$CDM, normalized to $1$ today. In figure \ref{fig:w} we have plotted the effective equation of state parameter $w$ (\ref{eq:w}), but since the results overlap too much at $x = 0$ we have also plotted the difference with $\La$CDM in figure \ref{fig:dw} to get a cleaner picture. In figure \ref{fig:w0} we have potted today's values of $w$ with respect to $\xi$. In figures \ref{fig:roDE} and \ref{fig:wDE} we have plotted the effective dark energy component $\hat{\ro}_{\rm DE}$ and the corresponding equation of state $w_{\rm DE}$, respectively. In figures \ref{fig:phi} and \ref{fig:psi} we have $\ph$ and $\psi$, where we must stress that the former is a different non-local functional of $R$ in each model. Moreover, in the action-based model it is $\psi$ that controls the dark energy component $\ro_{\rm DE}$, while in the projector-based model it is $\ph$. In figure \ref{fig:si} we have plotted the $\si$ and $\ti{\si}$ quantities of the action-based model which correspond to the (dimensionless) effective Newton's constant (\ref{eq:siroDE}) in the modified Friedman equation (\ref{eq:modFried}) and the effective Newton's constant (\ref{eq:tisi}) in the localized action (\ref{eq:locxiMM}), respectively. Finally, in table \ref{tab:mu2ws} we have given the numerical values of $\mu^2$, $w_0$ and $w_{{\rm DE},0}$.

\subsubsection{Analysis}

A first general remark is that, by increasing $\xi$ we get arbitrarily close to $\La$CDM, as anticipated in section \ref{sec:deSittersol}. More precisely, note how, as $\xi$ increases, the dark energy component $\hat{\ro}_{\rm DE}$ tends to behave more and more like a cosmological constant, both in the future and past around $x = 0$, \footnote{Although it is forced to be zero during RD.} (figure \ref{fig:roDE}), while the effective Newton's constants of the action-based model $\si$ and $\ti{\si}$ tend towards one (figure \ref{fig:si}). 

\begin{figure}[h!]
\begin{center}
\includegraphics[width=16cm]{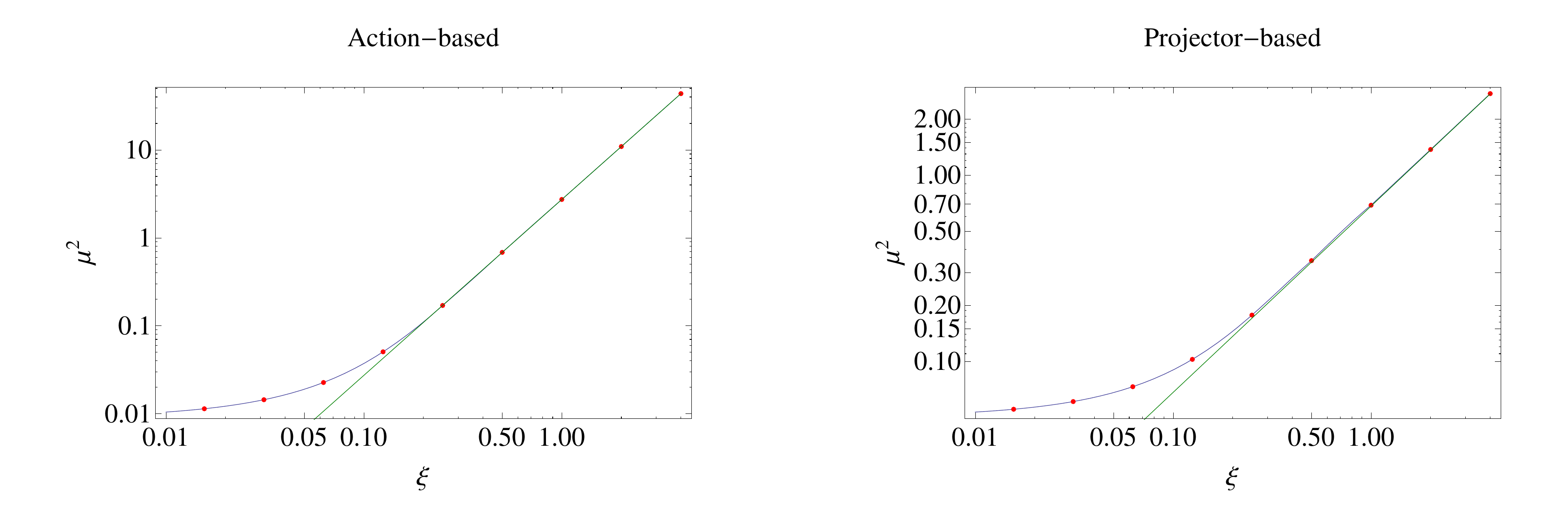} 
\caption{The $\mu^2(\xi)$ relation which gives $h_0 = 1$ (red dots) along with an interpolation (blue line) and the de-Sitter solution relation (\ref{eq:mu2ofxi}) (green line).}
\label{fig:mu2ofxi} 
\end{center}
\end{figure}

Thus, for large enough $\xi$, one should find the $\mu^2(\xi)$ relation of the de-Sitter solutions (\ref{eq:dSab}) and (\ref{eq:dSpb}) with the $\La$ of $\La$CDM, i.e. $\hat{\ro}_{\La} \equiv 1 - \hat{\ro}^0_R - \hat{\ro}^0_M \approx 0.6824$. More precisely, defining
\beq
\hat{\ro}_{\La} \equiv \frac{\La}{3H_0^2} \, ,
\eeq
we have that (\ref{eq:dSab}) and (\ref{eq:dSpb}) give
\beq \label{eq:mu2ofxi}
\mu^2 = 4 \, \hat{\ro}_{\La}\, \xi^2  \, , \hspace{1cm} \mu^2 = \hat{\ro}_{\La} \xi \, ,
\eeq
respectively. In figure \ref{fig:mu2ofxi} this relation corresponds to the green line and we see that the dots follow that trend indeed, for already small $\xi$ values. For very small $\xi$ we have that the transition to the dS phase is not complete yet at $x = 0$ and thus (\ref{eq:mu2ofxi}) does not hold. 

\begin{figure}[h!]
\begin{center}
\includegraphics[width=16cm]{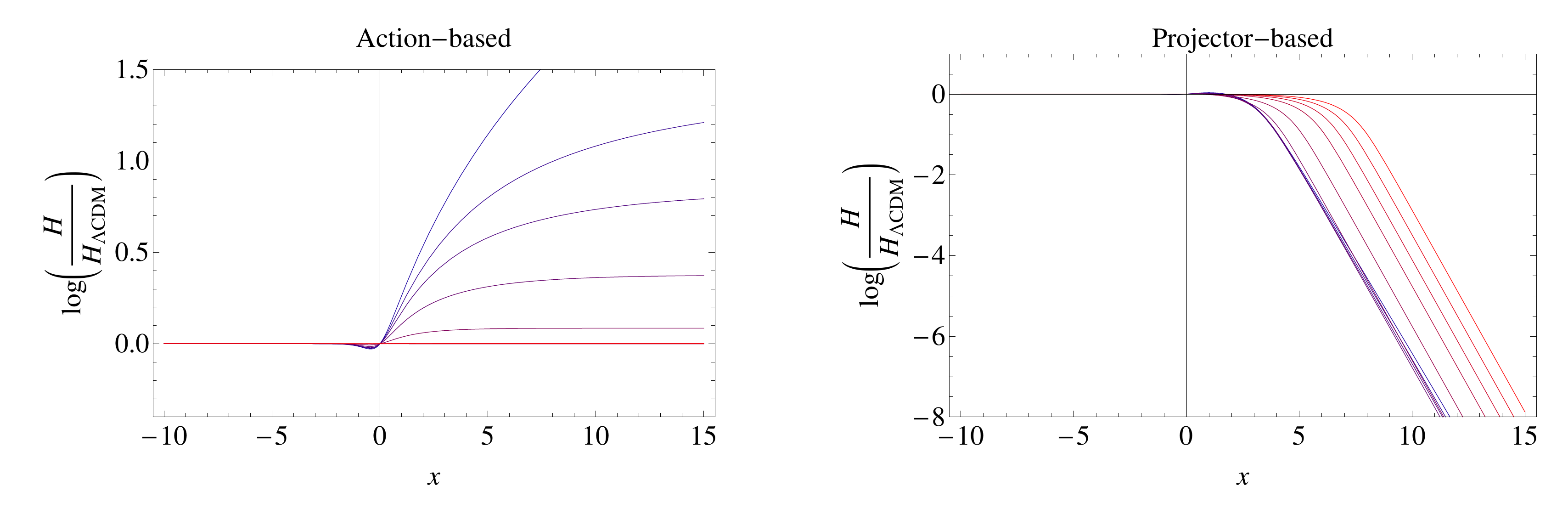} 
\caption{The logarithmic departure from the Hubble parameter of $\La$CDM.}
\label{fig:dlogH} 
\end{center}
\end{figure}

\begin{figure}[h!]
\begin{center}
\includegraphics[width=16cm]{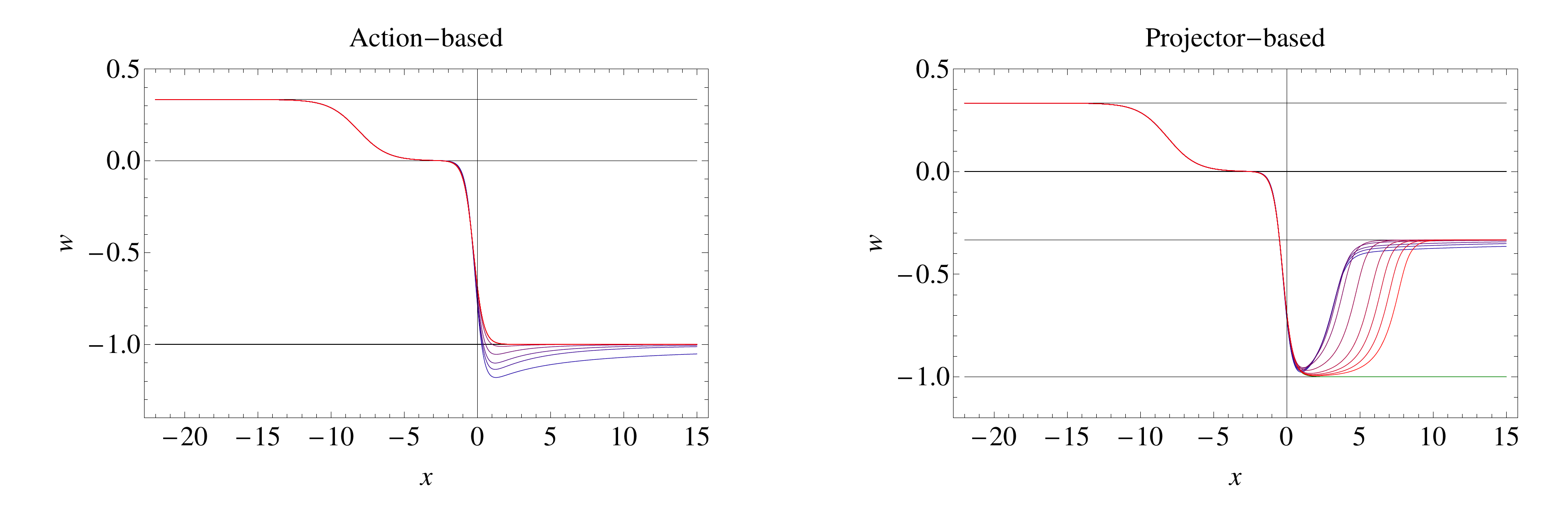} 
\caption{The effective equation of state parameter $w$.}
\label{fig:w} 
\end{center}
\end{figure}

For the action-based model we have that the de-Sitter solution in the $\xi > 0$ case is an attractor, since the universe reaches that state asymptotically (see figures \ref{fig:w} and \ref{fig:roDE}). The acceleration is faster than in $\La$CDM (see figure \ref{fig:dlogH}), but one tends towards $H_{\La{\rm CDM}}$ with increasing $\xi$. For the projector-based model that solution is unstable and the universe is rather attracted towards a $w = -1/3$ phase after the de-Sitter one, for all values of $\xi$. Increasing $\xi$ however makes the de-Sitter phase last longer (see figures \ref{fig:dlogH}, \ref{fig:roDE} and \ref{fig:wDE}), as could be expected by the fact that in the $\xi \to \infty$ limit one recovers $\La$CDM. A $w = -1/3$ value is interesting since it implies zero acceleration $\ddot{a} = 0$, and therefore $a \sim t$. Thus, although the dark energy component tends to zero as $t \to \infty$, it dominates over matter at late times.

\begin{figure}[h!]
\begin{center}
\includegraphics[width=16cm]{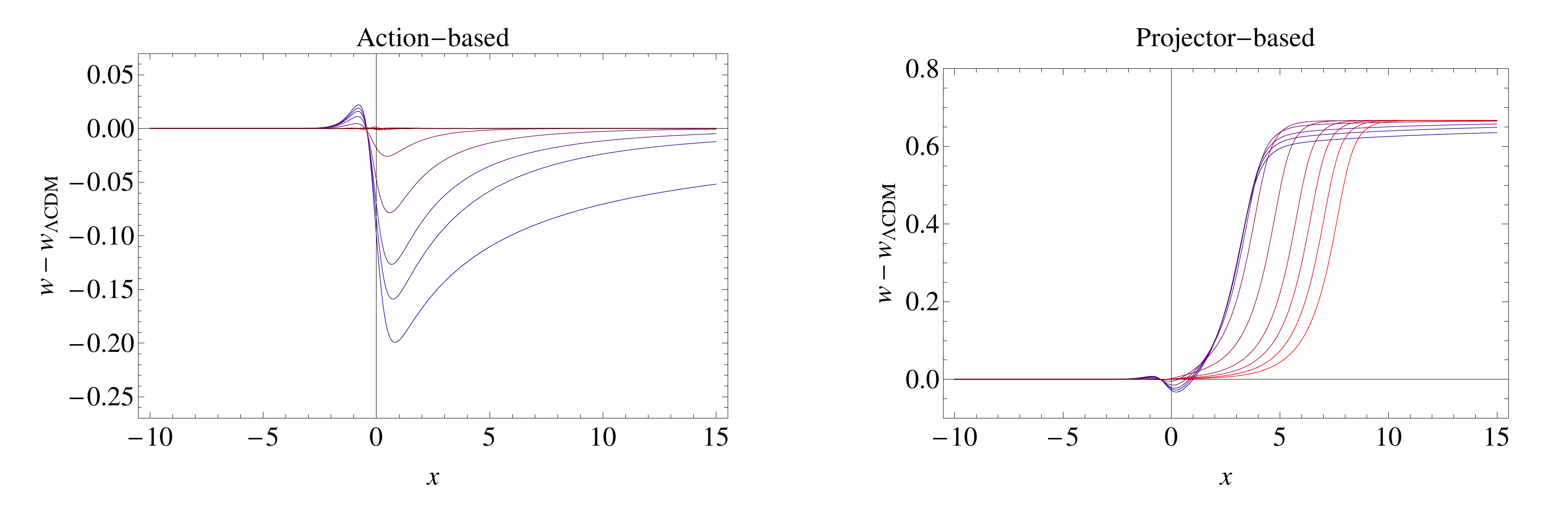} 
\caption{Departure from the equation of state parameter of $\La$CDM.}
\label{fig:dw} 
\end{center}
\end{figure}

\begin{figure}[h!]
\begin{center}
\includegraphics[width=15cm]{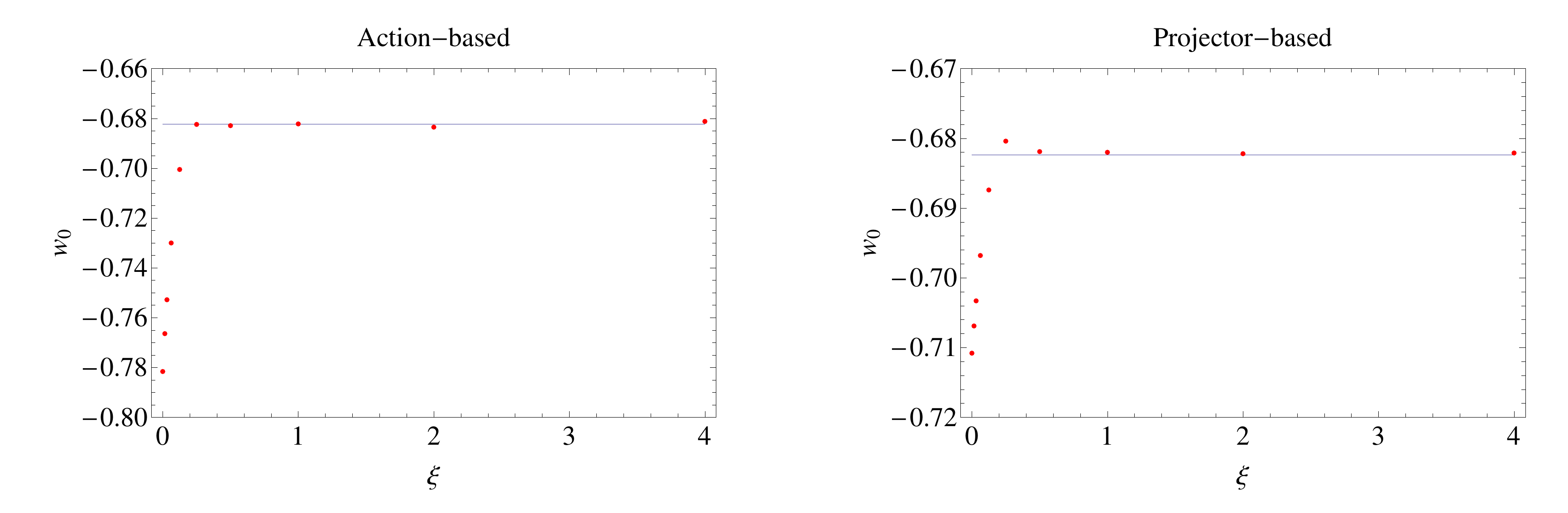} 
\caption{The effective equation of state parameter today $w_0$ (red dots) with the $\La$CDM result (blue line).}
\label{fig:w0} 
\end{center}
\end{figure}

\begin{figure}[h!]
\begin{center}
\includegraphics[width=16cm]{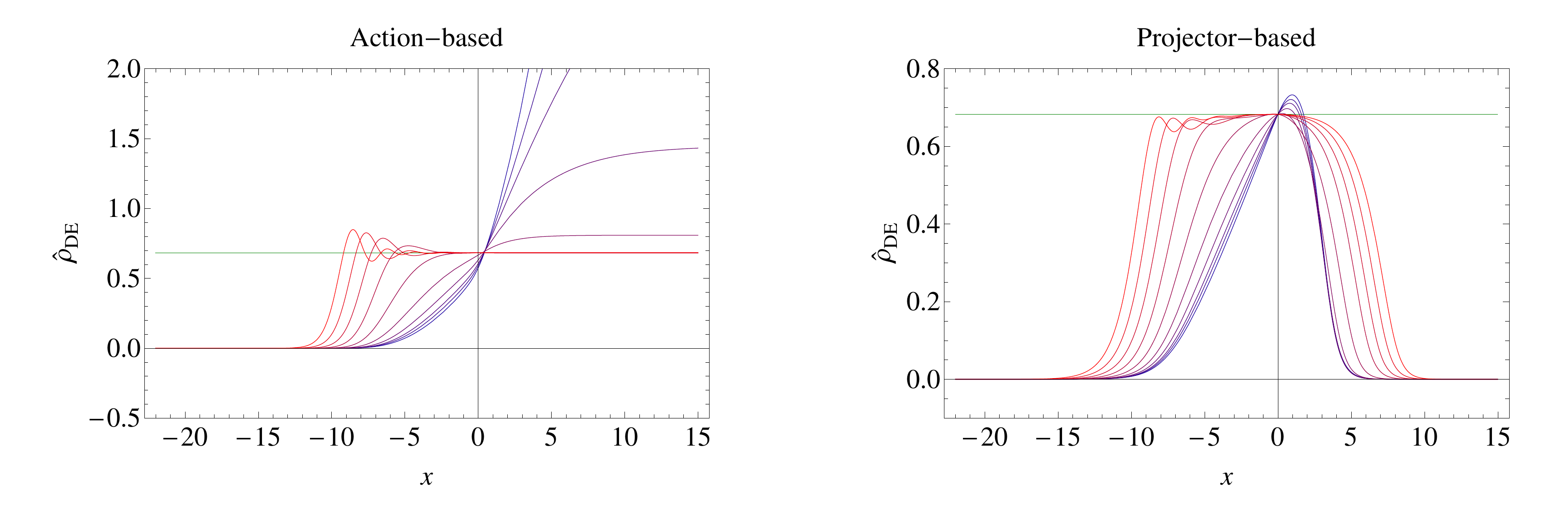} 
\caption{The dimensionless effective dark energy component $\hat{\ro}_{\rm DE}$.}
\label{fig:roDE} 
\end{center}
\end{figure}

\begin{figure}[h!]
\begin{center}
\includegraphics[width=16cm]{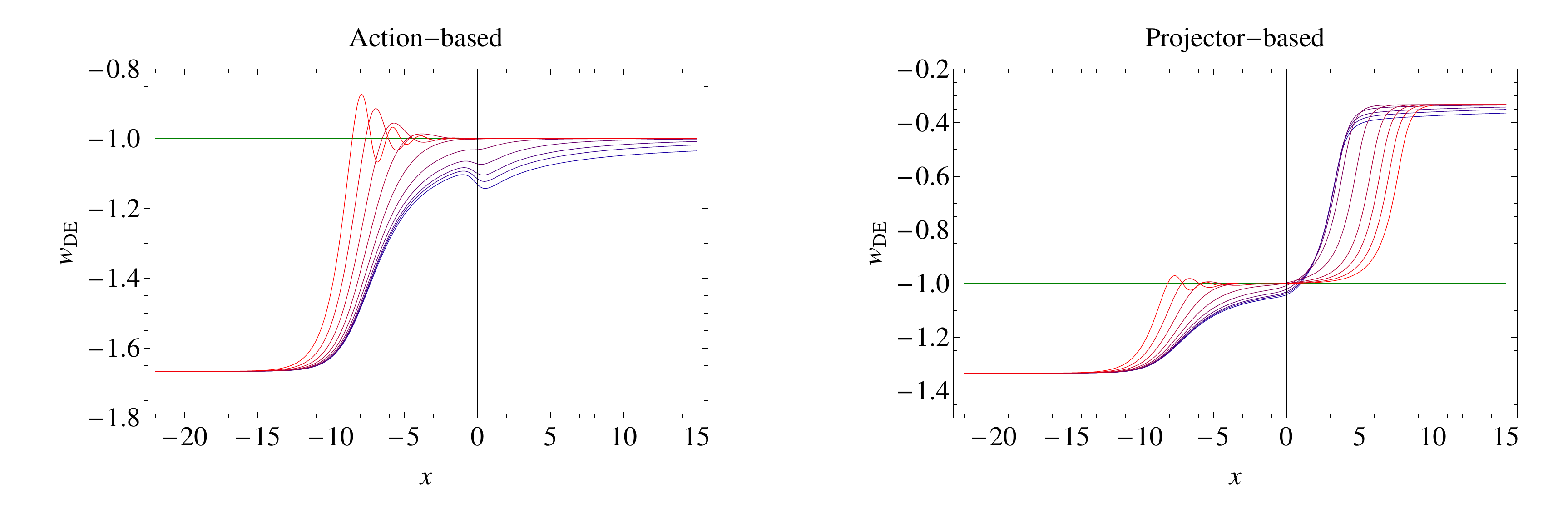} 
\caption{The dark energy effective equation of state parameter $w_{\rm DE}$.}
\label{fig:wDE} 
\end{center}
\end{figure}

Another noteworthy feature is that the observable departure from GR (figure \ref{fig:dlogH}, \ref{fig:w} and \ref{fig:dw}) starts roughly around today, i.e. when the curvature $\sim H^2$ approaches the $m^2$ scale. On the other hand, the dark energy component $\hat{\ro}_{\rm DE}$ starts being non-zero as we enter the MD era, i.e. roughly around $x_{\rm eq} \approx -8$, since this is when $R$ ``wakes-up''. 

Finally, the fact that the dark energy component starts from zero and then grows, i.e. $\ro_{\rm DE} > 0$ and $\dot{\ro}_{\rm DE} > 0$ at the beginning, implies that $w_{\rm DE}$ starts below $-1$ because of (\ref{eq:conseqw}). Thus, non-local dark energy models have this in common that their equation of state starts on the phantom side.

\pagebreak

\subsection{Plots \& tables} \label{sec:plottab}

\vspace{-0.2cm}

\vspace{-0.2cm}

\pagebreak

\vspace{-0.2cm}

\vspace{-0.2cm}

\pagebreak

\begin{figure}[h!]
\begin{center}
\includegraphics[width=16cm]{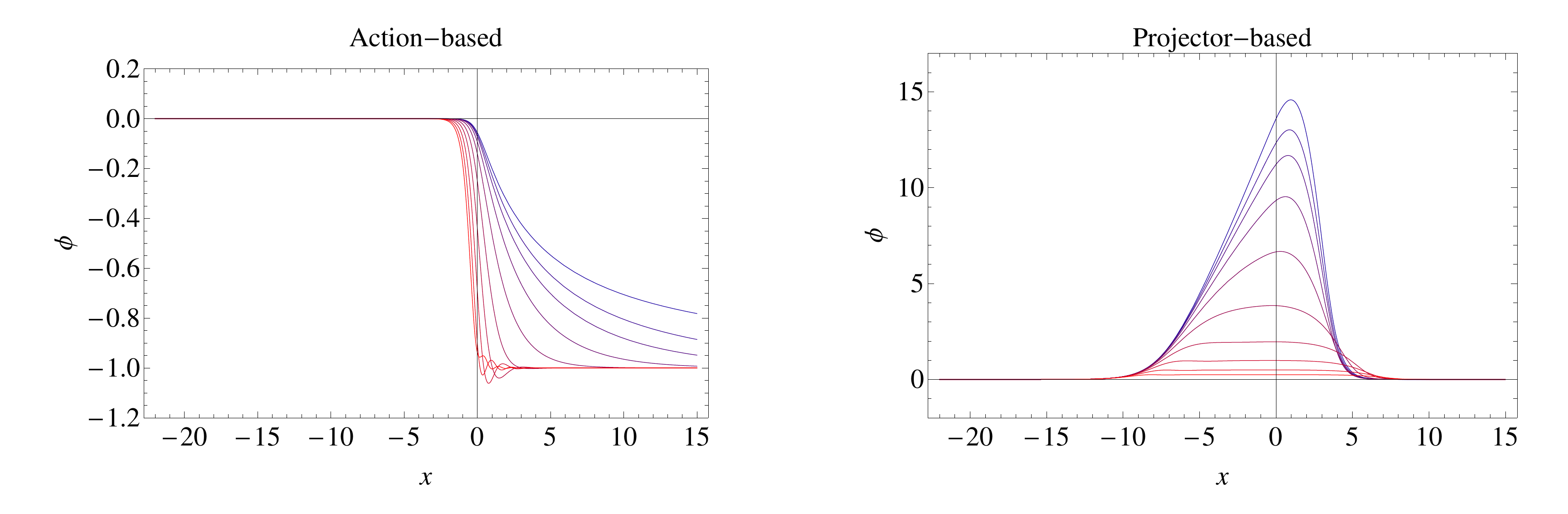} 
\caption{The dimensionless localizing field $\ph$.}
\label{fig:phi} 
\end{center}
\end{figure}

\vspace{-0.2cm}

\begin{figure}[h!]
\begin{center}
\includegraphics[width=16cm]{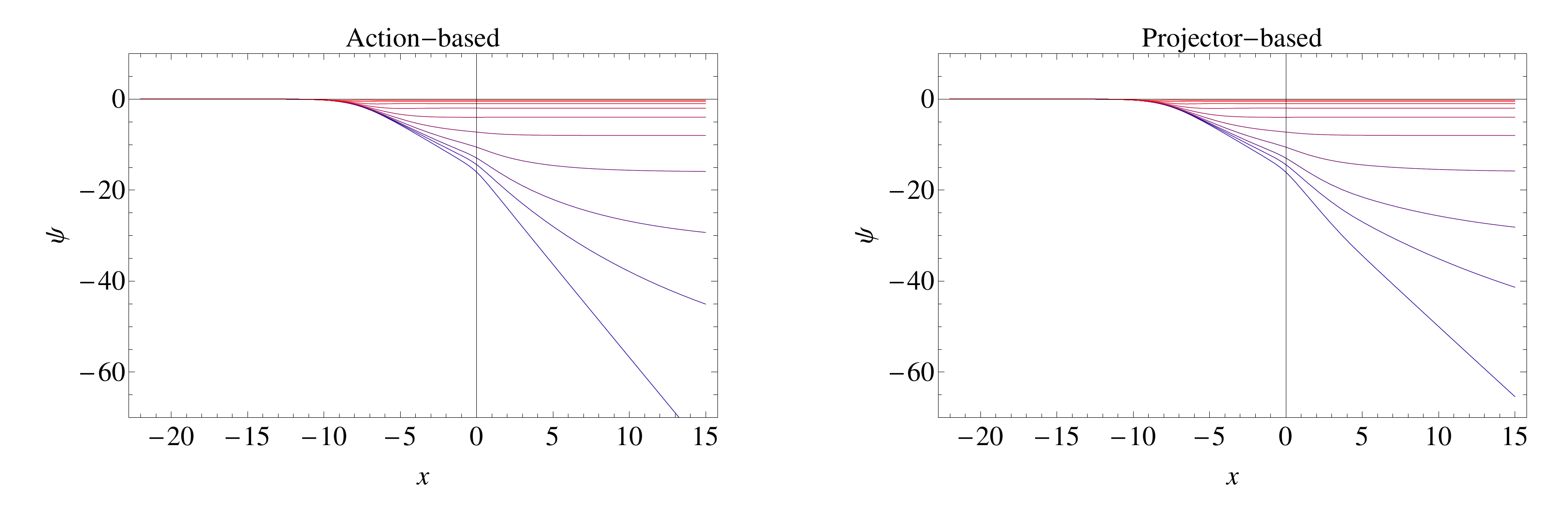} 
\caption{The dimensionless localizing scalar $\psi$.}
\label{fig:psi} 
\end{center}
\end{figure}

\vspace{-0.2cm}

\begin{figure}[h!]
\begin{center}
\includegraphics[width=16cm]{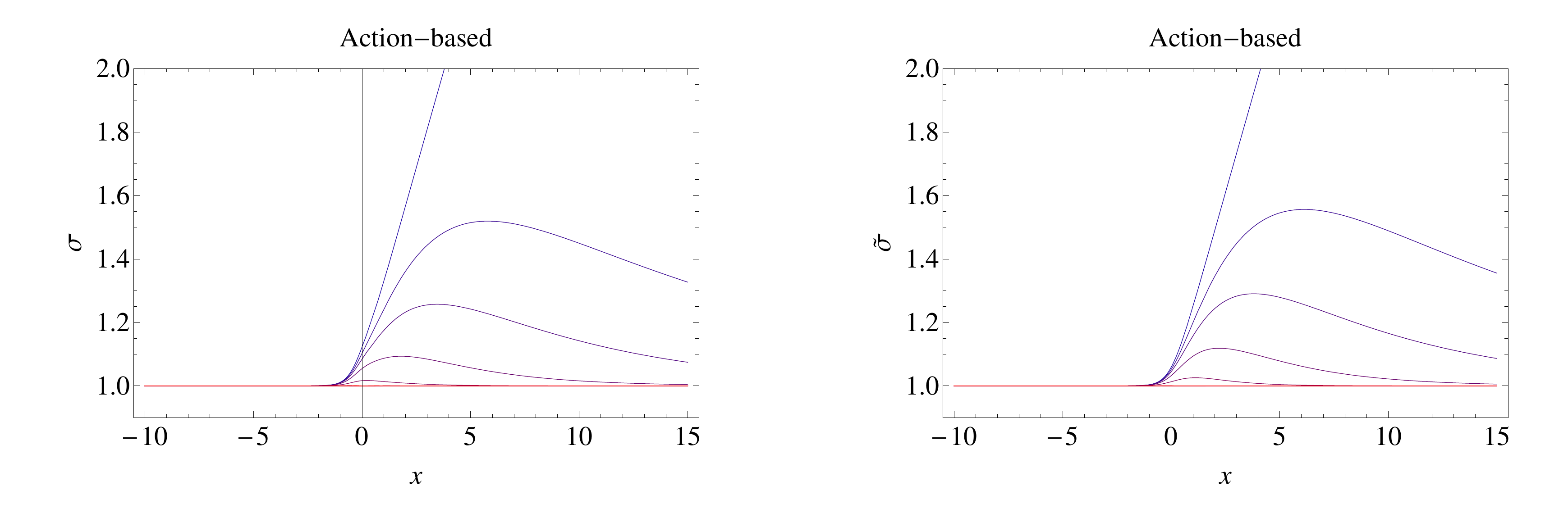} 
\caption{The effective Newton's constants $\si$ and $\ti{\si}$ of the action-based model.}
\label{fig:si} 
\end{center}
\end{figure}

\pagebreak

\begin{table}[h!]
\centering
\begin{tabular}{|c||c|c|c||c|c|c|} 
 \hline
  & \multicolumn{3}{c|}{Action-based} & \multicolumn{3}{|c|}{Projector-based} \\
 \hline
 $\log_2 \xi$  &  $\mu^2$  &  $w_0$  &  $w_{{\rm DE},0}$ &  $\mu^2$  &  $w_0$  &  $w_{{\rm DE},0}$  \\
 \hline 
 $-\infty$ & $0.0089235$ & $-0.7816$ & $-1.1307$ & $0.050252$ & $-0.7108$ & $-1.0417$ \\
 $-6$ & $0.0113795$ & $-0.7664$ & $-1.1144$ & $0.055373$ & $-0.7069$ & $-1.0359$ \\
 $-5$ & $0.0144145$ & $-0.7528$ & $-1.0992$ & $0.060895$ & $-0.7033$ & $-1.0306$ \\
 $-4$ & $0.022624$ & $-0.7300$ & $-1.0720$ & $0.073170$ & $-0.6968$ & $-1.0212$ \\
 $-3$ & $0.050545$ & $-0.7005$ & $-1.0314$ & $0.10260$ & $-0.6874$ & $-1.0074$ \\
 $-2$ & $0.17022$ & $-0.6824$ & $-0.9994$ & $0.17733$ & $-0.6804$ & $-0.9971$ \\
 $-1$ & $0.68365$ & $-0.6829$ & $-1.0011$ & $0.34773$ & $-0.6819$ & $-0.9993$ \\
 $0$ & $2.7328$ & $-0.6822$ & $-1.0007$ & $0.69005$ & $-0.6820$ & $-0.9994$ \\
 $1$ & $10.9275$ & $-0.6835$ & $-1.0004$ & $1.3739$ & $-0.6822$ & $-0.9997$ \\
 $2$ & $43.687$ & $-0.6812$ & $-1.0001$ & $2.7408$ & $-0.6821$ & $-0.9996$ \\
 \hline
 $\La$CDM & $-$ & $-0.6824$ & $-1$ & $-$ & $-0.6824$ & $-1$ \\
 \hline
\end{tabular}
\caption{The values of the mass parameter and today's effective equation of state parameters.}
\label{tab:mu2ws}
\end{table}

\section{Analytic approximations}

Now that we have some concrete insight into the physics, let us try to reproduce the essence of the numerical results through analytic approximations. The equations of motion can be solved analytically if we assume that $w$, or alternatively $\ze$, is constant, which is the case when we are well-inside a definite phase of the universe's history (\ref{eq:zeval}). Here we know that the solutions admit such plateau values (see figure \ref{fig:w}), but even if we did not, we could assume they exist and check the consistency of the solutions afterwards.

We start by solving for $\psi$ (\ref{eq:psicEOM}), which obeys the same equation in both models. For $\xi = 0$ we get
\beq \label{eq:solpsixi0}
\psi = - 6 \, \frac{2 + \ze}{3 + \ze}\, x + a_1 + a_2 \exp \[ -\(3+\ze\) x \] \, ,
\eeq
while for $\xi \neq 0$ we get
\bea
\psi & = & -\frac{1}{\xi} + a_1 \exp \[ -\frac{1}{2}\, x \( 3 + \ze - \sqrt{\( 3 + \ze \)^2 - 24\xi \( 2 + \ze \)} \) \] \nn \\
 & & + a_2 \exp \[ -\frac{1}{2}\, x \( 3 + \ze + \sqrt{\( 3 + \ze \)^2 - 24\xi \( 2 + \ze \)} \)  \] \, . \label{eq:attractpsi}
\eea
These have the same form only in RD where $\ze = -2$
\beq
\psi = C + a_2 e^{-x} \, .
\eeq
For more general $\ze$, the exponentials are decaying if $\ze > -3$ (and thus $w < 1$), which is the case in all phases of interest (\ref{eq:zeval}), so these solutions are stable. In the $\xi = 0$ case we then have a linear evolution, while in the $\xi \neq 0$ case we have an attractor behaviour towards $-1/\xi$. This is confirmed in figure \ref{fig:psi}, although the convergence is quite slow for low $\xi$. In RD, which is where we begin, the integration constants are fixed by the choice of initial conditions. Remember that these are theory-level data, i.e. different choices correspond to different definitions of $\bo^{-1}$ and thus to different theories. Here the data (\ref{eq:initdata}) translate into $C = a_2 = 0$, thus giving\footnote{Considering a non-zero integration constant in RD corresponds to a different theory, namely, the one where the inversion of $\ti{\bo}$ is affine, i.e. it is of the form $\psi \sim f + \ti{\bo}^{-1}_{\rm r} R$,
where $f$ is a homogeneous solution $\ti{\bo} f = 0$. This extension has been studied in \cite{FoffaMaggioreMitsou2} for the projector-based model with $\xi = 0$. Since $f$ is made of a constant part and a decaying exponential, the non-trivial part of the modification is $f = {\rm const}$ and this simply amounts to adding an $m$-dependent cosmological constant in the equation. Indeed, since $g_{\mu\nu}$ is trivially transverse, we have 
\beq
m^2 \( \bo^{-1} R\)^{\rm T} = m^2 g_{\mu\nu} f + m^2 \( \bo_{\rm r}^{-1} R\)^{\rm T} \, .
\eeq 
The effect on cosmology is similar to the one of $\xi$, as it bridges the Maggiore model with $\La$CDM.}
\beq \label{eq:psiRD}
\psi_{\rm RD} = 0 \, .
\eeq
For $\xi \neq 0$, we have that during the MD and DE phases $\psi = -1/\xi$ because of the attractor behaviour (\ref{eq:attractpsi}). So let us focus on $\xi = 0$ where the solution takes the form (\ref{eq:solpsixi0}). In the simplest approximation, the beginning of the MD phase $\ze = -3/2$ occurs at matter-radiation equality $x_{\rm eq} = \log \hat{\ro}^0_R / \hat{\ro}^0_M$, so this is where $\psi$ should start being non-zero. This gives
\beq \label{eq:psiMD}
\psi_{\rm MD} \approx  - 2 \( x - x_{\rm eq} \) \, , \hspace{1cm}  x > x_{\rm eq} \, .
\eeq
Then, considering $x = 0$ as the transition from MD to de-Sitter $\ze \approx 0$, and matching with the above result, we get that
\beq \label{eq:psiDE}
\psi_{\rm DE} \approx - 2 \( 2 x - x_{\rm eq} \)  \, , \hspace{1cm}  x > 0 \, .
\eeq
Indeed, in figure \ref{fig:psi} we see that the slope increases (from $2$ to $4$) after MD and as a further check we can verify that $\psi(0) \approx 2 x_{\rm eq} \approx -16$ seems correct. In the projector-based case, the slope then decreases again in the future because we pass from the quasi-de-Sitter phase $\ze = 0$ to the ultimate $w = -1/3$ phase, giving $\ze = -1$, and thus a slope of $3$. 

Let us now look at each model separately.

\subsection{Action-based model}

We wish to solve (\ref{eq:phcEOM}) for $\ph$ by analytic approximations. To do so, we first need to solve for $h$ with a constant $\ze$
\beq \label{eq:hsolzeconst}
h' = \ze h \, ,   \hspace{1cm} \Rightarrow \hspace{1cm} h \sim e^{\ze x} \, .
\eeq
Then, we start by computing the solution for RD where $\psi_{\rm RD} = 0$ to get
\beq
\ph = b_1 + b_2 e^{-x} \to b_1 \, ,
\eeq
whatever the value of $\xi$, so this result is stable. With vanishing initial conditions we have
\beq
\ph_{\rm RD} = 0 \, .
\eeq
For the subsequent phases we must consider the $\xi > 0$ and $\xi = 0$ cases separately.

\subsubsection{The case $\xi > 0$}

Using (\ref{eq:hsolzeconst}) and $\psi = -1/\xi$ the equation of $\ph$ gives
\bea
\ph & \sim & - \frac{3\mu^2}{2 \xi \( \( \ze - 3 \) \ze + 3 \xi \( 2 + \ze \) \)}\, e^{-2\ze x} + {\rm hom.}  \, ,
\eea
where the homogeneous part is the same as for $\psi$ (\ref{eq:attractpsi}) since their equations differ only through their sources. Therefore, the homogeneous solutions of $\ph$ are stable as well. We can thus focus on the inhomogeneous part which is diverging for MD $\ze = -3/2$. Indeed, the $\ph$ profile in the interval $x < 0$ of figure \ref{fig:phi} is exactly the one of an exponential with a negative $\Ord(1)$ factor in front. In the de-Sitter phase however, the solution is attracted towards a constant. The de-Sitter solutions are known exactly (\ref{eq:dSab}) and coincide with the observed value of $-1$. With this behaviour for $\ph$, and $\psi = -1/\xi$, we have that $\si$ (\ref{eq:siroDE}) is also constant at late times and thus so is $H^2$.

\subsubsection{The case $\xi = 0$}

\begin{figure}[h]
\begin{center}
\includegraphics[width=16cm]{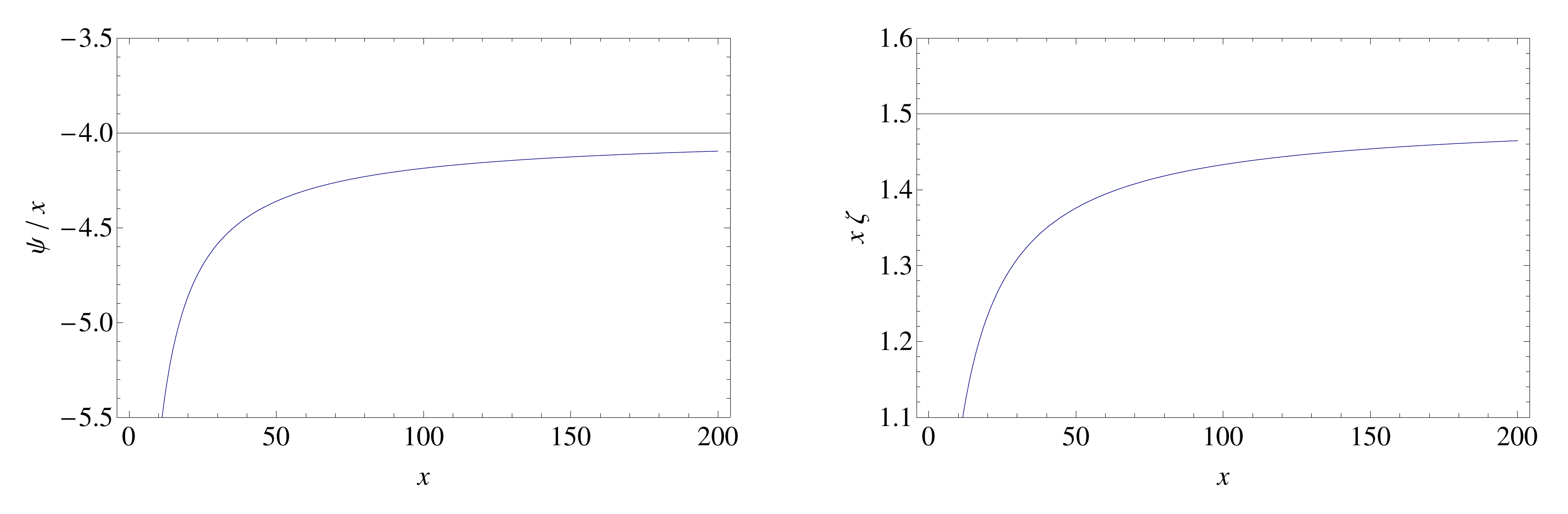} 
\caption{The functions $\psi/x$ and $x \ze$ in the action-based $\xi = 0$ model tending towards the values $-4$ and $3/2$, respectively.}
\label{fig:psizeasymp} 
\end{center}
\end{figure}

To get the MD solution here we have to use (\ref{eq:psiMD})
\beq 
\phi \sim -\frac{4}{81}\, \mu^2 \( 9 \( x - x_{\rm eq} \) - 5 \) e^{3x} + b_1 + b_2 e^{-\frac{3}{2}\, x} \, ,
\eeq
which is again unstable and fits with figure \ref{fig:phi}. In the $\ze = 0$ case we have to use the (\ref{eq:psiDE}) solution to get
\beq
\ph \sim - \frac{2}{3}\, \mu^2 \( 3 x - 2 \) x + b'_1 + b'_2 e^{-3x} \, .
\eeq
Surprisingly, this is not at all the kind of behaviour we observe since $\ph$ is constant at late times. This implies that the assumption $\ze = 0$ is not valid, i.e. $\ze$ tends towards zero as $x \to \infty$ but too slowly. We therefore need a more precise ansatz for $\ze$ and we thus proceed perturbatively from infinity. Using the leading order solutions
\beq
\psi_{\rm DE} \approx -4 x \, , \hspace{1cm} \ph_{\rm DE} = -1 \, ,
\eeq
we have that (\ref{eq:Friedab}) and (\ref{eq:zeab}) give 
\beq
h^2 \approx \frac{4 \mu^2 x}{1+\ph} \, , \hspace{1cm} \ze \approx \frac{6 \mu^2 h^{-2}}{1+\ph} \, ,
\eeq
and thus imply 
\beq
\ze \approx \frac{3}{2x} \, .
\eeq
As a check, in the left panel of figure \ref{fig:psizeasymp} we have plotted $\psi/x$ and $x \ze$ to see that they tend indeed towards $-4$ and $3/2$, respectively.
We can then solve $h' = \ze h$ to find $h \sim x^{3/2}$. Now that we have the more precise profiles $h^2 \sim x^3$ and $\ze \sim 3/2x$ for large $x$, we can plug them in the equation of $\ph$ and solve. The result is a combination of a Meijer G-function, an error function and a decaying exponential, whose $x \to \infty$ limit is an integration constant, consistent with the numerical result \ref{fig:phi}. 

A growing Hubble parameter at late times is more violent than the constantly accelerated expansion of a de-Sitter phase, so let us see what it implies for the fate of the universe.

\subsubsection{Big rip singularity}

We have $H = (2/T)\, x^{3/2}$, for some positive constant $T$ with dimensions of time. To estimate the latter, we try to guess the asymptotic value of $x^{-3/2} h$ by going at large $x$ and find a good estimate in $x^{-3/2} h \to 0.09$, so we have that $T \approx 22 H_0^{-1}$. The equation for $a(t)$ is then
\beq
\dot{a} = H a = \frac{2}{T} \( \log a \)^{3/2} a \, ,
\eeq
whose solution is
\beq \label{eq:DEera}
a(t) = \exp \[ \frac{T^2}{\( t_{\rm rip} - t \)^2} \] \, .
\eeq
This is an example of the so-called ``big rip'' singularity, i.e. the divergence of the scale factor and the Hubble parameter at finite time
\beq \label{eq:bigrip}
\lim_{t \to t^-_{\rm rip}} a(t) = \infty \, , \hspace{1cm} \lim_{t \to t^-_{\rm rip}} H(t) = \infty \, .
\eeq
In our case this occurs far in the future since $T$ corresponds to several times the age of the universe. Moreover, we must not forget that, since $H$ is growing in the DE, the curvature $R$ will eventually reach an energy scale where this effective description ceases to be valid, so the region close to the singularity cannot be trusted.

It turns out that a big rip is a usual consequence of phantom equation of state parameters $w_{\rm DE} < -1$. Indeed, the phenomenology of such types of dark energy was first considered in \cite{C,CKW}\footnote{For a study of the type of future singularities caused by phantom dark energy see \cite{NOT} and for the case where this occurs with the Deser-Woodard type of non-locality see \cite{BNOS}.} where it was realized that $w < -1$ in GR would generically imply a future singularity at a finite time (\ref{eq:bigrip}). For constant $w$ this is easy to show. The continuity and first Friedmann equations read
\beq
\dot{\ro} + 3 H \( 1 + w \) \ro = 0 \, , \hspace{1cm} \dot{a} = a \sqrt{\frac{8\pi G}{3}\, \ro}  \, .
\eeq
The first gives $\ro = \ro_0 a^{-3(1+w)}$ and, plugging this in the second, we get
\beq
\dot{a} = H_0 \, a^{-\frac{3}{2} (1+w)+1} \, .
\eeq
The solution can be written as
\beq \label{eq:aofconstw}
a(t) = \[ - \frac{3}{2}\,H_0 \(1+w\) \( t_{\rm rip} - t \) \]^{\frac{2}{3(1+w)}} \, ,
\eeq  
where $t_{\rm rip}$ is the integration constant. Since $1+w < 0$, the bracket is positive, while the power is negative and we thus have (\ref{eq:bigrip}) indeed. In our case we have that $w < -1$, but tends towards $-1$ as time passes. Thus, whether there will be a big rip or not depends on how fast this convergence is. We now know that for $\xi > 0$ there is no big rip, but rather an eternal de-Sitter phase, while for $\xi = 0$ no de-Sitter solution exists and we have a big rip. This feature can be traced back to the discontinuity of the asymptotic behaviour of $\psi$ as $\xi \to 0$. For $\xi > 0$ we have that $\psi$ tends to the constant value $-1/\xi$, while for $\xi = 0$ it goes like $\sim -4 x$.

\subsection{The projector-based model}

We now focus on (\ref{eq:phicEOMpb}) which does not depend explicitly on $\xi$, although $\psi$ does. In the RD phase we have $\psi_{\rm RD} = 0$ so $\ph$ has only a homogeneous solution, which is decaying, and thus the $\ph_{\rm RD} = 0$ solution is stable. We then enter MD, where the choice of $\xi$ is relevant.

\subsubsection{The case $\xi > 0$}  

With $\psi = -1/\xi$ we can solve (\ref{eq:phicEOMpb})
\beq
\ph = \frac{1}{\xi} + b_1 \exp \[ -\frac{1}{2}\, x \( 3 - \ze - \sqrt{21 + \ze \( 6 + \ze \)} \) \] + b_2 \exp \[  -\frac{1}{2}\, x \( 3 - \ze + \sqrt{21 + \ze \( 6 + \ze \)} \) \] \, .
\eeq
In MD $\ze = -3/2$ the exponentials decay and we are attracted towards the constant solution $1/\xi$ as can be checked in figure \ref{fig:phi}. In de-Sitter $\ze = 0$ however we have a diverging mode $\sim \exp \( \(\sqrt{21} - 3\)x/2 \)$ in the homogeneous solution, so this phase is unstable. This leads us to the final stage of the universe's history which is a $\ze = -1$ phase ($w = -1/3$), in which case
\beq
\ph = b'_1 + b'_2 e^{-4x} \, ,
\eeq
so this phase is stable. From figure \ref{fig:phi} we see that $b'_1 = 0$.

\subsubsection{The case $\xi = 0$}

To get the MD solution here we have to use (\ref{eq:psiMD}) to get
\beq 
\phi = -2 + 2\( x - x_{\rm eq} \) + {\rm hom}.  \, ,
\eeq
where the homogeneous part decays. This is indeed what we observe in \ref{fig:phi}, i.e. a linear trend which cuts the $x = 0$ axis at approximately $\ph(0) \approx -2(1 + x_{\rm eq}) \approx 14$. Then, for $\ze = 0$, using (\ref{eq:psiDE}) we get again the same kind of diverging mode in the homogeneous solution as in the $\xi > 0$ case. We must thus finally consider the case $\ze = -1$, and $\psi \sim -3 x$, where the solution is
\beq
\ph = \frac{9}{4}\, x + b_1 + b_2 e^{-4x} \, .
\eeq
As in the $\xi = 0$ action-based model, this final trend is not at all the behaviour we observe, which means that $\ze$ does not tend fast enough to $-1$. 
Here however we have a simpler way to deduce $\ph$ at large $x$. Indeed, since here $\si = 1$, we have that $w = w_{\rm DE}$ at late times so we can use (\ref{eq:wDEpb}) and (\ref{eq:w}) to get 
\beq
\ze \approx \frac{1}{2}\, \frac{\ph'}{\ph} \, .
\eeq
Then, using the lowest order result $\ze = -1$ the above equation gives $\ph \approx e^{-x/2}$, which is indeed the behaviour we observe \ref{fig:phi}. Note that this technique would not have worked in the action-based model because there $\si \gg 1$ in the future (see figure \ref{fig:si}). Indeed, had we used $w = w_{\rm DE}$ and (\ref{eq:wDEab}) and (\ref{eq:w}), we would have rather found $\ze = 1/x$ instead of $3/2x$. Thus, $w$ and $w_{\rm DE}$ tend to the same value but at different paces.

\section{Stability}

As we have already argued in section \ref{eq:classeffthstab}, the diverging modes of a non-tachyonic ghost, which is what we have here, should manifest themselves at time scales of the order of the mass scale. This implies that the background solutions we have studied above are potentially unstable under linear perturbations, but this does not necessarily spoil the viability of the cosmological history. Indeed, since $m \sim H_0$ the typical time interval for the divergence to become notable is of the order of the age of the universe $\De t \sim m^{-1} \sim H_0^{-1}$.

The linear perturbations of the $\xi = 0$ models have been studied in \cite{DFKKM}, where it was shown that there are indeed no notable divergences up until today. As already mentioned, these models have even been studied with a full Boltzmann/MCMC code and found to be statistically equivalent to $\La$CDM \cite{DFKMP}, with respect to the present data precision. We know that with large enough values of $\xi$ we can approach GR with a cosmological constant with arbitrary precision. At the level of the cosmological background evolution, we have verified indeed that $\xi$ interpolates between the $\xi = 0$ models and $\La$CDM. There is therefore no reason why this should not be the case in general, and we thus we expect the $\xi$ extensions to be equally viable at the level of the perturbations as well. 

\begin{figure}[h!]
\begin{center}
\includegraphics[width=14cm]{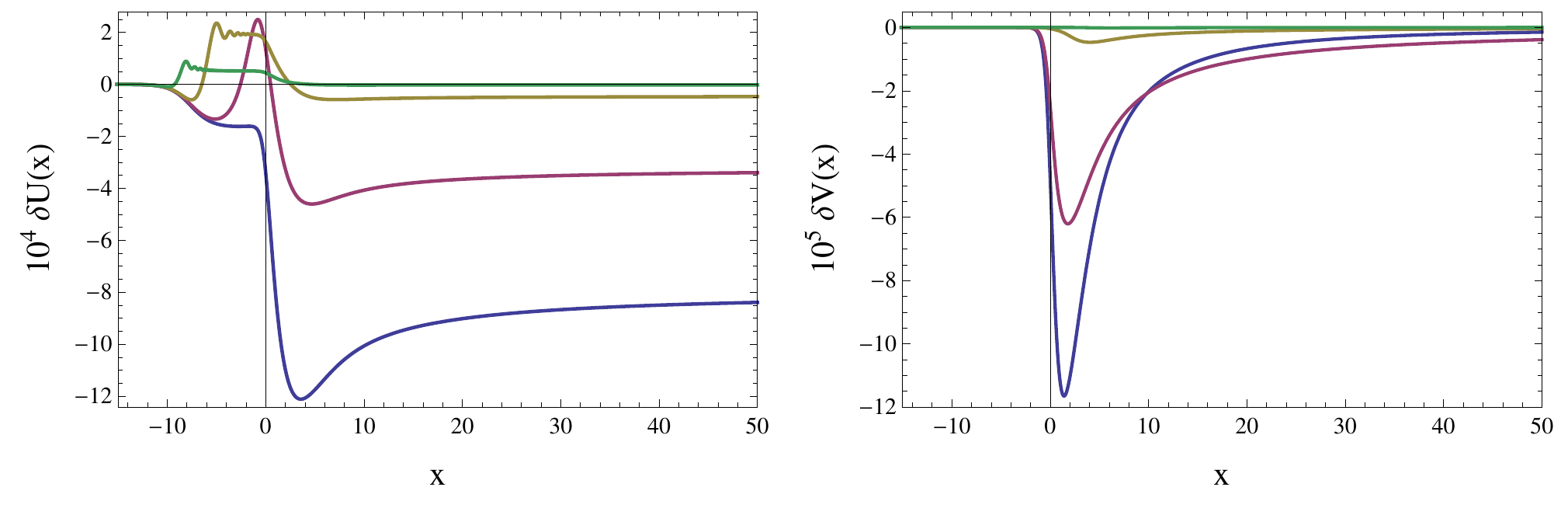} 
\caption{The linear perturbations of $U \equiv -\psi$ and $V \equiv -\mu^{-2}/3 \ph$ as a function of $x$ for the modes $\ka = 5 \times 10^{-3}$ (blue), $\ka = 5 \times 10^{-2}$ (purple), $\ka = 5 \times 10^{-1}$ (brown), $\ka = 5$ (green) in the MM model.}
\label{fig:dUdV} 
\end{center}
\end{figure}

An interesting fact regarding the perturbations is that in the action-based model they are actually even bounded. The perturbations of the two auxiliary scalar modes are given in figure\footnote{Courtesy of Yves Dirian.}  \ref{fig:dUdV}. We have plotted several different values of comoving wave-number $\ka \equiv k / k_{\rm eq}$, where $k_{\rm eq} = a_{\rm eq} H_{\rm eq}$ is the comoving wave-number corresponding to the horizon scale at matter-radiation equality\footnote{For the numerical integration the set-up of \cite{DFKKM} has been used.}.
Since $k_{\rm eq} \approx 42 H_0$, we have that the displayed choices of $\ka$ range from sub-horizon to super-horizon modes today and all of them tend to a constant for large $x$. Incidentally, the same holds with respect to cosmic time $t$ and, in particular, they are smooth in the $t \to t_{\rm rip}^-$ limit. We see that the large wave-length modes tend to diverge, as expected, soon after $x = 0$, but are then quickly tamed towards a constant evolution. We can now understand this as the consequence of Hubble friction. Indeed, if $H$ admits a singularity at finite time, the big rip, then by continuity the Hubble friction term $\sim H \dot{\ph}$ in the equations of the scalars will inevitably dominate at some point over any other term, i.e. even over the tendency of ghost modes to diverge\footnote{In \cite{DFKKM}, instead of focusing on the scalar modes themselves, the authors have chosen to treat the deviation from GR as an effective dark energy fluid and thus focused on the effective quantities $\ro_{\rm DE}, p_{\rm DE}, \te_{\rm DE}$ and $\si_{\rm DE}$ that are the energy density, pressure, velocity and anisotropic stress scalars, respectively. The conservation equation then leads to the evolution equation for the contrast $\de_{\rm DE} \equiv \de \ro_{\rm DE}/ \bar{\ro}_{\rm DE}$ which is eq. (6.9) of \cite{DFKKM}. In this description, the ``wrong'' relative sign appears in the fact that the sound speed squared $c_s^2$ is negative at all times, as shown in figure 16 of \cite{DFKKM}. The fact that $\de_{\rm DE}$ tends to zero in the future (figure 14 of \cite{DFKKM}) had already made the authors of \cite{DFKKM} deduce that the Hubble friction dominated the dynamics.}. 

One could still be worried by the small window where $\de \ph, \de \psi$ grow significantly around $x = 0$, especially in the case of large scales where the effect is the strongest. However, as shown in \cite{DFKKM}, this has no notable effect in the evolution of observable quantities such as the dark matter energy density or the Bardeen potentials. 

Finally, note that the $w_{{\rm DE},0}$ values found here, which range between $-1.13$ and $-1$, are consistent with the present observational data  \cite{DFKMP}, but nevertheless give different predictions than $\La$CDM, \footnote{Note that, {\it if one assumes a constant} $w_{\rm DE}$ for the background, then the present data narrow the result down to $w_{\rm DE} = -1.00 \pm 0.05$, \cite{SNe}, which therefore excludes part of the models we have considered here. However, in these models $w_{\rm DE}$ is not constant and a full comparison with the data has proved their viability, even if $w_{{\rm DE},0}$ can go down to $-1.13$.}. Future missions such as the Dark Energy Survey \cite{DES} and EUCLID \cite{EUCLID} are expected to measure $w_{{\rm DE},0}$ with a percent precision and will thus allow to discriminate these models from $\La$CDM. The $\xi$-parametrization we proposed, which is an original feature of the present thesis, allows more flexibility for matching the desired value, since it covers all values of $w_{{\rm DE},0}$ from the one of the MM model $w_{{\rm DE},0} \approx -1.13$ up to the one of $\La$CDM $w_{{\rm DE},0} = -1$. Of course this lowers the predictive power of the model, but we see that the predictions remain quite sharp.

\chapter{Conclusions}

In this thesis we have elaborated on the formulation, properties and phenomenology of some non-local theories of gravitation containing a fixed mass parameter, with the ultimate aim being of providing a viable dark energy model.

\subsubsection{Linear massive gauge theories}

We have started our investigation by trying to understand, under several viewpoints, the properties of linear massive gauge theories in order to prepare the ground for their non-local formulations and generalizations. We have found that performing a $d+1$ harmonic decomposition of the fields, whether in the Lagrangian or Hamiltonian formalisms, provides a very transparent understanding of the dynamical content of these theories. In particular, this decomposition reveals the structure of the spin-2 theory with generic mass term. Once the non-dynamical fields have been integrated-out, the $d$-scalar sector (\ref{eq:SscalhG}), which is the interesting one, can be neatly represented by two fields, one of which is a ghost 
\bea
S_{\rm scal.} & = & \frac{d-1}{d} \int \ed^D x \[  - \frac{1}{2}\, \pa_{\mu} \ti{\Phi} \pa^{\mu} \ti{\Phi}  - \frac{1}{2}\, m^2 \ti{\Phi}^2  - \ti{\Phi} \( \ro - \De \si \)  \right.  \\
 & & \hspace{2.2cm} \left. + \frac{1}{m^4} \( \frac{1}{2}\, \pa_{\mu} G \pa^{\mu} G  - \frac{1}{2}\, m_{\rm ghost}^2 G^2 - \frac{m^2}{d-1}\,G \( \ro - d p \) \)\] \, ,  \nn
\eea
with
\beq
m_{\rm ghost}^2 \equiv \frac{1 + d\( 1+ 1/\al \)}{d-1}\, m^2 \, ,
\eeq
where $\ti{\Phi} \equiv \Phi + m^{-2} G$ and both $\Phi$ and $G$ are analytic in $m$. From this the dependence of the physics on the $(m^2, \al)$ parameters is clear. The Fierz-Pauli point $\al = 0$ is the only ghost-free theory, but it is also the only one which is discontinuous in the $m \to 0$ limit, since $\Phi$ survives in the action. Remarkably, $\Phi$ is a gauge-invariant combination, under the gauge symmetry of the massless theory. This implies that, although the massive action is not gauge-invariant, the physics is invariant under a $2$-parameter subset of gauge transformations, so that this could be called a ``hidden'' symmetry.

Moreover, this property is preserved on a de-Sitter background as well, but then the $d$-scalar field $\Om$ is a combination of the $h_{\mu\nu}$ fields that is non-local in time, and actually quite ugly (\ref{eq:Omofh}). Again, integrating-out the non-dynamical fields, the $d$-scalar action reads (\ref{eq:SofOm})
\beq
S_{\rm scal.} = \frac{d-1}{d} \frac{M^2}{m^2} \int \ed^D x \, \sqrt{-g} \[ -\frac{1}{2}\, \pa_{\mu} \Om \, \pa^{\mu} \Om - \frac{1}{2}\, m^2 \Om^2 \] \, .
\eeq 
where $M^2 \equiv m^2- (d-1)H^2$. This reflects quite elegantly the dependence of the spectrum on the mass $m$ on a de-Sitter background, with the special case $M^2 = 0$ corresponding to the so-called ``partially massless theory''.

We have then moved on to the computation of the propagators of each theory and have discussed the St\"uckelberg formalism. Both approaches show how the apparent discontinuity in the degrees of freedom as $m \to 0$ can be understood as the smooth decoupling of some modes. Using the St\"uckelberg trick, we were able to reformulate the equations of motion in a gauge-invariant way, even in the presence of a mass, with the price to pay being the loss of locality. Nevertheless, locality is restored with the appropriate choice of gauge, which leads us to interpret the mass term as an obstruction to having both gauge-invariant and local representations of the theory. In the spin-2 case, we have also found that the non-local formulation of Fierz-Pauli theory actually has one more gauge symmetry than GR itself! This is the symmetry of linearized conformal transformations which is responsible for killing the ghost mode in this context.

A useful by-product of this construction are the transverse projectors ${\cal P}$, that is, non-local operators which make the gauge-field transverse and gauge-invariant and thus allow a straightforward construction of massive gauge-invariant theories. In the spin-1 case, only one such projector exists and the only gauge-invariant quadratic theory one can construct is nothing but the non-local formulation of the Proca action of massive electrodynamics. In the spin-2 case however, because the subspace of transverse tensors splits into traceless and pure-trace parts, there are two projectors and thus one has access to more models than the ones that are equivalent to the local theory. These are therefore genuinely non-local, i.e. they are non-local whatever the gauge we choose. We have thus considered these models, and in particular (\ref{eq:projeq})
\beq 
\( \bo - m_g^2 \) {}_0 {\cal P}_{\mu\nu\ro\si} h^{\ro\si} + \( z \bo - m_s^2 \) {}_s {\cal P}_{\mu\nu\ro\si} h^{\ro\si} = -T_{\mu\nu} \, ,
\eeq
which, on top of a massive graviton, contains an extra propagating scalar mode corresponding to the trace $h$. In the local theory, this mode is either non-dynamical (the Fierz-Pauli mass term), or it is ghost-like (all other mass terms). In the above non-local theory it is both dynamical and healthy if $z > 0$. Finally, the projectors have also simplified our computation of propagators in the non-local setting thanks to their nice algebraic properties (\ref{eq:Paverprod}).

We have also shown in more than one way an important aspect of local linear gauge theories, which is that the constraint structure which is due to gauge symmetry is such that it guarantees the rule $N_{\rm f} = 2 N_{\rm d}$, i.e. that there are always twice as many degrees of freedom as there are dynamical fields. This is important to note because it does not hold in the case of non-local field theories in general.

\subsubsection{Non-local subtleties}

We have then paused to discuss some peculiar aspects of non-local field theory. We have mentioned that the usual variational principle applied on some non-local action cannot yield causal equations of motion, but that there exists a modification of that principle which respects causality. The construction is inspired by the ``in-in'' formalism for the quantum effective action $\Ga$, i.e. the action which controls the dynamics of some expectation value of the field operator. The corresponding variational principle requires initial data to be imposed, instead of boundary data, and thus provides an action-based description of irreversible systems. This is for example the case of non-local field theories, where the combination of non-locality and causality privileges the past with respect to the future and thus implies an arrow of time.

We have also discussed the localization procedure which turns non-local equations into local ones by integrating-in auxiliary fields, and thus allow us to see the dynamical content of the theory. The auxiliary fields have constrained initial conditions, because this data corresponds to the fixed choice of $\bo$ inverse we do in the non-local theory. However, they obey dynamical equations of motion, so that $N_{\rm f} \leq 2 N_{\rm d}$ in general. The only exceptions to this rule are the non-local formulations of local theories, where the auxiliary fields correspond to St\"uckelberg fields and are thus pure-gauge. 

The presence of dynamical fields that have constrained initial conditions forbids any quantum interpretation of genuinely non-local theories, since one cannot implement these constraints at the quantum level, in terms of constraints on the Hilbert space, without spoiling unitarity. Thus, genuinely non-local theories are necessarily classical effective theories. 

We have then addressed the important issue of classical stability. Indeed, non-local theories often contain ghost-like or tachyonic dynamical fields, that are only seen in the localized theory. In the literature their constrained status has often been invoked in order to minimize their impact on stability. We have argued that, on the contrary, they should be considered on the same footing as regular dynamical fields in a stability analysis, i.e. they are very capable of destabilizing a given solution of interest. This is because, for genuinely non-local theories, these fields interact non-linearly and are thus excited whatever their initial conditions, making the initial data constraints irrelevant in a stability analysis. The latter must therefore be performed just as a in the case of unconstrained dynamical fields to decide whether some solution is stable or not.

\subsubsection{Non-local gravity and cosmology}

Armed with what we have learned in the previous chapters, we finally went on to construct generally-covariant non-local theories of gravity, massive or not. We saw two ways to proceed, the action-based one and the projector-based one, in order to guarantee the transversality of our equations. Having constructed a class of models, some simple phenomenological constraints have narrowed it down to two models, the $\xi$-M projector-based model (\ref{eq:xiM}) 
\beq 
G_{\mu\nu} - \frac{d-1}{2d} m^2 \( g_{\mu\nu} \ti{\bo}^{-1} R \)^{\rm T} = 8 \pi G T_{\mu\nu} \, ,
\eeq
and
the $\xi$-MM action-based model (\ref{eq:xiMM})
\beq
S = \ti{M}^2  \int \ed^D x\, \sqrt{-g} \[R - \frac{d-1}{4 d}\, m^2 R \ti{\bo}^{-2} R \]   \, ,  
\eeq
where $\ti{\bo} \equiv \bo - \xi R$. These are not theories of massive gravity, since the tensor modes are massless, although for the $\xi$-MM one could add a Weyl-squared term $W \ti{\bo}^{-2} W$ to make them massive without spoiling background cosmology. These are one-parameter extensions of the models proposed by Maggiore \cite{Maggiore2} and Maggiore - Mancarella \cite{MaggioreMancarella}, corresponding to the case $\xi = 0$. 

In the limit $\xi \to \infty$ one obtains GR with a cosmological constant $\La \sim m^2$, so the phenomenology of these models should lie between the $\xi = 0$ ones and $\La$CDM. We have confirmed this for the cosmological background through both a numerically analysis and analytic approximations. For $\xi > 0$ we found that both models admit de-Sitter solutions, although they are unstable in the projector-based case. Indeed, there the future universe ultimately leaves the de-Sitter phase to settle on a $w = -1/3$ phase.   

These theories share the same linearized limit and contain a scalar ghost. However, the latter is ultra-light and the divergence is expected to manifest itself only at cosmological time-scales of the order of the age of the universe. This has been confirmed by a recent study of the perturbations for $\xi = 0$ \cite{DFKKM}, i.e. the divergence is too slow to spoil the observational tests. In the $\xi = 0$ action-based model, the ghost dynamics are even bounded, which is explained by a big rip singularity in the future. It implies that at some point Hubble friction will dominate, thus diluting the perturbations, and it appears that this domination occurs already shortly after today. The $\xi = 0$ models have both been recently found to be consistent with the present data, and as privileged as $\La$CDM, through a full Boltzmann/MCMC analysis \cite{DFKMP}. This should therefore also hold for the $\xi > 0$ models since they lie somewhere in-between. Although considering one more parameter ($\xi$) for models that already work perfectly well can only lower their predictive power, we find interesting to have a parameter that continuously bridges to GR with $\La > 0$.

\appendix

\chapter{Bi-tensors} \label{sec:bitensors}

In this appendix we define the notion of bi-tensor, the mathematical structure behind generally covariant Green's functions, and discuss some properties that are going to be useful for our purposes.

\section{Definition}

Just as higher-rank tensors are constructed using the tensor products (in the sense of fibre bundle theory) of vectors and covectors, bi-tensors can be constructed through some other type of tensor product of ordinary tensors. In order to formalize this construction, it is convenient to first remind some properties of ordinary tensors and of the corresponding tensor product. 

\subsection{Tensors}

\subsubsection{Manifold \& scalars}

We start with a $D$-dimensional real differentiable manifold $\cal M$ with atlas $A_{\cal M}$, i.e. a set of pairs $\( U_i, f_i \)$ of open sets $U_i \subset {\cal M}$ and homeomorphisms
\bea
f_i & : & U_i \to \Rs^D \nn \\
 & & p \mapsto x_i^{\mu} \, , \hspace{1cm} \mu = 0, 1, \dots, d \, ,
\eea
such that the $U_i$ cover all of $\cal M$ and the transition functions from $\Rs^D$ to $\Rs^D$
\beq
f_{ij} \equiv f_i \circ f_j^{-1} : f_j\(U_i \cap U_j\) \to f_i\( U_i \cap U_j \) \, ,
\eeq
are smooth. Any continuous map $\ph : {\cal M} \to \Rs$ can then be represented by functions $\ph_i$ from $\Rs^D$ to $\Rs$ by pulling it back along some $f_i^{-1}$
\beq
\ph_i \equiv \ph \circ f_i^{-1} : f_i (U_i) \to \Rs \, .
\eeq
A scalar field is then defined as such a map for which all $\ph_i$ are smooth. Inverting we get $\ph = \ph_i \circ f_i$, so on $U_i \cap U_j$ we have
\beq
\ph_i \circ f_i = \ph_j \circ f_j \, , \hspace{1cm} \Rightarrow \hspace{1cm}  \ph_j = \ph_i \circ f_{ij} \, ,
\eeq
which is nothing but the transformation rule for a scalar function 
\beq
\ph_i(x_i) = \ph_j(x_j) \, ,
\eeq
under the coordinate transformation $x_i = f_{ij}(x_j)$. Since the $U_i$ cover $\cal M$ and the $f_i$ are homeomorphisms, we have that the $\ph_i$ functions fully determine $\ph$. Finally, we note that the set of scalar fields, denoted by $C^{\infty}({\cal M})$, forms an algebra whose addition and multiplication operations are the ordinary point-wise addition and multiplication in the target space $\Rs$.

\subsubsection{Tangent bundle}

We then consider the tangent bundle $T^1{\cal M}$. This is a $2D$-dimensional differentiable manifold along with a continuous surjective map $\pi : T^1{\cal M} \to {\cal M}$, such that $\pi^{-1}(p) \simeq \Rs^D$ for all $p \in {\cal M}$. In fibre bundle language, $T^1{\cal M}$ is the total space, $\cal M$ is the base and $\Rs^D$ is the fibre. This structure means that $T^1{\cal M}$ {\it locally} looks like ${\cal M} \times \Rs^D$, i.e. every point of $\cal M$ has a neighbourhood $U_i \subset {\cal M}$ such that $\pi^{-1}(U_i) \simeq U_i \times \Rs^D$. As a matter of fact, once $\pi$ is given, we restrict the atlas of $\cal M$ to the charts whose open set $U_i$ is small enough to satisfy this condition, i.e. to the sets which ``trivialize'' the fibre bundle. The atlas of the tangent bundle $A_{T^1{\cal M}}$ is then constructed out of $A_{\cal M}$ as follows. For every chart $\( U_i, f_i \) \in  A_{\cal M}$ we pick an open set $V_i \in T^1 {\cal M}$ and an homeomorphism
\bea
g_i & : & V_i \to \Rs^{2D} \nn \\
 & & q \mapsto \( x_i^{\mu}, k_i^{\nu} \) \, , 
\eea
such that 
\beq
\pi(V_i) = U_i \, , \hspace{1cm}  \( f_i \circ \pi \circ g_i^{-1}\)(x_i,k_i) = x_i \, , \hspace{1cm} \bigcup_i V_i = T^1{\cal M} \, ,
\eeq
i.e. $g_i$ is such that the function associated to the projection map is the trivial projection onto the base coordinates\footnote{The fact that we can cover $T^1{\cal M}$ with as many $V_i$ as there are $U_i$ is possible because we have demanded that $\pi^{-1}(U_i) \simeq U_i \times \Rs^D$.}. Moreover, the set of charts $\(V_i, g_i\)$ must be such that the corresponding transition functions 
\beq
g_{ij} \equiv g_i \circ g_j^{-1} : g_j\(V_i \cap V_j\) \to g_i\( V_i \cap V_j \)  \, ,
\eeq
read
\beq \label{eq:transfunc}
g_{ij}(x_j, k_j) = \( f_{ij}^{\mu}(x_j), \frac{\pa f_{ij}^{\nu}}{\pa x_j^{\ro}} (x_j) \, k_j^{\ro} \)  \, .
\eeq
A set of such pairs $\( V_i, g_i \)$ constitutes an atlas $A_{T^1{\cal M}}$ for $T^1{\cal M}$. The appearance of the transition functions of $\cal M$ in the transformation of the fibre coordinates in (\ref{eq:transfunc}) shows that the structure of $T^1{\cal M}$ is naturally induced by the one of $\cal M$. 

A vector field $X$ is a section of this bundle, i.e. a continuous map $X : {\cal M} \to T^1{\cal M}$ that is a right-inverse of the projection $\pi \circ X = {\rm id}_{\cal M}$. It can be expressed through local $\Rs^D \to \Rs^D$ functions, i.e. on $U_i$ we define its pullback $X_i \equiv X \circ g_i^{-1}$, which by the property $\pi \circ X = {\rm id}_{\cal M}$ has the form
\bea
X_i & : & f_i(U_i) \to g_i(V_i) \nn \\
 & & x_i^{\mu} \mapsto \( x_i^{\mu}, X_i^{\nu}(x_i) \) \, , \label{eq:tiXmap}
\eea
and the $X_i^{\mu}(x_i)$ are required to be smooth. As for the scalars, the full set of $X_i$ functions fully determines $X$. Since the projection map is trivial, the relevant information ultimately lies in the functions $X_i^{\mu}(x)$ that are what one usually refers to as ``the local components of the vector field'' on $U_i$, \footnote{The advantage of the section representation is that it is global and thus unique, while the $X_i^{\mu}(x)$ information is local and contains as many functions as the number of $U_i$ that are needed to cover $\cal M$.}. As in the case of scalar fields, we can invert $X = X_i \circ g_i$ and have that on $U_i \cap U_j$
\beq
X_i \circ g_i = X_j \circ g_j \, , \hspace{1cm} \Rightarrow \hspace{1cm}  X_j = X_i \circ g_{ij} \, ,
\eeq
which, given (\ref{eq:transfunc}), translates into the well-known rule
\beq \label{eq:Xtrans}
X_i^{\mu}(x_i) = \frac{\pa x_i^{\mu}}{\pa x_j^{\nu}}(x_j) \, X_j^{\nu}(x_j)  \, ,
\eeq
under the coordinate transformation $x_i = f_{ij}(x_j)$. The set of sections, denoted by $\Ga(T^1{\cal M})$, forms an $C^{\infty}({\cal M})$-vector space, whose addition and multiplication by a $\ph \in C^{\infty}({\cal M})$ operations are defined on each $U_i$ through the functions $X_i^{\mu}$, which then determine the resulting vector field\footnote{Indeed, we cannot define these operations using directly the maps $X$ and $Y$ because their target space is not a space of numbers.}. We have that if $X^{\mu}_i$, $Y_i^{\mu}$ and $\ph_i$ are the local functions associated to $X$, $Y$ and $\ph$, respectively, then the local functions of $X + Y$ and $\ph X$ are given by $X^{\mu}_i + Y^{\mu}_i$ and $\ph_i X^{\mu}_i$, \footnote{Note that scalar fields can also be expressed in this fibre bundle language as sections of $T^0 {\cal M}$. The base is still $\cal M$, the fibre is just $\Rs$, the transition maps are trivial $\xi_{ij}(x_i,k_j) = ( f_{ij}^{\mu}(x_i), k_j)$ and the scalar fields are sections which in local coordinates are given by functions $x_i^{\mu} \mapsto \( x_i^{\mu}, \ph_i(x_i) \)$. The addition and multiplication operations on $\Ga(T^0{\cal M})$ must then be defined through the local functions.}. 

At this point we can make contact with the alternative definition of a vector field which is as a derivation on $C^{\infty}({\cal M})$, i.e. an $\Rs$-linear operator $D_X : C^{\infty}({\cal M}) \to C^{\infty}({\cal M})$ obeying the Leibniz rule
\beq
D_X \(\al \ph + \be \ph'\) = \al D_X \ph + \be D_X \ph' \, , \hspace{1cm} D_X \(\ph \ph'\) = \( D_X \ph \) \ph' + \ph D_X \ph' \, ,
\eeq 
where $\al,\be$ are real constants. Indeed, these properties fully determine $D_X$: if $\ph_i$ denotes the local functions of $\ph$ then the local functions of $D_X \ph$ are $X_i^{\mu} \pa_{\mu} \ph_i$, for some functions $X_i^{\mu}$ which we can identify with the fibre components of a section (\ref{eq:tiXmap}). Indeed, the fact that $D_X \ph \in C^{\infty}({\cal M})$ implies 
\beq \label{eq:Xtrans2}
X_i^{\mu}(x_i) \frac{\pa}{\pa x_i^{\mu}} = X_j^{\nu}(x_j) \frac{\pa}{\pa x_j^{\nu}} \, ,
\eeq
which is precisely (\ref{eq:Xtrans}). It is a common abuse of terminology to call this derivation the ``vector field'', in which case the $\pa_{\mu}$ form a basis of vector fields. 

Finally, anticipating the generalization to tensors, we must look for yet another operator interpretation of vector fields. To that end we can define the cotangent bundle $T_1{\cal M}$ following the same steps as we did for $T^1{\cal M}$, only this time with coordinates $\( x_i^{\mu}, k_{i\nu} \)$ and with transition functions obeying
\beq 
g_{ij}(x_j, k_j) = \( f_{ij}^{\mu}(x_j),  \frac{\pa f_{ji}^{\ro}}{\pa x_i^{\nu}}  (f_{ij}(x_j)) \, k_{\ro} \) \, .
\eeq
A covector $\al$ is then a section of $T_1{\cal M}$, and has a natural action as a linear functional $\al: \Ga(T^1{\cal M}) \to C^{\infty}({\cal M})$. Indeed, its associated local functions $\al_i \equiv \al \circ \psi_i^{-1}$
\bea
\al_i & : & f_i(U_i) \to g_i(V_i) \nn \\
 & & x_i^{\mu} \mapsto \( x_i^{\mu}, \al_{i\nu}(x_i) \) \, ,
\eea
transform as
\beq \label{eq:altrans}
\al_{i\mu}(x_i) = \frac{\pa x_j^{\nu}}{\pa x_i^{\mu}}(x_i(x_j)) \, \al_{j\nu}(x_j) \, ,
\eeq
under the coordinate transformation $x_i = f_{ij}(x_j)$, and thus 
\beq
\ph_i(x_i) \equiv \al_{i\mu}(x_i) X_i^{\mu}(x_i) = \al_{j\mu}(x_j) X_j^{\mu}(x_j) \equiv \ph_j(x_j)
\eeq
transforms as the local function on $U_i$ of a scalar. This defines the interior product $X \cdot \al \in C^{\infty}({\cal M})$. Just as $\pa_{\mu}$ provides a basis for vector fields, because of its transformation properties, so does the differential $\ed x^{\mu}$ provide a basis for covectors
\beq
\ed x_i^{\mu} = \frac{\pa x_i^{\mu}}{\pa x_j^{\nu}} \, \ed x_j^{\nu} \, ,
\eeq
and we have the analogue of (\ref{eq:Xtrans2})
\beq 
\al_{i\mu}(x_i)\, \ed x_i^{\mu} = \al_{j\nu}(x_j)\, \ed x_j^{\nu} \, .
\eeq
Alternatively, the vectors can also be interpreted as linear functionals on $\Ga(T_1({\cal M}))$. It is this dual linear operator interpretation that generalizes straightforwardly to the case of higher-rank tensors.

\subsubsection{Tensor bundle}

Having defined $T^1{\cal M}$ and $T_1{\cal M}$ we can construct the tensor product  bundle
\beq
T^n_m {\cal M} \equiv T_1 {\cal M}\, \us{m \,\,{\rm times}}{\underbrace{\otimes \dots \otimes}} \,T_1{\cal M} \otimes T^1 {\cal M}\, \us{n \,\,{\rm times}}{\underbrace{\otimes \dots \otimes}} \,T^1{\cal M} \, .
\eeq
The $\otimes$ operation means that one takes the tensor product of the fibres at each point $p \in {\cal M}$, but keeps the same base manifold $\cal M$. The fibre coordinates will therefore take values in the vector space generated by $k^1_{\mu_1} \dots k^m_{\mu_m} k_1^{\nu_1} \dots k_n^{\nu_n}$, thus corresponding to $k_{\mu_1 \dots \mu_m}^{\nu_1 \dots \nu_n}$ coordinates. So $T^n_m {\cal M}$ is a $(D+D^{n+m})$-dimensional differentiable manifold with a projection map $\pi: T^n_m{\cal M} \to {\cal M}$ and fibre $\pi^{-1}(p) \simeq \Rs^{D^{n+m}}$. The set of charts $( g_i, V_i )$
\bea
g_i & : & V_i \to \Rs^{D+D^{n+m}} \nn \\
 & & q \mapsto \( x_i^{\mu}, k_{i\mu_1 \dots \mu_m}^{\,\,\nu_1 \dots \nu_n} \) \, ,  
\eea
is such that 
\beq
\pi(V_i) = U_i \, , \hspace{1cm}  \( f_i \circ \pi \circ g_i^{-1}\)(x_i,k_i) = x_i \, , \hspace{1cm} \bigcup_i V_i = T^n_m{\cal M} \, ,
\eeq
and the transition functions $g_{ij} \equiv g_i \circ g_j^{-1}$ are of the form
\beq \label{eq:transfunc2}
g_{ij}(x_j, k_j) = \( f_{ij}^{\mu}(x),  \frac{\pa f_{ji}^{\al_1}}{\pa x_i^{\mu_1}}  (f_{ij}(x_j))  \dots  \frac{\pa f_{ji}^{\al_m}}{\pa x_i^{\mu_m}}  (f_{ij}(x_j)) \, \frac{\pa f_{ij}^{\nu_1}}{\pa x_j^{\be_n}} (x_j) \dots  \frac{\pa f_{ij}^{\nu_n}}{\pa x_j^{\be_n}} (x_j)  \, k_{\al_1 \dots \al_m}^{\be_1 \dots \be_n} \) \, . 
\eeq
A tensor of rank $(n,m)$ is then a section of $T^n_m{\cal M}$, i.e. a map $T : {\cal M} \to T^n_m{\cal M}$ that is a right-inverse of the projection $\pi \circ T = {\rm id}_{\cal M}$. Thus, defining the local functions $T_i \equiv T \circ f_i^{-1}$ we have
\bea
T_i & : & f_i(U_i) \to g_i(V_i) \nn \\
 & & x_i^{\mu} \mapsto \(x_i^{\mu}, T_{i\mu_1 \dots \mu_m}^{\,\,\nu_1 \dots \nu_n}(x_i) \) \, ,
\eea
and the fibre components $T_{i\mu_1 \dots \mu_m}^{\,\,\nu_1 \dots \nu_n}(x_i)$, given (\ref{eq:transfunc2}), transform as
\beq
T_{i\mu_1 \dots \mu_m}^{\,\,\nu_1 \dots \nu_n}(x_i) = \frac{\pa x_j^{\al_1}}{\pa x_i^{\mu_1}}  (x_i(x_j))  \dots  \frac{\pa x_j^{\al_m}}{\pa x_i^{\mu_m}} (x_i(x_j))  \frac{\pa x_i^{\nu_1}}{\pa x_j^{\be_n}} (x_j) \dots  \frac{\pa x_i^{\nu_n}}{\pa x_j^{\be_n}} (x_j) \, T_{j\al_1 \dots \al_m}^{\,\,\be_1 \dots \be_n}(x_j) \, ,
\eeq
under the coordinate transformation $x_i = f_{ij}(x_j)$. As in the case of (co-)vectors, by a slight abuse of language, one usually calls $T_{i\mu_1 \dots \mu_m}^{\,\,\nu_1 \dots \nu_n}(x_i)$ the components of the tensor field. The addition and multiplication by a scalar operations are defined through the local functions just as in the case of vectors. We can now include the tensor product among the operations of interest, which is also defined through the local functions. If $T \in \Ga(T^n_m{\cal M})$ and $S \in \Ga(T^s_r{\cal M})$, then $T \otimes S \in \Ga(T^{n+s}_{m+r}{\cal M})$ is given by $(T \otimes S)_i \equiv (T \otimes S) \circ f_i^{-1}$
\beq
(T \otimes S)_i (x_i) = \(x_i^{\mu}, T_{i\mu_1 \dots \mu_m}^{\,\,\nu_1 \dots \nu_n}(x_i) S_{i\mu_{m+1} \dots \mu_{m+r}}^{\,\,\nu_{n+1} \dots \nu_{n+s}}(x_i) \) \, .
\eeq
Finally, using the $X = X^{\mu} \pa_{\mu}$ and $\al = \al_{\mu} \ed x^{\mu}$ interpretation of (co-)vectors, the ``basis of $\Ga(T^n_m{\cal M})$'' in this case is the tensor product 
\beq \label{eq:tensbasis}
\ed x^{\mu_1} \otimes \dots \otimes \ed x^{\mu_m} \otimes \pa_{\nu_1} \otimes \dots \otimes \pa_{\nu_n} \, ,
\eeq
where here $\otimes$ means ``multiplication and evaluation at the same point of $\cal M$'', so that
\beq
T_{i\mu_1 \dots \mu_m}^{\,\,\nu_1 \dots \nu_n}(x_i) \, \ed x_i^{\mu_1} \otimes \dots \otimes \ed x_i^{\mu_m} \otimes \frac{\pa}{\pa x_i^{\nu_1}} \otimes \dots \otimes \frac{\pa}{\pa x_i^{\nu_n}}  = T_{j\mu_1 \dots \mu_m}^{\,\,\nu_1 \dots \nu_n}(x_j) \, \ed x_j^{\mu_1} \otimes \dots \otimes \ed x_j^{\mu_m} \otimes \frac{\pa}{\pa x_j^{\nu_1}} \otimes \dots \otimes \frac{\pa}{\pa x_j^{\nu_n}} \, .
\eeq

\subsection{Bi-tensors}

\subsubsection{Bi-manifold \& bi-scalars}

We now wish to construct tensor-like fields that depend on two points of $\cal M$. We therefore begin by defining the Cartesian product ${\cal M}^2 \equiv {\cal M}_{\rm L} \times {\cal M}_{\rm R}$, where these are two copies of $\cal M$ that we will call the ``left'' and ``right'' ones. In the above product it is understood that ${\cal M}^2$ has the product topology and atlas $A_{{\cal M}_{\rm L}} \times A_{{\cal M}_{\rm R}}$, i.e. the one made of the pairs 
\beq
\( U_{i|j}, f_{i|j} \) \equiv \( U_i \times U_j, f_i \times f_j \) \, .
\eeq
Thus, a chart on ${\cal M}^2$ is a pair of open sets, one on ${\cal M}_{\rm L}$ and one on ${\cal M}_{\rm R}$, followed by a pair of functions that coordinatize each open set independently. The product topology gives 
\beq
U_{i|j} \cap U_{k|l} \equiv \( U_i \times U_j \) \cap \( U_k \times U_l \) = \( U_i \cap U_k \) \times \( U_j \cap U_l \) \, ,
\eeq
and the same for the union operation, and the transition functions decompose
\beq
f_{ik|jl} \equiv f_{i|k} \circ f_{j|l}^{-1} = \( f_i \circ f_k^{-1}\) \times \( f_j \circ f_l^{-1} \)\, ,
\eeq
so that these two manifolds do not ``see'' each other, i.e. one can perform coordinate transformations on each one of them independently. A bi-scalar field is a map $\ph : {\cal M}^2 \to \Rs$ such that the functions 
\bea
\ph_{i|j} \equiv \ph \circ f_{i|j}^{-1} & : & f_i(U_i) \times f_j(U_j) \to \Rs \nn \\
 & & \( x_i^{\mu}, y_j^{\nu} \) \mapsto \ph_{i|j}\( x_i, y_j \) \, ,
\eea 
are smooth in both arguments. Following the same steps as for the ordinary scalar field, its transformation under {\it independent} coordinate transformations $x_i = f_{ik}(x_k)$ and $y_j = f_{jl}(y_l)$ is thus
\beq
\ph_{i|j}(x_i,y_j) = \ph_{k|l}(x_k,y_l) \, .
\eeq

\subsubsection{Bi-tensor bundle}

We can now define the bi-tensor bundle $B^n_m|^s_r {\cal M}$ as follows. It is a differentiable fibre bundle of dimension $2D+D^{n+m+r+s}$, based on ${\cal M}^2$, with projection map $\pi_B : B^n_m|^s_r{\cal M} \to {\cal M}^2$ and fibre $\pi_B^{-1}(p_{\rm L}, p_{\rm R}) \simeq \Rs^{D^{n+m+r+s}}$. Its atlas $A_{B^n_m|^s_r{\cal M}}$ is constructed as follows. For every pair $\( U_{i|j}, f_{i|j} \) \in A_{{\cal M}^2}$, we pick an open set $V_{i|j} \subset B^n_m|^s_r {\cal M}$ and a homeomorphism
\bea
g_{i|j} & : & V_{i|j} \to \Rs^{2D+D^{n+m+r+s}} \nn \\
 & & q \mapsto \( x_i^{\mu}, y_j^{\nu}, k_{i\mu_1 \dots \mu_m}^{\,\,\nu_1 \dots \nu_n}|_{j\ro_1 \dots \ro_r}^{\,\,\,\si_1 \dots \si_s} \) \, ,  
\eea
such that 
\beq
\pi_B (V_{i|j}) = U_{i|j} \, , \hspace{1cm}  \( f_{i|j} \circ \pi_B \circ g_{i|j}^{-1}\)\(x_i,y_j,k_{i|j}\) = (x_i,y_j) \, , \hspace{1cm} \bigcup_{i,j} V_{i|j} = B^n_m|^s_r {\cal M} \, ,
\eeq
and the transition functions $g_{ik|jl} \equiv g_{i|j} \circ g_{k|l}^{-1}$ are of the form
\bea
g_{ik|jl}\( x_k, y_l, k_{k|l} \) & = & \( f_{ik}^{\mu}(x_k), f_{jl}^{\nu}(y_l),  \frac{\pa f_{ki}^{\al_1}}{\pa x_i^{\mu_1}}  (f_{ik}(x_k))  \dots  \frac{\pa f_{ki}^{\al_m}}{\pa x_i^{\mu_m}}  (f_{ik}(x_k))   \right. \nn \\
 & & \left. \frac{\pa f_{ik}^{\nu_1}}{\pa x_k^{\be_n}} (x_k) \dots  \frac{\pa f_{ik}^{\nu_n}}{\pa x_k^{\be_n}} (x_k) \, \frac{\pa f_{lj}^{\ga_1}}{\pa y_j^{\ro_1}}  (f_{jl}(y_l))  \dots  \frac{\pa f_{lj}^{\ga_m}}{\pa y_j^{\ro_m}}  (f_{jl}(y_l)) \right. \nn \\
 & & \left. \frac{\pa f_{jl}^{\si_1}}{\pa y_l^{\de_n}} (y_l) \dots  \frac{\pa f_{jl}^{\si_n}}{\pa y_l^{\de_n}} (y_l)  \, k_{k\al_1 \dots \al_m}^{\,\,\,\be_1 \dots \be_n}|_{l\ga_1 \dots \ga_r}^{\,\,\de_1 \dots \de_s} \) \, . \nn \label{eq:transfunc3} \\
\eea
Note that we have used a column to distinguish between the two types of indices, i.e. the ``left'' ones mixing with Jacobians evaluated at the left point $x$, and the ``right'' ones mixing with Jacobians evaluated at the right point $y$. A bi-tensor $G$ would then be a section of $B^n_m|^s_r{\cal M}$, i.e. a continuous map $G : {\cal M}^2 \to B^n_m|^s_r{\cal M}$ that is a right-inverse for the projection map $\pi_B \circ G = {\rm id}_{{\cal M}^2}$. Thus, defining the functions $G_{i|j} \equiv G \circ f_{i|j}^{-1}$, in local coordinates
\bea
G_{i|j} & : & f_{i|j}(U_{i|j}) \to g_{i|j}(V_{i|j}) \nn \\
 & & \( x_i^{\mu}, y_j^{\nu} \) \mapsto \( x_i^{\mu}, y_j^{\nu}, G_{i\mu_1 \dots \mu_m}^{\,\,\nu_1 \dots \nu_n}|_{j\ro_1 \dots \ro_r}^{\,\,\,\si_1 \dots \si_s}(x_i,y_j) \) \, ,
\eea
the local components $G_{i\mu_1 \dots \mu_m}^{\,\,\nu_1 \dots \nu_n}|_{j\ro_1 \dots \ro_r}^{\,\,\,\si_1 \dots \si_s}(x_i,y_j)$, given (\ref{eq:transfunc3}), transform as
\bea
G_{i\mu_1\dots \mu_m}^{\,\,\nu_1 \dots \nu_r}|_{j\ro_1 \dots \ro_r}^{\,\,\,\si_1\dots \si_s} (x_i, y_j) & = &  \frac{\pa x_k^{\al_1}}{\pa x_i^{\mu_1}}(x_i(x_k)) \dots \frac{\pa x_k^{\al_m}}{\pa x_i^{\mu_m}}(x_i(x_k)) \frac{\pa x_i^{\nu_1}}{\pa x_k^{\be_1}}(x_k) \dots \frac{\pa x_i^{\nu_n}}{\pa x_k^{\be_n}}(x_k) \nn \\
 & & \frac{\pa y_l^{\ga_1}}{\pa y_j^{\ro_1}}(y_j(y_l)) \dots \frac{\pa y_l^{\ga_r}}{\pa y_j^{\ro_r}}(y_j(y_l)) \frac{\pa y_j^{\si_1}}{\pa y_l^{\de_1}}(y_l) \dots \frac{\pa y_j^{\si_s}}{\pa y_l^{\de_s}}(y) \nn \\
 & &  \times G_{k\al_1\dots \al_m}^{\,\,\,\be_1 \dots \be_r}|_{l\ga_1 \dots \ga_r}^{\,\,\,\de_1\dots \de_s}(x_k,y_l) \, ,  \label{eq:bitenstrans}
\eea
under the independent coordinate transformations $x_i = f_{ik}(x_k)$ and $y_j = f_{jl}(y_l)$. In order to express such an object in the notation (\ref{eq:tensbasis}) we need to define a new kind of product. We thus use the notation
\bea
& & G_{\mu_1\dots \mu_m}^{\nu_1 \dots \nu_r}|_{\ro_1 \dots \ro_r}^{\si_1\dots \si_s} (x, y) \, \( \ed x^{\mu_1} \otimes \dots \otimes \ed x^{\mu_m} \otimes \frac{\pa}{\pa x^{\nu_1}} \otimes \dots \otimes \frac{\pa}{\pa x^{\nu_n}} \) \nn \\
 & & \hspace{3cm}  \otimes_B \( \ed y^{\ro_1} \otimes \dots \otimes \ed y^{\ro_m} \otimes \frac{\pa}{\pa y^{\si_1}} \otimes \dots \otimes \frac{\pa}{\pa y^{\si_s}} \) \, ,
\eea
and dub $\otimes_B$ the ``bi-tensor'' product, which means that the tensors on each side are evaluated on independent points of $\cal M$. This notation is again consistent with the transformation rule (\ref{eq:bitenstrans}) given the way the basis transforms. It is then straightforward to generalize the concept to bi-tensor densities and also to ``tri-tensors'', ``quadri-tensors'' etc, by taking more and more bi-tensor products.

\subsection{Bi-tensor calculus}

\subsubsection{Differentiation}

Since a bi-tensor basically ``lives'' on two points of the manifold, and their corresponding tangent tensor spaces, it can be covariantly differentiated at each point separately. Indeed, the transformation (\ref{eq:bitenstrans}) implies that one can apply covariant derivatives at each point separately, and with respect to the corresponding indices only, because $x$ and $y$ are independent. One must simply let the notation reflect the choice of point, so we will use $\na_{\rm L}$ and $\na_{\rm R}$ for the operators on bi-tensors, while we will use $\na_{\mu}|$ and $\na|_{\mu}$ for their representation on the bi-tensor components. For example, given $G \in \Ga(B^1_0|^1_1{\cal M})$,
\beq
\na_{\mu}| \, G^{\nu}|^{\ro}_{\si}(x,y) \equiv \frac{\pa}{\pa x^{\mu}}\, G^{\nu}|^{\ro}_{\si}(x,y) + \Ga^{\nu}_{\,\,\,\al\mu}(x) \, G^{\al}|^{\ro}_{\si}(x,y)  \, , 
\eeq
are the local components of $\na_{\rm L} G \in \Ga(B^1_1|^1_1{\cal M})$, while
\beq
\na|_{\mu} \, G^{\nu}|^{\ro}_{\si}(x,y) \equiv \frac{\pa}{\pa y^{\mu}}\, G^{\nu}|^{\ro}_{\si}(x,y) + \Ga^{\ro}_{\,\,\,\al\mu}(y) \, G^{\nu}|^{\al}_{\si}(x,y) - G^{\nu}|^{\ro}_{\al}(x,y) \Ga^{\al}_{\,\,\,\si\mu}(y)  \, ,
\eeq 
are the local components of $\na_{\rm R} G \in \Ga(B^1_0|^1_2 {\cal M})$. Pay attention to the various dependencies and index contractions. With this additional information the commutator of covariant derivatives generalizes accordingly. We have for instance
\bea
\[ \na_{\mu}|, \na|_{\nu} \] G^{\ro}|^{\si}_{\ta}(x,y) & = & 0  \, , \\
\[ \na_{\mu}|, \na_{\nu}| \] G^{\ro}|^{\si}_{\ta}(x,y) & = & R^{\ro}_{\,\,\,\al\mu\nu}(x) \, G^{\al}|^{\si}_{\ta}(x,y) \, , \\
\[ \na|_{\mu}, \na|_{\nu} \] G^{\ro}|^{\si}_{\ta}(x,y) & = & R^{\si}_{\,\,\,\al\mu\nu}(y) \, G^{\ro}|^{\al}_{\ta}(x,y) - G^{\ro}|^{\si}_{\al}(x,y) R^{\al}_{\,\,\,\ta\mu\nu}(y) \,  \, .
\eea

\subsubsection{Integration on ${\cal M}$}

Remember that integration is defined on manifolds by splitting the integral through a partition of unity subordinate to the open cover $U_i$ and evaluating the integral on each $U_i$ using the local functions. More precisely, let us denote by $I$ the set of indices indexing the open sets $U_i$ of $A_{\cal M}$. We can then pick a locally finite covering $I' \subset I$, i.e. a subset $\left\{ U_i \right\}_{i \in I'}$ that still covers $\cal M$ but such that for every $p \in {\cal M}$ there exists only a finite number of $U_i$ for which $p \in U_i$. Smooth manifolds which admit such locally finite refinements are called ``paracompact''. Then, a partition of unity subordinate to $\{U_i\}_{i\in I'}$ is the attribution of a scalar $\ro_i$ to each $U_i$ with $i \in I'$ such that
\begin{itemize}
\item
${\rm supp}(\ro_i) \subset U_i$,

\item
$\sum_{i \in I'} \ro_i = 1$.

\end{itemize}
The integral of a scalar field $\ph$ over $\cal M$ is then defined as follows. One first needs to define a measure, i.e. a $D$-form $\om$ such that the local density functions
\beq
\om_{i\mu_1 \dots \mu_D}(x_i) = \om_i(x_i) \vep_{\mu_1 \dots \mu_D} \, , 
\eeq
have positive definite sign $\om_i(x_i) > 0$. Given a metric tensor $g$, the physically sensible choice is $\om_i(x_i) = \sqrt{- g_i(x_i)}$ where $g_i(x_i)$ is the determinant of $g_{i\mu\nu}(x_i)$. We then have that the integral of $\ph$ is given by
\beq
\int_{\cal M} \om\, \ph \equiv \sum_{i \in I'} \int \ed^D x_i\, \ro_i(x_i)\om_i(x_i) \ph_i(x_i) \, .
\eeq
where $\ro_i, \om_i$ and $\ph_i$ are the local functions of $\ro, \om$ and $\ph$ on $U_i$, respectively. The sum in the right-hand side is well defined because for each $i \in I'$ only but a finite number of elements are non-zero. 

The generalization to bi-tensors is straightforward. It relies on the fact that if $\ro^{\rm L}_i$ and $\ro^{\rm R}_i$ form partitions of unity of ${\cal M}_{\rm L}$ and ${\cal M}_{\rm R}$ subordinate to their respective atlases, then $\ro^{\rm L}_i \ro^{\rm R}_j$ forms a partition of unity of ${\cal M}^2$ subordinate to $\left\{ U_{i|j} \right\}_{(i,j) \in I'^2}$. As for differentiation, one can then define the integration on ${\cal M}_{\rm L}$ and ${\cal M}_{\rm R}$ independently. For example, given $G \in \Ga(B^n_m|^0_0{\cal M})$, which is a scalar on ${\cal M}_{\rm R}$, one can define
\beq
\int_{{\cal M}_{\rm R}} \om\, G \, ,
\eeq
by specifying the local functions on $U_i$
\beq
\( \int_{{\cal M}_{\rm R}} \om\, G\){}_{i\mu_1 \dots \mu_m}^{\,\,\nu_1 \dots \nu_n}(x_i) \equiv \sum_{j \in I'} \int \ed^D y_j\, \ro_j^{\rm R}(y_j) \, \om(y_j)\, G_{i\mu_1 \dots \mu_m}^{\,\,\nu_1 \dots \nu_n}|_j(x_i,y_j) \, . 
\eeq
Given the independence of the two space-time points, the above object is clearly an element of $\Ga(T^n_m {\cal M}_{\rm L})$. As is usual in the literature, we will use a slightly less rigorous notation to describe such integrals, i.e. one that does not care about how the integral is partitioned or about the fact that usually several coordinate charts are needed. In this case for instance we can write
\beq
\( \int_{{\cal M}_{\rm R}} \om\, G\){}_{\mu_1 \dots \mu_m}^{\nu_1 \dots \nu_n}(x) \equiv \int_{\cal M} \ed^D y\, \om(y)\, G_{\mu_1 \dots \mu_m}^{\nu_1 \dots \nu_n}|(x,y) \, , 
\eeq
so that one can see with respect to which manifold we are integrating. Finally, we define the following notations. For $T \in \Ga(T^n_m{\cal M})$ and $T' \in \Ga(T^m_n{\cal M})$,
\bea
\(G \cdot_{\om} T \)_{\mu_1 \dots \mu_m}^{\nu_1 \dots \nu_n}(x) & = & \int_{\cal M} \ed^D y\, \om(y)\, G_{\mu_1 \dots \mu_m}^{\nu_1 \dots \nu_n}|_{\ro_1 \dots \ro_n}^{\si_1 \dots \si_m}(x,y)\, T^{\ro_1 \dots \ro_n}_{\si_1 \dots \si_m}(y) \, , \\
\( T' \cdot_{\om} G \)_{\nu_1 \dots \nu_n}^{\mu_1 \dots \mu_m}(y) & = & \int_{\cal M} \ed^D x\, \om(x)\, T'^{\si_1 \dots \si_m}_{\ro_1 \dots \ro_n}(x)\, G_{\si_1 \dots \si_m}^{\ro_1 \dots \ro_n}|_{\nu_1 \dots \nu_n}^{\mu_1 \dots \mu_m}(x,y) \, ,
\eea
are also elements of $\Ga(T^n_m{\cal M})$ and $\Ga(T^m_n{\cal M})$, respectively. We thus have that, for any measure $\om$, the elements of $\Ga(B^n_m|^m_n{\cal M})$ can be thought of as left-$\Rs$-linear endomorphisms of $\Ga(T^n_m{\cal M})$ and right-$\Rs$-linear endomorphisms of $\Ga(T^m_n{\cal M})$. Finally, since in the physically relevant cases $\om(x) = \sqrt{-g(x)}$, a dot without argument means $\cdot_{\sqrt{-g}}$.

\section{Bi-tensor distributions}

The notion of bi-tensor combined with the notion distribution, ultimately allows to define the notion of functional analysis on manifolds. The most interesting cases for us are the Dirac delta bi-tensor and the Green's bi-tensor. Disclaimer: here we will only focus on the aspects of the generalization of these notions to curved space-time. We will not concern ourselves with the functional analysis side of the field, i.e. we will not care about domains, continuity and convergence issues that are nevertheless crucial aspects of the theory of distributions.

\subsection{The Dirac delta bi-tensor}

The Dirac delta bi-tensor is defined, as the ordinary Dirac delta, by its distributional properties. The ${n \choose m}$-Dirac delta bi-tensor associated to the measure $\om$ is the bi-tensor $\De \in \Ga(B^n_m|^m_n {\cal M})$ satisfying
\beq
\De \cdot_{\om} T = T \, , \hspace{1cm} T' \cdot_{\om} \De = T' \, ,
\eeq 
for all $T \in \Ga\( T^n_m {\cal M} \)$ and $T' \in \Ga\( T^m_n {\cal M} \)$. This uniquely determines its associated local functions, which are of course going to be related to the Dirac delta function. The latter transforms as a scalar density of weight $-1$ under a diffeomorphism $x' = f(x)$. Indeed,
\beq
1 \equiv \int \ed^D x' \, \de^{(D)}(x') = \int \ed^D x \, \det \[ \frac{\pa f}{\pa x}(x) \] \, \de^{(D)}(f(x)) =  \int \ed^D x \, \de^{(D)}(x) \, ,
\eeq
so
\beq
\de^{(D)}(f(x)) = \det \[ \frac{\pa f}{\pa x}(x) \]^{-1} \de^{(D)}(x)  \, .
\eeq
Thus the combination $\de^{(D)}(x)/ \om(x)$ is a scalar. Repeating with the shifted diffeomorphism $f(x) \to f(x) - f(y)$, we get
\beq
\de^{(D)}(f(x)-f(y)) = \det \[ \frac{\pa f}{\pa x}(x) \]^{-1} \de^{(D)}(x-y)  \, ,
\eeq
so that
\beq
\frac{\de^{(D)}(x-y)}{\om(x)} = \frac{\de^{(D)}(x-y)}{\om(y)} = \frac{\de^{(D)}(x-y)}{\sqrt{\om(x)}\sqrt{\om(y)}} \, ,
\eeq
are all the same scalar {\it when $x$ and $y$ are coordinates of the same chart}, and thus transform together under the same transition functions. We can now make the link with the Dirac delta bi-tensor. To fully determine the latter it suffices to determine its local functions $\De_{i|j} \equiv \De \circ f_{i|j}^{-1}$ on the open sets $U_{i|j}$. We then have that 
\beq
\De_{i\mu_1 \dots \mu_m}^{\,\,\nu_1 \dots \nu_n}|^{\,\,\,\si_1 \dots \si_m}_{j\ro_1 \dots \ro_n} (x_i,y_j) = 0 \, , \hspace{1cm} U_i \cap U_j = \emptyset \in {\cal M} \, ,
\eeq
while, if $U_i \cap U_j$ is non-empty in $\cal M$,
\beq
\De_{i\mu_1 \dots \mu_m}^{\,\,\nu_1 \dots \nu_n}|^{\,\,\,\si_1 \dots \si_m}_{j\ro_1 \dots \ro_n} (x_i,y_j) = \frac{\pa f_{ji}^{\si_1}}{\pa x_i^{\mu_1}}(f_{ij}(y_j)) \dots \frac{\pa f_{ji}^{\si_m}}{\pa x_i^{\mu_m}}(f_{ij}(y_j)) \frac{\pa f_{ij}^{\nu_1}}{\pa y_j^{\ro_1}}(y_j) \dots \frac{\pa f_{ij}^{\nu_n}}{\pa y_j^{\ro_n}}(y_j) \, \frac{\de^{(D)}(x_i- f_{ij}(y_j))}{\om_i(x_i)} \, .
\eeq
The latter is obtained by considering the case $i = j$
\beq
\De_{i\mu_1 \dots \mu_m}^{\,\,\nu_1 \dots \nu_n}|^{\,\,\,\si_1 \dots \si_m}_{i\ro_1 \dots \ro_n} (x_i,y_i) = \de_{\mu_1}^{\si_1} \dots \de_{\mu_m}^{\si_m} \de_{\ro_1}^{\nu_1} \dots \de_{\ro_n}^{\nu_n} \, \frac{\de^{(D)}(x_i- y_i)}{\om_i(x_i)} \, ,
\eeq
and transforming the right coordinate $y_i$ to $y_j$ using the transition function $f_{ij}$. An important property for what follows is the one involving the left and right-differentiations
\beq \label{eq:derDe}
\na_{\rm L} \De = - \na_{\rm R} \De \, , \hspace{1cm} \De \cdot_{\om} \na T  = - (\na T) \cdot_{\om} \De \, ,
\eeq
which is proved by convolution with test tensors and integration by parts.

\subsection{Bi-tensor Green's functions} \label{sec:Greensfunc}

Let $L[\na]$ denote a covariant differential operator acting on $\Ga(T^n_m {\cal M})$, i.e. the space of ${n \choose m}$-tensors. In terms of local components we thus have\footnote{The bi-tensor notation here might appear misleading since $L$ is made of differential operators acting on a single space-time point, but since it is an endomorphism on $\Ga(T^n_m {\cal M})$, we can express it as the convolution with a bi-tensor indeed. We just need to rewrite
\beq
\( L T \)_{\mu_1 \dots \mu_m}^{\nu_1 \dots \nu_n}(x) = \int \ed^D y \, \sqrt{-g(y)} \, \De_{\mu_1 \dots \mu_m}^{\nu_1 \dots \nu_n}|_{\nu'_1 \dots \nu'_n}^{\mu'_1 \dots \mu'_m}(x,y) L_{\mu'_1 \dots \mu'_m}^{\nu'_1 \dots \nu'_n}|_{\ro_1 \dots \ro_n}^{\si_1 \dots \si_m} T^{\ro_1 \dots \ro_n}_{\si_1 \dots \si_m} (y) \, ,
\eeq
and then integrate by parts the covariant derivatives in $L$ so that they act on $\De$. The boundary terms drop because of the Dirac delta in $\De$ and the result is the convolution of $T$ with a bi-tensor. }
\beq
\( L T \)_{\mu_1 \dots \mu_m}^{\nu_1 \dots \nu_n} \equiv L_{\mu_1 \dots \mu_m}^{\nu_1 \dots \nu_n}|_{\ro_1 \dots \ro_n}^{\si_1 \dots \si_m} T^{\ro_1 \dots \ro_n}_{\si_1 \dots \si_m} \, .
\eeq
Note that because of the derivatives the kernel ${\rm Ker}[L]$ is non-zero, i.e. there exists $T$ such that $L T = 0$. There are therefore, roughly speaking, as many inverses of $L$ as there are elements in ${\rm Ker}[L]$. A Green's function for $L$ is a bi-tensor $G \in \Ga(B^n_m|^m_n {\cal M})$ such that its local functions satisfy
\beq \label{eq:Gdef}
L_{\mu_1 \dots \mu_m}^{\nu_1 \dots \nu_n}|_{\ro_1 \dots \ro_n}^{\si_1 \dots \si_m}[\na_{\rm L}](x)\, G_{\si_1\dots \si_m}^{\ro_1 \dots \ro_n}|_{\nu'_1\dots \nu'_n}^{\mu'_1 \dots \mu'_m}(x,y) = \De_{\mu_1 \dots \mu_n}^{\nu_1 \dots \nu_n}|_{\nu'_1 \dots \nu'_n}^{\mu'_1 \dots \mu'_m}(x,y)  \, ,
\eeq
where here it is the Dirac delta bi-tensor associated with $\sqrt{-g}$ that is being used. Given such a bi-tensor $G$, we have that the operator
\beq
L_G^{-1} T \equiv G \cdot T  \, , 
\eeq
is a {\it right}-inverse of $L$, i.e.
\beq
L L^{-1}_G = {\rm id}_{\Ga(T^n_m{\cal M})} \, .
\eeq
In this thesis, we will only focus on right-inverses that are $\Rs$-linear operators
\beq
L^{-1} \( \al T + \al' T' \) = \al L^{-1} T + \al' L^{-1} T' \, , \hspace{1cm} \al, \al' \in \Rs, {\rm constant}
\eeq
and which can therefore be expressed as the convolution with a Green's bi-tensor\footnote{These must be contrasted with the more general case where the right-inverse is given by an affine operator
\beq 
L^{-1}_{h,G}(T) \equiv h + L_G^{-1} T \, ,
\eeq
with $h \in {\rm Ker}[L]$ a homogeneous solution of $L$ that is independent of $T$.}. On flat space-time we have that the bi-tensor structure of $G$ simplifies considerably. For the local functions corresponding to the same charts on ${\cal M}_{\rm L}$ and ${\cal M}_{\rm R}$, i.e. when $x$ and $y$ are in the same coordinate chart, the converse property (\ref{eq:convGdef}) along with Poincar\'e covariance imply that all Green's bi-tensors can be expressed in terms of a Green's function
\beq
G_{\mu_1 \dots \mu_m}^{\nu_1 \dots \nu_n}|_{\nu'_1 \dots \nu'_n}^{\mu'_1 \dots \mu'_m}(x,y) = \de_{\mu_1}^{\mu'_1} \dots \de_{\mu_m}^{\mu'_m} \de_{\nu'_1}^{\nu_1} \dots \de_{\nu'_n}^{\nu_n}  G(x-y)  \, ,
\eeq
where
\beq
(L G)(x) = \de^{(D)}(x) \, .
\eeq
We will use the ``r'' subscript when referring to retarded Green's functions, i.e. those that obey
\beq \label{eq:retG}
G_{{\rm r}, \dots}^{\,\,\,\dots}|_{\dots}^{\dots}(x,y) = 0 \, , \hspace{1cm} \mbox{unless $y$ is in the past light-cone of $x$} \, .
\eeq 
On flat-space time this condition uniquely determines $G$ because it totally determines the initial conditions
\beq
\lim_{x^0 \to - \infty} \pa_{x^0}^n G(x) = 0 \, , 
\eeq
where $n$ goes from $0$ to the degree of $L$. For instance, the retarded Green's function of $L = \bo - m^2$ reads
\beq
G_{\rm r}(x) \equiv \lim_{\ep \to 0^+} \int \frac{\ed^D k}{(2\pi)^D} \, \frac{\exp \[ i\et_{\mu\nu} k^{\mu} x^{\nu} \]}{\( k^0 + i \ep \)^2 - \vec{k}^2 - m^2} \, ,
\eeq
and in $D = 4$ takes the simple form
\bea  \label{eq:Gr4DMink}
G_{\rm r}(x) & = & -\frac{1}{2\pi}\, \te(x^0)  \[ \de(|x|^2) - \te(|x|^2) \frac{m J_1 \( m |x| \)}{2|x|} \] \nn \\
 & = & -\frac{1}{4\pi} \[ \frac{\de(x^0 - |\vec{x}|)}{|\vec{x}|} - \te(x^0)\, \te(|x|^2)\frac{m J_1 \( m |x| \)}{|x|} \] \, , 
\eea
where
\beq
|x| \equiv \sqrt{-\et_{\mu\nu} x^{\mu} x^{\nu}} \, , \hspace{1cm} |\vec{x}| \equiv \sqrt{\de_{ij} x^i x^j} \, ,
\eeq
and $J_1$ is a Bessel function of the first kind. We see that $G_{\rm r}(x-y)$, seen as a function of $y$, has a singular part which is supported only {\it on} the past light-cone of $x$ and a non-singular part which is supported on the {\it inside} of the cone. The latter vanishes in the $m \to 0$ limit, consistent with the fact that the information then propagates only at the speed of light and its trajectory is thus stuck on the cone. Finally, note that the domain of definition of $L_{\rm r}^{-1}$ are the tensors that vanish sufficiently fast at past infinity for $m \neq 0$ and past null infinity for $m = 0$. 

The generalization to curved space-time presents the following subtleties. First of all, the retarded Green's bi-tensor of $\bo - m^2$ is still supported inside the past light-cone, it is just that the latter is now non-trivial. Indeed, there might be more than one geodesic linking a given pair of points, the most striking example being the gravitational lensing effect. Second, one needs to impose global hyperbolicity on the pair $\( {\cal M}, g \)$ in order to have a causal space-time with a past that extends to infinity, and in which case the past light-cone would also extend to the infinite past. In that case, the domain of definition of $L_{\rm r}^{-1}$ are the tensors that vanish sufficiently fast at past infinity. More precisely, since $\bo - m^2$ is second-order, taking $t$ to denote the global time coordinate (Geroch's theorem), we need
\beq \label{eq:pastlimitT}
\lim_{t \to - \infty} T = 0 \, , \hspace{1cm} \lim_{t \to -\infty} \na_N T = 0 \, ,
\eeq
for any past-pointing time-like $N$ (light-like for $ m =0$). Since in practical calculations one may have other differential operators acting on $T$ before $L_{\rm r}^{-1}$, imposing the above condition will not suffice in general, so we will need to be more conservative. If ${\cal C}_x$ denotes the interior of the past light-cone of $x$, then we will demand that $\overline{{\rm supp}(T) \cap {\cal C}_x}$ is compact for all $x$ and will refer to such tensors as tensors with ``finite past''.

\section{Green's bi-tensor properties}

\subsection{Converse of (\ref{eq:Gdef})} \label{sec:convGdef}

Here we show that (\ref{eq:Gdef}) holds also when one acts on the point $y$ instead of $x$
\beq \label{eq:convGdef}
L_{\ro_1 \dots \ro_n}^{\si_1 \dots \si_m}|_{\mu_1 \dots \mu_m}^{\nu_1 \dots \nu_n}[\na_{\rm R}](y)\, G_{\nu'_1\dots \nu'_n}^{\mu'_1 \dots \mu'_m}|_{\si_1\dots \si_m}^{\ro_1 \dots \ro_n}(x,y) = \De_{\nu'_1 \dots \nu'_n}^{\mu'_1 \dots \mu'_m}|_{\mu_1 \dots \mu_n}^{\nu_1 \dots \nu_n}(x,y)  \, .
\eeq
Indeed, acting with $L[\na_{\rm R}](y)$ on (\ref{eq:Gdef}) and using (\ref{eq:derDe}) we get
\bea
L_{\mu'_1 \dots \mu'_m}^{\nu'_1 \dots \nu'_n}|_{\ka_1 \dots \ka_n}^{\la_1 \dots \la_m}[\na_{\rm R}](y)\,  L_{\mu_1 \dots \mu_m}^{\nu_1 \dots \nu_n}|_{\ro_1 \dots \ro_n}^{\si_1 \dots \si_m}[\na_{\rm L}](x)\, G_{\si_1\dots \si_m}^{\ro_1 \dots \ro_n}|_{\nu'_1\dots \nu'_n}^{\mu'_1 \dots \mu'_m}(x,y) \nn \\
 = L_{\mu'_1 \dots \mu'_m}^{\nu'_1 \dots \nu'_n}|_{\ka_1 \dots \ka_n}^{\la_1 \dots \la_m}[\na_{\rm R}](y) \De_{\mu_1 \dots \mu_n}^{\nu_1 \dots \nu_n}|_{\nu'_1 \dots \nu'_n}^{\mu'_1 \dots \mu'_m}(x,y) \nn \\
 = L_{\mu'_1 \dots \mu'_m}^{\nu'_1 \dots \nu'_n}|_{\ka_1 \dots \ka_n}^{\la_1 \dots \la_m}[ -\na_{\rm L}](x) \De_{\mu_1 \dots \mu_n}^{\nu_1 \dots \nu_n}|_{\nu'_1 \dots \nu'_n}^{\mu'_1 \dots \mu'_m}(x,y) \, . \label{eq:Ly1}
\eea
Thus, the convolution with a test tensor on ${\cal M}_{\rm L}$, using $\[ \na_{\rm L}, \na_{\rm R} \] = 0$, gives
\bea
\int \ed^D x\, \sqrt{-g(x)} \,  L_{\mu_1 \dots \mu_m}^{\nu_1 \dots \nu_n}|_{\ro_1 \dots \ro_n}^{\si_1 \dots \si_m}[\na_{\rm L}](x) \[ L_{\mu'_1 \dots \mu'_m}^{\nu'_1 \dots \nu'_n}|_{\ka_1 \dots \ka_n}^{\la_1 \dots \la_m}[\na_{\rm R}](y)\,  G_{\si_1\dots \si_m}^{\ro_1 \dots \ro_n}|_{\nu'_1\dots \nu'_n}^{\mu'_1 \dots \mu'_m}(x,y) \,T^{\mu_1 \dots \mu_n}_{\nu_1 \dots \nu_n}(x) \] \nn \\
 \os{(\ref{eq:Ly1})}{=} \int \ed^D x\, \sqrt{-g(x)} \, L_{\mu'_1 \dots \mu'_m}^{\nu'_1 \dots \nu'_n}|_{\ka_1 \dots \ka_n}^{\la_1 \dots \la_m}[ -\na_{\rm L}](x) \De_{\mu_1 \dots \mu_n}^{\nu_1 \dots \nu_n}|_{\nu'_1 \dots \nu'_n}^{\mu'_1 \dots \mu'_m}(x,y) \, T^{\mu_1 \dots \mu_n}_{\nu_1 \dots \nu_n}(x) \nn \\
 \os{\rm i.b.p.}{=} \int \ed^D x\, \sqrt{-g(x)} \, \De_{\mu_1 \dots \mu_n}^{\nu_1 \dots \nu_n}|_{\nu'_1 \dots \nu'_n}^{\mu'_1 \dots \mu'_m}(x,y) L_{\mu'_1 \dots \mu'_m}^{\nu'_1 \dots \nu'_n}|_{\ka_1 \dots \ka_n}^{\la_1 \dots \la_m}[\na_{\rm L}](x) \, T^{\mu_1 \dots \mu_n}_{\nu_1 \dots \nu_n}(x) \nn \\
 = \( LT \)_{\ka_1 \dots \ka_n}^{\la_1 \dots \la_m}(y) \, , \nn \\
\eea
where in the second step we have integrated by parts and the boundary terms have dropped because of the Dirac delta. Comparing the first line with the last we get that the term in square brackets obeys the distributional definition of $\De$, i.e.  (\ref{eq:convGdef}).

\subsection{Conditions for also being a left-inverse} \label{sec:leftinv}

In general $L^{-1}$ is not a left-inverse $L^{-1} L \neq {\rm id}$, as it is most obvious when acting on $h \in {\rm Ker}[L]$
\beq
L^{-1} L h = 0 \, . 
\eeq
The most general statement is rather
\beq \label{eq:notleftinverse}
\( L^{-1} L - {\rm id} \) T \in {\rm Ker}[L] \, , \hspace{1cm} \forall T \in \Ga(T^n_m{\cal M}) \, ,
\eeq
since applying $L$ from the left will give zero. Note that in general the resulting element of ${\rm Ker}[L]$ will depend on $g$, because $L$ does, and is obviously also $\Rs$-linear in $T$. Indeed, because of the very existence of non-zero elements in ${\rm Ker}[L]$, left-inverses generically do not exist. To understand this intuitively consider for instance the operator $\pa_t^2$ in one dimension and the following acausal Green's function
\beq
G(t,t') = \te(t-t') \te(t'-t_0) (t-t')  - \te(t_0 - t') \te(t' - t) (t' - t) \, ,
\eeq
so that the inverse operation is
\beq
(\pa^{-2} f)(t) \equiv \int_{-\infty}^{\infty} \ed t' \, G(t,t') f(t') = \int_{t_0}^t \ed t' (t - t') f(t') \, , 
\eeq
and we get 
\beq
(\pa^{-2} \pa^2 f)(t) - f(t) = - f(t_0) - f'(t_0)(t-t_0) \in {\rm Ker}\[\pa^2\] \, .
\eeq
It is clear that with this definition, $\pa^{-2}$ is a left inverse $\pa^{-2} \pa^2 = {\rm id}$ only on the subspace of functions obeying $f(t_0) = f'(t_0) = 0$. Moreover, we see that the resulting element of the kernel is determined by the boundaries of the convolution, i.e. the support of the Green's function with respect to the second argument. In the retarded case where $t_0 \to -\infty$ the integral makes sense only for functions that decrease sufficiently fast at infinity, i.e.
\beq
\lim_{t \to -\infty} f(t) = 0 \, , \hspace{1cm} \lim_{t \to -\infty} \dot{f}(t) = 0 \, ,
\eeq 
and then $\pa^{-2}$ is a left-inverse. This is actually the case in any dimension and on arbitrary geometries, i.e. the obstruction to being a left-inverse is generated by non-trivial boundaries of the support of $G$. Now that we have understood this using the simplest example, let us consider the case $L = \bo$ which is the one of interest in this thesis, for arbitrary dimension and for globally hyperbolic $\( {\cal M}, g \)$ so that the past light-cones extend to past infinity. We have
\bea
\bo_{\rm r}^{-1}  \bo  T_{\mu_1 \dots \mu_m}^{\nu_1 \dots \nu_n}(x) & = & \int_{\cal M} \ed^D y \, \sqrt{-g(y)} \, G_{{\rm r}\mu_1 \dots \mu_m}^{\,\,\,\nu_1 \dots \nu_n}|_{\ro_1 \dots \ro_n}^{\si_1 \dots \si_m}(x,y) \bo T_{\si_1 \dots \si_m}^{\ro_1 \dots \ro_n}(y)  \nn \\
 & = & \int_{\cal U} \ed^d y \, \sqrt{-g(y)} \, G_{{\rm r}\mu_1 \dots \mu_m}^{\,\,\,\nu_1 \dots \nu_n}|_{\ro_1 \dots \ro_n}^{\si_1 \dots \si_m}(x,y)\, N^{\mu}(y) \na_{\mu} T_{\si_1 \dots \si_m}^{\ro_1 \dots \ro_n}(y) \nn \\
 & & - \int_{\cal M} \ed^d y \, \sqrt{-g(y)} \, \na|^{\mu} G_{{\rm r}\mu_1 \dots \mu_m}^{\,\,\,\nu_1 \dots \nu_n}|_{\ro_1 \dots \ro_n}^{\si_1 \dots \si_m}(x,y)\, \na_{\mu} T_{\si_1 \dots \si_m}^{\ro_1 \dots \ro_n}(y) \nn \\
 & = & \int_{\cal U} \ed^d y \, \sqrt{-g(y)} \, G_{{\rm r}\mu_1 \dots \mu_m}^{\,\,\,\nu_1 \dots \nu_n}|_{\ro_1 \dots \ro_n}^{\si_1 \dots \si_m}(x,y)\, N^{\mu}(y) \na_{\mu} T_{\si_1 \dots \si_m}^{\ro_1 \dots \ro_n}(y) \nn \\
 & & - \int_{\cal U} \ed^d y \, \sqrt{-g(y)} \, N^{\mu}(y) \na|_{\mu} G_{{\rm r}\mu_1 \dots \mu_m}^{\,\,\,\nu_1 \dots \nu_n}|_{\ro_1 \dots \ro_n}^{\si_1 \dots \si_m}(x,y)\, T_{\si_1 \dots \si_m}^{\ro_1 \dots \ro_n}(y) \nn \\
 & & + \int_{\cal M} \ed^D y \, \sqrt{-g(y)} \, \bo_y G_{{\rm r}\mu_1 \dots \mu_m}^{\,\,\,\nu_1 \dots \nu_n}|_{\ro_1 \dots \ro_n}^{\si_1 \dots \si_m}(x,y)\,  T_{\si_1 \dots \si_m}^{\ro_1 \dots \ro_n}(y) \nn \\
 & \os{(\ref{eq:convGdef})}{=} & \int_{\cal U} \ed^d y \, \sqrt{-g(y)} \, W_{x,\mu_1 \dots \mu_m}^{\,\,\,\,\nu_1 \dots \nu_n}(y) + T_{\mu_1 \dots \mu_m}^{\nu_1 \dots \nu_n}(x)  \, , \label{eq:notleftinverseex}
\eea 
where $N$ is the normal vector to ${\cal U}$ and
\beq \label{eq:Wronskian}
W_{x,\mu_1 \dots \mu_m}^{\,\,\,\,\nu_1 \dots \nu_n}(y) \equiv G_{{\rm r}\mu_1 \dots \mu_m}^{\,\,\,\nu_1 \dots \nu_n}|_{\ro_1 \dots \ro_n}^{\si_1 \dots \si_m}(x,y)\, \overleftrightarrow\na_N \, T_{\si_1 \dots \si_m}^{\ro_1 \dots \ro_n}(y) \, ,
\eeq
is the Wronskian of $G(x,y)$ and $T(y)$ with respect to the derivative operator $\na_N$ acting on $y$. The question now is: what is $\cal U$? If the integrand we started with was smooth, then by Stokes' theorem we would have that ${\cal U}$ is the boundary of the support of the integrand. However, since $G_{\rm r}(x,y)$ is non-zero only when $y$ is on the past light-cone of $x$, we have that it is actually a distribution, just like in the flat space-time case (\ref{eq:Gr4DMink}). Thus, the integration by parts has to be understood in the way it is used for distributions: the boundary term is supported on the boundary of the support of the distribution. Since in our case the integrand is supported on the past light-cone ${\cal L}_x$ of $x$, the integral of the Wronskian is actually supported on $\pa {\cal L}_x$ which lies at past (null) infinity. Thus, we have that the conditions that one must impose on $T$ for $\bo_{\rm r}^{-1}$ to be a left inverse are (\ref{eq:pastlimitT}), i.e. precisely the ones for which $\bo_{\rm r}^{-1}$ is defined anyway. We thus have that $\bo_{\rm r}^{-1}$ is also a left inverse on the domain of $\Ga(T^n_m{\cal M})$ where it is defined. 

It is quite interesting to see how this computation goes through in the massive case $L = \bo - m^2$ since then the support of the Green's function is inside the past light-cone so that $\pa {\cal U} = {\cal L}_x$. For simplicity let us work on flat space-time, since in that case we have an explicit result (\ref{eq:Gr4DMink}). We then see that we have the singular part of $\bo^{-1}$, which is treated as before and thus gives an integral supported at past infinity. The smooth part which is supported on the inside of the cone however has a non-zero limit $|x| \to 0$ from the inside of the cone
\beq 
\lim_{|x|\to 0^+} G_{\rm r}(x) = -\frac{1}{4\pi} \[ \frac{\de(x^0 - |\vec{x}|)}{|\vec{x}|} - \frac{m^2}{2}\,\te(x^0) \] \, ,
\eeq
so the corresponding Wronskian boundary term is not zero and lies on ${\cal L}_x$, not on $\pa{\cal L}_x$. Therefore, if we wanted this to be zero for all $x$ then we would need to impose $T = 0$. However, what we see is that the smooth part of $G_{\rm r}(x)$ is actually constant on the light-cone, so that the Wronskian (\ref{eq:Wronskian}) is a total derivative. Thus, by Stokes' theorem it also amounts to an integral that is supported on $\pa {\cal L}_x$ at past infinity. For generic space-times we would need to know the limiting behaviour of the Green's function on the light-cone to answer the question of left-inversion.

\subsection{Commutation relations of $L^{-1}$} \label{sec:commrel}

We are now interested in understanding the commutator $\[ M, L^{-1} \]$ where $M[\na]$ is some differential operator. To do so we can simply act with the derivation $\[ M, \cdot\, \]$ on the equation $L L^{-1} = {\rm id}$ to get
\beq
\[ M, L \] L^{-1} + L \[ M, L^{-1} \] = 0 \, .
\eeq
Isolating $\[ M, L^{-1} \]$ would require the use of a left-inverse which, as we have seen in the previous section, does not exist when acting on generic functions. We can make use of the weaker equation (\ref{eq:notleftinverse}) to get the most conservative statement
\beq
\[ M, L^{-1} \] T = - L^{-1} \[ M, L \] L^{-1} T + X  \, , \hspace{1cm} X \in {\rm Ker}[L] \, . 
\eeq 
where $X$ is $\Rs$-linear in $\[ M, L \] L^{-1} T$. For instance, in the case $L = \bo$ and $M = \na_{\mu}$, we get the following rule for the retarded inverse on a scalar field of finite past
\beq \label{eq:nabocom}
\[ \na_{\mu}, \bo_{\rm r}^{-1} \] \ph = \bo_{\rm r}^{-1} \( R_{\mu}^{\nu} \na_{\nu} \bo_{\rm r}^{-1} \ph\) \, ,
\eeq
i.e. there is no $X$ part precisely because then $\bo_{\rm r}^{-1}$ is also a left inverse. Isolating $\bo_{\rm r}^{-1} \na_{\mu}$ and restricting to an Einstein space-time $R_{\mu\nu} = \ka\, g_{\mu\nu}$, where $\ka$ is constant, we get 
\beq
\bo_{\rm r}^{-1} \na_{\mu} = \( 1 - \ka \bo_{\rm r}^{-1} \) \na_{\mu} \bo_{\rm r}^{-1} \, .
\eeq
Inverting the operator in the bracket in a causal way, we get
\beq \label{eq:Einteincomm}
\na_{\mu} \bo_{\rm r}^{-1} = \( \bo - \ka \)_{\rm r}^{-1} \na_{\mu} \, .
\eeq

\subsection{Displacing the indices of Green's bi-tensors}

Since we only use metric compatible covariant derivatives $\[ \na,  g \] = 0$, we have that $\[ g, L \] = 0$ for any differential operator $L$. At the level of the Green's bi-tensors we have that the isomorphism $g$ between $\Ga(T^n_m{\cal M})$ and $\Ga(T^{n-1}_{m+1}{\cal M})$ 
\beq
 T_{\mu_1 \dots \mu_{m+1}}^{\nu_1 \dots \nu_{n-1}} =  g_{\mu_{m+1} \nu_n} T_{\mu_1 \dots \mu_m}^{\nu_1 \dots \nu_n} \, ,
\eeq
induces an isomorphism between the Green's functions of $L$ in $\Ga(B^n_m|^m_n{\cal M})$ and the ones in $\Ga(B^{n-1}_{m+1}|^{m+1}_{n-1}{\cal M})$ which is found through
\bea
\( G \cdot T \)_{\mu_1 \dots \mu_m}^{\nu_1 \dots \nu_n}(x) & \equiv & \int \ed^D y \, \sqrt{-g(y)}\, G_{\mu_1 \dots \mu_m}^{\nu_1 \dots \nu_n}|_{\ro_1 \dots \ro_n}^{\si_1 \dots \si_m}(x,y)\, T_{\si_1 \dots \si_m}^{\ro_1 \dots \ro_n}(y) \\
 & = & \int \ed^D y \, \sqrt{-g(y)}\, G_{\mu_1 \dots \mu_m}^{\nu_1 \dots \nu_n}|_{\ro_1 \dots \ro_n}^{\si_1 \dots \si_m}(x,y)\, g^{\ro_n \si_{m+1}} (y)\, T_{\si_1 \dots \si_{m+1}}^{\ro_1 \dots \ro_{n-1}}(y) \nn \\
 & \equiv & g^{\mu_{m+1} \nu_n}(x)\int \ed^D y \, \sqrt{-g(y)}\, G_{\mu_1 \dots \mu_{m+1}}^{\nu_1 \dots \nu_{n-1}}|_{\ro_1 \dots \ro_{n-1}}^{\si_1 \dots \si_{m+1}}(x,y)\,  T_{\si_1 \dots \si_{m+1}}^{\ro_1 \dots \ro_{n-1}}(y) \, ,\nn 
\eea
so that 
\beq
G_{\mu_1 \dots \mu_{m+1}}^{\nu_1 \dots \nu_{n-1}}|_{\ro_1 \dots \ro_{n-1}}^{\si_1 \dots \si_{m+1}}(x,y) = g_{\mu_{m+1} \nu_n}(x)\, G_{\mu_1 \dots \mu_m}^{\nu_1 \dots \nu_n}|_{\ro_1 \dots \ro_n}^{\si_1 \dots \si_m}(x,y)\, g^{\ro_n \si_{m+1}} (y) \, .
\eeq
Indeed, the latter trivially obeys $L[\na_{\rm L}] G = \De$ since $[g, L] = 0$.

\end{document}